\documentclass[11pt,a4paper]{article}
\usepackage{jheppub,amsmath, amsthm, amssymb,slashed,url,diagram,bm}
\usepackage{graphicx}
\usepackage{epstopdf}
\def\t{\widetilde} 

\def\btau{{\bm \tau}}

\def\NSthree{{\mathrm{NS}}^3}
\def\NSRR{{\mathrm{NS}}\cdot{\mathrm{R}}^2}
\def\be{\begin{equation}}
\def\ee{\end{equation}}

\def\frakl{{\mathfrak L}}
\def\eusmn{{\eusm N}}

\def\PGL{{\mathrm{PGL}}}
\def\OSp{{\mathrm{OSp}}}
\def\hat{\widehat}
\def\tilde{\widetilde}
\def\frak{\mathfrak}
\def\D{{\mathcal D}}
\def\S{{\mathcal S}}
\def\RP{{\Bbb{RP}}}
\def\NS{{\mathrm{NS}}}
\def\Ra{{\mathrm{R}}}
\def\W{{\mathcal W}}

\def\MM{\mathfrak M}

\def\O{{\mathcal O}}
\def\V{{\mathcal V}}

\def\w{{u}}
\def\dzzt{\D(\t z,z|\theta)}
\def\dzztt{\D(\t z,z|\t\theta,\theta)}
\def\Bbb{\mathbb}
\def\SIgma{\Sigma}
\def\red{{\mathrm{red}}}

\def\d{{\mathrm d}}

\def\R{{\mathbb R}}
\def\RR{{\mathcal R}}
\def\C{{\mathbb C}}
\def\U{{\mathcal U}}
\def\D{{\mathcal D}}

\def\G{{\mathcal G}}
\def\[{\bigl [}
\def\Spin{{\mathrm{Spin}}}
\def\]{\bigr ]}
\def\CP{{\mathbb{CP}}}
\def\N{{\mathcal N}}
\def\T{{\mathcal T}}

\def\Z{{\mathbb Z}}

\def\half{{\frac{1}{2}}}
\def\CC{{\mathcal C}}

\def\L{{\eusm L}}
\def\h{\hat }
\def\h{\widehat}

\def\V{{\mathcal V}}
\def\J{{\mathcal J}}

\def\M{{\mathcal M}}
\def\W{{\mathcal W}}

\def\Y{\mathcal Y}

\def\l{\langle}
\def\r{\rangle}

\def\epsilon{\varepsilon}

\def\sf{{\mathfrak s}}

\def\i{{\mathbf i}}
\def\spin{{\mathrm{spin}}}
\def\m{\cmmib m}
\def\tilde{\widetilde}
\def\bar{\overline}
\def\neg{\negthinspace}
\def\Ber{{\mathrm {Ber}}}
\def\BBer{{\text{\it{Ber}}}}

\font\teneurm=eurm10 \font\seveneurm=eurm7 \font\eighteurm=eurm8 \font\fiveeurm=eurm5
\newfam\eurmfam
\textfont\eurmfam=\teneurm \scriptfont\eurmfam=\seveneurm
\scriptscriptfont\eurmfam=\fiveeurm

\font\teneusm=eusm10 \font\seveneusm=eusm7 \font\fiveeusm=eusm5
\newfam\eusmfam
\textfont\eusmfam=\teneusm \scriptfont\eusmfam=\seveneusm
\scriptscriptfont\eusmfam=\fiveeusm
\def\eusm#1{{\fam\eusmfam\relax#1}}
\font\tencmmib=cmmib10 \skewchar\tencmmib='177
\font\sevencmmib=cmmib7 \skewchar\sevencmmib='177
\font\fivecmmib=cmmib5 \skewchar\fivecmmib='177
\newfam\cmmibfam
\textfont\cmmibfam=\tencmmib \scriptfont\cmmibfam=\sevencmmib
\scriptscriptfont\cmmibfam=\fivecmmib

\def\F{\eusm F}
\def\Pi{\varPi}
\def\g{\text{{\teneurm g}}}
\def\sg{\text{{\eighteurm g}}}

\def\n{\text{{\teneurm n}}}
\def\sn{\text{{\eighteurm n}}}
\def\ssn{\text{{\seveneurm n}}}
\def\m{\text{{\teneurm m}}}
\def\sm{\text{{\eighteurm m}}}

\title{Notes On Super Riemann Surfaces And Their Moduli}

 \author{Edward Witten}
\affiliation{School of Natural Sciences, Institute for Advanced Study,\\ 1 Einstein Drive, Princeton, NJ 08540 USA}
\abstract{These are notes on the theory of super Riemann surfaces and their moduli spaces, aiming 
to collect results that are useful for a better understanding of superstring perturbation theory in the RNS
formalism.  
}

\begin{document} \maketitle

\section{Introduction}\label{intro}   

What are now understood as superstrings -- string theories with spacetime supersymmetry -- can be formulated
in terms of a worldsheet action with worldsheet supersymmetry \cite{GeSa,bvh,dz}, naturally coupled to two-dimensional supergravity \cite{H,Ga}.
This leads to a description of superstring scattering
amplitudes in terms of integration over a suitable class of two-dimensional supergeometries, an early reference being \cite{Ma}.
The relevant  supergeometries  are best understood  as super Riemann surfaces,
which were  initially defined in \cite{BaSo,BaS,Fr,YMa}.
The subsequent literature, which includes papers  such as \cite{RSV, BMFS,CR,LR,RT} 
 developing the theory of super Riemann surfaces and papers such as  \cite{MNP,AGNSV,RSVtwo,DPhagain,DPh,DPhtwo,DPhthree,Nelson,GN,Giddings,Martwo,EHV,Cohn,Bel,Beltwo} with applications to string theory,
 is too vast to be fully cited here.

The present notes aim to give a relatively understandable account  of aspects of super Riemann surface theory that are
particularly relevant for superstring perturbation theory. 
A reconsideration of superstring perturbation theory will appear elsewhere \cite{Witten}.
A companion set of notes gives an introduction to supermanifolds and integration \cite{Wittennotes}, including 
some concepts used here.  The reader
 will also require some familiarity with ordinary Riemann surfaces and complex manifolds, and some general 
 familiarity with superstring theory.  We have tried
 to structure these notes so that many of the more technical parts can be omitted on first reading.

The material described here is mostly standard.  One  point on which we differ from some of the literature  is that in 
describing super Riemann surfaces,
we assume no relation between holomorphic and antiholomorphic odd variables, since no such relation
is natural for superstrings.  In particular, in the framework assumed here, one is not allowed to take the complex conjugate of an odd variable.  
Much of the literature emphasizes the case of  a super Riemann surface with a real structure for the odd variables, as is appropriate
 for the nonsupersymmetric (and tachyonic) Type 0 string theory.    On some points, we have filled in what appear to be missing
details in the literature or have attempted to simplify previous arguments.

We introduce super Riemann surfaces from a purely holomorphic point of view in section \ref{bintro}, and from a smooth 
point of view in section \ref{sufsmooth}.
 Neveu-Schwarz and Ramond punctures are introduced in section \ref{nsrpunctures}.  Some examples 
 of low genus are the topic of section \ref{examples}.
In section \ref{infinity}, we describe the behavior of supermoduli space at infinity; this is 
crucial input for superstring perturbation theory.  Up to this point,
we consider only oriented super Riemann surfaces without boundary.  
The generalizations appropriate  for open and/or unoriented strings can be found in section \ref{oundor}.
In section \ref{contourintegrals}, we describe the super analog of the period matrix of a Riemann surface.  

In the bulk of this paper, all super Riemann surfaces  are $\N=1$ super Riemann surfaces, with holomorphic odd dimension  1.
This is the relevant case for the usual supersymmetric string theories.  However, in section \ref{duality}, we describe some rather pretty facts 
\cite{Delone,DRS,BR,RoRa} about
$\N=2$ super Riemann surfaces and generic complex supermanifolds of dimension $1|1$.  These facts  
have perhaps not yet been fully incorporated
in string theory. 

We generally write $\Sigma$ for a super Riemann surface, $\Sigma_0$ for an ordinary Riemann surface, and $\Sigma_\red$ for
the reduced space of $\Sigma$.
Also, $\M$ is generally the moduli space of ordinary Riemann surfaces and $\MM$ is  the moduli space of super Riemann 
surfaces. Further details like the genus, the number and types of punctures, or (in the bosonic case) the 
specification of a spin structure are indicated in an obvious way.
 Many statements apply either to the ordinary moduli
spaces that parametrize smooth Riemann surfaces or super Riemann surfaces, or to the Deligne-Mumford compactifications 
that parametrize also certain singular
surfaces (see section \ref{infinity}).  To refer specifically to the compactifications, we write $\h\M$ or $\h\MM$.  Integration cycles for string
perturbation theory are generically called $\Gamma$ for bosonic string theory and $\varGamma$ for superstring theory; their compactifications are
$\hat\Gamma$ and $\hat\varGamma$.  

\section{Super Riemann Surfaces From A Holomorphic Point Of View}\label{bintro} 

\subsection{Complex Supermanifolds And Super Riemann Surfaces}\label{holp}

We begin by describing\footnote{As noted in the introduction, we consider only $\N=1$ super Riemann surfaces except in section \ref{duality}.}
 a super Riemann surface $\Sigma$ as a complex supermanifold.  As such $\Sigma$ has dimension $1|1$, so locally 
 it is isomorphic to $\C^{1|1}$
and can be described by bosonic and fermionic local coordinates $z$ and $\theta$.  Its tangent bundle $T\Sigma$ and 
its cotangent bundle $T^*\Sigma$ are both of rank $1|1$.
For example, the cotangent bundle has a basis of 1-forms $\d z$ and $\d\theta$; the tangent bundle has the dual 
basis $\partial_z$ and $\partial_\theta$. 

So far, $\Sigma$ could be any complex supermanifold of dimension $1|1$.  To make $\Sigma$ into a super 
Riemann surface, we need one more piece of structure, which is
a subbundle $\D\subset T\Sigma$ of rank $0|1$, which is required to be ``completely nonintegrable'' in the sense 
that if $D$ is a nonzero section of $\D$ (in some open set
$U\subset\Sigma$), then  $D^2=\frac{1}{2}\{D,D\}$ is nowhere proportional to $D$ (we describe this by saying that $D^2$ 
is nonzero mod $D$).  A typical example is
\begin{equation}\label{poky}D_\theta=\frac{\partial}{\partial\theta}+\theta\frac{\partial}{\partial z}.\end{equation}
For this choice, we see that $D_\theta^2=\partial/\partial z$, which indeed is nowhere a multiple of $D_\theta$; indeed, 
$D_\theta$ and $D_\theta^2$ are everywhere a basis of $T\Sigma$.

To see that this example is typical, consider a section of $\D$ of the general form
\begin{equation}\label{oky} D=a \frac{\partial}{\partial\theta} + b\frac{\partial}{\partial z}.\end{equation}
Here $a$ is an even function and $b$ is an odd function.  (In general, $a$ and $b$ depend on $z$, $\theta$, 
and possibly on the even and odd moduli of $\Sigma$.)
In supermanifold theory, to say that a quantity is ``nonzero'' means that it is nonzero after reducing modulo the 
odd variables (that is, setting them to zero).  The odd function
$b$ is certainly proportional to odd variables, so for $D$
to be nonzero means that $a$ is nonzero and thus invertible.  The condition that $D^2$ is not a multiple of $D$ is 
invariant under $D\to f D$ for any nonzero function $f$,  and by taking $f=a^{-1}$
and redefining $b$,
we can reduce to $D=\partial/\partial\theta+b\partial/\partial z$.  We expand $b$ in powers of $\theta$:
\begin{equation}\label{practo}D=\frac{\partial}{\partial\theta}+(b_0+b_1\theta)\frac{\partial}{\partial z}.
\end{equation}
The condition that $D^2$ is not a multiple of $D$ implies that $b_1$ is nonzero, so we can
replace $\theta$ by $\theta^*=b_1^{-1/2}(b_0+b_1\theta)$.  After also rescaling $D$ by $b_1^{1/2}$, we reduce 
to $D=\partial_{\theta^*}+\theta^*\,\partial_z$,
showing that locally one can always pick holomorphic  coordinates in which $D$ has the form indicated in (\ref{poky}).  
We call these superconformal coordinates.  
A local superconformal coordinate system $z|\theta$ determines the section $D_\theta$ of $\D$ defined in (\ref{poky}).  
(To explain the notation, it can be shown -- for instance by using
eqn. (\ref{zobo}) below -- that $D_\theta$ is completely determined by the choice of $\theta$.)

In many ways, super Riemann surfaces have simpler properties than generic $1|1$ supermanifolds, but describing 
them explicitly can be more difficult.
For example, to describe a $1|1$ supermanifold,
 one may start with $\CP^{2|1}$, with homogeneous coordinates $z_1 \dots z_3|\theta$, and impose a homogeneous 
 equation $G(z_1\dots z_3|\theta)=0$.
 But generically in this way one obtains a complex supermanifold that is not a super Riemann surface, since its tangent bundle 
 does not have an appropriate
 subbundle $\D$.  It is actually quite tricky to determine which $1|1$ complex
 supermanifolds possess the additional structure corresponding to a super Riemann surface.  The gluing construction to which 
 we turn presently is one of the few general
 ways to construct super Riemann surfaces. 

\subsubsection{Superconformal Transformations And Primary Fields}\label{supertrans}

Let $z|\theta$ be local superconformal coordinates.
A vector field $W=\alpha \,\partial_\theta+ \beta\,\partial_z$ is said to generate an infinitesimal superconformal transformation or simply to be superconformal if it preserves the subbundle $\D\subset T\Sigma$,
which is equivalent to saying that for $D_\theta=\partial_\theta+\theta\partial_z$, $[W,D_\theta\}$ is proportional to $D_\theta$.  (The symbol $[~,~\}$ denotes a commutator or anticommutator depending on the statistics of the objects
in question.)  A short computation reveals that a basis of superconformal vector fields is given by the odd vector fields
 \begin{equation}\label{zerb}\nu_f=f(z)\left(\frac{\partial}{\partial\theta}-\theta\frac{\partial}{\partial z}\right), \end{equation}
and the even ones
\begin{equation}\label{terb} V_g=g(z)\frac{\partial}{\partial z}+\frac{\partial_zg(z)}{2}\theta\frac{\partial}{\partial\theta}.\end{equation}
 $f$ and $g$ are even functions that depend on $z$ and perhaps on moduli, but not on $\theta$.     
 Explicitly, \begin{equation}\label{latef}\{\nu_f,D_\theta\}=(\theta \partial_zf)D_\theta, ~~~[V_g,D_\theta]=-(\partial_z g/2)D_\theta.\end{equation}
Note that $\nu_f^2=V_g $ with $g=-f^2$.  If we set $g(z)=z$, we get the superconformal vector field $z\partial_z+\frac{1}{2}\theta\partial_\theta$,
which generates the scaling
\begin{equation}\label{merb} z\to \lambda z, ~~\theta\to\lambda^{1/2}\theta,~~\lambda\in\C^*,\end{equation}
and we see that  $\theta$ scales with dimensions of $(\mathrm{length})^{1/2}$ or $(\mathrm{mass})^{-1/2}$, while $D_\theta$ scales with mass dimension $1/2$.

If $z|\theta$ and $\h z|\h \theta$ are two local superconformal coordinate systems, then $D_\theta=\partial_\theta+\theta\partial_z$ and $D_{\h\theta}=\partial_{\h\theta}+\h\theta
\partial_{\h z}$ are two local nonzero sections of $\D$, and so are related by $D_\theta= F\,D_{\h\theta}$ for some nonzero function $F$.  Acting with this formula
on the function $\h \theta$, we determine $F$:
\begin{equation}\label{mirto} D_\theta=(D_\theta\h \theta)D_{\h\theta}. \end{equation}
On the other hand, for any two coordinate systems $z|\theta$ and $\h z|\h\theta$, we can use the chain rule of calculus to compute
\begin{equation}\label{irto}D_\theta=(D_\theta\h \theta)D_{\h \theta}+(D_\theta\h z-\h \theta D_\theta\h \theta)\partial_{\h z}, \end{equation}
so if $z|\theta$ is a superconformal coordinate system, then the condition that $\h z|\h \theta$ is also a superconformal coordinate system can be stated \cite{Fr}
\begin{equation}\label{brito} D_\theta\h z-\h \theta D_\theta \h \theta = 0. \end{equation}
It is actually possible from this to deduce a general formula for a superconformal change of coordinates. 
Let us try \cite{CR} a general ansatz
\begin{align}\h z & = u(z)+\theta \zeta(z)\cr \h\theta & = \eta(z)+\theta v(z).\end{align}
Then (\ref{brito}) gives $\zeta=v\eta$, $v^2=u'+\eta\eta'$, so the relation between the two coordinate systems is
\begin{align}\label{zobo} \h z & = u(z)+\theta\eta(z)\sqrt{u'(z)} \cr \h\theta& = \eta(z)+\theta\sqrt{u'(z)+\eta(z)\eta'(z)}.\end{align}

In any superconformal coordinate system $z|\theta$, the object $D_\theta$ gives a local trivialization of the line bundle $\D$.  A global section of $\D$
assigns to each superconformal coordinate system $z|\theta$ a function $A(z|\theta)$ such that the vector field $A(z|\theta)D_\theta$ is independent of the choice of superconformal
coordinates.  Recalling (\ref{mirto}), this means that the functions $A(z|\theta)$ and $\h A(\h z|\h\theta)$ in two superconformal coordinate systems are related by
\begin{equation}\label{retro}A =  F^{-1} \h A, ~~ F = D_\theta\h\theta. \end{equation}
A field $A$ with this property is called a superconformal primary field of dimension $-1/2$. (The normalization comes from the fact that $D_\theta$ scales with
mass dimension $1/2$.)   More generally a section of $\D^n$, for any $n$, corresponds to a field that transforms under a change of superconformal
coordinates  as
\begin{equation}\label{etro}A=F^{-n}\h A.\end{equation}
Such a field is a superconformal primary of dimension $-n/2$.  As an example of this definition, if $\phi$ is a globally-defined function on $\Sigma$,
then since
\begin{equation}\label{betrom}D_\theta\phi = F D_{\h\theta}\phi,\end{equation}
$D_\theta\phi$ is a section of $\D^{-1}$ or in other words a superconformal primary of dimension $1/2$.  

If $A$ is a superconformal primary of some dimension $h$, then locally we can expand
\begin{equation}\label{zetro}
A=a_0+\theta a_1,\end{equation}
where $a_0$ and $a_1$ are conformal primaries of dimensions $h$ and $h+1/2$.

\subsubsection{Gluing and Split Super Riemann Surfaces}\label{spinc}

With the help of (\ref{zobo}), one can construct all possible super Riemann surfaces by gluing together open sets 
$U_\alpha\subset \C^{1|1}$ by means of superconformal
automorphisms.  One simply imitates the gluing procedure that builds an ordinary manifold by gluing 
together small open sets.  All super Riemann surfaces can be built in
this way.  The only problem is that the formulas quickly get complicated.

An important special case of the gluing 
procedure arises if we glue by superconformal automorphisms with $\eta=0$.  Then (\ref{zobo}) 
simplifies and the gluing of superconformal coordinates in
open sets $U_\alpha$ and $U_\beta$ reduces to
\begin{align}\label{robo} z_\alpha=& u_{\alpha\beta}(z_\beta)\cr
                                             \theta_\alpha= &(u'_{\alpha\beta}(z_\beta))^{1/2}\theta_\beta.\end{align}                                
We can consistently forget the $\theta$'s, because (\ref{robo}) says that $z_\alpha$ is a function of $z_\beta$ only, independent of $\theta$.  This would not be true for more
general superconformal gluing with $\zeta\not=0$. When we forget the $\theta$'s, the functions $u_{\alpha\beta}(z_\beta)$ are gluing functions that build an ordinary
Riemann surface $\Sigma_\red$ (known as the reduced space of $\Sigma$) by gluing small open sets in $\C$.  (To build the super Riemann surface $\Sigma$ takes a little more information, as we discuss shortly, since the gluing relation for the $\theta$'s depends on the
signs of the square roots $(u'_{\alpha\beta})^{1/2}$.)
 Forgetting the $\theta$'s  amounts to a projection $\pi:\Sigma\to\Sigma_\red$.
  The fibers of $\pi$ are linear spaces of dimension $0|1$, parametrized by the odd coordinates $\theta_\alpha$.  Thus
$\Sigma$ is the total space of a line bundle $V\to \Sigma_\red$, where the fibers of $V$ are fermionic.

To identify $V$, we observe that the gluing law for the  holomorphic differential $\d z_\alpha$                                     
 is $\d z_\alpha=u'_{\alpha\beta}(z_\beta)\,\d z_\beta$, so that (\ref{robo}) says that $\theta_\alpha$ transforms as $(\d z_\alpha)^{1/2}$.
 Now $\d z_\alpha$ is a local section of a line bundle, the canonical or cotangent bundle\footnote{For any space $X$, we generically write $T^*X$ for the cotangent bundle to $X$.
 So we would naturally write $T^*\Sigma_\red$ for the cotangent bundle to $\Sigma_\red$.  However, the abbreviation $K$ is traditional and convenient.} $K$  of the ordinary Riemann surface $\Sigma_\red$.  So $(\d z_\alpha)^{1/2}$ would
 be a section of $K^{1/2}$, and thus $\theta_\alpha$ transforms as a (fermionic) section of $K^{1/2}$.  If a linear function on the fibers of a line bundle
 $V$ is a section of $K^{1/2}$, we must identify $V$ as the dual line bundle $K^{-1/2}$.
 
 So we conclude that gluing laws of the particular form (\ref{robo}) describe a super Riemann surface $\Sigma$ that is the total space of the line bundle  
 ${\Pi} K^{-1/2}$ over $\Sigma_\red$.  Here the symbol ${\Pi}$
 simply represents reversal of ``parity'' or statistics; it is a reminder that the fibers of the fibration $\Sigma\to\Sigma_\red$ are fermionic.   
 
 In this construction, $K^{1/2}$ may be any line bundle over $\Sigma_\red$ whose square is isomorphic to $K$.  If $\Sigma_\red$ has genus $\g$ 
  (in which case we also say that $\Sigma$ has genus $\g$), then up to isomorphism,
 there are $2^{2\sg}$ choices of $K^{1/2}$.  The choice of $K^{1/2}$ enters in the choices of signs of the square roots $(u'_{\alpha\beta})^{1/2}$ that appear in the superconformal
 gluing relations (\ref{robo}).  A choice of $K^{1/2}$ is called a spin structure.
 
 A super Riemann surface $\Sigma$ with a projection to $\Sigma_\red$ as described above  is called a split super Riemann surface.
  The reduced space $\Sigma_\red$  is what we will call a spin curve -- an ordinary Riemann surface together    with a choice of spin structure.
  Topologically, spin structures are naturally classified as being even or odd depending on whether the space $H^0(\Sigma_\red,K^{1/2})$ of global holomorphic sections of the line bundle $K^{1/2}$ has even or odd dimension;
   this is the only property of a spin structure that is invariant under all diffeomorphisms, including those that are not continuously connected to the identity.
The moduli space $\M_\spin$ of spin curves accordingly has two connected components, which we will call $\M_{\spin,\pm}$, parametrizing respectively
  a Riemann surface with a choice of even or odd spin structure.  $\M_{\spin,\pm}$ is an unramified cover of $\M$,
  the moduli space of Riemann surfaces without a choice of spin structure; the degree of the cover is the number   
  of even or odd spin structures, namely $\half(2^{2\sg}\pm 2^\sg)$.
  
  To get $\MM$, the moduli space of super Riemann surfaces, we must allow more general gluing with $\zeta\not=0$.  Since the odd parameters are nilpotent,
  this does not change the topology of the situation, so again $\MM$ has two connected components $\MM_\pm$.  $\MM_\pm$ reduces to $\M_{\spin,\pm}$ if
  we set the odd moduli (and therefore $\zeta$) to zero.  So $\M_{\spin,\pm}$ is the reduced space of $\MM_\pm$.

 \subsection{Moduli Of Super Riemann Surfaces}\label{otorg}  

We will now try to understand more systematically the even and odd gluing parameters that arise in 
building a super Riemann surface $\Sigma$.

Recall that $\Sigma$ is built out of small open sets $U_\alpha$ that are glued together on intersections 
$U_\alpha\cap U_\beta$.   So a first-order
deformation of the gluing data is given by a superconformal vector field $\phi_{\alpha\beta}$ defined in each intersection 
$U_\alpha\cap U_\beta$.  The idea is that before gluing $U_\beta$ to $U_\alpha$, we transform by 
$1+w \phi_{\alpha\beta}$, with $w$ an infinitesimal parameter.
 The $\phi_{\alpha\beta}$ are  subject to a constraint
\begin{equation}\label{tumbo}\phi_{\alpha\beta}+\phi_{\beta\gamma}+\phi_{\gamma\alpha}=0 \end{equation}
in each triple intersection $U_\alpha\cap U_\beta\cap U_\gamma$.  This constraint ensures that  
the gluing data remain consistent after being modified by the $\phi_{\alpha\beta}$.
The $\phi_{\alpha\beta}$ are also subject to an equivalence relation
\begin{equation}\label{rumbo} \phi_{\alpha\beta} \cong \phi_{\alpha\beta}+\phi_\alpha-\phi_\beta,\end{equation}
where each $\phi_\alpha$ is a superconformal vector field defined in $U_\alpha$.  The meaning of 
this equivalence is that $\Sigma$ is unchanged if we transform each $U_\alpha$
by a symmetry generated by  $\phi_\alpha$ before gluing the $U_\alpha$ together.  
All this is precisely analogous to the deformation theory of ordinary complex manifolds.  The constraint (\ref{tumbo}) means that
we  should interpret $\phi_{\alpha\beta}$ as a one-cocycle on $\Sigma$ with values in 
the sheaf\footnote{This sheaf assigns to any small open set $U\subset \Sigma$
the space of superconformal vector fields on $U$.}  $\S$ of superconformal vector fields on $\Sigma$.  
Modulo the equivalence relation (\ref{rumbo}), such a cocycle
determines an element of $H^1(\Sigma,\S)$.

Let $T\MM$ be the tangent bundle to $\MM$ and let $T\MM|_\Sigma$ be its fiber at the point in $\MM$ corresponding to $\Sigma$.
What is explained in the last paragraph implies that $T\MM|_\Sigma=H^1(\Sigma,\S)$. 

While $H^1(\Sigma,\S)$ has this interpretation,  the corresponding space of global holomorphic sections
 $H^0(\Sigma,\S)$ has a more elementary interpretation: it classifies superconformal vector fields on $\Sigma$ that are holomorphic everywhere,
and so determine infinitesimal automorphisms of $\Sigma$.  So $H^0(\Sigma,\S)$ is the Lie algebra of the supergroup $\G$ of superconformal automorphisms of $\Sigma$.  For a super Riemann surface $\Sigma$, the higher sheaf cohomology
groups $H^k(\Sigma,\S)$, $k>1$, are always zero,\footnote{This statement can be deduced from the analogous statement for the ordinary
Riemann surface $\Sigma_\red$.  As $\Sigma_\red$ has
complex dimension 1, its  cohomology with values in any coherent sheaf
vanishes above degree 1.} so we only have to consider the cases $k=0,1$.

In superstring perturbation theory, at least for closed oriented superstrings, one normally encounters
only values of the genus $\g$ of $\Sigma$ and the number of punctures such that there are no infinitesimal automorphisms, $H^0(\Sigma,\S)=0$.    
For the time being, we consider Riemann surfaces without punctures (they will be incorporated in section \ref{nsrpunctures}).
With no punctures, just as for
ordinary Riemann surfaces,  infinitesimal automorphisms are absent precisely for  $\g\geq 2$, so the cases $\g=0,1$ are exceptional.
  (For those exceptional cases, the moduli spaces
can be described by hand; see section \ref{examples}.)  

We would like to calculate the even and odd dimension of $T\MM|_\Sigma$. For $\g\geq 2$, 
this dimension does not depend on the odd moduli of $\Sigma$, for the following reason.  Because the odd moduli
are infinitesimal, they could affect the cohomology only by a ``jumping'' process in which cohomology classes of neighboring degree pair up and disappear under an infinitesimal
perturbation.  For the present problem, with $\g\geq 2$, since the cohomology is nonzero only in degree 1, 
jumping cannot occur and the odd moduli do not affect the dimension of the cohomology.
It turns out that there is also no jumping for $\g=0$ or for $\g=1$ with even spin structure (for the case of $\g=1$ with odd spin structure, see the end of section \ref{ryon}).

So to determine the dimension of $T\MM|_\Sigma$, we can make a convenient choice of the odd moduli.  The most convenient choice is to set them to zero,
that is, to take $\Sigma$ to be a split super Riemann surface.
In this case,  we have a decomposition $\S=\S_+\oplus \S_-$
where $\S_+$ consists of  vector fields 
\begin{equation}\label{oco}V_g=g(z)\partial_z+\half g'(z) \theta\partial_\theta\end{equation}
with $\g$ even,
 and $\S_-$ consists of vector fields \begin{equation}\label{zond}\nu_f=f(z)(\partial_\theta-\theta\partial_z).\end{equation}
with $f$ odd.
Similarly, for $\Sigma$ split, there is a natural decomposition $T\MM|_\Sigma=T_+\MM|_{\Sigma}\oplus T_-\MM|_{\Sigma}$ in even and odd subspaces, and
\begin{equation}\label{tromob}T_\pm\MM|_{\Sigma}=H^1(\Sigma,\S_\pm).\end{equation}

To evaluate (\ref{tromob}), we can express the cohomology computation as a computation on the reduced space $\Sigma_\red$.
We can associate to $V_g$ the ordinary vector field $g(z)\partial_z$ on  $\Sigma_\red$; it is a section of $T\Sigma_\red$, the tangent bundle to $\Sigma_\red$.
(This is the same as $K^{-1}$, where $K$ is the canonical bundle of $\Sigma_\red$.)  So $\S_+$ is the sheaf of sections of $T\Sigma_\red$ and
\begin{equation}\label{romort}T_+\MM|_{\Sigma}=H^1(\Sigma_\red,T\Sigma_\red).\end{equation}
The right hand side is the tangent space to $\M$, the moduli space of ordinary Riemann surfaces, at the point corresponding to $\Sigma_\red$.  For $\Sigma_\red$ of genus $\g\geq 2$,
it has dimension $3\g-3$.    One route to this result is the Riemann-Roch theorem. For any line bundle $\L$
of degree $n$, this theorem asserts that
\begin{equation}\label{plo}\dim\,H^0(\Sigma_\red,\L)-\dim\,H^1(\Sigma_\red,\L)=1-\g+n.\end{equation}
Also, $H^0(\Sigma_\red,\L)=0$ if $n<0$.  For $\L=T\Sigma_\red$, we have $n=2-2\g$, which is negative for $\g\geq 2$, so the Riemann-Roch formula gives
$\dim\,H^1(\Sigma_\red,T\Sigma_\red)=3\g-3$.

Similarly, we can associate to $\nu_f$ the object $f(z)\partial_\theta$, which we view as an odd vector field along the fibers of $\Sigma\to\Sigma_\red$,
or in other words as a section of $K^{-1/2}=T\Sigma_\red^{1/2}$.  So $\S_-$ is the sheaf of sections of $T\Sigma_\red^{1/2}$ and
\begin{equation}\label{yomort}T_-\MM|_{\Sigma} ={\Pi} H^1(\Sigma_\red,T\Sigma_\red^{1/2}),\end{equation}
where the symbol ${\Pi}$ is meant to remind us to view this space as being fermionic.  
The line bundle $T\Sigma_\red^{1/2}$ has degree $1-\g$, and
 for $\g\geq 2$, the Riemann-Roch formula gives $\dim\,H^1(\Sigma_\red,T\Sigma_\red^{1/2})=2\g-2$.
 
 Thus for $\g\geq 2$, the dimension of the moduli space of super Riemann surfaces of  genus $\g$ with no punctures is
 \begin{equation}\label{zomort}\dim\,\MM_\sg= 3\g-3|2\g-2.\end{equation}
The general formula that holds for all $\g$ includes a contribution in the Riemann-Roch formula from $H^0(\Sigma,\S)$, which
is the Lie algebra of the supergroup $\G$ of automorphisms of $\Sigma$.  The general formula is
\begin{equation}\label{womort}\dim\,\MM_\sg-\dim\,\G=3\g-3|2\g-2.\end{equation}
For example, for $\g=0$, $\MM_0$ is a point, of dimension $0|0$, while $\G$ is the supergroup
$\mathrm{OSp}(1|2)$, of dimension $3|2$.

\subsubsection{Odd Moduli And Spin Structures}\label{oddspin}

The fact that the odd moduli take values in ${\Pi} H^1(\Sigma_\red,T\Sigma_\red^{1/2})$ means that once odd moduli are turned on, it does not make
sense to claim that two super Riemann surfaces are the same except for a different choice of spin structure.  The odd moduli take values in a space whose
definition depends on a choice of $T\Sigma_\red^{1/2}$ and thus on a choice of spin structure, so odd moduli with different spin structures cannot be compared.  
This means that what happens in superstring perturbation
theory in genus 1 is atypical.  In genus 1, with an even spin structure (and no punctures), there are no odd moduli.  It therefore makes sense to sum over
spin structures before performing any other integration, and this leads to the classic proof of vanishing of the one-loop contribution to the cosmological constant,
via an identity whose use in this context goes back to \cite{GSO}.  When odd moduli are present, the closest analog of this procedure
is to first integrate over the odd moduli, to reduce an integral over $\MM_\sg$ to an integral over $\M_\sg$. After this, it makes sense to sum
over spin structures before performing the remaining integrals.  This can be elegantly done
for $\g=2$ at least for the vacuum amplitude \cite{DPhthree,Wittenother}, though as far as is known, there is no such natural procedure in general. 

\subsubsection{Deformations As A Complex Supermanifold}\label{defc}

In this analysis, we have used a gluing construction to interpret first-order deformations of $\Sigma$ as a super Riemann surface in terms
of the sheaf cohomology group $H^1(\Sigma,\S)$.

Alternatively, we could simply deform $\Sigma$ as a complex supermanifold, not requiring the deformations to preserve a super Riemann surface
structure.  This is much easier to analyze.  In this case, the vector fields $\phi_{\alpha\beta}$ by which we perturb can be any holomorphic
sections of the tangent bundle $T\Sigma$, and hence the first-order deformations of $\Sigma$ as a complex supermanifold are parametrized
by $H^1(\Sigma,T\Sigma)$.  The same cohomology group
parametrizes the first-order deformations of an ordinary complex manifold, for essentially the same reasons.

\subsubsection{$\MM_\sg$ as an Orbifold}\label{efc}

The moduli space $\M_\sg$ of ordinary Riemann surfaces $\Sigma_0$ of genus $\g$ is not a manifold but an orbifold.
This is so because $\Sigma_0$ may have automorphisms.  Usually in perturbative string theory, one considers a situation
in which $\Sigma_0$ has no continuous automorphisms (either because $\g\geq 2$, or because $\Sigma_0$ is endowed with
a sufficient number of punctures, as described in section \ref{nsrpunctures}), but it still may have a finite group of automorphisms.

Orbifold singularities of $\M_\sg$ arise when $\Sigma_0$ has extra automorphisms compared to a generic surface with the same
genus.  For example, a generic surface of genus at least $3$ has no nontrivial automorphisms at all, so for $\g\leq 3$,
$\M_\sg$ has an orbifold singularity whenever $\Sigma_0$ has a nontrivial automorphism group.  For $\g=2$, the generic
automorphism group is $\Z_2$, and orbifold singularities occur when it becomes larger.

Super moduli space $\MM_\sg$ is similarly an orbifold.  Indeed, its orbifold structure is more prominent than that of $\M_\sg$,
for the following simple reason.  
Every super Riemann surface $\Sigma$ is infinitesimally close to a surface with enhanced symmetry.  That is so because there
is always an enhanced symmetry when we turn off the odd moduli; a split super Riemann surface $\Sigma$
always has a $\Z_2$ symmetry $\btau$
that acts trivially on $\Sigma_\red$ and acts as $-1$ on the fibers of the fibration $\Sigma\to\Sigma_\red$. (In eqn. (\ref{robo}),
this is the symmetry that acts as $\btau:z_\alpha|\theta_\alpha\to z_\alpha|-\theta_\alpha$ for all $\alpha$.)  The odd moduli are
odd under this symmetry.
So the locus of enhanced symmetries is dense in $\MM_\sg$, while in $\M_\sg$ it has positive (bosonic) codimension.

In the present notes, we will not go quite deeply enough for 
the orbifold nature of $\MM_\sg$  to play a major role, though the $\Z_2$ automorphism group of a split super Riemann
surface will make an occasional appearance.  

The fancy way to describe the orbifold nature of $\MM_\sg$ is
to refer to it as the moduli ``stack'' of super Riemann surfaces rather than the moduli ``space.''

\subsection{Superconformal Vector Fields Reconsidered}\label{reconsider}

In the definition of a super Riemann surface $\Sigma$, we postulated the existence of a subbundle $\D\subset T\Sigma$
of rank $0|1$.  As $T\Sigma$ has rank $1|1$, the quotient $T\Sigma/\D$ has rank $1|0$.  

On the other hand, we also postulated that if $D$ is a local nonzero section of $\D$, then $D^2$ is everywhere nonzero mod $D$.
This means that $D^2$ has an everywhere nonzero projection to $\L=T\Sigma/\D$.  The existence of a natural map from a nonzero
section $D$ of $\D$ to a nonzero section $D^2$ of $\L$ (a quadratic map in the sense that if $f$ is a nonzero function, then
$(f D)^2=f^2 D^2$ mod $\D$)
implies that $\L\cong \D^2$.

Thus the natural exact sequence $0\to \D\to T\Sigma\to T\Sigma/\D\to 0$ becomes
\begin{equation}\label{omurk} 0 \to \D\to T\Sigma\to \D^2\to 0.  \end{equation}
We also sometimes need the dual sequence:
\begin{equation}\label{zomurk} 0\to \D^{-2}\to T^*\Sigma\to \D^{-1}\to 0.\end{equation}
Here one must recall that to dualize an exact sequence of vector bundles, one dualizes each bundle involved and reverses the
direction of the maps.  To get (\ref{zomurk}), we just need to know that the
 dual of $T\Sigma$ is the cotangent bundle $T^*\Sigma$ and the dual of $\D^h$ is $\D^{-h}$. 
 Just as $T\Sigma$ has a distinguished
subbundle of rank $0|1$, namely $\D$, (\ref{zomurk}) says that  $T^*\Sigma$ has a distinguished subbundle of rank $1|0$, isomorphic to $\D^{-2}$.
Concretely, what singles out the subbundle  $\D^{-2}\subset T^*\Sigma$ is that, under the duality
between $T^*\Sigma$ and $T\Sigma$, it is normal to $\D\subset T\Sigma$.
Explicitly, in local superconformal coordinates $z|\theta$,  the subbundle $\D\subset T\Sigma$ is generated by the vector field $D_\theta=\partial_\theta+\theta\partial_z$, so
$\D^{-2}\subset T^*\Sigma$ is generated by the 1-form 
\begin{equation}\label{amurk}\varpi=\d z-\theta\d\theta,\end{equation}
 whose contraction with $D_\theta$ vanishes.

We would like to give a new description of the space of superconformal vector fields.  The formulas
(\ref{zerb}) and (\ref{terb}) for superconformal vector fields are clear enough, and not difficult to verify.  But
they have the peculiar property that the functions $f(z)$ and $g(z)$ depend on $z$ only and not $\theta$.
Can we combine $f$ and $g$ to a superfield?

In \cite{RSV}, it is shown how to do this.  In terms of a superfield $\V(z|\theta)=g(z)+2\theta f(z)$, a general superconformal vector
field is 
\begin{equation}\label{orod}\W= \V(z,\theta)\partial_z+\half D_\theta \V\, D_\theta. \end{equation}
In fact, upon expanding the right hand side in terms of $\partial_z$ and $\partial_\theta$, one finds
\begin{equation}\label{rod} \W=V_g+\nu_{f},\end{equation}
with $V_g$ and $\nu_f$ as defined before.

Since (\ref{orod}) has been written in local superconformal coordinates $z|\theta$,
the global nature of the superfield $\V(z|\theta)$ is not immediately obvious.  To understand it, we observe that a superconformal
vector field $\W$ is in particular a vector field, and thus a section of $T\Sigma$.  We can project $\W$ from $T\Sigma$ to $T\Sigma/\D$
by dropping the $D_\theta$ term.  In other words, mod $D_\theta$, $\W$ is equivalent to $\V(z,\theta)\partial_z$, where we view
$\partial_z$ as giving a basis for $T\Sigma/\D\cong \D^2$.   So  globally  the superfield $\V$ is a section of $\D^2$, as stated
in \cite{RSV}.

The map from superconformal vector fields to sections of $\D^2$ is one-to-one, as is clear from (\ref{rod}).
This helps in understanding the superconformal ghost fields of string theory.  These fields start life as a superconformal vector
field with reversed statistics.  So they can be combined to a superfield $C$ that is a section of ${\Pi} \D^2$, where the symbol ${\Pi}$
tells us that $C$ has the opposite to usual statistics (its lower component is odd).  Locally, we  can expand
\begin{equation}\label{fryt} C=c+\theta\gamma,\end{equation}
where $c$ is an odd section of $T\Sigma_\red$, and $\gamma$ is an even section of $T\Sigma_\red^{1/2}$.

 The equivalence of $\S$ with the sheaf of sections of $\D^2$ has an application that we will explain later: it leads to
 a convenient description of the dual space to $H^1(\Sigma,\S)$.   
This equivalence also has one obvious drawback:  
superconformal vector fields have a natural graded Lie algebra structure, which is not particularly visible in the description via
sections of $\D^2$.

\subsection{Holomorphic Volume Forms}\label{holvol}

Now we will discuss holomorphic volume forms on $\Sigma$.  Like a holomorphic differential $a(z) \d z$ on an ordinary Riemann
surface, a holomorphic volume form is the right object for a contour integral -- an integral on a submanifold of $\Sigma$ of real codimension 1.
To introduce volume forms suitable for ``bulk'' integrals over $\Sigma$ (analogous to a 2-form $a(\bar z, z)\d\bar z\wedge \d z$ on an ordinary
Riemann surface), we will need the smooth description of super Riemann surfaces that we introduce in section \ref{colzo}.


In general, on a complex supermanifold $X$ of dimension $p|q$, one defines  a holomorphic line 
bundle $\BBer(X)$  
(known as the Berezinian of $X$) of holomorphic densities on $X$.  The most elementary definition of $\BBer(X)$ is as follows  
(see, for example, sections 3.1 and  5.3.1 of \cite{Wittennotes}, as well as appendix \ref{obz} below).  Given
any local  trivialization of $T^*X$ by basis elements $d z^1\dots |\dots \d\theta^q$ of $T^*X$, there is a corresponding local trivialization of   
$\BBer(X)$ by a basis element  $[\d z^1\dots|\dots \d\theta^q]$.  Under a change of basis, the symbol $[\d z^1\dots|\dots\d\theta^q]$
transforms as one would expect a density to transform (that is, it transforms by the Berezinian -- the superanalog of the determinant -- of the
matrix giving the change of basis).   $\BBer(X)$ is the analog for complex supermanifolds of what for an ordinary
complex manifold is called the determinant line bundle.  A holomorphic section of $\BBer(X)$ is the superanalog of a holomorphic $p$-form
on an ordinary  complex manifold of dimension $p$.

We claim that in the case of a super Riemann surface $\Sigma$, $\BBer(\Sigma)$ is naturally isomorphic to $\D^{-1}$.   So sections of
$\D^{-1}$, which correspond to superconformal primaries of dimension 1/2, are equivalent to sections of $\BBer(\Sigma)$.  For example,
in discussing eqn. (\ref{betrom}), we showed that if $\phi$ is a function on $\Sigma$, then there is a section of $\D^{-1}$ that in any local
superconformal coordinate system $z|\theta$ can be represented by $D_\theta\phi$.  We must therefore have a section of $\BBer(\Sigma)$
given by
\begin{equation}\label{torox}\sigma=[\d z|\d\theta]\,D_\theta\phi,\end{equation}
and in particular we claim that  $\sigma$ does not depend on the choice of the superconformal coordinate system $z|\theta$.
Clearly $\phi$ plays no essential role here.  The assertion that $\BBer(\Sigma)\cong \D^{-1}$ is equivalent to the statement that
the expression $[\d z|\d\theta]\,D_\theta$ is independent of the choice of superconformal coordinates.

 Thus if $\h z|\h \theta$ is any other superconformal coordinate system, we claim that
\begin{equation}\label{oxon}[\d \h z|\d \h \theta]\, D_{\h\theta}=[\d  z|\d \theta]\,D_\theta.   \end{equation} 
For a first orientation to this statement, suppose that the two coordinate systems differ by scaling
$\h z |\h \theta=\lambda z|\lambda^{1/2}\theta$.  Then $[\d\h z|\d\h\theta]=\lambda^{1/2}[\d z|\d\theta]$ (where a factor of $\lambda$ comes
from scaling of $z$ and a factor of $\lambda^{-1/2}$ from scaling of $\theta$), while $D_{\h\theta}=\lambda^{-1/2}D_\theta$.

The general proof follows by a direct calculation \cite{GN}.  By definition of the Berezinian,
\begin{equation}\label{oxpon}[\d\h z|\d\h\theta]=[\d z|\d \theta]\,\Ber(M),\end{equation}
where
\begin{equation}\label{xonson}M=\begin{pmatrix} \partial_z\h z & \partial_z \h\theta\cr \partial_\theta \h z & \partial_\theta\h\theta\end{pmatrix}.\end{equation}
With the help of (\ref{brito}), one may verify that
\begin{equation}\label{bosons} M=\begin{pmatrix} 1 & 0 \cr -\theta & 1\end{pmatrix}
\begin{pmatrix} \partial_z\h z+\h\theta\partial_z\h\theta &~~ \partial_z\h\theta\cr 0 & ~~D_\theta\h\theta\end{pmatrix}\begin{pmatrix}1 & 0 \cr \h\theta & 1 \end{pmatrix},\end{equation}
from which it follows that 
\begin{equation}\label{berz}\Ber(M)=\frac{\partial_z\h z+\h\theta\partial_z\h\theta}{D_\theta\h\theta}=D_\theta\h\theta.\end{equation}
In the last step, we used the relation
\begin{equation}\label{erz} \partial_z \h z+\h\theta\partial_z\h\theta-(D_\theta\h\theta)^2=0,\end{equation}
which follows upon applying the operator $D_\theta$ to eqn. (\ref{brito}).   Using this result for $\Ber(M)$ together with (\ref{mirto}),
we find that (\ref{oxpon}) implies the desired result (\ref{oxon}).

A less computational proof that $\BBer(\Sigma)\cong\D^{-1}$ can be found in appendix \ref{obz}.

\subsubsection{What is this Good For?}\label{whatfor}

The most important application of the isomorphism $\BBer(\Sigma)\cong \D^{-1}$ will be as a tool in writing Lagrangians.
Another important application will involve duality on a super Riemann surface.

But here we will explain an application involving conserved currents and contour integrals.
In superconformal field theory, a superconformal primary $\sigma$ of dimension $1/2$ is the superanalog of a conserved (holomorphic)
current.  Hence for every codimension 1 cycle $\gamma$, there is a conserved charge $q_\gamma=\int_\gamma\sigma$.  The statement that $q_\gamma$
is ``conserved'' means that it depends only on the homology class of $\gamma$.
From the point of view of super Riemann surfaces, $\sigma$ corresponds to a holomorphic section of $\D^{-1}$ and hence of $\BBer(\Sigma)$.
The fact that such a $\sigma$ can be integrated over a codimension 1 homology cycle $\gamma$ is  a special case of a general fact about
complex supermanifolds. 
As explained for example in section 5.3.1 of \cite{Wittennotes}, on a complex manifold $X$ of dimension $p|q$, a holomorphic section
$\sigma$ of $\BBer(X)$ can be naturally integrated over a cycle $\gamma\subset X$ of real codimension $p|0$, with a result
that only depends on the homology class of $\gamma$.   Moreover, up to homology, a codimension $p|0$ cycle $\gamma\subset X$ is naturally
determined by a corresponding ordinary codimension $p$ homology cycle $\gamma_\red\subset X_\red$.  In the super Riemann surface context,
that means that an ordinary one-cycle $\gamma_\red\subset \Sigma_\red$ determines the charge $q_\gamma=\oint_\gamma\sigma$.

An example of a holomorphic section $\sigma$ of $\BBer(X)$ is $\sigma=D_\theta\phi$, where $\phi$ is a  holomorphic function.  But in this
case, the conserved charges vanish.  It is instructive to explain this using the interpretation of  $\BBer(X)$ as the space of holomorphic
integral forms on $X$ of top degree. (See \cite{Wittennotes} for an explanation of the relevant concepts.)  
Given local superconformal coordinates $z|\theta$, we can
define the section $[\d z|\d \theta]$ of $\BBer(X)$ and also the vector field $D_\theta$.  Each of these separately depends on the choice of coordinates,
but as we have seen the product $[\d z|\d\theta]\,D_\theta$ does not.  The contraction operator $\i_{D_\theta}$ transforms like $D_\theta$,
so $\i_{D_\theta}[\d z|\d\theta]$, which we understand as a holomorphic integral form of codimension 1, does not depend on the choice of coordinates.
More generally, given a function $\phi$, we can define the codimension 1 holomorphic integral form  $\lambda=\phi\,\i_{D_\theta}[\d z|\d\theta]$.
Explicitly, the  contraction operator is
\begin{equation}\label{domox}\i_{D_\theta}=\frac{\partial}{\partial\d\theta}+\theta\frac{\partial}{\partial\d z},\end{equation}
and as an integral form, $[\d z|\d\theta]$ is naturally written $\delta(\d z)\delta(\d\theta)$, so 
\begin{equation}\label{omox}\lambda= \phi \left(\frac{\partial}{\partial\d\theta}+\theta\frac{\partial}{\partial\d z}\right)\delta(\d z)\delta(\d\theta).\end{equation}
The exterior derivative is
\begin{equation}\label{pomox}\d=\d z\frac{\partial}{\partial z}+\d\theta\frac{\partial}{\partial \theta}. \end{equation}
Using the fact that $x\delta'(x)=\mp\delta(x)$ for an even or odd variable $x$, we find
\begin{equation}\label{momox}\d\lambda= -D_\theta\phi\,\delta(\d z)\delta(\d\theta)= [\d z|\d\theta]\,D_\theta\phi.   \end{equation}
So $\oint_\gamma [\d z|\d\theta]\,D_\theta\phi=\oint_\gamma\d\lambda=0$, by the superspace version of Stokes's theorem.   For another explanation of this result, see eqn. (\ref{domoxot}).

\section{Super Riemann Surfaces From A Smooth Point Of View}\label{sufsmooth} 

So far, we have considered a super Riemann surface purely as a complex supermanifold of dimension $1|1$ with some additional structure.
This is not really the right structure for a string worldsheet, as on the string worldsheet, there are both holomorphic and antiholomorphic
degrees of freedom.  
For string theory, we have to know how to go over to a smooth description, in which one can discuss functions on $\Sigma$ that are not necessarily
holomorphic.  This will enable us to discuss topics such as fields and Lagrangians on $\Sigma$, deformations of $\Sigma$ as described by fields,
Riemannian geometry on $\Sigma$, etc.

\subsection{String Worldsheets And Their Parameter Space}\label{colzo}

We will adopt the viewpoint of \cite{Wittennotes}, section 5.5.  A string worldsheet $\Sigma$ is a smooth supermanifold that
is embedded in a product $\Sigma_L\times \Sigma_R$ of holomorphic Riemann surfaces or super Riemann surfaces.  (The notation $\Sigma_L$ and
$\SIgma_R$ is meant to evoke left- and right-moving degrees of freedom in string theory.)
For both heterotic and Type II superstrings, $\Sigma_R$ is a super Riemann surface.  For the heterotic string, $\Sigma_L$ is an ordinary Riemann surface,
but for Type II, $\Sigma_L$ is another super Riemann surface. 

What we will do with $\Sigma$ is to use it as an integration cycle: we will define
the worldsheet action of string theory by integrating a closed form -- actually a holomorphic section of $\BBer(\Sigma_L\times \Sigma_R)$ -- over
$\Sigma$.  For this purpose,
small deformations of the embedding of $\Sigma$ in $\Sigma_L\times \Sigma_R$ do not matter.  The basic example is that the reduced
spaces of $\Sigma_L$ and $\Sigma_R$ are complex conjugates
(the complex conjugate of a complex manifold $X$ is the same space with opposite complex structure) and $\Sigma_\red$ is
the diagonal in $\Sigma_{L,\red}\times \Sigma_{R,\red}$.  Then $\Sigma$ is obtained from $\Sigma_\red$ by a slight thickening in the fermionic
directions, as explained in \cite{Wittennotes}, section 2.2.  This last operation is not completely natural, but it is natural up to homology, which will be good enough.\footnote{If one picks $\Sigma_\red$ to be the diagonal in $\Sigma_{L,\red}\times \Sigma_{R,\red}$, then one can restrict
the homologies in question to be infinitesimal ones that act trivially on $\Sigma_\red$.  We call these fermionic homologies.
In any event,  $\Sigma$ is also unique up to isomorphism
as a smooth supermanifold and its holomorphic and antiholomorphic structures are both also unique.  Only the relation between
the holomorphic and antiholomorphic structures depends on how $\SIgma$ is embedded in $\SIgma_L\times \SIgma_R$.}
 More generally, it suffices if $\SIgma_L$ is sufficiently close to the complex
conjugate of $\Sigma_R$ and $\Sigma_\red$ is close to the diagonal in $\Sigma_{L,\red}\times \SIgma_{R,\red}$.  For more on all this,
see section 5 of \cite{Wittennotes}.

 The odd dimension of $\Sigma$ is the same as that of $\Sigma_L\times \Sigma_R$, and its even dimension
(as a smooth supermanifold) is 2. For the heterotic string,  $\Sigma_L\times \Sigma_R$ is a complex supermanifold of dimension $2|1$,
so  $\Sigma\subset \Sigma_L\times \Sigma_R$ is a smooth supermanifold of dimension $2|1$.  For Type II,
$\Sigma_L$ and $\Sigma_R$ are both super Riemann surfaces and $\Sigma$ is a smooth supermanifold of dimension $2|2$.

By definition, a holomorphic function on $\Sigma$ is a holomorphic function on $\Sigma_R$, restricted to $\Sigma$.  If $z|\theta$ are local superconformal
coordinates on $\Sigma_R$, then their restrictions to $\Sigma$ are local holomorphic functions on $\Sigma$ and locally any  holomorphic function on 
$\Sigma$ is $f(z|\theta)$.  Similarly, an antiholomorphic function on $\Sigma$ is by definition the restriction to $\Sigma$ of a holomorphic function on 
$\Sigma_L$.   For the heterotic string, if $\t z$ is a local holomorphic function on $\Sigma_L$, its restriction to $\Sigma$ is a local antiholomorphic function and locally 
any antiholomorphic function is a function $g(\t z)$.     Given such local coordinates on $\Sigma_L$ and $\Sigma_R$,  with $\t z$ 
sufficiently close\footnote{For example, $\t z$ may be the complex conjugate of $z$ if the odd variables including
odd moduli 
are set to zero, though one does not have to limit oneself to this case.} to the complex conjugate of $z$,  we call these functions -- or more precisely their restrictions to $\Sigma$ -- a standard coordinate system $\t z;\neg z|\theta$
on $\Sigma$.  (Our convention will generally be to list antiholomorphic coordinates before holomorphic ones, separating them by the
semicolon.)   For Type II, we take local superconformal coordinates $\t z|\t\theta$ and $z|\theta$ on $\Sigma_L$ and $\SIgma_R$ respectively, with $\t z$ sufficiently
close to the complex conjugate of $z$, to define a standard local coordinate system $\t z;\neg z|\t \theta;\neg\theta$.

Since $\Sigma_L\times \Sigma_R$ is a product, its Berezinian is $\BBer(\Sigma_L\times \Sigma_R)\cong \BBer(\Sigma_L)\otimes \BBer(\SIgma_R)$.
A holomorphic section $\sigma$ of $\BBer(\Sigma_L\times\Sigma_R)$ can be integrated over $\Sigma$, with a result that is invariant
under small deformations of $\SIgma$ within $\Sigma_L\times \Sigma_R$.  This is explained in \cite{Wittennotes}, section 5.3.1.  One
way to understand the statement is to observe that $\sigma$ can be understood as a codimension 2 integral form on $\Sigma_L\times \Sigma_R$,
so it can be integrated over the codimension $2|0$ cycle $\SIgma$, with a result that only depends on the homology class of $\Sigma$.  
Since the ability to integrate a section of $\BBer(\Sigma_L\times \Sigma_R)$ over $\SIgma$ is important, we will
explain it in another way.  We consider the heterotic string as an example. As $\Sigma$ is a smooth supermanifold, it has a Berezinian line bundle $\Ber(\Sigma)$ in the smooth sense, whose sections
can be integrated; a trivialization of $\Ber(\SIgma)$
 in any standard coordinate system is given by the symbol $[\d\t z;\neg\d z|\d\theta]$.  $\BBer(\Sigma_L) $ and $\BBer(\Sigma_R)$ 
 are likewise trivialized by the symbols $[\d\t z]$ and $[\d z|\d\theta]$, and so their tensor product is trivialized by the tensor product
 $[\d\t z]\otimes [\d z|\d\theta]$.  The definitions  ensure that the map
 from $\BBer(\SIgma_L\times\Sigma_R)$ to $\Ber(\Sigma)$ that takes $[\d\t z]\otimes [\d z|\d\theta]$ to $[\d\t z;\neg\d z|\d\theta]$ does
 not depend on the choice of coordinates.  
So we get a natural isomorphism
\begin{equation}\label{oturg}\BBer(\Sigma_L\times\SIgma_R)|_\Sigma\cong \Ber(\Sigma),\end{equation}
and once again a section of $\BBer(\Sigma_L\times \Sigma_R)$ can be integrated over $\Sigma$.
When focusing on the holomorphic structure of $\Sigma$, we sometimes write simply $\BBer(\Sigma)$ for $\BBer(\Sigma_R)$.

For the heterotic string, $\Sigma_L$ is an ordinary Riemann surface, so $\BBer(\Sigma_L)$ is simply the space of $(1,0)$ forms
on $\Sigma_L$ (these are regarded as $(0,1)$-forms on $\Sigma$ since a holomorphic function $\t z$ on $\Sigma_L$ is
regarded as an antiholomorphic function on $\Sigma$).  So if $\phi$ is a function on $\SIgma$ and we define the operator $\t\partial$ by                          
\begin{equation}\label{dofog}\tilde\partial\phi=\d\tilde z\,\frac{\partial\phi}{\partial \t z},\end{equation}
then $\t\partial\phi$ is a section of $\BBer(\Sigma_L)$, pulled back to $\Sigma$.  
 This is a more elementary analog of the fact that, because of the isomorphism $\BBer(\Sigma_R)\cong \D^{-1}$,
the expression
\begin{equation}\label{tomong}[\d z|\d\theta]\,D_\theta\phi\end{equation}
makes sense as a section of $\BBer(\Sigma_R)$.  Multiplying the two constructions, we get a section of $\BBer(\Sigma_L)\otimes \BBer(\Sigma_R)\cong \Ber(\Sigma)$:
\begin{equation}\label{romong}\tilde\partial\phi\, [\d z|\d\theta] \,D_\theta\phi =[\d\t z;\negthinspace\d z|\d\theta]\,\partial_{\t z}\phi\, D_\theta\phi. \end{equation}
This expression does not depend on the choice of standard local coordinates and can be integrated over $\Sigma$.  
If $\phi$ is real-analytic, it can be extended to a holomorphic function on $\Sigma_L\times \SIgma_R$, defined in a neighborhood of $\Sigma$.
Then the expression in \ref{romong}) is a holomorphic section of $\BBer(\Sigma_L)\otimes \BBer(\SIgma_R)$, defined in that neighborhood.
We defer the Type II analog of this to section \ref{tanalog}.

\subsection{Integration Cycles Of Superstring Perturbation Theory}\label{parspace}

Having explained what we mean by a superstring worldsheet, we 
can ask what is the space parametrizing such worldsheets over which we should integrate
in superstring perturbation theory.  This is more subtle than one might think based on experience with the bosonic string.
In bosonic string theory, the parameter space over which one wants to integrate to compute perturbative scattering amplitudes
is a canonically defined moduli space of conformal
structures on an ordinary Riemann surface $\Sigma_0$.  In superstring theory, the worldsheet $\Sigma$ is only defined up to
homology, and there does not seem to be a useful notion of the moduli space of such worldsheets. Instead, one defines
the relevant parameter space $\varGamma$  up to homology by a procedure similar to the one used to define $\Sigma$.
  (In each of the two cases, there
are natural choices of $\Sigma_\red$ and $\varGamma_\red$, and if one makes those choices, then one can restrict to infinitesimal
homologies that act trivially on the reduced spaces.) 

We follow the viewpoint of \cite{DEF}, p. 95 (see also \cite{Wittennotes}, section 5.6), and define the desired parameter space as a smooth
supermanifold $\varGamma$ embedded in  the moduli space $\MM_L\times \MM_R$ that 
parametrizes independent deformations of $\Sigma_L$ and $\Sigma_R$.
First we define a submanifold $\varGamma_\red$ of the reduced space 
$(\MM_L\times \MM_R)_{\red}=\MM_{L,\red}\times \MM_{R,\red}$ that is characterized by saying
that the reduced spaces $\Sigma_{L,\red}$ and $\Sigma_{R,\red}$ are complex conjugate (with no condition on
the spin structures).    
Then we thicken $\varGamma$ in the fermionic directions, in a way that is unique up to homology,
 to make a smooth supermanifold embedded
in $\MM_L\times \MM_R$ with the same  odd dimension as $\MM_L\times \MM_R$.  
  Thus the relation of $\varGamma$ to $\MM_L\times\MM_R$ is very similar to the relation of $\SIgma$ to $\Sigma_L\times \Sigma_R$ in section \ref{colzo}.

The superstring path integral constructs a section $\Xi$ of $\BBer(\MM_L\times \MM_R)$ that is holomorphic in a neighborhood of $\varGamma_\red$.
This section can be integrated over $\varGamma$ by virtue of the same arguments that we used in section \ref{colzo} to show that a section of $\BBer(\Sigma_L\times \Sigma_R)$
can be integrated over $\Sigma$.  
And by the same arguments as before, subject to a caveat
mentioned shortly, the integral of $\Xi$ over $\varGamma$ is invariant under small deformations of $\varGamma$.  Perturbative
superstring scattering amplitudes are computed via such integrals.

A caveat is needed because $\MM_L$, $\MM_R$, and $\varGamma$ are all not compact and the integrals required in superstring perturbation theory  have a delicate behavior
at infinity.  A condition is needed on just how to define $\varGamma$ at infinity.  This is described in section 6.5 of \cite{Witten}.

We have described the construction of $\varGamma$ in a way that applies uniformly to heterotic and Type II superstrings.
For the heterotic string, $\MM_R$ is the moduli space of super Riemann surfaces, with reduced space $\M_{\spin}$, which parametrizes an ordinary Riemann surface $\Sigma_{R,\red}$
with a choice of spin structure.  $\MM_L$  parametrizes the ordinary Riemann surface $\Sigma_L=\Sigma_{L,\red}$, so $\MM_L$ and its reduced space are both the moduli
space $\M$ of Riemann surfaces.  $\varGamma_\red\subset \M\times \M_{\spin}$ is defined
by the condition that $\Sigma_L$ is the complex conjugate of $\Sigma_{R,\red}$ (here we place no condition on the spin structure of the latter).  This
condition means that a point in $\varGamma_\red$ is determined by its projection to the second factor of $ \M\times \M_{\spin}$
and  $\varGamma_\red$ is actually a copy of 
$\M_\spin$. The dimension of $\varGamma$ is $6g-6|2\g-2$.

For Type II, $\MM_L$ and $\MM_R$ are both copies of the moduli space of super Riemann surfaces, and the reduced spaces are both copies of $\M_{\spin}$.
$\varGamma_\red\subset \M_{\spin}\times \M_{\spin}$ is defined by the condition that $\Sigma_L$ and $\Sigma_R$, ignoring their spin structures, are complex conjugate.  The two
spin structures can vary independently.
The dimension of $\varGamma$ is $6g-6|4g-4$. 

Henceforth, until section \ref{tanalog}, we mostly concentrate on the heterotic string.

\subsection{Lagrangians}\label{lagrangians}

To formulate the heterotic string on $\R^{10}$,  four contributions to the Lagrangian are important.  We will now see
that they all make sense in the present context.  

 In writing Lagrangians, we make the following definition
\begin{equation}\label{pogy}\dzzt=-i[\d\t z;\d z|\d\theta] \end{equation}
both as an abbreviation and to make it a little easier to compare to standard formulas.
This is a superanalog of the following.  In the bosonic world, the differential form $\d\bar z\wedge \d z$ is imaginary (since if $z=x+iy$,
then $\d\bar z\wedge\d z=2i\d x \wedge \d y$), and sometimes one defines the real form $\d^2 z = -i \d\bar z\wedge \d z$.  Our expression 
$\dzzt$ is analogous. 

The first possible Lagrangian
has essentially been described already in eqn. (\ref{romong}).  A map from the string worldsheet to $\R^{10}$ is described
by fields $X^I(\t z; \negthinspace z|\theta)$, $I=1\dots 10$.  The kinetic energy for these fields is 
\begin{equation}\label{bomong}I_X=\frac{1}{2\pi\alpha'}\int \dzzt\,\sum_{IJ}\eta_{IJ}\partial_{\t z}X^I D_\theta X^J, \end{equation}
where $\eta_{IJ}$ is the metric tensor of $\R^{10}$.  This expression does not depend on the choice of local coordinates since as already
explained, the integrand is a section of $\BBer(\Sigma_L\times \Sigma_R)$.  If the functions $X^I(\t z; \negthinspace z|\theta)$ are real-analytic,
the expression $I_X$ is also invariant under small displacements of $\Sigma$ in $\Sigma_L\times \Sigma_R$.

To describe the current algebra degrees of freedom of the heterotic string, we will use the description by fermions, since this is convenient
for writing an action.  We first pick a square root $\L$ of the line bundle $\BBer(\Sigma_L)=K_{\Sigma_L}$.   
Next we introduce 32 fields $\Lambda_a$, $a=1\dots 32$, valued in ${\Pi}\L$ (that is, they are fermionic fields valued in $\L$).  
These fields are often called current algebra fermions and the choice of $\L$ is called a choice of spin structure for 
them.\footnote{For brevity we will consider the $\Spin(32)/\Z_2$ heterotic string; in constructing the $E_8\times E_8$ heterotic
string, one divides the $\Lambda_a$  in two sets of 16 with a separate choice of $\L$ for each set.}  The expression
$D_\theta \Lambda_a$ makes sense as a section of $\L\otimes \D^{-1}$, because the line bundle  $\L$ is
antiholomorphic -- that is, it is a pullback from $\Sigma_L$, like its square $\BBer(\Sigma_L)$, and
 can be constructed using gluing functions that are functions of $\t z$ only and so commute with $D_\theta$.
Accordingly, $\sum_a\Lambda_a D_\theta\Lambda_a$ is a section of $\L^2\otimes \D^{-1}\cong \BBer(\Sigma_L\times \Sigma_R)$, restricted to $\Sigma$.
So the expression
\begin{equation}\label{zotto}I_\Lambda= \frac{1}{2\pi}\int \dzzt \sum_a  \Lambda_a D_\theta\Lambda_a \end{equation}
does not depend on local coordinates and can be integrated.  

What remain are the kinetic energies for the ghosts and antighosts.  The holomorphic ghosts are a section $C$ of ${\Pi}\S={\Pi} \D^{2}$, as explained
in section \ref{reconsider}.  Here $\S\cong \D^2$ is the sheaf of superconformal vector fields on $\Sigma_R$ (which we pull back to $\Sigma$).  
As this is a holomorphic line bundle, the expression $\t\partial C$ makes sense as a section of $\BBer(\Sigma_L)\otimes
{\Pi}\D^2$.  The holomorphic antighosts are a section $B$ of ${\Pi} \D^{-3}$.    Hence $B\tilde \partial C$
is a section of ${\BBer}(\Sigma_L)\otimes \D^{-1}\cong \BBer(\Sigma_L\times\Sigma_R)$, so it too can be integrated:
\begin{equation}\label{otto}I_{B,C}=\frac{1}{2\pi}\int \dzzt\,B\t \partial  C. \end{equation}
Finally, the antiholomorphic antighosts and ghosts are sections $\t B$ and $\t C$ of ${\Pi} \BBer(\Sigma_L)^2$ and ${\Pi}{\BBer}(\Sigma_L)^{-1}$, respectively,
so $\t B D_\theta \t C$ is  again a section of  $\BBer(\Sigma_L)\otimes\D^{-1}\cong \BBer(\Sigma_L\times\Sigma_R)$, leading to one last term in the action:
\begin{equation}\label{zottor}I_{\t B,\t C}=\frac{1}{2\pi}\int \dzzt\, \t B D_\theta\t C.\end{equation}

In superstring perturbation theory, it is important to understand the ghost and antighost zero-modes.
The $C$ zero-modes obey $\t\partial C=0$, so they are globally-defined superconformal vector fields, with parity reversed.  The $\t C$ zero-modes are globally-defined antiholomorphic
vector fields, also with parity reversed.  For this statement, one uses the fact that $D_\theta \t C=0$ implies that also $0=D_\theta^2\t C=\partial_z\t C$.
 Usually in superstring perturbation theory, there are no global $C$ or $\t C$ zero-modes because the genus is too large or there are too many punctures.  To understand the $B$ and $\t B$
zero-modes, we will need Serre duality (section \ref{motorg}).

The local expansions of the fields introduced above are as follows.  One has
\begin{equation}\label{oryt}X=x+\theta\psi, \end{equation}
where $x$ has antiholomorphic and holomorphic conformal dimension $(0,0)$ and $\psi$ has dimension $(0,1/2)$;
\begin{equation}\label{zoryt}\Lambda=\lambda+\theta F ,\end{equation}
where $\lambda$ has dimension $(1/2,0)$ and $F$ has dimension $(1/2,1/2)$; 
\begin{equation}\label{ghosts} B=\beta+\theta b,~~   C=c+\theta\gamma,\end{equation}
where $\beta$ and $\gamma$ are even fields of respective dimensions $(0,3/2)$ and $(0,-1/2)$, and $b $ and $c$ are odd fields
of respective dimensions $(0,2)$ and $(0,-1)$; and finally
\begin{equation}\label{tghosts} \t B=\t b+\theta \t f,~~\t C=\t c+\theta \t g,\end{equation}
where $\t b$ and $\t c$ are odd fields of dimensions $(2,0)$ and $(-1,0)$, while $\t f$ and $\t g$ are even fields of dimensions $(2,1/2)$ and $(-1,1/2)$.
The fields $F$, $\t f$, and $\t g$ are all auxiliary fields that (after performing the integral over $\theta$) appear quadratically in the action,
without derivatives.  They can hence be ``integrated out,'' and the worldsheet action of the heterotic string is often written in terms of only
the remaining fields.

\subsection{Cohomology and Duality}\label{motorg} 

In the present section, we emphasize the holomorphic structure of $\Sigma$ and we write $\BBer(\Sigma)$ for
what we have been calling $\BBer(\Sigma_R)$, the Berezinian of $\Sigma$ in the holomorphic sense.   And we interpret a section $g(\t z;\neg z|\theta)\d\t z$ of $\BBer(\Sigma_L)$ as a $(0,1)$-form on $\Sigma$.
What can be integrated on $\Sigma$ is therefore a $(0,1)$-form with values in $\BBer(\Sigma)$.  What follows is formulated
in terms of objects defined on the smooth supermanifold $\Sigma$ without regard to the embedding in $\Sigma_L\times \Sigma_R$.

The action (\ref{otto}) for the fields $B$ and $C$ would make sense more generally if $C$ is a section of an arbitrary holomorphic line bundle
$\RR$ and $B$ is a section of $\RR^{-1}\otimes \BBer(\Sigma)$.  The pair of line bundles $\RR$ and $\RR^{-1}\otimes\BBer(\Sigma)$ has a particular
significance related to duality.  

First let us review how duality works on an ordinary compact Riemann surface $\Sigma_0$. What can be integrated over $\Sigma_0$ is a $(1,1)$-form $\alpha=f(\bar z,z)\d \bar z\,\d z$.
From a holomorphic point of view, a $(1,0)$-form is best understood as a section of $K=T^*\Sigma_0$, the canonical bundle of $\Sigma_0$, and a $(1,1)$-form
is best understood as a $(0,1)$-form with values in $K$.

Now let $u$ be a $(0,k)$-form with values in some line bundle $\RR$ (where $k=0$ or 1), and let $v$ be a $(0,1-k)$-form valued in $\RR^{-1}\otimes K$.  Then the product $uv$ is naturally a $(0,1)$-form with values in $K$, so it can be integrated, to give
\begin{equation}\label{doofus}\Phi(u,v)=\int_{\Sigma_0} uv. \end{equation}
 Letting $\Omega^k(\Sigma_0,\T)$ denote
the space of $(0,k)$-forms on $\Sigma_0$ valued in any line bundle $\T$, $\Phi$ is best understood as a natural duality
\begin{equation}\label{wongo} \Phi: \Omega^k(\Sigma_0,\RR)\times \Omega^{1-k}(\Sigma_0,\RR^{-1}\otimes K) \to \C.\end{equation}
Calling $\Phi$ a duality means that it is a nondegenerate pairing that 
establishes the two vector spaces $\Omega^k(\Sigma_0,\RR)$ and $\Omega^{1-k}(\Sigma_0,\RR^{-1}\otimes K)$ as each other's dual spaces. 
(Because of the infinite-dimensionality of the spaces involved, there are some technicalities in formulating this precisely,
but these need not concern us here.)

To state the implications of this for cohomology, we need to define the Dolbeault cohomology groups
$H^k(\Sigma_0,\RR)$, $k=0,1$ of $\Sigma_0$ with values in a holomorphic line bundle $\RR$.  For this, one introduces the operator
\begin{equation}\label{okno}\bar\partial=\d\bar z\frac{\partial}{\partial \bar z} \end{equation}
mapping $(0,0)$-forms valued in $\RR$ to $(0,1)$-forms valued in $\RR$.  (We have assumed a local holomorphic trivialization of $\RR$ in this
simple way of writing the $\bar\partial$ operator.)  Its kernel -- the space of holomorphic sections of $\RR$ -- is defined to 
be $H^0(\Sigma_0,\RR)$.  On the other hand, $H^1(\Sigma_0,\RR)$ is defined to be the cokernel of the operator $\bar\partial$ or in other words the space
of all $\RR$-valued $(0,1)$-forms  modulo those that are $\bar\partial$-exact.

Because the duality (\ref{wongo}) between forms is invariant $u\to u+\bar\partial u$ or $v\to v+\bar\partial v$ (assuming that $\Sigma_0$ is
compact\footnote{Otherwise one should constrain $u$ (or $v$) to have compact support and one arrives at a duality between the ordinary
$\bar\partial$ cohomology groups and the corresponding cohomology groups with compact support.  The same statement holds later
when we discuss the super case.})
it induces a pairing between cohomology groups:
\begin{equation}\label{zongo}\Psi: H^k(\Sigma_0,\RR)\times H^{1-k}(\Sigma_0,\RR^{-1}\otimes K)\to \C.\end{equation}
For $\Sigma_0$  a compact Riemann surface, these cohomology groups are finite-dimensional, and Serre duality is the statement
that the pairing $\Psi$ is nondegenerate, establishing a duality between the cohomology groups in question.

All this has an analog on a super Riemann surface $\Sigma$.  The analog of a $(0,1)$-form is a 1-form proportional to $\d\t z$,
and the analog of the $\bar\partial$ operator is the operator
\begin{equation}\label{barp}\t\partial=\d\t z\frac{\partial}{\partial\t z} \end{equation}
from sections of a line bundle $\T$ to $(0,1)$-forms  valued in $\T$.  Again one defines $H^0(\Sigma,\T)$ as the kernel of $\t\partial$
and $H^1(\Sigma,\T)$ as its cokernel.  (One important subtlety is that in general, when one varies the odd moduli of $\Sigma$, dimensions of cohomology groups can jump.  We will formulate the following only in the absence of such jumping.)

  We again define $\Omega^k(\Sigma,\T)$, $k=0,1$,
 as the space of $(0,k)$-forms on $\Sigma$ with
values in a holomorphic line bundle $\T$.  The product of a $(0,k)$-form valued in $\RR$ (a line bundle whose fibers we allow to be either even or odd)
with a $(0,1-k)$-form valued in $\RR^{-1}\otimes \BBer(\Sigma)$ is a $(0,1)$-form
valued in $\BBer(\Sigma)$. So rather as before we have a natural duality, defined by integration. There are two differences from the bosonic case:  

(1) Since $\BBer(\Sigma)$ is itself a fermionic
line bundle (a typical local section being the odd object $[\d z|\d\theta]$), if one wants an integral $\int uv$ to be bosonic,
where $u$ is a section of $\Omega^k(\Sigma,\RR)$ and $v$ is a section of $\Omega^{1-k}(\Sigma,\RR^{-1}\otimes \BBer(\Sigma))$, then
either $u$ or $v$ must be odd.  Accordingly, the duality is usually formulated as a pairing between $\Omega^k(\Sigma,\RR)$ and
${\Pi} \Omega^{1-k}(\Sigma,\RR^{-1}\otimes \BBer(\Sigma))$.  (It does not much matter which of the two sides of the pairing one takes $\Pi$ to act on,
since one is free to replace $\RR$ by $\Pi\RR$.)

(2) The second point is a little more subtle.  An ordinary Riemann surface has moduli, but these moduli are bosonic and can be set
equal to complex values.  That is what we have effectively done in assuming above that the integral $\int_{\Sigma_0} uv$ is $\C$-valued, rather than being a function of
the moduli.  However, super Riemann surfaces have odd moduli, which cannot be set to nonzero complex values. So unless one wants to simply set
the odd moduli to zero, one should work not over $\C$ but over some ring $\C'$ that is generated over $\C$ by odd elements.  Then one does
not quite set the even and odd moduli of $\Sigma$ to complex constants; one sets them to even and odd elements of $\C'$. 
(For a relatively elementary account of introducing a ring such as $\C'$ in the context of field theory, see \cite{DeWitt}.)  In practice, this is somewhat like
speaking in prose; if one does what comes naturally, one never has to think about it, though a technical description of what is involved may cause confusion.

With these points understood, the superanalog of the duality (\ref{wongo}) reads
\begin{equation}\label{widongo}\Phi: \Omega^k(\Sigma,\RR)\otimes {\Pi}\Omega^{1-k}(\Sigma,\RR^{-1}\otimes \BBer(\Sigma))\to \C'.\end{equation}
Again the definition is simply
\begin{equation}\label{idongo} \Phi(u,v)=\int_\Sigma uv. \end{equation}

As in the bosonic case, the fact that $\Phi(u,v)$ is invariant under $u\to u+\t\partial u$ or $v\to v+\t\partial v$ means that it determines a pairing between
cohomology groups:
\begin{equation}\label{ritongo}\Psi:H^k(\Sigma,\RR)\times {\Pi} H^{1-k}(\Sigma,\RR^{-1}\otimes \BBer(\Sigma))\to \C'.\end{equation}
The superanalog of Serre duality is the statement that this is a nondegenerate pairing.  If $\Sigma$ is split, this statement reduces to ordinary Serre duality
(separately for the even and  odd parts of the cohomology groups).  Our no-jumping hypothesis ensures that the nondegeneracy of the pairing persists when
infinitesimal fermionic moduli are turned on.
We have formulated all this for a holomorphic line bundle $\RR$, but all statements have immediate analogs if $\RR$ is a holomorphic 
vector bundle of any rank $a|b$ and $\RR^{-1}$ is replaced by the dual bundle $\RR^*$.

As an example of the use of this duality, we know that the tangent space to the moduli space $\MM$ of super Riemann surfaces at the
point corresponding to $\Sigma$ is
$H^1(\SIgma,\S)=H^1(\SIgma,\D^2)$, where $\S\cong\D^2$ is the sheaf of superconformal vector fields.  (A derivation of this
fact from the Dolbeault point of view is in section \ref{defsmooth}.) So the cotangent space
to $\MM$ is $T^*\MM=H^0(\Sigma,\D^{-2}\otimes \BBer(\Sigma))=H^0(\Sigma,\D^{-3})$, where we use the isomorphism $\BBer(\Sigma)\cong \D^{-1}$.    
This is relevant to superstring perturbation theory, because the zero-modes of the antighost field $B$
are holomorphic sections of $\D^{-3}$ or in other words elements of $H^0(\SIgma,\D^{-3})$.

The analogous statements
for ordinary Riemann surfaces are perhaps more familiar.  The tangent space to the moduli space $\M$ of ordinary Riemann 
surfaces at the point corresponding to $\Sigma_0$ is $H^1(\Sigma_0,T)$, so the cotangent space to $\M$ at the
same point is $H^0(\Sigma_0,T^{-1}\otimes K)=H^0(\Sigma_0,K^2)$.  Here we use Serre duality (\ref{zongo}); $T\cong K^{-1}$ is the tangent bundle of $\Sigma_0$.  In bosonic string theory, the zero-modes of the antighost field $b$ are elements of $H^0(\SIgma_0,K^2)$.

\subsection{Deformation Theory}\label{defsmooth}

\subsubsection{Deformation Theory  Via Embedding}\label{descemb}

We now return to the integration cycle $\varGamma\subset\MM_L\times \MM_R$ of superstring perturbation theory, as defined in
section \ref{parspace}.  It is a smooth cs supermanifold (a concept described in \cite{Wittennotes}, section 2.1), 
meaning that  the odd coordinates of $\varGamma$ have no real structure.
Certain general remarks here are applicable to both heterotic and Type II superstrings.

The most natural notion of the tangent or cotangent bundle of a cs supermanifold is the analog of what for an ordinary
smooth manifold would be the complexified tangent or cotangent bundle.  The reason for this is that if $M$ is a cs manifold,
there is no real structure for its odd tangent or cotangent vectors, so it is unnatural to try to impose one for even
tangent and cotangent vectors.  Instead we simply define the tangent and cotangent
bundles $TM$ and $T^*M$ as ($\Z_2$-graded) complex vector bundles.

With this understood, the tangent bundle $T\varGamma$ of $\varGamma$ is the direct sum $T\MM_L\oplus T\MM_R$ (restricted to $\varGamma$), and similarly
$T^*\varGamma=T^*\MM_L\oplus T^*\MM_R$.    This is an analog of the fact that if $Y$ is an ordinary complex manifold that is the complexification of a middle-dimensional submanifold $N$,
then the complexified tangent space to $N$ is the tangent space to $Y$, restricted to $N$.  This fact is relevant because 
we have defined $\varGamma$ so that its complexification is $\MM_L\times \MM_R$.

We can regard $T^*\MM_R$ as the holomorphic cotangent space to $\varGamma$
and $T^*\MM_L$ as its antiholomorphic cotangent space.   So far our remarks have been general; now we return to the heterotic 
string.  Comparing to the Lagrangian (\ref{otto})
for holomorphic ghost and antighost fields, and also recalling the description of $T^*\MM$ from the end of section \ref{motorg},
we see now that the space of $B$ zero-modes is $\Pi T^*\MM_R$, the holomorphic cotangent bundle to $\varGamma$ with parity reversed.
Similarly, comparing to the Lagrangian (\ref{zottor}) for antiholomorphic ghost and antighost fields, and recalling the description of
$T^*\MM_L$, we see that the space of
$\tilde B$ zero-modes is $\Pi T^*\MM_L$, the antiholomorphic cotangent bundle to $\varGamma$ with parity reversed. 

Of course, we define the tangent and cotangent bundles of the cs supermanifold $\Sigma\subset \Sigma_L\times \Sigma_R$ in the
same way, so $T\Sigma=(T\Sigma_L\oplus T\Sigma_R)|_\SIgma$, $T^*\Sigma=(T^*\Sigma_L\oplus T^*\SIgma_R)|_\Sigma$.  
We abbreviate the summands of $T\SIgma$ as $T_L\Sigma$ and $T_R\Sigma$ and call them
 the antiholomorphic and holomorphic tangent bundles to $\SIgma$.  We use
analogous notation and terminology for the summands of $T^*\SIgma$.  The familiar structures that are present because $\Sigma_R$
is a super Riemann surface can all be restricted to $\SIgma$.  For instance, $T_R\SIgma$ has a subbundle $\D$ of rank $0|1$ with
an exact sequence $0\to \D\to T_R\SIgma\to \D^2\to 0$.    The sheaf of sections of $T_R\SIgma$
 has a subsheaf $\S$ of superconformal vector fields.  $\S$ is a sheaf of graded Lie algebras; if we forget that structure, there
 is an isomorphism $\S\cong \D^2$.  

\subsubsection{Deformation Theory Via Fields: Review}\label{defreview}

In section \ref{lagrangians}, we described the worldsheet Lagrangian for the heterotic string on the super Riemann surface $\Sigma$.
This can be the starting point for  heterotic string perturbation theory.  To develop this perturbation theory, we must understand
the deformations of $\Sigma$ in terms of fields on $\Sigma$, rather than via embeddings as in (\ref{descemb}) or  cutting and gluing 
as  in section \ref{otorg}.  This will be our next objective.  (The analog for Type II superstring
theory is similar and is briefly described in section \ref{newdef}.)

First, we recall the appropriate constructions
on an ordinary Riemann surface $\Sigma_0$, with notation chosen in anticipation of the
superstring case (which we study starting in section \ref{deffields}).  The 
complexified
tangent bundle of $\Sigma_0$ is $T\Sigma_0=T_L\SIgma_0\oplus T_R\SIgma_0$, where $T_R\SIgma_0$ and $T_L\SIgma_0$ are respectively the holomorphic and antiholomorphic subbundles of $T\SIgma_0$.  If $z$ and $\t z$ are local holomorphic and antiholomorphic coordinates on $\Sigma_0$, then $T_R\SIgma_0$ and $T_L\SIgma_0$ are generated respectively by
$\partial_{z}$ and by $\partial_{\t z}$.  A function $f$ on $\SIgma_0$ is holomorphic if it is annihilated
by $\partial_{\t z}$, and antiholomorphic if it is annihilated by $\partial_z$.  To deform the
holomorphic structure of $\SIgma$, we perturb $\partial_{\t z}$ by
\begin{equation}\label{doofog}\partial_{\t z}\to \partial'_{\t z}=\partial_{\t z}+ h_{\t z}^z\partial_z.\end{equation}
An additional 
 deformation $\partial_{\t z}\to \partial_{\t z}+ u_{\t z}^{\t z}\partial_{\t z}$ is not interesting, since it
does not deform the condition for a function to be holomorphic.  
Multiplying  $\partial_{\t z}$  by a nonzero function does not affect the 
subbundle $T_L\SIgma_0$ of $T\Sigma_0$ that is generated by $\partial_{\t z}$, 
so we allow a gauge transformation
\begin{equation}\label{zorby}\partial_{\t z}\to e^{\t\varphi} \partial_{\t z},\end{equation}
for any function $\t\varphi$.  This gauge-invariance can be used to remove the 
$u_{\t z}^{\t z}$ deformation.  Conversely, the holomorphic structure of $\Sigma_0$ is deformed by
\begin{equation}\label{orby}\partial_z\to \partial'_z=\partial_z+h_z^{\t z}\partial_{\t z}. \end{equation}
Again, a shift in $\partial_z$ by a multiple of itself is uninteresting, as it does not affect
the criterion for a function to be antiholomorphic; we are interested in $\partial_{z}$ only
up to a gauge transformation
\begin{equation}\label{zorbya}\partial_{ z}\to e^{\varphi} \partial_{z}.\end{equation}
In a diffeomorphism-invariant theory, we  only want to consider the deformations 
$h_{\t z}^z$ and $h_z^{\t z}$ modulo deformations
generated by a vector field
\begin{equation}\label{moryba}q=q^{\t z}\partial_{\t z}+q^z\partial_z.\end{equation}
Modulo (\ref{zorby}) and (\ref{zorbya}), the transformations
generated by the commutator of $\partial_{\t z}$ or $\partial _z$ with $q$ are
\begin{align}\label{meldoc}h_{\t z}^z&\to h_{\t z}^z+\partial_{\t z}q^z \cr
                                           h_z^{\t z}&\to h_z^{\t z}+\partial_z q^{\t z}.\end{align}
The field $h_{\t z}^z$, modulo this equivalence, defines an element of the holomorphic
sheaf cohomology
group $H^1(\Sigma_0,T_R\SIgma_0)$, and the field $h_z^{\t z}$, modulo this equivalence,
defines an element of the analogous antiholomorphic sheaf cohomology group, which we will
call $\t H^1 (\Sigma_0,T_L\SIgma_0)$.

These are the standard answers, although on an ordinary Riemann surface, since it is obvious
from the beginning that antiholomorphic deformations are complex conjugates of holomorphic
ones, one does not always write out the two cases in such detail. We have presented
the analysis this way as preparation for  the superstring
case, where it is not true that the antiholomorphic deformations are complex conjugates
of holomorphic ones. 

 Notice that, perhaps counterintuitively, a first-order holomorphic deformation is
made by deforming the embedding in $T\Sigma_0$
of the antiholomorphic tangent bundle $T_L\SIgma_0$, which is generated
by $\partial_{\t z}$, without changing $T_R\SIgma_0$,
and a first-order antiholomorphic deformation is made by deforming the embedding in $T\SIgma_0$ of
the 
holomorphic tangent bundle $T_R\SIgma_0$, which is generated by $\partial_z$, without changing
$T_L\SIgma_0$.   However, the deformation from $\partial_{\t z}$ to $\partial'_{\t z}$
affects the condition for a section of $T_R\SIgma_0$ to be holomorphic; and similarly an antiholomorphic
deformation affects the condition for a section of $T_L\Sigma$ to be antiholomorphic.

Finally, 
the fields $h_{\t z}^z$ and $h_z^{\t z}$ can be regarded
as either deformations of the metric of $\Sigma_0$ modulo Weyl transformations, or equivalently
as deformations of its complex structure, which we call $J$. We will briefly describe the second
point of view.   $J$ is a linear transformation of $T\Sigma_0$
that obeys $J^2=-1$ and acts as $i$ and $-i$ on $T_R\Sigma_0$ and $T_L\SIgma_0$, respectively.
One can deform $\SIgma_0$ as a complex manifold by deforming $J$.  The condition $J^2=-1$ sets $\delta J_{\t z}^{\t z}=\delta J_z^z=0$,
so the deformation only involves $\delta J_{\t z}^z$ and $\delta J_z^{\t z}$, which can be identified with 
$h_{\t z}^z$ and $h_z^{\t z}$ in the above formulas. The
derivation shows that $\delta J_{\t z}^z$ is a deformation of type $(1,0)$ on the space of complex
structures on $\Sigma$  (it represents a first-order deformation
of the holomorphic structure of $\SIgma_0$) while $\delta J_z^{\t z}$ is of type $(0,1)$ (it represents a first-order
deformation of the antiholomorphic structure).  In complex dimension greater
than 1, the deformation $\delta J$ would be subject to an integrability condition, but in complex dimension 1, this is trivial.
For more on such matters, see section   \ref{compstructures}.      

\subsubsection{Superstring Deformation Theory Via Fields}\label{deffields}

The case of deforming a heterotic string worldsheet $\Sigma$ is similar, with a few inevitable differences.
First we restate in terms of objects defined on $\Sigma$ some concepts that we previously
described in terms of the embedding $\Sigma\subset \Sigma_L\times \Sigma_R$.  
A function $F(\t z;\neg z|\theta)$ defined in some open set $U\subset \Sigma$ is holomorphic if it is annihilated by vector fields valued in
the antiholomorphic tangent bundle $T_L\Sigma$ of $\Sigma$ and antiholomorphic if it is annihilated by those valued in the holomorphic tangent bundle $T_R\Sigma$.
In a standard local coordinate system $\t z;z|\theta$,   $F$ is holomorphic if it obeys
\begin{equation}\label{torob} \partial_{\tilde z} F = 0,\end{equation}
Similarly the condition for antiholomorphy is
\begin{equation}\label{prorob}D_\theta F=0. \end{equation}
This condition of course implies that $F$ is also annihilated by $\partial_z=D_\theta^2$, so that it is annihilated
by all vector fields valued in  $T_R\Sigma$.

In deforming a heterotic string worldsheet $\Sigma$, there are potentially three types of
deformation to consider: {\it (i)} deformations of the holomorphic structure of $\SIgma$,
which mean deformations of the embedding of $T_L\SIgma $ in $T\SIgma$; {\it (ii)} deformations
of the antiholomorphic structure of $\Sigma$, which mean deformations of the embedding of
$T_R\SIgma$ in $T\SIgma$; and {\it (iii)} deformations of the superconformal structure (rather than
the holomorphic or antiholomorphic structure) of
$\Sigma$ -- in other words deformations
of the embedding of $\D$ in $T_R\SIgma$.

In a diffeomorphism-invariant theory, we are only interested in first-order deformations of $\Sigma$ modulo
those that are generated by a vector field.  An arbitrary vector field $v$ on $\Sigma$ can, of
course, be written \begin{equation}\label{melfog}
v=w^{\t z}\partial_{\t z}+w^z\partial_z+w^\theta\partial_\theta,\end{equation} with coefficients
$w^{\t z}$, $w^z$, and $w^\theta$ that are functions of $\t z;\neg z|\theta$.  However, it turns out
that a different expansion is more illuminating:
\begin{equation}\label{plonzo}v=q^{\t z}\partial_{\t z}+\left(q^z\partial_z+\frac{1}{2}D_\theta q^z
D_\theta\right)+q^\theta D_\theta.\end{equation}
This way of making the expansion is useful largely because if $q^z$ is holomorphic (annihilated by
$\partial_{\t z}$) then $q^z\partial_z+\frac{1}{2}D_\theta q^z
D_\theta$ is a superconformal vector field (see eqn. (\ref{orod})).

The simplest deformations to analyze are actually those of type {\it (iii)} in which  the
superconformal structure is changed. These are deformations in which a generator $D_\theta$ of $\D$ is shifted
by a section of $T_R\SIgma$ that is not proportional to $D_\theta$ (a transformation
$D_\theta\to e^f D_\theta$ does not affect the subbundle of $T_R\Sigma$ generated by $D_\theta$).
Such a deformation has the form
\begin{equation}\label{lonzo}D_\theta\to D_\theta+r_\theta^z\partial_z. \end{equation}
However, conjugating by a vector field $q^\theta D_\theta$ transforms $D_\theta$ by
$D_\theta\to D_\theta+[D_\theta,q^\theta D_\theta]$.  Modulo a multiple of $D_\theta$, this
is $D_\theta\to D_\theta -2q^\theta\partial_z$.  So by taking $q^\theta=-r^z_\theta/2$, we can in
a unique way eliminate deformations of type {\it (iii)} while also eliminating the equivalence
by vector fields of the form $q^\theta D_\theta$.

Now we turn to deformations of type {\it (i)}.
To deform the holomorphic structure of $\Sigma$, we make a first-order perturbation of the
condition of holomorphy, deforming the condition $\partial_{\t z}F=0$ to $\partial'_{\t z}F=0$ with
\begin{equation}\label{worb}\partial_{\t z}'= \partial_{\t z} +
h_{\t z}^z\partial_z+\chi_{\t z}^\theta \partial_\theta. \end{equation}
As in the bosonic case, there is no point in perturbing $\partial_{\t z}$ by an additional term $u_{\t z}{}^{\t z}\partial_{\t z}$, since this would not affect the condition for a function to be 
holomorphic.  Thus we allow gauge transformations
\begin{equation}\label{otorb}\partial_{\t z}\to e^{\t \varphi}\partial_{\t z}.\end{equation}   On
the fields $h_{\t z}^z$ and $\chi_{\t z}^\theta$,
we want to impose the equivalence relation of ignoring trivial deformations $[\partial_{\t z},v]$
for any vector field $v$.  It is illuminating to do this first ignoring the condition that the deformation
is supposed to preserve the holomorphic
superconformal structure of $\Sigma$.  In this case, one would conveniently use the generic expansion (\ref{melfog})
of a vector field $v$ to generate gauge invariances
\begin{equation}\label{omorb} h_{\t z}^z\to  h_{\t z}^z+ \partial_{\t z}w^z,~~ \chi_{\t z}^\theta\to \chi_{\t z}^\theta+\partial_{\t z}w^\theta.\end{equation}
(The commutator with $v$ also shifts $\partial_{\t z}$ by a term $(\partial_{\t z}w^{\t z})\partial_{\t z}$, which we remove via (\ref{zorby}).) 

We can make this look more familiar if we introduce a 1-form $\d\t z$ and combine the fields $h_{\t z}^z$ and $\chi_{\t z}^\theta$ to
a 1-form with values in $T_R\Sigma$:
\begin{equation}\label{pico}\CC= \d\t z\left(h_{\t z}^z\partial_z+\chi_{\t z}^\theta\partial_\theta\right).\end{equation}
Then the equivalence relation (\ref{omorb}) amounts to
\begin{equation}\label{ico}\CC\to \CC +\t\partial v',\end{equation}
where 
$\t\partial=\d\t z\,\partial_{\t z}$ was introduced in eqn. (\ref{barp}) to define Dolbeault cohomology,
and $v'$ is a 0-form with values in $T_R\Sigma$:
\begin{equation}\label{rico} v'=w^z \partial_z+w^\theta\partial_\theta. \end{equation}
The equivalence classes form by definition the Dolbeault cohomology group $H^1(\Sigma,T_R\Sigma)$.

We have arrived from a new vantage point at the result of section \ref{defc}: the first-order deformations of $\Sigma$ as a complex supermanifold
are parametrized by $H^1(\Sigma,T_R\Sigma)$.  
The only difference is that in this derivation, the cohomology appears via the field $\CC$, while the previous approach was based on gluing functions and Cech cohomology.   
The same arguments as on an ordinary complex manifold show 
that the two types of cohomology are equivalent, but the description by fields is a useful starting point for superstring perturbation theory.  

However, so far we did not impose the condition that we want deformations 
that preserve the fact that from a holomorphic point of view,
$\Sigma$ is a super Riemann surface, rather than a more general complex supermanifold
of dimension $1|1$.  The super Riemann surface structure is defined by the 
subbundle $\D\subset T_R\Sigma\subset T\Sigma$.  So we want to deform
$\Sigma$ preserving the fact that $T_R\Sigma$ has the holomorphic subbundle $\D$.
A first-order deformation of the holomorphic structure of $\Sigma$ 
will leave fixed $T_R\Sigma$ as a subbundle
of $T\Sigma$  (for the same reasons as in the discussion in section \ref{defreview} of deformation of an ordinary
Riemann surface $\Sigma_0$).
We look for deformations that also do not alter the
embedding of $\D$ in $T_R\Sigma\subset T\Sigma$.   (Deformations that do alter the embedding
of $\D$ in $T_R\Sigma$ are the type {\it (iii)} deformations that we already studied above.)  However, again
as on an ordinary Riemann surface,
the first-order holomorphic deformation will modify the condition for a section of $\D$ to be
holomorphic.
Any section of $\D$ is locally of the form  $e^ \mu D_\theta$  for some function $ \mu$;
the condition for this section to be holomorphic after the deformation is that 
it should commute with the modified operator $\partial_{\t z}'$:
\begin{equation}\label{tosco}[\partial_{\t z}',e^ \mu D_\theta]=0.\end{equation}
The condition that a function $ \mu$ should exist locally obeying this equation is that the perturbation in (\ref{worb}) should take
values not in $T\Sigma$ but in its subsheaf $\S$ of superconformal vector fields.  Thus, recalling from  eqns. (\ref{oco})
and (\ref{zond}) the general form of a superconformal vector field, we specialize (\ref{worb}) to perturbations of the following
kind:
\begin{equation}\label{zorb}\partial_{\t z}'=\partial_{\t z}+
\left(h_{\t z}^z(\t z;\neg z)\partial_z+\half \partial_z h_{\t z}^z(\t z;\neg z)\theta\partial_\theta\right)
+\chi_{\t z}^\theta(\t z;\neg z) \left(\partial_\theta-\theta\partial_z\right) .\end{equation}
A key difference from (\ref{worb}) is that the fields $h$ and $\chi$ defined in this 
new way do not depend on $\theta$.
Precisely for such perturbations, one can locally find a function $ \mu$ such 
that (\ref{tosco}) holds to first-order in the
perturbation.\footnote{In verifying this, one uses the relations (\ref{latef}), 
which do not require holomorphy of $f$ and $g$, only the fact that they do
not depend on $\theta$.} 
Of course, $ \mu$ is uniquely
determined only modulo the possibility of adding a function that 
commutes with the unperturbed operator $\partial_{\t z}$.  What we are constructing
is the sheaf of holomorphic sections of $\D$ with its perturbed complex 
structure, not a particular holomorphic section of this sheaf.

It is again convenient to multiply eqn. (\ref{zorb}) by the $(0,1)$-form $\d\t z$.  
Then the perturbation of interest is a $(0,1)$-form on $\Sigma$
valued in $\S$.  As before, we need to impose an equivalence relation on this $(0,1)$-form,
because perturbations that are generated by vector fields are uninteresting.  
Now it is best to use the expansion (\ref{plonzo}) of a  vector field
that is better adapted to the superconformal structure.  Vector fields  
$q^\theta D_\theta$ have already been used to remove deformations of type {\it (iii)},
and vector fields $q^{\t z}\partial_{\t z}$ leave $\partial_{\t z}$ fixed, modulo a transformation 
(\ref{otorb}).   Finally, vector fields $q^z\partial_z+\frac{1}{2}D_\theta q^z
D_\theta$ generate the expected gauge transformations of $h_{\t z}^z$ and $\chi_{\t z}^\theta$,
\begin{equation}\label{pojj}h_{\t z}^z\to h_{\t z}^z+\partial_{\t z}a^z,~~\chi_{\t z}^\theta
\to \chi_{\t z}^\theta+\partial_{\t z} \eta^\theta,\end{equation}
where $q^z=a^z+2\theta \eta^\theta$.

 The result is that one should
classify the  perturbations by $h_{\t z}^z$ and $\chi_{\t z}^\theta$ up to $\t\partial$-exact forms.  Thus equivalence classes now give a
Dolbeault
description of $H^1(\Sigma,\S)$.  We have arrived at the same description as in section \ref{otorg}
of the space of first-order deformations of $\Sigma$ as a super Riemann surface, but now in terms of fields rather than Cech cocycles.  The fields appearing in (\ref{zorb})
have a familiar interpretation: $h_{\t z}^z$ is usually called a metric perturbation, and $\chi_{\t z}^\theta$ is called a gravitino field.

Now let us  perturb the antiholomorphic structure of $\Sigma$.  As already noted, prior to any perturbation, antiholomorphic
functions on $\Sigma$ are precisely the functions annihilated by $D_\theta$.  So now we want to perturb the operator $D_\theta$.  The most general perturbation
that we have to consider is to replace $D_\theta$ by
\begin{equation}\label{zoffo}D_\theta'=D_\theta+ c_\theta^{\t z}\partial_{\t z}.\end{equation}
There is no point in perturbing $D_\theta$ by a multiple of itself, since this does not affect the condition for a function to be antiholomorphic,
that is, annihilated by $D_\theta$.  For deforming the antiholomorphic structure of $\SIgma$,
there is also no point in perturbing $D_\theta$ by a multiple of $\partial_z=D_\theta^2$, since any
function annihilated by $D_\theta$ is annihilated by $\partial_z$. So (\ref{zoffo}) is the most general possible perturbation of the antiholomorphic
structure.  (A shift of $D_\theta$ by a multiple of $\partial_z$ is a type {\it (iii)} deformation
that we have already considered above.)
Again, we are only interested in this perturbation modulo perturbations $[D_\theta,v]$ induced by vector fields $v$ on $\Sigma$.  In perturbing
the antiholomorphic structure of $\Sigma$, the important vector fields are of the form $v=q^{\t z}\partial_{\t z}$ (since contributions of
other vector fields to $[D_\theta,v]$  are proportional to $D_\theta$ or $\partial_z$), 
and in the commutator $[D_\theta,v]$, we only care about the 
 term $D_\theta q^{\t z}\,\partial_{\t z}$. 
So the relevant  equivalence relation  on the field $c_\theta^{\t z}$ that appears in (\ref{zoffo}) is
\begin{equation}\label{rotok} c_\theta^{\t z}\cong c_\theta^{\t z}+D_\theta q^{\t z}. \end{equation}
Let us expand these functions in powers of $\theta$, 
\begin{align}\label{otok} c_\theta^{\t z}(\t z;\neg z|\theta)=&e_\theta^{\t z}(\t z;\neg z)+\theta h_z^{\t z}(\t z;\neg z),\cr  q^{\t z}(\t z;\neg z|\theta)
=&t^{\t z}(\t z;\neg z)+\theta u_\theta^{\t z}(\t z;\neg z). \end{align} 
We find that  the gauge-equivalence (\ref{rotok}) reads
\begin{align}\label{toxco}e_\theta^{\t z}& \cong e_\theta^{\t z}+ u_\theta^{\t z}\cr 
    h_z^{\t z}&\cong h_z^{\t z} +\partial_z t^{\t z}. \end{align}
Clearly we can set $u_\theta^{\t z}=- e_\theta^{\t z}$, eliminating  $e_\theta^{\t z}$ and completely fixing 
the gauge-invariance generated by $u_\theta^{\t z}$.  But the space of fields
$h_z^{\t z}$ modulo the equivalence relation in (\ref{toxco}) is a standard description of a nontrivial sheaf cohomology group -- or more
exactly it is standard except that the complex structure has been reversed.
We will  call this group $\tilde H^1(\Sigma,T_L\Sigma)$, the sheaf cohomology of $\Sigma$ with values in the sheaf of antiholomorphic sections of
$T_L\Sigma$.  (The tilde is meant as a reminder that this is cohomology with values in a sheaf of antiholomorphic sections.)
The field $h_z^{\t z}$ is again known as a metric perturbation.

To summarize, the first-order holomorphic deformations of the smooth supermanifold $\Sigma$ preserving its relevant structures
are given by the holomorphic sheaf
cohomology $H^1(\Sigma,\S)$, where $\S\cong \D^2$ is the sheaf of (holomorphic) superconformal vector fields.  And its first-order
antiholomorphic deformations
are given by the antiholomorphic sheaf cohomology $\t H^1(\Sigma,T_L\Sigma)$,  the cohomology with values in the sheaf of antiholomorphic
vector fields.  These are the answers expected from a description by Cech cocycles and gluing, or from the embedding in $\Sigma_L\times
\Sigma_R$, but here we have obtained 
an equivalent description via fields on $\Sigma$.  This is a useful starting point for superstring perturbation theory.

\subsubsection{Complex Structures}\label{compstructures}

Now we will  briefly explain how to state our results in terms of complex structures on $\SIgma$
and their deformations.

   In general, consider a smooth cs supermanifold $M$
of dimension $2p|q$.  An almost complex structure is an even  endomorphism $\J$ of the complexified
tangent bundle $TM$ of $M$ that obeys $\J^2=-1$.  Given an almost complex structure,
we define $T_RM$ and $T_LM$ to be the subbundles of $TM$ on which $\J$ acts as $i$
and $-i$, respectively.  We require that after reducing modulo the odd variables, $T_LM$ is
close to the complex conjugate of $T_RM$.   This implies in particular that the even
parts of the ranks of $T_LM$ and $T_RM$ are equal (they 
have ranks $p|a$ and $p|b$ for some $a$ and $b$ with $a+b=q$).

$\J$ is said to be integrable if the sections of $T_LM$ form a Lie algebra and likewise
the sections of $T_RM$ form a Lie algebra.  
Let us first consider the case of an ordinary complex manifold, meaning that $M$ has dimension
$2p|0$ for some $p$.
For $p=1$,
integrability of $\J$ is a trivial condition, because $T_LM$ (for example) is generated by
a single vector field $w$ (for example $w=\partial_{\t z}+h_{\t z}^z\partial_z$), and given
any one vector field $w$,  the vector fields $e^{\t \varphi}w$ always form a Lie algebra.  By
contrast, for $p>1$, integrability is a severe constraint; this is because if $T_LM$ is
generated by two vector fields $w$ and $w'$, we have to ask whether $[w,w']$ can
be expressed as a linear combination of $w$ and $w'$.  

Now let us consider the case that $M$ has dimension $2|1$.  Without essential loss
of generality, we can consider the case that $T_LM$ has rank $1|0$ and $T_RM$
has rank $1|1$.  Integrability of $T_LM$ is now trivial, because $T_LM$ is generated
by a single vector field; but integrability of $T_RM$ is non-trivial, since $T_RM$
is generated by two vector fields, namely an odd one and an even one.

However, integrability of $T_RM$ becomes trivial again if we are given that $M$ has
a holomorphic superconformal structure, meaning that $T_RM$ has a rank
$0|1$ subbundle $\D$ and is generated by $D_\theta$ and $D_\theta^2$, for some  section
(and in fact for any generic section)
$D_\theta$ of $\D$.  This makes integrability of $T_RM$ trivial, because no matter
what odd vector field $D_\theta$ we choose, the vector fields of the form $e^\varphi D_\theta$
and $e^{\h\varphi} D_\theta^2$ always form a Lie algebra.  

The case described in the last paragraph is of course the case that $M=\Sigma$ is the
worldsheet of a heterotic string.  Now let us describe deformation theory of $\Sigma$ in this
language. We deform the complex structure $\J$ of $\Sigma$ under the condition that $\J^2=-1$
and that $\Sigma$ has a holomorphic superconformal structure, in other words
$T_R\Sigma$ has a distinguished subbundle $\D$ of rank $0|1$.  The deformation can be described,
just as on an ordinary complex manifold, by a tensor $\delta \J$ that maps $T\SIgma$ to itself.
The condition $\J^2=-1$ means that $\delta\J_{\t z}^{\t z}=\delta\J_z^z=\delta\J_z^\theta=\delta\J_\theta^z
=\delta\J_\theta^\theta=0$.  The superconformal structure determines $\delta\J_{\t z}^\theta$ in terms of 
$\delta\J_{\t z}^z$, as exhibited in eqn. (\ref{zorb}).  The fact that $T_R\Sigma$ is generated by
$D_\theta$ and $D_\theta^2=\partial_z$ means that there is no need to deform $\partial_z$ independently of
$D_\theta$, so  $\delta\J_z^{\t z}$ is determined in terms of $\delta\J_\theta^{\t z}$.   
As we have seen in analyzing the deformations of type {\it (iii)} in section \ref{deffields}, nothing new comes
from varying $\D$ within $T_R\Sigma$.  

Though this involves jumping ahead of our story slightly, there is one more important case
in which integrability is trivial.  This is the case of the worldsheet of a Type II superstring.
Here $M$ has dimension $2|2$ and $T_LM$ and
$T_RM$ each has rank $1|1$, and each is endowed with a superconformal structure.  This means
that $T_R\Sigma$ is generated by $D_\theta$ and $D_\theta^2$ where $D_\theta$ is a section of
a distinguished rank $0|1$ subbundle $\D\subset T_RM$, and similarly $T_L\Sigma$ is generated by
$D_{\t \theta}$ and $D_{\t\theta}^2$, where $D_{\t\theta}$ is a section of a distinguished rank $0|1$ subbundle
$\t\D\subset T_LM$.  Deformations of the almost complex structure of $M$ that preserve such a structure
are automatically integrable, since vector fields $e^\varphi D_\theta$ and $e^{\h\varphi}D_\theta^2$ (and similarly
vector fields $e^{\t \varphi}D_{\t \theta}$ and $e^{\h{\t\varphi}}D_{\t\theta}^2$) always form a Lie algebra.

\subsection{Relation To Supergravity}\label{relsup}

Many approaches to two-dimensional supergravity
 can be found in the literature, for example \cite{H,Ga,GN,NM}.  We will aim for a shortcut here, introducing only
the necessary definitions and minimizing the number of equations. In contrast to much of the literature,
we start by describing the conformally-invariant structure, which after all is the main structure of importance
for string theory.   Then we describe the super version of a Riemannian metric.
This has both intrinsic interest and some applications in string theory.

We focus here on describing the worldsheet of a heterotic string in the language of supergravity.
For the  Type II
analog, see section \ref{newsuper}.

\subsubsection{The Conformally Invariant Case}\label{elsup}

In a standard coordinate system $\t z;\neg z|\theta$, we can define the 1-forms
\begin{align}\label{zonko}\t E_0& = \d \t z \cr E_0& = \d z-\theta\d\theta \cr
        F_0 & = \d\theta ,\end{align}
obeying
\begin{align}\label{onko}\d\t E_0 & = 0 \cr \d E_0+F_0\wedge F_0&=0 \cr \d F_0&=0.\end{align}
 To understand the significance of $E_0$, recall that $T_R^*\Sigma$, which is generated by $E_0$ and $F_0$,
  appears in an exact
sequence (\ref{zomurk}):
\begin{equation}\label{zomur} 0 \to \D^{-2}\to T_R^*\Sigma\to \D^{-1}\to 0,\end{equation}
where $\D^{-2}$ is generated by $\varpi=\d z-\theta\d\theta$, which we have taken for $E_0$.

From a conformally invariant point of view, we are interested not in $E_0$ and $\t E_0$, but in the associated holomorphic
and superconformal structures.   These are invariant under what we might call Weyl transformations.  We set
\begin{equation}\label{dolfus} \t E=e^{\t \varphi}\t E_0 ,\end{equation}
with an arbitrary function $\t\varphi$.  So $\t E$ obeys
\begin{equation}\label{olfus}\d\t E=0~~\mod~\t E.\end{equation}
This means that $\d\t E=\alpha\wedge \t E$ for some 1-form $\alpha$ (in fact $\alpha=\d\t\varphi$).  Similarly we set 
\begin{equation}\label{lfus}E=e^\varphi E_0,\end{equation}
again with an arbitrary function $\varphi$.  Now there is an odd 1-form $F$, unique up to sign (the sign is unique if
one asks for $F$ to vary continuously with $\varphi$ and to equal $F_0$ at $\varphi=0$) such that
\begin{equation}\label{ombus}\d E+F\wedge F = 0 ~~\mod \t E\wedge E. \end{equation}
We leave the reader as an exercise to verify that this is true and to determine $F$.  We remark only that $F$ is a linear combination of
$E_0$ and $F_0$:
\begin{equation}\label{zeldo}F=aF_0+b E_0,~~a\not=0. \end{equation}

From the conditions stated in the last paragraph, one can reconstruct the complex and superconformal structure of $\Sigma$,
and moreover, these are Weyl-invariant, that is, 
independent of $\t\varphi$ and $\varphi$.  The decomposition $T^*\SIgma=T_L^*\SIgma\oplus T_R^*\SIgma$
of $T^*\SIgma$ in antiholomorphic and holomorphic summands is given by declaring that $T_L^*\Sigma$ is generated by
$\t E$, while $T_R^*\Sigma$ is generated by\footnote{Eqn. (\ref{zeldo}) implies that the linear span of $E$ and $F$ is Weyl-invariant,
so that the definition of $T_R^*\SIgma$ is Weyl-invariant.} $E$ and $F$.    Finally, the line bundle $\D^{-2}\subset T_R^*\SIgma$ is the subbundle generated
by $E$.  (The decomposition $T\Sigma=T_L\SIgma\oplus T_R\SIgma$ is the dual of the decomposition of $T^*\Sigma$, and in
particular $\D\subset T\Sigma$ is the subbundle orthogonal to both $E$ and $\t E$.)  Thus the conditions of the last paragraph
capture the entire structure of a heterotic string worldsheet, expressed in supergravity language. 

More explicitly, given 1-forms $\t E$, $E$, and $F$ obeying (\ref{olfus}) and (\ref{ombus}), one can locally fix the Weyl factors
$\t\varphi$ and $\varphi$ so that (\ref{olfus}) and (\ref{ombus}) reduce to (\ref{onko}), and then one can introduce standard coordinates
$\t z;\neg z|\theta$ so that $\t E$, $E$, and $F$ take the form given in (\ref{zonko}).   So again, eqn. (\ref{olfus}) and (\ref{ombus})
contain the full structure of a heterotic string worldsheet.

We have achieved much greater brevity than can sometimes be found in the literature, because we have started with the conformally
invariant case.  Also, we have presented only a minimum set of necessary equations.  For example, we have written no equation for
$\d F$; none is needed.   This will remain so when we introduce the analog of a Riemannian metric.

\subsubsection{More On Deformation Theory}\label{defagain}

It is interesting to consider deformation theory from this point of view.   
The analysis will be equivalent to that of section \ref{deffields}, but in a dual language.
We will consider first-order deformations of $\t E$ and $E$,
modulo both Weyl rescaling and deformations generated by a vector field
\begin{equation}\label{oplonzo}v=q^{\t z}\partial_{\t z}+\left(q^z\partial_z+\frac{1}{2}D_\theta q^z
D_\theta\right)+q^\theta D_\theta.\end{equation}
We parametrize the vector field as in eqn. (\ref{plonzo}).
There is no need to consider deformations of $F$, since $F$ is uniquely determined (up to sign) 
in terms of $E$ and $\t E$ by eqn. (\ref{ombus}).

First we consider deformations of $E$.  Deformations of $E$ that are proportional to itself can be removed by a Weyl transformation,
and deformations of $E$ that are proportional to $F$ can be removed in a unique fashion using the $q^\theta D_\theta$
term in $v$.  So we can forget about $q^\theta$ and consider only deformations of $E$ that are proportional to $\t E$: 
\begin{equation}\label{emobo}E\to E+h_{\t z}^z \t E. \end{equation}
The notation $h_{\t z}^z$ is motivated by the way that  this field transforms under a holomorphic reparametrization of $z$ and $\t z$.
Geometrically,  $h_{\t z}^z\t E$ is a $(0,1)$-form with values in  $\D^{2}$.  (Up to a Weyl transformation, $E$ is any section of $\D^{-2}$,
and the deformation $E\to E+  h_{\t z}^z \t E$  followed by the projection to the second term, namely $h_{\tilde z}^z\t E$,
 is a map from sections of $\D^{-2}$ to $(0,1)$-forms, or equivalently a $(0,1)$-form with
values in $\D^2$.)  
   Often, one omits $\t E$ and says informally that $h_{\t z}^z$ is a $(0,1)$-form with values in $\D^{2}$. 
$h_{\t z}^z$ is subject to the gauge-invariance 
\begin{equation}\label{arox} h_{\t z}^z\to h_{\t z}^z+\partial_{\t z}q^z,\end{equation}
which reflects the transformation of $E$ generated by\footnote{
The transformation of a 1-form $E$ generated by a vector field $v$ is in general given by ${\mathfrak L}_v(E)$,
where $\mathfrak L_v=\i_v \d+\d \i_v$ is the Lie derivative.} the vector field $v$ of eqn (\ref{oplonzo}).  So the deformations of the holomorphic structure of $\Sigma$, modulo trivial ones, are given by $H^1(\Sigma,\D^{2})$, a familiar
answer.

In deforming $\t E$, we can again disregard deformations of $\t E$ that are proportional to itself, as these can be removed by
a Weyl transformation.  So we consider a deformation $\t E\to \t E+u^{\t z}_zE+c_\theta^{\t z} F$.  
However, the requirement that $\d\t E=0$ $\mod~\t E$
determines $u$ in terms of $c$.  In standard coordinates $\t z;\neg z|\theta$ (that is, with $\t E$, $E$, and $F$ as in eqn.
(\ref{zonko})), the relation is $u^{\t z}_z=D_\theta c_\theta^{\t z}$.  Geometrically, $c_\theta^{\t z}$ is a section of $\D^{-1}\otimes T_L\Sigma$.  
The equivalence relation on $c_\theta^{\t z}$ that comes from deformations by the vector field $v$ is
\begin{equation}\label{oxo}  c_\theta^{\t z}\to c_\theta^{\t z}+D_\theta q^{\t z}. \end{equation}
This description of antiholomorphic deformations of $\Sigma$  is familiar from eqn. (\ref{rotok}), and, as explained in that connection, it
amounts to a description of the space of antiholomorphic first-order deformations of $\Sigma$ via the antiholomorphic sheaf cohomology group 
$\t H^1(\Sigma,T_L\Sigma)$.

\subsubsection{The Super Analog Of A Riemannian Metric}\label{anamet}

Our remaining goal along these lines is to describe the super analog of a Riemannian metric.  
Before considering super Riemann surfaces, 
let us review Riemannian geometry  on an ordinary Riemann surface $\Sigma_0$.  We write $T_\C^*\Sigma_0$ for the complexified 
cotangent bundle of the smooth two-manifold $\Sigma_0$.  (We complexify the cotangent bundle of $\Sigma_0$ to match the
way we defined the cotangent bundle of a smooth cs supermanifold.)  To match our notation in the super case, we write
  $T_\C^* \Sigma_0=T_L^*\Sigma_0\oplus T_R^*\Sigma_0$, where the two summands are respectively 
the spaces of $(0,1)$-forms and of $(1,0)$-forms on $\Sigma_0$.
The additional structure that leads to Riemannian geometry can be formulated in many ways.  For example, one can introduce a 
hermitian metric on the line bundle $T_R^*\Sigma_0$;
this is equivalent to a Riemannian metric on $\Sigma_0$.  In the spirit of supergravity, it is more useful to take for the
basic data  a ``vierbein,'' which 
in the present context  means locally a complex conjugate pair of nonzero sections $\bar E$ and $E$ of $T_L^*\SIgma_0$
and $T_R^*\SIgma_0$,
subject to the gauge-invariance
\begin{equation}\label{homb} E\to e^{i\w }E,~~\bar E\to e^{-i\w }\bar E,\end{equation}
with $\w $ a real-valued function.
The quantity $E\otimes \bar E$ is gauge-invariant, and we can think of it as the Riemannian metric on $\Sigma_0$.
Explicitly, if $z$ is a local complex coordinate on $\Sigma_0$, then as $E$ is supposed to be of type $(1,0)$ and $\bar E$ is its complex conjugate, we have
\begin{equation}\label{omb} E=e^\varphi \d z,~~\bar E=e^{\bar \varphi}\d\bar z,\end{equation}
for some complex-valued function $\varphi$.  The Riemannian metric is
\begin{equation}\label{comb}\d s^2=E\otimes \bar E=e^{2\mathrm{Re}\,\varphi}|\d z|^2. \end{equation}
With the aid of the gauge-invariance (\ref{homb}), we can take $\varphi$ to be real, if we so desire.
In that case, the vierbein reads
\begin{equation}\label{romb}E=e^\phi\,\d z,  ~~~\bar E=e^\phi \d\bar z,\end{equation}
with real-valued $\phi$.
 The metric is then $e^{2\phi}|\d z|^2$;
the function $e^{2\phi}$ is often called the Weyl factor.

The next step is to define the Levi-Civita connection on 
$T_\C^*\Sigma_0$; it is uniquely characterized by being metric-compatible and torsion-free.
It is a 
connection $\omega$ with structure group $U(1)$.  We represent $\omega$ by a real 1-form on $\Sigma$. Under the gauge transformation
(\ref{homb}), it transforms as
\begin{equation}\label{tri}\omega\to\omega+\d u.\end{equation}
This ensures gauge covariance of the following extension of the exterior derivative:
\begin{equation}\label{ritz}DE=(\d-i\omega)E,~~D\bar E=(\d+i\omega)\bar E.\end{equation}
The Levi-Civita connection is defined by requiring
\begin{equation}\label{zitz}DE=D\bar E=0.\end{equation}
Writing $\omega = \d z \,\omega_z+\d \bar z\,\omega_{\bar z}$, a very short computation reveals that
\begin{equation}\label{witz}\omega_{\bar z}=-i\partial_{\bar z} \varphi,~~\omega_z=i\partial_z\bar\varphi.\end{equation}

The sections $\bar E$ and $E$ of $T_L^*\Sigma_0$ and $ T_R^*\Sigma_0$ 
cannot be defined globally (unless the Euler characteristic of $\Sigma_0$ vanishes).  
Globally, a better formulation is to think of $E$ as an isomorphism between $T_R^*\Sigma$ and a line bundle $\mathcal U$ (on which
$\omega$ is a connection), while $\bar E$ is an isomorphism between the complex conjugate line bundles.
 A similar remark applies in the superstring context that we consider next.

Now let $\Sigma$ be a heterotic string worldsheet.         Then 
$T^*\Sigma= T_L^*\Sigma\oplus T_R^*\Sigma$, where  $T_L^*\Sigma$ is of rank $1|0$ and $T_R^*\Sigma$ is of rank $1|1$.
Since $T_L^*\Sigma$ has rank $1|0$, it is a direct analog of $T_L^*\SIgma_0$ in the bosonic case.  As for
 $T_R^*\Sigma$, it has rank $1|1$, but it has a subbundle $\D^{-2}$ of rank $1|0$, and this
 will play the role of $T_R^*\Sigma_0$ in the bosonic case. 

We now can introduce what we claim is the appropriate analog on a super Riemann surface of a Riemannian structure on an ordinary 
Riemann surface.  The appropriate structure, locally,
is the choice of a nonzero section $\t E$ of $T_L^*\Sigma$ and of a nonzero section $E$ of $\D^{-2}\subset T_R^*\Sigma$, subject to a gauge invariance that will be specified shortly.  As in section \ref{elsup},
in standard local coordinates $\t z;\neg z|\theta$, we have 
\begin{equation}\label{byan} E=e^{\varphi}(\d z-\theta\d\theta)=e^\varphi \varpi,~~\t E=e^{\t \varphi}\d\t z.\end{equation}
  We also introduce a connection $\omega$ on the line bundle $\D^{-2}$.  We assume the gauge-invariance 
\begin{equation}\label{zyme} E\to e^{i\w }E,~~\t E\to e^{-i\w }\t E,~~~\omega\to \omega+\d u.\end{equation}
The gauge-covariant exterior derivatives of $E $ and $\t E$ are
\begin{align}\label{namo} D E & = (\d -i\omega)E\cr
                                                 D\t E & = (\d+i\omega)\t E ,\end{align}
                                                 where $\d$ is the ordinary exterior derivative.
The strongest reality condition that makes sense for this data is the following.  If $\Sigma$ has been defined so that its reduced space is
the diagonal in $\Sigma_{L,\red}\times \Sigma_{R,\red}$, then we can ask that when all odd variables ($\theta$ and the odd moduli)
vanish, $\t E$ is the complex conjugate of $E$, $\omega$ is real, $\t\varphi$ is the complex conjugate of $\varphi$, and $u$ is also real.                                                 
                                                 
Locally, the gauge-invariance can be used to reduce to the case
$\varphi=\phi(\t z;\neg z)+\theta\chi(\t z;\neg z)$, $\tilde\varphi=\phi$, where $\phi$ is a 
 field quite analogous to the one that appears in (\ref{romb}), and $\chi$ is a fermionic partner of $\phi$.
With just one even field and one odd one, this
 is the  smallest set of independent fields that one can possibly hope for in a super extension of Riemannian geometry.
If the reality condition of the last paragraph is imposed, then $\phi$ is real when the odd variables vanish.

What remains is to specify the conditions analogous to those that in the ordinary case determine  the Levi-Civita connection.  These conditions are extremely simple:
\begin{equation}\label{delico} D\t E=0,~~ DE\cong 0\, \,{\mathrm{mod}}\,\, T_R^*\SIgma\otimes T_R^*\Sigma.\end{equation}
The condition on $E$ means that the part of $DE$ proportional to $\varpi\wedge \d\t z$ or $\d\theta\wedge \d\t z$ vanishes, but $DE$ may have terms
proportional to $\varpi \d\theta$ or $(\d\theta)^2$.
A very short computation suffices to show that  the conditions (\ref{delico}) do uniquely determine $\omega$.
We start by writing the exterior derivative in a convenient fashion:
\begin{equation}\label{horsefly}\d=\d\t z\,\partial_{\t z}+\d z\partial_z+\d\theta\,\partial_\theta =\d\t z \partial_{\t z}+\varpi \partial_z+\d\theta D_\theta.\end{equation}
(In verifying this, recall that the quantity $\d\theta$ is even.)
Similarly for the 1-form $\omega$, we write
\begin{equation}\label{orsef}\omega=\d
\t z \omega_{\t z}+\varpi  \omega_z+ \d\theta\omega_\theta. \end{equation}
Now explicitly we compute
\begin{equation}\label{rosef} D\t E=e^{\t\varphi}\varpi\wedge \d\t z\bigl(\partial_z\t\varphi+i\omega_z\bigr)-
e^{\t\varphi}\d\theta \wedge \d\t z\bigl(D_\theta\t\varphi+i\omega_\theta\bigr).\end{equation} 
Setting this to zero, we determine part of the connection:
\begin{equation}\label{bosef}\omega_z=i\partial_z\t\varphi,~~\omega_\theta=iD_\theta\t\varphi.\end{equation}
Similarly, we evaluate $D E$:
\begin{equation}DE=e^{\varphi}\d\t z\wedge \varpi\bigl(\partial_{\t z}\varphi-i\omega_{\t z}\bigr)+e^\varphi \varpi\wedge \d\theta\bigl(D_\theta \varphi
-D_\theta\t\varphi\bigr)-e^\varphi \d\theta\wedge \d\theta.\end{equation}
Setting to zero the coefficient of $\d\t z\wedge\varpi$ (which is the part not valued in $T_R^*\Sigma\otimes T_R^*\SIgma$),
we complete the determination of the connection:
\begin{equation}\label{yotso} \omega_{\t z}=-i\partial_{\t z}\varphi.
\end{equation}

In this presentation, there is no need to mention the odd 1-form $F$ of section \ref{elsup}; it is determined in terms of $E$ by
eqn. (\ref{ombus}).  There is a variant of the above construction in which one does include $F$,
thus completing $\t E$ and 
$E$ to a basis $\t E,E,F$ of $T^*\SIgma$.
Gauge transformations act on $F$ by $F\to e^{iu/2}F$, and the constraint on $DE$ is replaced by 
$DE+F\wedge F=0$. There is no need
to state an independent condition on $DF$.

\subsection{Analogs For Type II}\label{tanalog}  

So far, we have mostly concentrated on the heterotic string.  Here we will much more briefly indicate
the analogs  for Type II superstrings.  

The starting point was already described in  sections \ref{colzo} and \ref{parspace}.  A Type II superstring worldsheet $\Sigma$ is embedded in a product
$\Sigma_L\times \SIgma_R$, with  $\Sigma_L$ and $\Sigma_R$ both  super Riemann surfaces.  
$\Sigma$ has codimension $2|0$ in $\Sigma_L\times\SIgma_R$.  We require that the reduced space $\SIgma_{L,\red}$ is the
complex conjugate of $\Sigma_{R,\red}$ (or sufficiently close to this) and that the reduced space of $\Sigma$ is the diagonal
in $\Sigma_{L,\red}\times \Sigma_{R,\red}$ (or, again, sufficiently close to this).

Deformations of $\Sigma_L\times \Sigma_R$ are parametrized by $\MM_L\times \MM_R$, where now $\MM_L$ and $\MM_R$
are both copies of the moduli space of super Riemann surfaces.  The integration cycle $\varGamma$ of Type II superstring perturbation
theory is a subsupermanifold $\varGamma\subset \MM_L\times\MM_R$ that satisfies the following two conditions:
(1) its reduced space is the subspace of $\MM_{L,\red}\times \MM_{R,\red}$ characterized by the condition that the reduced
spaces $\Sigma_{L,\red}$ and $\Sigma_{R,\red}$, with their spin structures ignored,
 are complex conjugate; (2) its odd dimension is the odd dimension of 
$\MM_L\times \MM_R$.  These conditions determine $\varGamma$ up to infinitesimal  homology 
and -- together with some discussion of
how $\varGamma$ behaves at infinity --  suffice for superstring perturbation theory.  

The tangent bundles $T\Sigma$ and $T\varGamma$ have decompositions familiar from section \ref{descemb}.  
We have $T\Sigma=T_L\SIgma\oplus T_R\Sigma$, where $T_L\Sigma$ and $T_R\SIgma$ are the restrictions to $\Sigma$
of $T\Sigma_L$ and $T\Sigma_R$.  $T_L\Sigma$ and $T_R\SIgma$ are both of rank $1|1$, with distinguished subbundles
$\t \D$ and $\D$ inherited from the super Riemann surface structures of $\Sigma_L$ and $\Sigma_R$.  These are summarized
by exact sequences
\begin{equation}\label{zorky} 0\to \t\D\to T_L\Sigma \to \t\D^2\to 0,\end{equation}
and
\begin{equation}\label{orky}0\to \D\to T_R\Sigma\to  \D^2\to 0\end{equation}
with the usual properties.   
Locally, one can choose coordinates  in which $\D$ is generated
by $D_\theta=\partial_\theta+\theta \partial_z$, and $\t\D$ is generated by $\t D_\theta=\partial_{\t\theta}+\t\theta\partial_{\t z}$.
Given such coordinates, with $\t z$ sufficiently close to the complex conjugate of $z$, we call $\t z; z|\t\theta;\theta$ a standard
local coordinate system.  
The sheaves of sections of $T_L\SIgma$ and $T_R\SIgma$ have subsheaves $\t\S$ and $\S$
of antiholomorphic and holomorphic superconformal vector fields.  
As sheaves (ignoring their graded Lie algebra structures), $\t\S$ is isomorphic to the sheaf of sections of $\t\D^2$ and $\S$ to the
sheaf of sections of $\D^2$.

The first-order 
deformations  of the holomorphic structure of $\Sigma$ are inherited from those of $\Sigma_R$ and are given by the holomorphic
sheaf cohomology $H^1(\Sigma,\S)$.  The first-order deformations of the antiholomorphic structure of $\Sigma$
are inherited from those of $\Sigma_L$ and are given by the antiholomorphic sheaf cohomology $\tilde H^1(\Sigma,\t\S)$.
The tangent space to $\varGamma$ has a decomposition $T\varGamma=\tilde H^1(\Sigma,\t\S)\oplus H^1(\SIgma,\S)$.  The cotangent
bundle to $\varGamma$ has a corresponding description using Serre duality.  

The setup is so similar to what it is for the heterotic string that most statements have fairly obvious analogs, so we will be brief
in what follows.

\subsubsection{Lagrangians}\label{latwo}

The Berezinian of the product $\Sigma_L\times \SIgma_R$ is a tensor product:
\begin{equation}\label{morno}\BBer(\Sigma_L\times \Sigma_R)=\BBer(\Sigma_L)\otimes \BBer(\Sigma_R).\end{equation}
   Because of the exact
sequences (\ref{zorky}) and (\ref{orky}), we have $\BBer(\Sigma_L)\cong \t\D^{-1}$, $\BBer(\Sigma_R)\cong \D^{-1}$ (as in
section \ref{holvol} or appendix \ref{obz}), 
so $\BBer(\SIgma_L\times \Sigma_R)\cong \t\D^{-1}\otimes \D^{-1}$.  
The Berezinian of the smooth supermanifold $\Sigma$ is the restriction to $\Sigma$ of $\BBer(\Sigma_L\times\SIgma_R)$,
for the same reason as for the heterotic string, so 
\begin{equation}\label{zibbo}\Ber(\Sigma)\cong \t\D^{-1}\otimes \D^{-1}.\end{equation}

Just as in section \ref{lagrangians}, this is the information that we need in order to construct Lagrangians.  
In doing this, we define
\begin{equation}\label{deloft}\dzztt = -i [\d\t z;\neg \d z|\d\t\theta;\neg\d\theta],\end{equation}
by analogy with eqn. (\ref{pogy}).
The usual matter fields of Type II  superstrings in $\R^{10}$  are scalar superfields $X^I(\t z|\t\theta;z|\theta)$.  Their kinetic energy is
\begin{equation}\label{kinex}I_X=\frac{1}{2\pi\alpha'}\int\dzztt\,\sum_{IJ}\eta_{IJ}D_{\t\theta}X^I D_\theta X^J.\end{equation}
This is well-defined because for a scalar field $X$, the expression $D_{\t\theta}X D_\theta X$ is naturally a section 
of $\t\D^{-1}\otimes \D^{-1}\cong \Ber(\Sigma^\star)$. 

Similarly, the holomorphic superghosts are a section $C$ of ${\Pi}\D^2$, and the corresponding antighosts are a section 
$B$ of ${\Pi} \D^{-3}$.  The expression $C D_{\t\theta} B$ is  a section
of $\t\D^{-1}\otimes \D^{-1}$, so again it can be integrated, giving the kinetic energy for the holomorphic ghost fields:
\begin{equation}\label{zinex}I_{B,C}=\frac{1}{2\pi}\int  \dzztt\, B D_{\t\theta}C.\end{equation}

The antiholomorphic superghosts are a section $\t C$ of ${\Pi}\t \D^2$, and the corresponding antighosts are a section 
$\t B$ of ${\Pi}\t\D^{-3}$.  In this case the action is
\begin{equation}\label{finex}I_{\t B,\t C}=\frac{1}{2\pi}\int \dzztt\,\t  B D_{\theta}\t C.\end{equation}

The $C$ and $\t C$ zero-modes are globally-defined holomorphic and antiholomorphic superconformal vector fields, with parity reversed.
(In most applications in superstring perturbation theory, there are no such global zero-modes.)
The $B$ and $\t B$ zero-modes are holomorphic and antiholomorphic cotangent vectors to the integration cycle $\varGamma$ of superstring perturbation theory.

\subsubsection{Deformations From A Smooth Point Of View}\label{newdef}

To discuss deformations of $\Sigma$ in terms of fields on $\Sigma$, rather than in terms of the embedding in $\Sigma_L\times \Sigma_R$, 
we start from the fact that holomorphic functions are those that are annihilated by $D_{\t\theta}$
(and hence also by $\partial_{\t z}=D_{\t \theta}^2$).  So to deform the holomorphic structure of $\Sigma$, we must, 
rather as in section \ref{defsmooth}, deform the operator
$D_{\t\theta}$ by adding perturbations proportional to $\partial_z$ and $D_\theta$.  To preserve the super Riemann surface 
structure of the holomorphic variables,
the perturbation must take values not in arbitrary sections of $T\Sigma$ but in the subsheaf $\S$.  The resulting analysis is similar to that
of section \ref{defsmooth}.  It gives the expected result that the first-order deformations of the holomorphic structure of $\Sigma$ are
parametrized by
$H^1(\Sigma,\S)$.

Reciprocally, to deform the antiholomorphic structure of $\Sigma$, we deform the $D_\theta$ operator by sections of $\t S$.  
An analysis similar to that of section \ref{defsmooth} shows that first-order deformations of the antiholomorphic structure of  $\Sigma$ are 
parametrized by $\t H^1(\Sigma,\t S)$.

\subsubsection{Supergravity}\label{newsuper}

To describe a Type II worldsheet in the language of supergravity, we proceed as in section \ref{relsup}.  First
we describe the conformally invariant structure.

  In standard local coordinates $\t z;\neg z|\t\theta;\theta$,
the distinguished subbundle $\t\D^{-2}$ of $T_L^*\SIgma$ is generated by $\t\varpi=\d \t z-\t\theta\d\t\theta$, and the distinguished 
subbundle $\D^{-2}$ of $T_R^*\SIgma$ is generated
by $\varpi=\d z-\theta\d\theta$.  
We can extend $\t\varpi$ to a basis of $T_L^*\Sigma$
\begin{align}\label{zing} \t E_0 & = \t\varpi=\d\t z-\t\theta\d\t\theta \cr \t F_0 & = \d\t\theta ,\end{align}
obeying
\begin{align}\label{ming} \d\t E_0+\t F_0\wedge \t F_0 & = 0 \cr \d \t F_0 & =0,\end{align}
and similarly we can extend $\varpi$ to a basis of $T_R^*\Sigma$,
\begin{align}\label{zingo} E_0 & = \varpi=\d z-\theta\d\theta \cr  F_0 & = \d\theta ,\end{align}
obeying
\begin{align}\label{mingo} \d E_0+ F_0\wedge F_0 & = 0 \cr \d F_0 & =0.\end{align}

We are really interested in the line bundles $\t \D^{-2}$ and $\D^{-2}$ generated by $\t E_0$ and $E_0$, not
in $\t E_0$ and $E_0$ themselves.  So we allow what one might call Weyl transformations from $\t E_0$ and $E_0$
to
\begin{equation}\label{pingo} \t E= e^{\t\varphi}\t E_0,~~ E=e^\varphi E_0,\end{equation}
with arbitrary functions $\t\varphi$, $\varphi$.  The conditions obeyed by $\t E$ and $E$ are simply that
there exist odd 1-forms $\t F$ and $F$ (unique up to sign) such that
\begin{align}\label{lingo}\d \t E+\t F\wedge \t F& = 0~~\mod ~\t E\cr
                                             \d E+F\wedge F&=0  ~~\mod ~E.\end{align}
 Eqn. (\ref{lingo}) describes a Type II superstring worldsheet in supergravity language.  The antiholomorphic cotangent bundle $T^*_L\Sigma$
 is generated by $\t E$ and $\t F$, while the holomorphic cotangent bundle $T_R^*\Sigma$                                            
  is generated by $E$ and $F$.  Their distinguished subbundles $\t\D^{-2}$ and $\D^{-2}$ are generated, respectively, by $\t E$ and by $E$.
  
  What we have just described is the Type II analog of what was explained for the heterotic string in section \ref{elsup}.
  We leave it to the reader to adapt the reasoning of section \ref{defagain} and describe deformation theory from this point of view.
  Here we will generalize the construction of section \ref{anamet} and describe the analog of a Riemannian metric in this context.

To describe a Riemannian metric, we want to give $E$ and $\t E$ not up to arbitrary Weyl transformations (\ref{pingo})
but up to the smaller class of ``gauge transformations''
 \begin{equation}\label{zonox} E\to e^{iu}  E,~~~\t E\to e^{-iu}\t E. \end{equation}
 We introduce a  gauge connection $\omega$ that transforms as 
 \begin{equation}\label{ponox}\omega\to \omega+\d u.\end{equation}
 The gauge-covariant exterior derivatives are $DE=(\d-i\omega)E$, $D\t E=(\d+i\omega)\t E$.
 Under appropriate conditions ($z$ and $\t z$ and likewise $E$ and $\t E$ are complex conjugates when odd variables are set to zero)
 one can require that  $\omega$ and  $u$ are real
 when odd variables vanish.   
 
 The conditions
 \begin{align}\label{blimp} D\t E&=0\mod \,T_L^*\Sigma\otimes  T_L^*\SIgma, \cr DE &=0\mod \,T_R^*\Sigma\otimes T_R^*\SIgma\end{align}
 play the role of eqn. (\ref{delico}) and uniquely determine $\omega$.    These are the analogs of the
 conditions that in ordinary Riemannian geometry determine the Levi-Civita connection.   As  in the remark at the end of section \ref{anamet},
 there is also a variant in which $\t E$ is completed to a basis $\t E,\t F$ of $T_L^*\SIgma$ and $ E$ is completed to a basis $E, \,F$ of $T_R^*\SIgma$,
 the constraint equations then being $D\t E+\t F\wedge \t F=0=D E+ F\wedge  F$.                                  
                                              
\section{Punctures}\label{nsrpunctures}

We now return to the purely holomorphic theory of super Riemann surfaces as introduced in section \ref{bintro}.  We aim to describe the ``punctures'' at which
external vertex operators can be inserted on a super Riemann surface.  (Some of the topics have been treated in \cite{Gid}.)

For our purposes, on an ordinary Riemann surface $\SIgma_0$, a ``puncture'' is the same thing as a marked (or labeled) point; we sometimes
use the two terms interchangeably.\footnote{We do not  think of removing the ``puncture'' from $\Sigma_0$, except when this is stated.  The reason that we use the term ``puncture'' in this fashion is that in the superstring case, we want a term to apply uniformly for both NS and Ramond insertions.   As will become clear, ``marked point'' would not be a sensible term for
what we will call a Ramond puncture.  }  On a super Riemann surface,
 there are two kinds of puncture.  Bosonic vertex operators are inserted at Neveu-Schwarz (NS) punctures, 
 while fermionic vertex operators are inserted at Ramond (R) punctures.  

NS punctures are a fairly obvious idea, quite analogous to  punctures  on an ordinary Riemann surface.  But 
Ramond punctures are a sufficiently unusual idea that one may ask how one knows that this concept is necessary.  One answer is
that, pragmatically, fermion vertex operators of superstring theory are spin fields \cite{FMS} that are naturally inserted at Ramond punctures.  
Another answer will become clear in
section \ref{infinity}.  If we do not know about Ramond punctures already, then 
this notion is forced upon us when we contemplate the compactification of supermoduli space or equivalently
the infrared region of superstring theory.  One way that a super
Riemann surface can degenerate involves the appearance of Ramond punctures, and this makes it inevitable to consider them.

We always consider punctures to be distinguishable, since in string perturbation theory, one will generically insert a different type of vertex
operator at each puncture.  So for example $\M_{\sg,\sn}$ will be the moduli space of genus $\g$ surfaces with $\n$ labeled punctures.

\subsection{Basics}\label{ubasics}

\subsubsection{Definitions}\label{definitions}

 If a super Riemann surface $\Sigma$ is parametrized locally by coordinates $z|\theta$,
then an NS vertex operator can be inserted at any point
\begin{align} \label{zinf}z& = z_0\cr
                        \theta& = \theta_0.\end{align}
                        The choice of such a point is what we mean by an NS puncture.
The parameters $z_0|\theta_0$ are the moduli of the NS puncture.  So adding an NS puncture increases the dimension of supermoduli space by $1|1$, and the moduli space
of super Riemann surfaces of genus $\g$ with $\n_  {\NS}$  NS punctures has dimension $3\g-3+\n_  \NS|2\g-2+\n_  \NS$.  It makes sense to integrate over the insertion point of an NS vertex operator,
keeping fixed\footnote{This statement holds in the heterotic string for NS vertex operators, and in Type II superstrings for NS-NS vertex operators, that is vertex operators that are of NS
type both holomorphically and antiholomorphically.} the moduli of $\Sigma$.

In studying an NS puncture, it is sometimes useful to know that a point in a super Riemann surface, as opposed to a more general $1|1$ supermanifold,
determines a divisor through that point.  In a complex supermanifold in general, a divisor is a submanifold of codimension $1|0$, so in a super Riemann surface,
it has dimension $0|1$ and is isomorphic to $\C^{0|1}$.  The divisor associated to the point $z|\theta=z_0|\theta_0$ is simply the orbit through that point
generated by the odd vector field $D_\theta$.  (Replacing $D_\theta$ with another nonzero section of $\D$ would not change this orbit.)  Concretely, $D_\theta$
generates the coordinate transformation $\theta\to\theta+\alpha$, $z\to z+\alpha\theta$, so the orbit through $z|\theta=z_0|\theta_0$ is given in parametric
form by
\begin{align}\label{ortz} z=& z_0+\alpha\theta \cr \theta=&\theta_0+\alpha\end{align} 
or equivalently by the equation
\begin{equation}\label{bortz} z=z_0-\theta_0\theta.\end{equation}
The fact that we can associate to an NS puncture a divisor improves the analogy between NS punctures and  Ramond punctures, which are definitely associated to divisors
as we will see momentarily.

A Ramond puncture is a much more subtle concept than an NS puncture; it is a singularity in the super Riemann surface structure of $\Sigma$.  In the presence of a Ramond puncture, $\Sigma$ is still
a smooth complex supermanifold of dimension $1|1$, and the tangent bundle
$T\Sigma$ still has  a distinguished subbundle $\D$ of rank $0|1$.  But it is no longer true that $\D^2$ is everywhere linearly independent of $\D$.  Rather, this condition fails along
a divisor in $\Sigma$, that is, a submanifold of dimension $0|1$.  The local behavior near a Ramond puncture is that, in suitable local coordinates $z|\theta$, $\D$ has a section
\begin{equation}\label{dorg}D^*_\theta=\frac{\partial}{\partial\theta}+z\theta \frac{\partial}{\partial z}.\end{equation}
(We reserve the name $D_\theta$ for a section of $\D$ of the form $\partial_\theta+\theta\partial_z$ in some coordinate system, and write $D^*_\theta$
for a section of a more general form $\partial_\theta+w(z)\theta\partial_z$ as in (\ref{dorg}).)
Thus
\begin{equation}\label{mong}{D^*_\theta}^2=z\frac{\partial}{\partial z}. \end{equation}
We see that ${D^*_\theta}^2$ vanishes on the divisor $z=0$.   We call this kind of divisor a Ramond divisor and generically
denote it as $\F$.

\subsubsection{Conformal Mapping To A Tube}\label{conftube}

Before going farther, let us explain the relation  of we have called NS and R punctures with NS and R string states.

On an ordinary Riemann surface, the connection between a local operator that may be inserted at, say, the point $z=0$ and a string
state arises from the coordinate transformation $z=e^\varrho$.  This transformation maps the $z$-plane with the point $z=0$ omitted to the cylinder parametrized by 
$\varrho$ with the equivalence relation
\begin{equation}\label{ommy}\varrho\cong \varrho+2\pi \sqrt{-1}. \end{equation}
Conformal field theory on this cylinder describes a string of circumference $2\pi$ propagating in the $\mathrm{Re}\,\varrho$ direction.
In a sense, the two descriptions differ by whether one thinks of $z=0$ as a marked point (at which an operator is inserted) or a puncture
(around which a string propagates).

What is the analog for super Riemann surfaces?  First we consider an NS puncture at $z=\theta=0$ on $\C^{1|1}$.
According to  eqn. (\ref{bortz}),  the divisor in $\C^{1|1}$ determined by this point is just $z=0$.
Omitting this divisor, we map its complement to a supertube by $z=e^\varrho$, $\theta=e^{\varrho/2}\zeta$.  $\varrho$ and $\zeta$ are superconformal
coordinates, since $D_\theta = e^{-\varrho/2}(\partial_\zeta+\zeta\partial_{\varrho})$.  They are subject to the equivalence relation
\begin{equation}\label{zomy}\varrho\cong \varrho+2\pi \sqrt{-1}, ~~\zeta\to -\zeta.\end{equation}
The minus sign in the transformation of $\zeta$ means that in superstring theory on this supertube, the strings that  propagate in the $\mathrm{Re}\,\varrho$
direction will be in the Neveu-Schwarz sector.

Next we consider the model (\ref{dorg}) with a Ramond puncture at $z=0$.  In this case, we omit the Ramond divisor and map to the tube simply by $z=e^\varrho$, leaving $\theta$
alone; $\varrho$
and $\theta$ are superconformal coordinates, since $D^*_\theta=\partial_\theta+\theta\partial_{\varrho}.$  The equivalence relation is now
\begin{equation}\label{zoomy}\varrho\cong \varrho+2\pi \sqrt{-1}, ~~\theta\to+\theta,\end{equation}
where now because of the $+$ sign for $\theta$, the strings propagating in the $\mathrm{Re}\,\varrho$ direction will be in the Ramond sector.

What we have encountered in the last two paragraphs are the two possible spin structures on the purely bosonic cylinder defined in eqn. (\ref{ommy}).

\subsubsection{More On Ramond Punctures}\label{multiram}
Generalizing (\ref{dorg}), an example of a (noncompact) super Riemann surface  with any number $\n_  \Ra$ of Ramond punctures is given by $\C^{1|1}$,
parametrized by $z|\theta$, with superconformal structure defined by
\begin{equation}\label{torg} D^*_\theta =\frac{\partial}{\partial\theta}+w(z)\theta\frac{\partial}{\partial z}\end{equation}
and
\begin{equation}\label{zomon} w(z)=\prod_{i=1}^{\sn_\Ra}(z-z_i).\end{equation}
So ${D^*_\theta}^2=w(z)\partial_z$, and the usual claim that the tangent bundle is generated by $\D$ and $\D^2$ fails precisely
at the divisors $\F_i$ given by $z=z_i$.  We write $\F=\sum_i\F_i$ for the divisor on which $\D^2$ vanishes mod $\D$ and we say that the superconformal
structure of $\Sigma$ degenerates along $\F$.

In the presence of Ramond punctures, though the subbundle $\D\subset T\SIgma$ is still part of an exact sequence
$0\to \D\to T\Sigma\to T\Sigma/\D\to 0$, it is no longer true that $T\Sigma/\D$ is isomorphic to $\D^2$.  $T\Sigma/\D$ is generated
by $\partial_z$, but $\D^2$ is generated, in the above example, by $w(z)\partial_z$, which vanishes on  $\F$.
 The relation between $\D^2$ and $T\Sigma/\D$
is $\D^2\cong T\Sigma/\D\otimes \O(-\F)$ (this is  a fancy way to say that a section of $\D^2$ is a section of $T\SIgma/\D$ that vanishes
along $\F$), or equivalently $T\Sigma/\D\cong \D^2\otimes \O(\F)$.  So in the presence of Ramond punctures, the familiar
exact sequence becomes
\begin{equation}\label{oxx}0\to \D\to T\Sigma\to \D^2\otimes \O(\F)\to 0.\end{equation}
It is also convenient to describe this in a dual language.  Dualizing (\ref{oxx}), we get an exact sequence
\begin{equation}\label{zoxx}0\to \D^{-2}\otimes \O(-\F)\to T^*\Sigma\to \D^{-1}\to 0.\end{equation}
So $T^*\Sigma$ has a distinguished subbundle $\L\cong \D^{-2}\otimes \O(-\F)$.  Concretely, $\L$ is the subbundle of $T^*\Sigma$ that is orthogonal
to $\D$.  So if $\D$ is generated by $\partial_\theta+\theta z \partial_z$, then $\L$ is generated by $\d z-z\theta\d\theta$.  More generally, for the example (\ref{torg}) with several
Ramond punctures,
$\L$ is generated by
\begin{equation}\label{zeb}\varpi=\d z-w(z)\theta\d\theta.\end{equation}

It turns out that on a compact super Riemann surface, the number $\n_\Ra$ of Ramond punctures is always even.  One might anticipate this from the fact that Ramond punctures
are the locations at which vertex operators for spacetime fermions are inserted.  Including $\n_\Ra$ Ramond punctures increases the bosonic dimension of supermoduli space
by $\n_\Ra$, and the fermionic dimension by $\n_\Ra/2$.  The statement about the bosonic dimension is not surprising; in the simple example of eqn. (\ref{torg}), we see that the
positions of Ramond punctures can be varied independently.  More interesting is the statement that the contribution of $\n_\Ra$ Ramond punctures to the fermionic dimension of supermoduli space
is $\n_\Ra/2$. It means that the fermionic moduli are global in nature, not associated to any particular Ramond puncture.  There is no notion of moving a Ramond
puncture in the fermionic directions.

The claim that the Ramond punctures contribute $\n_\Ra | \n_\Ra/2$ to the dimension of supermoduli space will be justified in section \ref{modcount}.  Given this claim,
the dimension of the moduli space of super Riemann surfaces of genus $\g$ with $\n_\NS$ Neveu-Schwarz punctures and $\n_\Ra$ Ramond punctures is
\begin{equation}\label{gendim}\dim \MM_{\sg,\sn_\NS,\sn_\Ra}=(3\g-3+\n_\NS+\n_\Ra)|(2\g-2+\n_\NS+\n_\Ra/2).\end{equation}

In superconformal field theory, it is very useful to describe Ramond vertex operators as spin operators that are inserted at branch points
of the worldsheet fermions.  To get to this description, we should replace what we have called $\theta$ by a new variable $\theta'$ that has square root
branch points on the divisor $\F$.  In the simple example (\ref{torg}), we take
 $\theta'=\sqrt{w(z)}\theta$.  At
zeroes of $w(z)$,  $\theta'$ is not well-defined, but away from those
zeroes the coordinates $z|\theta'$ (with either choice of sign for the square root
in the definition of $\theta'$) are
local superconformal coordinates in which the line bundle $\D$ is generated by $D_{\theta'}=\partial_{\theta'}+\theta'\partial_z$.  
So away from zeroes of $w$, the ordinary formulas of superconformal field theory are valid.  At those zeroes, $\theta'$ and
all fields that carry odd worldsheet fermion number (that is, all fields that are odd under the GSO projection) have square root branch points.
This description is very useful for understanding local properties such as the operator product expansion of  fields.  But
the global geometry tends to be obscured by the use of a double-valued coordinate.
In superstring perturbation theory, it is useful to also have a global description via a smooth supermanifold $\Sigma$,
albeit one whose superconformal structure degenerates along the divisor $\F$.   Just as one application, this will
greatly facilitate the analysis of pictures in section \ref{pictures}.

\subsubsection{Implications}\label{implic}

The peculiar nature of Ramond punctures has important implications for superstring perturbation theory.
In contrast to the NS case, it does not make sense to integrate over the insertion point of a Ramond vertex operator while keeping fixed the other moduli of $\Sigma$.  The odd moduli of $\Sigma$ are properties of the superconformal structure of $\Sigma$  and cannot be defined
 independently of the singularities of that  structure, which are the divisors on which the Ramond vertex operators are inserted.  
 So it does not make sense to change the position of a Ramond vertex operator while keeping fixed the
odd moduli.  To integrate over the insertion location of a Ramond vertex operator, we have to integrate over all the odd moduli.
And since there is no known natural way in general to integrate over odd moduli without also integrating over even moduli,  the only known manageable way to integrate over the position of a Ramond vertex operator
is to integrate over the whole supermoduli space.  

This fact was one of the essential sources of complication in the superstring literature of the 1980's.  To prove
spacetime supersymmetry of perturbative scattering amplitudes, one would like \cite{FMS, Martinec}
to integrate over the position of a certain Ramond vertex operator, the spacetime supersymmetry generator. 
More generally, the proof of gauge invariance for fermion vertex operators involves integrating over the position at which a Ramond vertex operator is inserted. 
In these integrals one needs to be able to integrate by parts.
 But the integral over the location of a Ramond vertex operator really only makes sense when extended to an integral over the whole
supermoduli space.   So one really needs to integrate by parts on a rather subtle supermanifold.
The most relevant version of ``integration by parts'' is  the supermanifold version of Stokes's theorem
 (for example see \cite{DEF,Beltwo,Wittennotes}).  When one integrates by parts, one needs to analyze possible surface terms at infinity,
so one needs an understanding of the behavior of supermoduli space at infinity as described in section \ref{infinity} below.  The supermanifold
version of Stokes's theorem and an analysis of what happens at infinity will be the main ingredients in a reconsideration of superstring perturbation theory \cite{Witten}.

\subsection{Superconformal Vector Fields And Moduli}\label{modcount}

Here we will repeat some of the considerations of section \ref{bintro} in the presence of Ramond punctures.

\subsubsection{The Sheaf Of Superconformal Vector Fields}\label{sheafsup}

We first consider the simple example (\ref{torg}) of a superconformal structure on $\C^{1|1}$ with $\n_\Ra$ Ramond punctures.
A calculation similar to the one that led previously to (\ref{zerb}) and (\ref{terb})  shows that
a general odd superconformal vector field preserving this superconformal structure is
\begin{equation}\label{odz} \nu_f = f(z)\left(\partial_\theta-w(z)\theta{\partial_z}\right), \end{equation}
and a general even one is
\begin{equation}\label{bodz} V_g= w(z)\left(g(z)\partial_z+\frac{g'(z)}{2}\theta\partial_\theta\right).\end{equation}
As before, the functions $f$ and $g$ are holomorphic functions of $z$ only and not $\theta$.  A check on these formulas
is the following.  Since the divisor $\F$ along which $w=0$ is intrinsically defined by the condition that $\D^2$ is proportional to 
$\D$ along this divisor, any vector field that preserves $\D$ must, when restricted to $\F$, be tangent to $\F$, so as to
leave $\F$ fixed (not pointwise, but as a divisor).  The vector fields $\nu_f$ and $V_g$  have this property.  $V_g$ vanishes along $\F$, and $\nu_f$ when restricted
to $\F$ is proportional to $\partial_\theta$, so it is tangent to $\F$.   We define a sheaf $\S$ of  superconformal vector
fields whose sections are vector fields of the form $V_g+\nu_f$.

\subsubsection{Reduced Structure}\label{basic}

 We can count the moduli of $\Sigma$ in the presence of Ramond punctures by the same reasoning as in section \ref{otorg}.  For variety, we will present the arguments
 somewhat differently.
 It suffices to consider the case that $\Sigma$ is split, meaning that $\Sigma$ is fibered over its reduced space, which is an ordinary Riemann surface $\Sigma_\red$. 
 The fibers of this fibration are vector spaces of rank $0|1$; thus $\Sigma$ is the total space of a line bundle $V\to \Sigma_\red$ with fermionic fibers. We can think of $V$
 as the normal bundle to $\Sigma_\red$ in $\Sigma$.  As described in section \ref{efc},
 a split super Riemann surface has a $\Z_2$ symmetry $\btau:\theta\to -\theta$ that leaves fixed the reduced space $\Sigma_\red$ and
 acts as $-1$ on the normal bundle $V$.
 So when restricted to $\Sigma_\red$, the tangent bundle $T\Sigma$ can be decomposed in subspaces
 that are even and odd under $\btau$.  The even subspace\footnote{We are not simply expanding
 $T\Sigma$ as the sum of bosonic and fermionic subspaces. A super vector space $W$ in general has no such canonical decomposition; if $W$ has rank $a|b$,
 then the group $GL(a|b)$ of automorphisms of $W$ does not preserve any decomposition in even and odd subspaces.  Rather, we make the decomposition in eqn (\ref{oform})
 using the symmetry $\btau$.  If $\Sigma$ is not split, it does not have the symmetry $\btau$; one would have to define $\btau$ to reverse the odd moduli of $\Sigma$.} of $T\Sigma|_{\SIgma_\red}$
  is the tangent bundle $T\Sigma_\red$ to $\Sigma_\red$, and the odd subspace is $V$:
   \begin{equation}\label{oform}T\Sigma|_{\Sigma_\red}= T\Sigma_\red\oplus  V.\end{equation}
   Similarly, we can restrict the exact sequence (\ref{oxx}) to $\Sigma_\red$ and decompose it in even and odd parts.  This decomposition is very simple.
   When restricted to $\Sigma_\red$, $\D$ is odd under $\btau$ and $\D^2\otimes \O(\F)$ is even. 
The exact sequence simply identifies $\D$ and $\D^2\otimes \O(\F)$ with the odd and even parts of $T\Sigma_\red$:
\begin{equation}\label{zurky} \D  \cong V,~~~ \D^2\otimes \O(\F)\cong T\Sigma_\red. \end{equation}
 Moreover, when restricted to $\Sigma_\red$, $\F$ is just
   an ordinary divisor $\sum_ip_i$, where $p_i$ are points in $\Sigma_\red$ (in the example (\ref{torg}), the $p_i$ are the zeroes of the function $w$), so we can
   write $\O(\sum_ip_i)$ for $\O(\F)$.
The isomorphisms in (\ref{zurky}) combine to give an isomorphism 
\begin{equation}\label{loxc} V^2\cong T\Sigma_\red\otimes \O\left(-\sum_{i=1}^{\sn_\Ra}p_i\right).\end{equation}
For a line bundle $V$ with such an isomorphism to exist, the  line bundle $T\Sigma_\red\otimes\O(-\sum_{i=1}^{\sn_\Ra}p_i)$ must have even degree.
Since this degree is actually $2-2\g-\n_\Ra$, we  learn that $\n_\Ra$ must be even.  The degree of $V$ is
\begin{equation}\label{zolt}\mathrm{deg}\,V=1-\g-\frac{\n_\Ra}{2}.\end{equation}

\subsubsection{Counting Moduli}\label{countmod}

We can now easily count the odd and even moduli of $\Sigma$.  In the split case, we decompose the sheaf $\S$ of superconformal vector fields as $\S_+\oplus \S_-$,
where the summands are respectively even and odd under $\btau$.  $\S_-$ is spanned by the vector fields $\nu_f$ in (\ref{odz}), and $\S_+$ is spanned by the vector fields 
$V_g$ in (\ref{bodz}).  The even part of the tangent bundle to supermoduli space $\MM$  (at a point in $\MM$ corresponding to the split surface $\Sigma$) is $H^1(\Sigma_\red,\S_+)$.  And the odd part is $H^1(\Sigma_\red,\S_-)$.

To $\nu_f$, we associate the vector field $f\partial_\theta$ along the fibers of the fibration $V\to \Sigma_\red$.  So we can regard $f$ as a section of $V$, and therefore
$\S_-$ is the sheaf of sections of the line bundle $V$.  Given the formula (\ref{zolt}) for the degree of $V$, and the Riemann-Roch formula (\ref{plo}), the dimension of
$H^1(\Sigma_\red,\S_-) $ is $\g-1-{\mathrm{deg}}\,\S_-=2\g-2+\n_\Ra/2$.  This is the number of odd moduli if there are no NS punctures.  Adding $\n_\NS$ such punctures simply increases the
number of odd moduli by $\n_\NS$, one for each puncture.  So in general the number of odd moduli is $2\g-2+\n_\NS+\n_\Ra/2$, as claimed in eqn. (\ref{gendim}).

The counting of even moduli is more obvious because it does not depend on the identification of $V$.  From eqn. (\ref{bodz}), we see that when
 restricted to $\Sigma_\red$, an even superconformal vector
field $V_g$ is simply a vector field along $\Sigma_\red$ that vanishes at the points $p_i$.  So we identify $\S_+$ as $T\Sigma|_\red\otimes \O(-\sum_ip_i)$,
of degree $2-2\g-\n_\Ra$.  Hence the dimension of $H^1(\Sigma_\red,\S_+)$ is $3\g-3+\n_\Ra$.  After including the contributions of $\n_\NS$ NS punctures, we arrive
at the formula for the number of even moduli claimed in eqn. (\ref{gendim}).

Instead of including the NS punctures as an afterthought, a more principled way to proceed is as follows.  In the presence of NS punctures, we restrict the sheaf $\S$ of
superconformal vector fields to its subsheaf $\S'$ that leaves fixed the NS punctures.\footnote{Concretely, the condition that a general superconformal
vector field $g(z)\partial_z+\half g'(z)\theta\partial_\theta  +f(z)(\partial_\theta-\theta\partial_z)$ should leave fixed the point $z|\theta=z_0|\theta_0$ is  $g(z_0)-\theta_0 f(z_0)=0=
\half g'(z_0)\theta_0+f(z_0)$.  These conditions define the sheaf $\S'$ of superconformal vector fields that leave fixed an NS puncture at $z_0|\theta_0$.}
$\S'$ is the sheaf of automorphisms of $\Sigma$ with its specified NS punctures.
So  the deformation space of $\Sigma$ with those punctures is $H^1(\Sigma_\red,\S')$.  The restriction from $\S$ to $\S'$ has the effect of increasing the dimension
of the cohomology group by $\n_\NS|\n_\NS$. (This can be proved by an argument similar to the analysis of eqn. (\ref{zeldox}) below.)

The reason that it is up to us whether to add the contribution of the NS punctures as an afterthought or incorporate them in the definition of the sheaf of superconformal
vector fields is that an NS puncture is something that is added to a pre-existing super Riemann surface.  By constrast, a Ramond puncture is part of the structure of the super
Riemann surface so there is no way to add it as an afterthought.  This statement is closely related to the point made in section \ref{implic}.  

\subsubsection{Generalized Spin Structures}\label{compar}

Eqn. (\ref{loxc}) generalizes the fact that in the absence of Ramond punctures, a spin structure is part of the structure of a super Riemann surface.  If $\n_\Ra=0$,
eqn. (\ref{loxc}) says that $V\cong T\Sigma_\red^{1/2}$ and in other words $V$ is a choice of spin structure.  For $\n_\Ra>0$, a super Riemann surface is not endowed with
a spin structure, but rather with a square root of $T\Sigma_\red\otimes \O(-\sum_{i=1}^{\sn_\Ra}p_i)$.  It is convenient to call such a square root a generalized spin structure.

Regardless of the value of $\n_\Ra,$ for a fixed choice of the $p_i$,
there are $2^{2\sg}$ choices of $V$.    The case $\n_\Ra>0$ has one conspicuous difference from the case $\n_\Ra=0$.  For $\n_\Ra>0$, the $2^{2\sg}$ equivalence classes of
line bundle $V$ that admit an isomorphism (\ref{loxc}) are permuted transitively as the points $p_i$ move around.   By contrast, spin structures are of two types -- even and odd --
that do not mix as one varies the moduli of $\Sigma_\red$.

\subsubsection{Another Look At Superconformal Vector Fields}\label{analook}
 
As in section \ref{reconsider}, we can try to combine the functions $f$ and $g$ that generate superconformal vector fields to a superfield.  Differently put, we
can look for  an equivalence between the sheaf $\S$ of superconformal vector fields and the sheaf of sections
of a line bundle over $\Sigma$.   Given our previous experience, we can try to do this by  projecting the sections $V_g$ and $\nu_f$ of $\S$
 to $T\Sigma/\D$.   Again the projection of a vector field $W$ to $T\Sigma/\D$ can be made by expressing  $W$
 as a linear combination of $D^*_\theta$ (which generates $\D$) and $\partial_z$ (which generates $T\Sigma$ mod $\D$) and keeping the coefficient of $\partial_z$.
 In this process, $\nu_f$ projects to $-2 w(z)f(z)\theta\partial_z$, and $V_g$ projects to $w(z)g(z)\partial_z$.  
 So a general section $V_g+\nu_f$ of the sheaf $\S$ (with $g$ even and $f$ odd) projects to the section 
  $w(z)(g(z)+2\theta f(z))\partial_z$ of $T\Sigma/\D$.  Since $g(z)+2\theta f(z)$ can be any function of $z$ and $\theta$, what we get this
  way is any section of $T\Sigma/\D$ that is divisible by $w$ or in other words vanishes along $\F$.  So we can identify
  $\S$ as $(T\Sigma/\D)\otimes \O(-\F)$.    However, the exact sequence (\ref{oxx}) tells us that $T\Sigma/\D$ is isomorphic to $\D^2\otimes \O(\F)$,
  so $(T\Sigma/\D)\otimes \O(-\F)$ is isomorphic to $\D^2$.  Thus the sheaf $\S$ of superconformal vector fields is isomorphic to $\D^2$
  just as in the absence of Ramond punctures:
  \begin{equation}\label{rex} \S\cong \D^2.\end{equation}
  
  The superconformal ghosts are a section of $\S$ with parity reversed, so we can express them as a field $C$ valued in ${\Pi} \D^2$.  
  To understand where the antighost fields $B$ should live, we have to ask how to make sense of the Lagrangian
  \begin{equation}\label{orn}I_{BC}=\frac{1}{2\pi i}\int [\d\t z;\d z|\d\theta]\, B\partial_{\t z}C. \end{equation}
  (For brevity, we consider the ghost Lagrangian of the heterotic string, as in eqn. (\ref{otto}).)  For this to make sense, the product $BC$ must take values in the holomorphic 
  Berezinian $\BBer(\Sigma)$, so $B$ must take values in $\BBer(\Sigma)\otimes \D^{-2}$.  $\BBer(\SIgma)$ can be computed by the method of section
  \ref{holvol} or (with the help of the exact sequence (\ref{zoxx})) by the reasoning of appendix \ref{obz}, with the result
  \begin{equation}\label{yrem}\BBer(\Sigma)\cong\D^{-1}\otimes \O(-\F).\end{equation}
  So $B$ is valued in $\D^{-3}\otimes \O(-\F)$.

 \subsection{Pictures}\label{pictures}

By now, we understand that a Ramond ``puncture'' is really a divisor on a super Riemann surface $\Sigma$.  This divisor is a copy of $\C^{0|1}$.        It 
is one of the connected components of the divisor $\F=\sum_{i=1}^{\sn_\Ra}\F_i$ on which the superconformal structure of $\Sigma$  degenerates.

The understanding that Ramond ``punctures'' are really divisors and not points raises the following question.   Is the insertion of a Ramond vertex operator at one of the $\F_i$
associated to the whole divisor $\F_i$, or should the vertex operator be inserted at a point on $\F_i$?

In the presence of  Ramond punctures, there are two different things one might mean by the moduli space of super Riemann surfaces.  One moduli space  $\MM$ parametrizes $1|1$ supermanifolds $\Sigma$ with a superconformal structure that
becomes degenerate along $\n_\Ra $ minimal divisors $\F_i$.  Another moduli space $\MM_1$ parametrizes the same data as before but now with a choice of a point 
on each $\F_i$.  Clearly the relation between the two moduli spaces is that $\MM_1$ is a fiber bundle over $\MM$, the fiber being a copy of $\prod_{i=1}^{\sn_\Ra}\F_i$:
\begin{equation}\label{ymurky}\begin{matrix} \prod_{i=1}^{\sn_\Ra}\F_i & \rightarrow & \MM_1 \cr
                           & & \downarrow \cr
                            & & \MM. \end{matrix}\end{equation}
(Presently we will generalize the story, and then we will rename $\MM_1$ as $\MM_{1,\dots,1}$.)                            
                            
Since there are two plausible moduli spaces, one question that comes to mind is whether the dimension formula   (\ref{gendim}) applies to $\MM$ or $\MM_1$.  The answer
to this question is that this dimension formula applies to $\MM$.  This is so because when we described the sheaf of superconformal vector fields in the presence of Ramond
punctures, we did not require the function $f$ in equation (\ref{odz}) to vanish at the zeroes of $w$. That being so, we allowed as symmetries superconformal vector fields that generate
nontrivial shifts $\theta\to\theta+\eta$ (with $\eta$ an odd constant) along the divisors $\F_i$.     Since we allowed these as symmetries, the moduli space that we constructed
is the space $\MM$ in which the divisors $\F_i$ are not endowed with distinguished points.

Although we included superconformal vector fields that generate shift symmetries $\theta\to\theta+\eta$ of $\F_i$, there are no
superconformal vector fields that rescale the divisor $\F_i$ by $\theta\to\lambda\theta$.  This can be seen from the detailed form
of (\ref{odz}) and (\ref{bodz}), where the coefficient of $\theta\partial_\theta$ always vanishes at $w=0$.  It is also clear from the
fact that a superconformal transformation must map the object $\varpi$ defined in eqn. (\ref{zeb}) to a multiple of itself; a transformation that
rescales $\theta$ at $w=0$ (except by a factor $\pm 1$) would lack this property.
We write $\S\F_i$
for the sheaf of vector fields $\partial_\theta$ that can arise at $\F_i$ by restricting a superconformal vector field.  So the space
of global section of $\S\F_i$ is of dimension $0|1$, generated by $\partial_\theta$. And we define $\S\F=\oplus_i\S\F_i$.

To construct $\MM_1$, we should replace $\S$ by its subsheaf $\S_1$ consisting of superconformal vector fields that vanish when restricted to $\F$.  $\S_1$ sits
in an exact sequence
\begin{equation}\label{zeldox}0\to \S_1\xrightarrow{i} \S\xrightarrow{r} \S\F\to 0.\end{equation}
  The map $i$ is the natural inclusion of $\S_1$ in $\S$, and $r$ is the restriction of a superconformal
vector field to $\F$.  The exact sequence is just a fancy way to say that sections of $\S_1$ are sections of $\S$ whose restrictions to $\F$ vanish.
From (\ref{zeldox}), one deduces an exact sequence of cohomology groups\footnote{We have shortened the long exact cohomology sequence that one derives from (\ref{zeldox})
using the vanishing of some of the groups involved.  For instance,   $H^1(\SIgma,\S\F)=0$ because the support of $\S\F$ is on 
 $\F$, which  has bosonic dimension 0. } 
\begin{equation}\label{hoxo}0 \to H^0(\Sigma,\S\F)\to H^1(\Sigma,\S_1)\to H^1(\Sigma,\S)\to 0. \end{equation}
 The group $H^0(\Sigma,\S\F)$ has dimension $0|\n_\Ra$ (with one basis vector $\partial_\theta$ for each of the $\F_i$), and the sequence (\ref{hoxo}) implies that the dimension of $H^1(\SIgma,\S_1)$ exceeds that of $H^1(\SIgma,\S)$ by
 $0|\n_\Ra$.  Of course,  we could have anticipated this result given the fibration (\ref{ymurky}).
 Since $H^1(\Sigma,\S_1)$ is the tangent space to $\MM_1$ at the point corresponding to $\Sigma$, while $H^1(\SIgma,\S)$ is the tangent space
 to $\MM$,  we deduce with the help of (\ref{gendim}) that the dimension of $\MM_1$ is
 \begin{equation}\label{endim}\dim \MM_{1;\sg,\sn_\NS,\sn_\Ra}=(3\g-3+\n_\NS+\n_\Ra)|(2\g-2+\n_\NS+\frac{3}{2}\n_\Ra).\end{equation}

 Now let us return to the question: should superstring perturbation theory be understood as an integral over $\MM$ or over $\MM_1$?  A little thought shows that if either one of these
 answers is correct, then it must be possible to construct perturbative superstring amplitudes as an integral over $\MM$.  Indeed, since $\MM_1$ admits a natural
 map to $\MM$,  if there is some way to compute superstring scattering ampltudes by integrating over $\MM_1$, then by first integrating over the fibers of the projection $\MM_1\to \MM$,
 we would get a recipe to compute the scattering amplitudes by integration over $\MM$.  
 
 In fact, it is up to us whether we want to use $\MM$ or $\MM_1$.   In constructing superstring vertex operators \cite{FMS}, one runs into a peculiar phenomenon of  ``picture number.''    (We will assume here a familiarity with the main ideas of that reference. The following
 remarks are presented in a somewhat heuristic way.  Some technical details can be found in appendix \ref{morep}.)  The
 picture number of an NS vertex operator takes values in $\Z$, while that of a Ramond vertex operator takes values in $\Z+1/2$.   Although all integer or half-integer values of the
 picture number are possible, the vertex operators have the most simple superconformal properties -- which usually
 is a welcome simplification --  only in the case of vertex 
 operators of ``canonical''
 picture number.  The canonical numbers are $-1$ for NS vertex operators and $-1/2$ for Ramond vertex operators.

 Not coincidentally, the negatives of these numbers, namely 1 and $1/2$, appear in the dimension formula (\ref{gendim}): the 
 contribution of $\n_\NS$ NS punctures and $\n_\Ra$
 Ramond punctures to the odd dimension of $\MM$ is $1\cdot \n_\NS+\frac{1}{2} \cdot \n_\Ra$.  In fact, the natural 
 interpretation of the picture number of a vertex operator is that it
 is minus the number of odd moduli associated to the presence of that operator.   The essential clue to this statement
 goes back to \cite{EHV}:  picture-changing, which is the operation introduced in \cite{FMS} that increases the picture number of a 
 vertex operator by 1, amounts to integration 
 over an odd modulus.  To integrate naturally over supermoduli space, each vertex operator should have a picture number that
 is minus its contribution to the odd dimension of supermoduli space.

 If we want to compute superstring scattering amplitudes using vertex operators in the canonical pictures, we should use the 
 moduli space $\MM$ where the number of odd moduli
 is 1 for each NS puncture and $1/2$ for each Ramond puncture.  However, the moduli space $\MM_1$ is a perfectly good 
 supermanifold with 1 odd modulus
  for each NS vertex operator and $3/2$ for each Ramond vertex operator.   If we want to compute scattering amplitudes by 
  integrating over $\MM_1$, we should use NS vertex operators
 of picture number $-1$ and Ramond vertex operators of picture number $-3/2$.  
 
 More generally, we can make a separate choice for each fermion vertex operator of whether we want its picture number to be $-1/2$ or $-3/2$.  We simply modify the definition
 of the sheaf $\S_1$ to say that it consists of superconformal vector fields that vanish on some chosen subset of the divisors $\F_i$.
 
 Can we define moduli spaces that are suitable for use with vertex operators with other values of the picture number?  This is no sooner said than done.  Suppose that
 we want the $i^{th}$ fermion vertex operator to have picture number $-1/2-k_i$, where the $k_i$ are nonnegative integers.  We introduce a more general
 sheaf $\S_{k_1,\dots,k_{\ssn_\Ra}}$ 
 by requiring the odd vector field $\nu_f$ to have a zero of order $k_i$ along each $\F_i$.  (A crucial fact is that $\S_{k_1,\dots,k_{\ssn_\Ra}}$ is a sheaf of super Lie algebras.  In constructing  
 a moduli space with $H^1(\Sigma,\S_{k_1,\dots,k_{\ssn_\Ra}})$ as its tangent space, one uses this fact. Roughly one wants to interpret $\S_{k_1,\dots,k_{\ssn_\Ra}}$ as generating the
 infinitesimal symmetries of the objects classified by the moduli space one is trying to define, so $\S_{k_1,\dots,k_{\ssn_\Ra}}$ has to be a sheaf of graded Lie algebras.)  The moduli spaces $\MM_{k_1,\dots,k_{\ssn_\Ra}}$ defined by this procedure have ``integration over the fiber'' maps that reduce the $k_i$.  These maps correspond to the picture-raising
 operations for the vertex operators.
 
 We can do something similar for NS vertex operators.    As already explained in section \ref{countmod}, the most natural way to incorporate an NS puncture in the supermoduli
 problem is to replace the sheaf  $\S$ by its subsheaf consisting of superconformal vector fields $\nu_f$ and $V_g$ such that $f$ and $g$ vanish at the puncture.
 This leads to the canonical moduli space $\MM_{\sg,\sn_{\NS},\sn_\Ra}$, suitable for NS vertex operators of picture number $-1$.  
 If we want the $i^{th}$ NS vertex operator to have picture number $-k_i$, we use the sheaf  of superconformal vector fields such that $g$ has a simple zero
 at the $i^{th}$ NS puncture but $f$ has a zero of order $k_i$.  (We must take $k_i>0$, since superconformal vector fields with $g$ having a prescribed zero and no constraint
 on $f$ do not form a graded Lie algebra.)           

What we have described so far is a framework to compute perturbative superstring amplitudes using vertex operators of arbitrary negative picture numbers.  Is there a similar
framework to use vertex operators of non-negative picture number?  The answer to this question in general appears to be ``no.''  One would need a moduli space with odd
dimension less than the canonical value given in eqn. (\ref{gendim}).  There does not seem to be a natural version of supermoduli space with a smaller odd dimension than is given
in this formula.  The question of existence of such a space is somewhat like the question of whether, 
starting with the canonical supermoduli space $\MM$, one can integrate over some or all of the odd moduli in a natural
way, without integrating over even moduli.

Alternatively, we could increase the bosonic dimension of supermoduli space by requiring higher order zeros of $g$, as 
well as $f$ (subject to the condition that the class
of superconformal vector fields considered must form a graded Lie algebra).  But nothing is known in superconformal 
field theory that would suggest how to construct
natural objects that can be integrated over supermoduli spaces with enhanced bosonic dimension.

\subsection{Punctures In Superstring Theory}\label{superpunctures}

\subsubsection{Reduced Spaces}\label{redsp}

Before describing superstring worldsheets and integration cycles in the presence of spin structures, we need to understand the reduced spaces of a super Riemann
surface with punctures and
of the corresponding moduli space $\MM_{\sg,\sn_\NS,\sn_\Ra}$.

  If  $\Sigma$ 
is a  super Riemann surface with punctures of either NS or R type, then its reduced space $\Sigma_\red$ is
 an ordinary Riemann surface with punctures.  Both NS and R punctures on $\Sigma$ become ordinary punctures on $\Sigma_\red$ (but they are still labeled as NS or R).  
 $\Sigma_\red$  is endowed with a generalized spin structure in the sense  
of section \ref{compar}.

The reduced space of $\MM_{\sg,\sn_\NS,\sn_\Ra}$ is the moduli space $\M_{\sg,\sn_\NS+\sn_\Ra,\spin}$ of Riemann surfaces with a generalized spin structure  and
a total of $\n_  \NS+\n_  \Ra$ punctures.  Equivalently,  $\MM_{\sg,\sn_\NS,\sn_\Ra,\red}=\M_{\sg,\sn_\NS+\sn_\Ra,\spin}$  parametrizes split super Riemann surfaces with the indicated number of punctures.
Here, a super Riemann surface $\Sigma$ with punctures is said to be split if it has the structure described in section \ref{basic}
and the NS punctures are contained in $\Sigma_\red\subset\Sigma$. 

\subsubsection{Worldsheets and Integration Cycles}\label{wint}

With this input, we can essentially repeat the description of superstring worldsheets and integration cycles that we gave in the unpunctured case.
A superstring worldsheet in the presence of punctures is
a smooth supermanifold $\Sigma$ that is embedded in a product $\Sigma_L\times\SIgma_R$ of Riemann surfaces or super Riemann surfaces with punctures.
For the heterotic string, $\Sigma_R$ is a super Riemann surface with punctures,   and $\Sigma_L$ is
an ordinary Riemann surface with punctures; for Type II superstrings, $\SIgma_L$ and $\SIgma_R$ are both super Riemann
surfaces with punctures.  The punctures in $\Sigma_L$ and $\Sigma_R$ will be the same in number, but there is no relation between their types.

As usual, the  basic example is the case that $\Sigma_{L,\red}$ and $\Sigma_{R,\red}$ are complex conjugate spaces,  $\Sigma_\red$ is
the diagonal in $\Sigma_{L,\red}\times \Sigma_{R,\red}$, and $\Sigma$ is obtained by thickening $\Sigma_\red $ in the odd directions.  For $\Sigma_{L,\red}$
and $\Sigma_{R,\red}$ to be complex conjugate means now that they are complex conjugate spaces with punctures at the same points (not necessarily of the same types).
For Type II, a puncture may be independently of NS or R type on $\Sigma_L$ and $\Sigma_R$, so there are four types of puncture, which one can label
as NS-NS, NS-R, R-NS, and R-R.

The moduli space of deformations of $\Sigma_L\times \Sigma_R$, with $\Sigma_L$ and $\SIgma_R$ allowed to deform
independently, is the obvious product $\MM_L\times \MM_R$.  As usual, we define an integration cycle $\varGamma\subset \MM_L\times
\MM_R$ whose odd dimension is that of $\MM_L\times \MM_R$ and whose reduced space is the diagonal in $\MM_{L,\red}\times \MM_{R,\red}$.  
As usual, the even dimension of $\varGamma$ is twice the complex dimension of 
 $\MM_L$ or equivalently of $\MM_R$.  From these facts, one can work out the dimension of $\varGamma$.  The most useful way to record the result seems
 to be to state the contribution of a puncture of a given type to the dimension of $\varGamma$. For Type II, these contributions are
\begin{align}\label{zobosot} \mathrm{NS-NS}:& ~~ 2|2 \cr
                                       \mathrm{NS-R}:&~~2|3/2\cr
                                               \mathrm{R-NS}:&~~2|3/2\cr
                                        \mathrm{R-R}:&~~2|1. \end{align}
For the heterotic string, the contributions of an NS or R puncture to the dimension of $\varGamma$ are $2|1$ and $2|1/2$, respectively.

As usual, we can deform $\Sigma\subset\Sigma_L\times \Sigma_R$ or $\varGamma\subset \MM_L\times \MM_R$ slightly away from the above-stated conditions
without changing the integrals that define the worldsheet action or the superstring scattering amplitudes.

The presence of punctures does not affect the definition of the worldsheet action.  At an NS puncture, superconformal primary fields are inserted in 
the standard way.  The R case is more subtle; see \cite{Witten}, section 5.

\section{Super Riemann Surfaces Of Low Genus}\label{examples}

Our goal in this section is to describe explicitly the moduli space of super Riemann surfaces of genus 0 and 1.   These are important special
cases and it helps to be familiar with them.

Genus 0 and 1 are  the  cases
in which the automorphism group $\G$ of a  super Riemann surface may have positive dimension.  The dimension formula for the moduli
space includes a contribution from $\G$:
\begin{equation}\label{kobo}\dim\,\MM_\sg-\dim\,\G= (3\g-3+\n_  \NS+\n_  \Ra)|(2\g-2+\n_  \NS+\frac{1}{2}\n_  \Ra).\end{equation}
Increasing the number of punctures makes $\G$ smaller, and even for $\g=0$ or 1, $\dim\,\G=0|0$ unless the number of punctures is very small.

\subsection{Genus Zero}\label{genzero}

\subsubsection{$\CP^{1|1}$ As A Super Riemann Surface}\label{srprof}

Let $\Sigma$ be a super Riemann surface of genus 0, initially without punctures.  We will prove that $\Sigma$ is split, but at first
let us just assume that this is the case.  The reduced space $\Sigma_\red$ of $\Sigma$ is an ordinary Riemann surface of genus 0.  
Since $\Sigma_\red$ is simply-connected and its tangent bundle $T\Sigma_\red$ is of even degree (namely degree 2), there is
up to isomorphism a unique square root of $T\SIgma_\red$, which we call $T\Sigma_\red^{1/2}$; it has degree 1.
The normal bundle to $\Sigma_\red$ in $\SIgma$ is isomorphic to ${\Pi} T\Sigma_{\red}^{1/2}$.  Odd deformations of $\Sigma_\red$ are
classified by $H^1(\SIgma_\red,{\Pi} T\SIgma_\red^{1/2})$, but this vanishes as $T\SIgma_\red$ has positive degree.  So $\Sigma$
cannot be deformed away from the split case, and actually is split.  

It is convenient to describe $\Sigma_\red$ by homogeneous complex coordinates $u,v$.  We write $\O(n)$ for the line bundle
over $\SIgma_\red$ whose sections are functions of $u,v$ homogeneous of degree $n$.  Since $T\Sigma_\red^{1/2}$ has degree 1,
the total space of the line bundle ${\Pi} T\Sigma_{\red}^{1/2}$ can be described by introducing an odd homogeneous coordinate $\theta$
that scales just like $u$ and $v$.  So a super Riemann surface $\Sigma$ of genus 0 (without Ramond punctures)
 is just a complex projective supermanifold $\CP^{1|1}$ with even and odd
homogeneous coordinates $u,\neg v|\theta$, all of degree 1.

 As a super Riemann surface, $\Sigma$ has additional structure that we have not yet described.  
We usually describe this additional structure as a subbundle $\D\subset T\SIgma$ of rank $0|1$.  However, it will here
be more convenient
to describe the dual picture, which is a line bundle $\D^{-2}\subset T^*\Sigma$ that appears in the dual exact sequence (\ref{zomurk}).

For a split super Riemann surface,
$\D$ is isomorphic to the pullback to $\Sigma$ of the line bundle $T\Sigma_\red^{1/2}\to \Sigma_\red$.  For $\Sigma$ of genus 0,
this pullback is isomorphic to the line bundle $\O(1)$ over $\Sigma=\CP^{1|1}$.  So $\D^{-2}\cong \O(-2)$.  Hence to turn the complex
supermanifold $\Sigma=\CP^{1|1}$ into a super Riemann surface, we need to give a subbundle of $T^*\Sigma$ isomorphic to $\O(-2)$,
or equivalently we need to specify a global section $\varpi$ of $T^*\Sigma\otimes \O(2)$.

Such a global section is a holomorphic 1-form on $\Sigma$ that is homogeneous of degree 2 in $u,\neg v|\theta$.  Explicitly we take
\begin{equation}\label{refto}\varpi =u\d v -v\d u-\theta\d\theta.\end{equation}
If we use the scaling symmetry of $\CP^{1|1}$ to set $u=1$, and write $z$ for $v$, then we get
\begin{equation}\label{beft} \varpi=\d z-\theta\d\theta.\end{equation}
$\varpi$ is orthogonal to the vector field $D_\theta=\partial_\theta+\theta\partial_z$, so $z|\theta$ are a system of superconformal
coordinates, away from the divisor $u=0$.

Purely as a complex supermanifold, $\Sigma$ would have the automorphism group $\PGL(2|1)$ of linear transformations of $u,\neg v|\theta$
modulo overall scaling. 
As a super Riemann surface, the automorphism group of $\Sigma$ is the subgroup of $\PGL(2|1)$ that leaves fixed $\varpi$.  
To identify this subgroup, an alternative description of $\varpi$ is useful.  We introduce the space $Y\cong \C^{2|1}$ with
coordinates $u,\neg v |\theta$ (so $\Sigma$ is the quotient by $\C^*$ of $Y$ minus the locus $u=v=0$).  We endow $Y$ with a nondegenerate bilinear form $\langle~,~\rangle$, skew-symmetric
in the $\Z_2$ graded sense, as follows.  	Writing $y=u,\neg v|\theta$ and $y'=u',\neg v'|\theta'$ for  two points $y,\neg y'\in Y$, we set
\begin{equation}\label{friendly} \langle y, y'\rangle =u  v' - v u' -\theta\theta'.\end{equation}
Every nondegenerate skew-symmetric form on $Y$ is equivalent to this one up to the action of $\PGL(2|1)$.  Any choice
of such a form -- such as the one we have given -- breaks $\PGL(2|1)$ to a subgroup\footnote{$\OSp(m|n)$ is the orthosymplectic group whose maximal bosonic
subgroup is $\mathrm{O}(m)\times \mathrm{Sp}(n)$; $n$ must be even.  $\PGL(n|m)$ is isomorphic to $\PGL(m|n)$, but $\OSp(m|n)$ has no such symmetry.}  $\OSp(1|2)$.    The dimension of
$\OSp(1|2)$ is $3|2$. 

The section $\varpi$ of $T^*\Sigma\otimes \O(2)$ can be defined as
\begin{equation}\label{uly}\varpi =\langle y,\d y\rangle,\end{equation}
so in particular its supergroup of automorphisms is $\OSp(1|2)$.    Actually, the supergroup
of symmetries that acts faithfully on $\SIgma\cong \CP^{1|1}$
is the connected subgroup of $\OSp(1|2)$, which is the quotient of $\OSp(1|2)$ by its center; the center
is generated by the transformation $(u,v|\theta)\to (-u,-v|-\theta)$, which acts trivially on $\CP^{1|1}$.
A maximal bosonic automorphism group of the super Riemann surface $\Sigma$
 is $\mathrm{SL}_2$ (or $\mathrm{Sp}(2)$), a double cover of the automorphism group $\PGL(2)$ of the
ordinary super Riemann surface $\Sigma_\red$.  
In terms of the superconformal coordinates $z|\theta$, the odd generators
of $\OSp(1|2)$ are explicitly $f(z)(\partial_\theta-\theta\partial_z)$, where $f(z)=\alpha+\beta z$.  They act by
\begin{align}\label{torcho}\theta\to \,& \theta+\alpha +\beta z\cr
z\to\, & z-(\alpha+\beta z)\theta.\end{align}
In terms of the same superconformal coordinates, the expression $\l y, y'\r$ becomes
\begin{equation}\label{indly}\l y,y'\r =z-z'-\theta\theta'.\end{equation}

The central element of the bosonic automorphism group $\mathrm{SL}_2$ acts
by $\btau':(u,v|\theta)\to (-u,-v|\theta)$. Modulo the center of $\OSp(1|2)$, this is equivalent to
\begin{equation}\label{zombo}\btau:(u,v|\theta)\to (u,v|-\theta), \end{equation}
which is
the universal symmetry of a split super Riemann surface (section \ref{efc}). 

\subsubsection{Adding NS Punctures}\label{addns}

Now we will add NS punctures.  Each puncture that is added can reduce the dimension of the automorphism group by 
at most $1|1$.  As the dimension of $\OSp(1|2)$ is $3|2$, it will take at least 3 NS punctures to remove all automorphisms and this
number is indeed sufficient.
Superstring perturbation theory is constructed primarily\footnote{There are a few exceptions to this statement in the context of open strings
and orientifolds.  See section \ref{undbe}.} by integrating over supermoduli spaces that do not have continuous automorphisms.
So in genus 0, if all punctures are of NS type, one is primarily interested in the case that the number of punctures is at least 3.

For a genus 0 super Riemann surface $\Sigma$ with 3 NS punctures, 
the dimension formula says that the dimension of supermoduli
space is $0|1$. So a set of 3 NS punctures should have 1 odd modulus.  
Explicitly, one can use $\OSp(1|2)$ and scaling
of the homogeneous coordinates of $\CP^{1|1}$ to map one puncture
to $u,\neg v|\theta=1,\neg 0|0$, a second to $0,\neg 1|0$, and the third to $1,\neg 1|\eta$, where $\eta$ is the modulus.  
From this, one might think that the moduli space $\MM_{0,3,0}$ would be a copy of $\C^{0|1}$, parametrized by $\eta$.
But one has to remember the universal automorphism $\btau$ (eqn. (\ref{zombo}))
 that changes the sign of all odd variables.
In the present context, it acts by $\btau:\eta\to -\eta$.  So the moduli space is actually the orbifold (or stack)
$\C^{0|1}/\Z_2$.  Since $\btau$ is an element of $\OSp(1|2)$ and is equivalent as a symmetry of $\Sigma$
 to the element $\btau'$ of the connected component of
 $\OSp(1|2)$, we see that although $\eta$ cannot be removed by an $\OSp(1|2)$ transformation, 
 it also cannot be defined by a formula invariant under  $\OSp(1|2)$ or even its connnected component.  
 Roughly speaking, $\eta^2$ is $\OSp(1|2)$-invariant; but $\eta^2=0$.

More generally, for any number $k$ of NS punctures, we can use 
$\OSp(1|2)$ and scaling of the homogeneous coordinates
to map the first three to $1,\neg 0|0$, $0,\neg1|0$, and $1,\neg 1|\eta$, and the others to 
$1,\neg z_i|\theta_i$, $i=4,\dots,k$.
The supermoduli space $\MM_{0,k,0}$ (genus 0, $k$ NS punctures and no Ramond punctures) 
can thus be parametrized by the $k-3$ even parameters $z_i$, $i=4,\dots,k$, 
along with the $k-2$ odd parameters $\theta_i$, $i=4,\dots,k$ and $\eta$.
This parametrization of the moduli space is used in practice in computing superstring scattering amplitudes in genus 0.  
For most computations of tree level scattering amplitudes (of NS fields) in superstring theory, one does not
need to know more about $\MM_{0,k,0}$ than we have just explained.   
The reason is that, as long as the external momenta
are generic, the subtleties of compactification of supermoduli space are not very important at tree level.

\subsubsection{Cross Ratios}\label{zddn}

However, it ultimately is helpful to understand a little more, so we will carry on.  To illustrate the main issues, it suffices
to consider the case $k=4$, where the dimension of the supermoduli space is $1|2$.  The reduced space of $\MM_{0,4,0}$
is the moduli space $\M_{0,4}$ of an ordinary Riemann surface of genus 0 with four punctures $z_1,\dots, z_4$.  In the present
section, there will be no Ramond punctures, so we will write just $\MM_{0,4}$ for $\MM_{0,4,0}$.

\begin{figure}
 \begin{center}
   \includegraphics[width=2.5in]{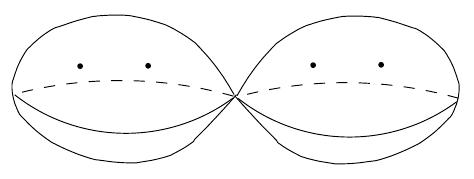}
 \end{center}
\caption{\small  
The moduli space $\M_{0,4}$ that parametrizes a genus 0 Riemann surface $\Sigma_0$ with four punctures $z_1,\dots,z_4$
is compactified
by adding ``points at infinity'' in which $\Sigma_0$ splits into a union of two components, each of which
contains two of the punctures.  There are three such points at infinity, depending on how $z_1,\dots,z_4$
are divided pairwise between the two components.    }
 \label{succinct}
\end{figure}
First of all, we should properly distinguish two related moduli spaces.  There is the moduli space $\M_{0,4}$ of four distinct
points in a genus zero surface $\Sigma_0$, and also its Deligne-Mumford compactification $\h\M_{0,4}$ in which two points are allowed to coincide, or more properly (fig. \ref{succinct}),
in which $\Sigma_\red$ may split into two components.\footnote{All this is more systematically
explained in section \ref{infinity}, which the reader may wish to consult first.
But hopefully the details are not needed for the moment.}  The following reasoning will
be more succinct if we concentrate on the compactification $\h\M_{0,4}$.   Similarly,
we write $\MM_{0,4}$ for the moduli space that parametrizes 4 distinct NS punctures in a genus 0 super Riemann surface,
and $\h\MM_{0,4}$ for its compactification.  

The compactified moduli space $\h\M_{0,4}$ can be parametrized by the $\PGL(2)$-invariant  cross ratio 
\begin{equation}\label{twelk} \Psi = \frac{(z_1-z_2)(z_3-z_4)}{(z_1-z_3)(z_4-z_2)}.\end{equation}
We include the value $\Psi=\infty$, which corresponds to one of the points at infinity in $\h\M_{0,4}$, where
either $z_1\to z_3$ or $z_2\to z_4$. The two possibilities are equivalent modulo the action of $\PGL(2)$ and
can be described more symmetrically by
 the degeneration shown in fig. \ref{succinct} with $z_1$ and $z_3$ on one side and
$z_2$ and $z_4$ on the other.  (The other points at infinity are $\Psi=0$, which corresponds to $z_1\to z_2$ or
$z_3\to z_4$, and $\Psi=-1$, which corresponds to $z_2\to z_3$ or $z_1\to z_4$.)  

A point worthy of note is that to define the parameter $\Psi$, we broke the permutation symmetry among $z_1,\dots,z_4$.  We could
alternatively have mapped $\h\M_{0,4}$ to $\CP^1$ via the $\PGL(2)$-invariant
\begin{equation}\label{welk}\Psi' =  \frac{(z_1-z_3)(z_4-z_2)}{(z_1-z_4)(z_2-z_3)}.\end{equation}
This would not make much difference.  Because of the identity
\begin{equation}\label{zelk}(z_1-z_2)(z_3-z_4)+(z_1-z_3)(z_4-z_2)+(z_1-z_4)(z_2-z_3)=0,\end{equation}
we have 
\begin{equation}\label{bombo} \Psi+\frac{1}{\Psi'}+1 = 0.\end{equation}
Thus the two maps from $\h\M_{0,4}$ to $\CP^1$ given by $\Psi$ or $\Psi'$ differ by a fractional linear transformation of $\CP^1$.

Now let us consider the supersymmetric case.  The supermanifold $\h\MM_{0,4}$ has dimension $1|2$.  Is this supermanifold
split? (See section 2.3.1 of \cite{Wittennotes} for an explanation of this concept.) 
 To get a ``yes,'' answer, we just need to find a holomorphic map from $\h\MM_{0,4}$ to $\CP^1$ that when the odd variables
vanish coincides with one of the above formulas.  To get such a map,
it suffices to have an $\OSp(1|2)$-invariant extension of the cross ratio 
that reduces to the cross ratio if the odd variables vanish.  There is no problem to find one.
For example, a super extension of the cross ratio $\Psi$
is given by the $\OSp(1|2)$ invariant $\hat\Psi=\l w_1,w_2\r \l w_3,w_4\r/\l w_1,w_3\r \l w_4,w_2\r$.  In view of (\ref{indly}), in the usual
superconformal coordinates, this becomes
\begin{equation}\label{forbly}\hat\Psi=\frac{(z_1-z_2-\theta_1\theta_2)(z_3-z_4-\theta_3\theta_4)}
{(z_1-z_3-\theta_1\theta_3)(z_4-z_2-\theta_4\theta_2)}.\end{equation}
Obviously, $\hat\Psi$ reduces to $\Psi$ if we set the $\theta$'s to 0.
The map from $\h\MM_{0,4}$ to $\h\M_{0,4}=\CP^1$  is
a splitting of $\h\MM_{0,4}$.  

There is a small catch. Obviously we could define another splitting by 
\begin{equation}\label{orbly}\hat\Psi'=\frac{(z_1-z_3-\theta_1\theta_3)(z_4-z_2-\theta_4\theta_2)}
{(z_1-z_4-\theta_1\theta_4)(z_2-z_3-\theta_2\theta_3)}.\end{equation}
In contrast to the bosonic case, the splittings given by $\hat\Psi$ and $\hat\Psi'$ (and a third  obtained from $\hat\Psi$ by exchanging $z_2|\theta_2$ with $z_4|\theta_4$) are not equivalent.  That is because the supersymmetric
extension of the identity (\ref{zelk}) does not hold:
\begin{equation}\label{orbom}(z_1-z_2-\theta_1\theta_2)(z_3-z_4-\theta_3\theta_4)
+(z_1-z_3-\theta_1\theta_3)(z_4-z_2-\theta_4\theta_2)+(z_1-z_4-\theta_1\theta_4)(z_2-z_3-\theta_2\theta_3)\not=0.\end{equation}
The  left hand side of (\ref{orbom}) is nilpotent, but that just implies that 
the splittings defined using $\hat\Psi$ and $\hat\Psi'$ are equivalent modulo nilpotent terms, which is a trivial statement, in the sense
that it holds by definition for all splittings.

So $\h\MM_{0,4}$ can be split in several fairly natural ways, but none of these is entirely natural. What property of a splitting
do we actually most care about? The answer to this question involves concepts explained in section \ref{infinity}, but we summarize
a few facts here.  A splitting is a holomorphic
projection $\pi:\h\MM_{0,4}\to \h\M_{0,4}$.  For $p$ a point in $\h\M_{0,4}$, $\pi^{-1}(p)$ is a divisor in $\h\MM_{0,4}$.
The subtleties of superstring perturbation theory come from the behavior at infinity, which means that in the present context,
we should focus attention on the distinguished points in $ \h\MM_{0,4}$ at which two of the $z_i$ coincide.  
Over the point  $p_{ij}\in\h \M_{0,4}$ given by $z_i-z_j=0$, there is a distinguished divisor  $\eusm D_{ij}\subset \h\MM_{0,4}$ given by
$z_i-z_j-\theta_i\theta_j=0$.  A good splitting is one such that for each $i$ and $j$, $\pi^{-1}(p_{ij})=\eusm D_{ij}$.  
By this criterion, none of the splittings that we have described is satisfactory at each of the divisors at infinity. The splitting given by $\hat\Psi$
behaves well for $z_1\to z_2$, since $\hat\Psi$ is explicitly proportional to $(z_1-z_2-\theta_1\theta_2)$.  It is equally good
for $z_1\to z_3$, in this case because $\hat\Psi^{-1}$ is proportional to a similar factor.  But what happens when $z_1\to z_4$?
The behavior we would like is
\begin{equation}\label{ofy}\hat\Psi= a+b(z_1-z_4-\theta_1\theta_4)+\O((z_1-z_4)^2,(z_1-z_4)\theta_1\theta_4), ~~b\not=0,\end{equation}
so that the equation $\hat\Psi-a=0$ will define the divisor $\eusm D_{ij}$.
But this is actually not the case.  So the splitting given by $\hat\Psi$ does not have the behavior one would wish 
at $z_1-z_4=0$, and similarly each of 
the other splittings fails to show the desired behavior at one of the divisors at infinity.

What we have said about $\h\MM_{0,4}$ extends to $\h\MM_{0,k}$ for all $k\geq 4$.  By making use of the supersymmetric
extension of the cross ratio, one can always define holomorphic splittings $\pi:\h\MM_{0,k}\to \h\M_{0,k}$.  But there are many
ways to do this and none is fully satisfactory.

The splittings of $\h\MM_{0,4}$ and their properties were first described in \cite{Ne}, with more detail in \cite{Gid}.  The relevance
to superstring perturbation theory is as follows \cite{GrSe}. Although tree level scattering amplitudes at generic values of the
external momenta do not have the subtleties that occur in higher genus, some of these subtleties do occur in tree level scattering amplitudes
at special values of the momenta (such as momentum zero). And moreover, as essentially also explained there, as well as in \cite{Ne},
in general a procedure to compute these amplitudes by first
integrating over the fibers of a projection $\pi:\h\MM_{0,4}\to \h\M_{0,4}$ will  only give the right answer if $\pi^{-1}(p_{ij})
=\eusm D_{ij}$.  Otherwise a correction has to be made at infinity.

In section \ref{infinity}, we will describe the genus $\g$ generalization of the divisor $\eusm D_{ij}$.  This
information  will be important input in a reconsideration
of superstring perturbation theory \cite{Witten},  not so much because one aims to find
 a good splitting -- supermoduli space is actually not holomorphically split in general \cite{DonW} -- but as part
of an explanation of how to handle the infrared region in the integration over supermoduli space.

To understand $\h\MM_{0,4}$ more deeply, one has to describe its orbifold structure.  Let us discuss what
happens along the reduced space $\h\MM_{0,4,\red},$ which roughly speaking (ignoring the orbifold structure)
 is the same as $\h\M_{0,4}$.  Generically, along $\h\MM_{0,4,\red}$, the automorphism group of 
$\Sigma$ is the universal $\Z_2$ symmetry of split super Riemann surfaces generated by $\btau:\theta\to-\theta$.
However, at the three points in $\h\MM_{0,4.\red}$ at which $\Sigma$ splits into two components, the automorphism
group is enhanced to $\Z_2\times \Z_2$, since one has a separate $\Z_2$ symmetry group on each component.
(This will be clearer in section \ref{infinity};  the  NS degeneration described in eqn. (\ref{ondo}) has separate $\Z_2$
symmetries $\theta\to -\theta$ and $\psi\to -\psi$ if $\varepsilon=0$, though for $\varepsilon\not=0$,
it has only a single $\Z_2$ symmetry.)  Thus, although
roughly speaking $\h\MM_{0,4,\red}$ is the same as $\h\M_{0,4}$ and is a copy of $\Bbb{CP}^1$ parametrized by the cross ratio, it is more accurate
to think of it as an orbifold version of $\Bbb{CP}^1$, with an automorphism group that is generically  $\Z_2$ and jumps
to $\Z_2\times \Z_2$ at the three points
at infinity.  The normal bundle to $\h\MM_{0,4,\red}$ in $\h\MM_{0,4}$ is, roughly speaking, a vector bundle over 
$\h\MM_{0,4,\red}$ of rank $0|2$, but actually this vector bundle must be defined in the orbifold sense \cite{Deltwo}.

\subsubsection{Ramond Punctures}\label{tryon}

We now will  describe a super Riemann surface $\Sigma$ of genus $0$ with $\n_  \Ra$ Ramond punctures.
We have already given such a description in affine coordinates in eqns. (\ref{torg}) and  (\ref{zeb}).  We want to rewrite these formula projectively, to better
see the behavior at infinity.  We actually can do so by hand.  In eqn. (\ref{refto}), we endowed $\CP^{1|1}$ with a superconformal
structure by picking a section $\varpi$ of $T^*\SIgma$ tensored with a suitable line bundle.  We simply need to modify
this formula so that in affine coordinates it will match with (\ref{zeb}):
\begin{equation}\label{helpme} \varpi=u\d v-v\d u-w(u,v)\theta\d\theta.\end{equation}
We take $w(u,v)$ to be homogeneous of degree $\n_  \Ra$ and to coincide with the function $w(z)$ in (\ref{zeb}) if we
set $u=1$, $v=z$.  The zeroes of this function are the Ramond divisors.
In order for $\varpi$ to transform homogeneously under the scaling of the homogeneous coordinates
$u,v$, we have to assume that $\theta$ scales with degree $1-\n_  \Ra/2$.  So $\SIgma$ is a weighted projective
superspace $\Bbb{WCP}^{1|1}(1,1|1-\n_  \Ra/2)$ with homogeneous coordinates $u,v|\theta$ whose weights are as indicated.

An important special case of this is that $\n_  \Ra=2$.  It is convenient then to put the Ramond punctures at 0 and $\infty$, which
we do by taking $w(u,v)=uv$.  Then in affine coordinates, we have
\begin{equation}\label{trymo} \varpi = \d z-z\theta\d\theta\end{equation}
with the Ramond punctures at $0,\infty$.  To verify that the structure at $z=\infty$
is equivalent to that at $z=0$, one may use the
 automorphism $z\to -1/z$, $\theta\to \pm \sqrt{-1}\,\theta$, which exchanges them, mapping $\varpi$ to a multiple
 of itself.  (Of course, the equivalence between $z=0$ and $\infty$ is obvious in the projective
 description.)
This super Riemann surface has no supermoduli, and has
an automorphism group $\G$ of dimension $1|1$, generated by the superconformal vector fields $z\partial_z$ and $\partial_\theta-\theta z\partial_z$,
which are regular both at 0 and $\infty$.  This is in keeping with the dimension formula (\ref{kobo}). 
A noteworthy fact is that the vector field $\partial_\theta-\theta z\partial_z$ is nonzero when restricted to $z=0$ (or $z=\infty$),
\begin{equation}\label{mogro}\left.\left(\partial_\theta-\theta z\partial_z \right)\right|_{z=0}=\partial_\theta.\end{equation} It
generates a shift symmetry of the Ramond divisors.

 If we add a single NS puncture at $z|\theta=z_0|\theta_0$,
then by the action of $\G$, we can uniquely map it to $z|\theta=1|0$.  In this way, we describe a Riemann surface 
of genus  0 with 1 NS puncture and 2 Ramond punctures.  It has no supermoduli and in particular is split, and its automorphism group is of dimension $0|0$, 
the only non-trivial element being the symmetry $\btau:\theta\to -\theta$ that is common
to all split super Riemann surfaces.   Of course, for $\n_  \NS=1$, $\n_  \Ra=2$, the dimension formula is consistent with $\dim\M=\dim\G=0|0$.

\subsection{Genus One}\label{ryon}

Here we will much more briefly consider the case that $\Sigma$ is a super Riemann surface of genus 1.  There are two cases; the spin structure of $\Sigma$ may be even or odd.  We only consider unpunctured surfaces.

Suppose first that the spin structure is even.  The reduced space of $\Sigma$ is an ordinary genus 1 Riemann surface, endowed
with a line bundle $T\Sigma_\red^{1/2}$ whose square is isomorphic to $T\Sigma_\red$.  If $\Sigma$ is split, then it is isomorphic
to the total space of the line bundle ${\Pi} T\Sigma_\red^{1/2}$ over $\Sigma_\red$.  The odd moduli that measure the departure of $\Sigma$
from splitness take values in $H^1(\Sigma_\red,T\Sigma_\red^{1/2})$, but this cohomology group vanishes.  So $\Sigma$ has no odd moduli
and cannot be deformed away from the split case.  So similarly to what we found for genus 0, a genus 1 super Riemann surface with an even
spin structure is automatically split.  

Such a surface can be explicitly described as the quotient of $\C^{1|1}$ with its standard supeconformal structure by the group $\Z\times\Z$ acting
by
\begin{equation}\label{trulyp}z\to z+1,~~\theta\to\theta\end{equation}
and
\begin{equation}\label{ruly}z\to z+\tau,~~\theta\to -\theta.\end{equation}
The moduli space $\MM$
 has dimension $1|0$, with $\tau$ as the only parameter.  The automorphism group $\G$ is also of dimension $1|0$, generated by $\partial_z$. 
Clearly these statements are consistent with the dimension formula (\ref{kobo}).

Now we consider the case of an odd spin structure (for much more, see \cite{JRa}).  In this case, $H^1(\Sigma_\red,T\SIgma_\red^{1/2})$ is of dimension 1, so there is 1 odd parameter
by which to deform away from a split situation.  Explicitly, one can describe the family of super Riemann surfaces as the quotient of $\C^{1|1}$ by
\begin{equation}\label{truly}z\to z+1,~~\theta\to\theta\end{equation}
and
\begin{equation}\label{rulyp}z\to z+\tau-\alpha\theta,~~\theta\to \theta+\alpha,\end{equation}
where $\alpha$ is the odd parameter.  At $\alpha=0$, the dimension formula is satisfied with $\dim\,\M=\dim\,\G=1|1$. $\G$ is generated
by $z\to z+b$, $\theta\to\theta$, along with $\theta\to\theta+\epsilon,~z\to z-\epsilon\theta$, with constants $b$ and $\epsilon$.  But
actually jumping of the cohomology occurs when $\alpha$ is turned on and this makes the proper formulation of the dimension formula subtle.
 Explicitly, when $\alpha\not=0$, a superconformal transformation $\theta\to\theta+\epsilon$, $z\to z-\epsilon\theta$ is not a symmetry
 of  (\ref{rulyp}), but shifts $\tau$ to $\tau-\alpha\epsilon$, so for $\alpha\not=0$, the odd automorphism of $\Sigma$
 is lost and the sense in which $\tau$ is a modulus is subtle.

A related jumping occurs as a function of $\alpha$ in the space of holomorphic differentials.  We explain this briefly because of its relevance 
to section \ref{peribasics}.  At $\alpha=0$, the space of closed holomorphic 1-forms on $\Sigma$ has dimension $1|1$, generated by
$\d z$ and $\d \theta$.  For $\alpha\not=0$, $\d\theta$ is still well-defined, but $\d z$ is not, since it is not invariant under (\ref{rulyp}).  Moreover,
there is no way to add an $\alpha$-dependent correction to $\d z$ to get an $\alpha$-dependent closed holomorphic 1-form that
equals $\d z$ at $\alpha=0$.  So the dimension of the space of closed holomorphic differentials jumps downward for $\alpha\not=0$.  As explained in section \ref{symmetry}, the space of closed holomorphic 1-forms on $\Sigma$
is naturally isomorphic to $H^0(\Sigma,\BBer(\Sigma))$,
so the same jumping occurs in that cohomology group.  In view of the Riemann-Roch theorem, the same jumping occurs in $H^1(\Sigma,
\BBer(\Sigma))$, as well.

\section{Behavior At Infinity}\label{infinity}

\subsection{Compactification Of The Bosonic Moduli Space}

\subsubsection{Introduction}
\begin{figure}
 \begin{center}
   \includegraphics[width=5in]{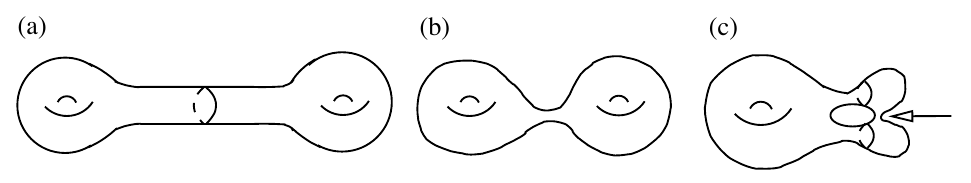}
 \end{center}
\caption{\small  
(a) A Riemann surface $\Sigma_0$ with a long tube.  (b) A conformally equivalent picture in which $\Sigma_0$ has a narrow neck. The example
sketched here represents a separating degeneration.  (c) A similar picture with a nonseparating degeneration, indicated by the arrow.}
 \label{surfbreak}
\end{figure}
The moduli space $\M$ of smooth Riemann surfaces is not compact, because a Riemann surface $\Sigma_0$ can develop a long tube, as in fig. \ref{surfbreak}(a).
A long tube is conformally equivalent (see the discussion of eqn. (\ref{varrz})) to a  narrow neck, which collapses when  the length of the long tube becomes infinite.  When this happens, $\Sigma_0$
may either divide into two topological components, as in fig. \ref{surfbreak}(b), or remain connected, as in fig. \ref{surfbreak}(c).  The two cases are called
separating and nonseparating degenerations, respectively. 

In either case, when a narrow neck collapses, $\Sigma_0$ separates  at least locally into two branches that meet at a point.  Let $x$ and $y$ 
be local parameters on the two branches.
Let us focus on the ``neck'' region  $N\subset \Sigma_0$ defined by $|x|,\,|y|<1$.  A local description of $\Sigma_0$ near 
its degeneration is given by the algebraic equation
\begin{equation}\label{onslo} xy = q, \end{equation}
where $q$ is a complex parameter that we consider small.  For $q=0$, the equation reduces to $xy=0$, and describes two 
branches -- with $x=0$ and $y=0$, respectively --
that meet at the point $x=y=0$.  This is a singular point, in some sense the simplest and most generic possible singularity of a Riemann surface.  It is known
as a node or ordinary double point. For small but nonzero $q$, this picture is modified in the region $|x|,|y|\lesssim q$ and 
the two branches are glued together through a narrow neck.

To develop more intuition, let us transform to variables that are adapted to the ``long tube'' picture of fig. \ref{surfbreak}(a).  
We do this by setting  \begin{equation}\label{varrz} x=e^\varrho,~~
y=q e^{-\varrho},\end{equation} where \begin{equation}\label{umber} \varrho=s+i\vartheta,\,~~~s,\vartheta\in \R.\end{equation}
This change of variables  actually make the conformal equivalence of a narrow neck and a long tube obvious; it  is a holomorphic
change of variables, so the metric $|\d x|^2+|\d y|^2$, which develops a narrow neck for $q\to 0$, is conformally equivalent to the metric $|\d\varrho|^2$,
which develops a long tube.
The conditions $|x|,\,|y|<1$ give $0>s > -\ln |q|^{-1}$.    So in terms of $\varrho$, the neck region $N$ is a tube of length 
\begin{equation}\label{proplength}\tau=\ln |q|^{-1}\end{equation}  (the distance along the tube is parametrized by $s$)  and of
circumference $2\pi$ (the angular variable being $\vartheta$).  In bosonic string theory, to compute a scattering 
amplitude, we have to integrate over $q$ (as well as the other
moduli of $\Sigma_0$), which means integrating over the proper length $\tau$ of the tube as 
well as over the angle $\mathrm{Arg}\,q$ (the latter parametrizes a relative twist
that can be introduced
in gluing the two sides of $\Sigma_0$).

We interpret $\tau$ as the elapsed proper time  for a closed string that is propagating down the tube.  
The meaning of the degeneration as $q\to 0$ is that the elapsed
proper time diverges.  The analog of $\tau$ in field theory is the proper time parameter in the Schwinger representation of the propagator of
a particle.  For example, the propagator of a scalar of mass $m$ is 
\begin{equation}\label{dofo}\frac{1}{p^2+m^2}=\int_0^\infty \d \tau\,\exp\left(-\tau(p^2+m^2)\right).\end{equation}
Here $p$ is the particle momentum and we write the formula in Euclidean signature so that the mass shell condition is $p^2+m^2=0$.
The pole of the propagator at $p^2+m^2=0$ comes from a divergence of the integral at $\tau\to\infty$.   In 
perturbative quantum field theory, infrared singularities
of Feynman amplitudes occur because of these poles.

In field theory, there is also an ultraviolet region, with $\tau\to 0$.  (In any Feynman diagram, one may use the proper time representation (\ref{dofo}) for 
all of the propagators, with a proper time parameter $\tau_i$ for each internal line.  Then to evaluate the diagram, one must integrate over the loop momenta and the  $\tau_i$.
The momentum integrals are always convergent as long as the $\tau_i$ are all positive, but divergences may occur as some or all of the $\tau_i$ vanish.  These are the ultraviolet
divergences of Feynman diagrams.)   There is
no analog in string theory of taking $\tau\to 0$.  This would mean taking $q\to\infty$, but the local description (\ref{onslo}) that we used 
in introducing $q$ is only
valid if $q$ is sufficiently small.  When $q$ ceases to be small, one must use a different description of $\Sigma_0$.

In the end, there is no ultraviolet region in the integral over $\M$.  
The precise statement of this is based on the fact that $\M$ has a natural Deligne-Mumford compactification $\h\M$ that parametrizes,
apart from smooth Riemann surfaces, only one additional type of object -- Riemann surfaces with the node or ordinary double point singularity modeled by
the equation $xy=q$.     $\h\M$ is built by adjoining to $\M$ certain  
``divisors at infinity'' $\eusm D_\lambda$ -- described more fully in section \ref{dmcomp} -- that
parametrize Riemann surfaces with nodes.   The phrase ``the region at infinity in $\M$'' refers to any small neighborhood of the  union of the $\eusm D_\lambda$.

 $\h\M$ is a smooth compact manifold (or more precisely a smooth compact orbifold).
 In general, there is no problem with integration of a smooth measure on a compact manifold (or orbifold).
The integrals required to 
compute perturbative bosonic string scattering are smooth except along the divisors $\eusm D_\lambda$.  In view of what has been just
explained, the singularities that occur along those divisors  are infrared effects.  The fact that the bosonic string measures
are smooth along $\M$ and that the $\eusm D_\lambda$  are  the only divisors that have to be added to $\M$ to achieve compactification   is a very precise
statement of the fact that in bosonic string perturbation theory, the only possible problems are infrared problems.  However, those infrared problems
 are serious: bosonic string theory has tachyon poles and dilaton tadpoles that do not have
a sensible physical interpretation, at least in the context of perturbation theory.

What about superstring theory?
The  moduli space $\MM$ of super Riemann surfaces has an analogous Deligne-Mumford compactification 
$\h\MM$ that is obtained by adjoining to $\MM$ certain divisors $\eusm D_\lambda$ that parametrize super Riemann surfaces with nodes.   We again call
these the divisors at infinity, and any small neighborhood of the union of these divisors is called ``the region at infinity in supermoduli space.''  $\h\MM$ is a smooth
compact supermanifold (or more precisely superorbifold).  To compute superstring scattering amplitudes, one must evaluate integrals over $\h\MM$ (or more precisely over certain integration cycles derived from $\h\MM$)
of measures whose
only singularities are along the divisors at infinity.
As in the bosonic string, the subtleties associated to the behavior at infinity are   infrared effects -- in which strings propagate for a long proper time near their mass shell. 
There is no potentially dangerous
ultraviolet region.  The difference between superstring theory and bosonic string theory is that in the superstring case, the infrared effects are harmless; there
are no tachyon poles, and the massless particle tadpoles are tamed by spacetime supersymmetry.  The infrared behavior is the same as one would expect in a field
theory with the same low energy content.  These matters will be reconsidered in \cite{Witten}.

\subsubsection{More Details On The Divisors At Infinity}\label{dmcomp}

\begin{figure}
 \begin{center}
   \includegraphics[width=4.5in]{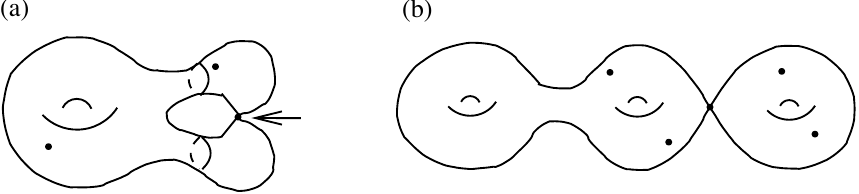}
 \end{center}
\caption{\small  
(a) Along the divisor $\eusm D_{\mathrm{nonsep}}\subset\h\M_{\sg,\sn}$ that parametrizes nonseparating degenerations of a Riemann surface $\Sigma_0$, the genus
of $\Sigma_0$ is reduced by 1, but the two points that are glued together at the node (marked by the arrow)
 count as additional punctures, one on each branch.  So this divisor
 is a copy of $\h\M_{\sg-1,\sn+2}$.    This is sketched here for $\g=\n=2$.
(b) Divisors in $\h\M_{\sg,\sn}$ that correspond
to separating degenerations correspond to decompositions $\g=\g_1+\g_2$, $\n=\n_1+\n_2$. The divisor corresponding to such a decomposition is a copy of
$\h\M_{\sg_1,\sn_1+1}\times \h\M_{\sg_2,\sn_2+1}$, where the points glued together count as one extra puncture on each side.  This is sketched here for $\g_1=2$, $\g_2=1$,
$\n_  1=\n_  2=2$.}
 \label{divisors}
\end{figure}

Now we will describe in more detail the divisors at infinity in the Deligne-Mumford compactification  of $\M_{\sg,\sn}$, the moduli
space that parametrizes a genus $\g$ surface $\Sigma_0$ with $\n$ punctures.   As we will see, the Deligne-Mumford compactification is defined
in such a way that the punctures always remain distinct as the moduli are varied.  When superficially it appears that two punctures would collide,
something else will happen instead. 

In what follows, we will exclude  a few exceptional cases with small values of $\g$ and $\n$.
The case $\g=1$, $\n=0$ is omitted because including it would create inconvenient exceptions to many statements. 
 The case $\g=0$, $\n<3$ can  then
be omitted for a reason explained below.  The discussion below will show that we can impose these restrictions not just on $\Sigma_0$ itself but on each of the
components to which it may degenerate.  The restrictions on $\g$ and $\n$ can be summarized by saying neither $\Sigma_0$ nor any of its components (if $\Sigma_0$
degenerates) has continuous symmetries.\footnote{In the Deligne-Mumford compactification of the moduli space of unoriented Riemann surfaces or Riemann surfaces
with boundary, one does have to include components with continuous symmetries, as we will see in section \ref{undbe}.  This turns out to be important
in the study of anomalies in superstring theory \cite{PolCai}, \cite{Witten}.}  Accordingly, for the values of $\g$ and $\n$ that we will discuss, the dimension of $\h\M_{\sg,\sn}$ is
always precisely
\begin{equation}\label{homer}\dim\,\h\M_{\sg,\sn}=3\g-3+\n.\end{equation}
There is no correction for the dimension of the automorphism group $\G$, since this always vanishes.

Now that these preliminaries are out of the way, let us consider a nonseparating degeneration (fig. \ref{divisors}(a)).  When
$\Sigma_0$ develops a node or double point, it is of course singular.  If, however, we simply unglue the two branches of $\Sigma_0$
that meet at the node, we get a smooth Riemann surface $\Sigma_1$ of genus $\g-1$ with $\n+2$ punctures, where the 2 extra punctures are the 2 points in $\Sigma_1$ that have to be
glued together to get $\Sigma_0$. ($\Sigma_1$ is sometimes called the normalization of $\Sigma_0$.)
 So the nonseparating divisor $\eusm D_{\mathrm{nonsep}}$ in $\h\M_{\sg,\sn}$ is a copy
of $\h\M_{\sg-1,\sn+2}$.  As a check on this, we count dimensions.  A quick computation using (\ref{homer}) shows that
$\dim\,\h\M_{\sg-1,\sn+2}=\dim\,\h\M_{\sg,\sn}-1$, consistent with the claim that $\h\M_{\sg-1,\sn+2}$ is a divisor at infinity in $\h\M_{\sg,\sn}$.

A separating degeneration occurs when $\Sigma_0$ decomposes into two components of genera $\g_1$ and $\g_2$ with $\g_1+\g_2=\g$.  In general, of the $\n$ punctures on $\Sigma_0$, we will have $\n_  1$ on one side and $\n_  2$ on the other side, with any $\n_  1,\n_  2$ such that  $\n_  1+\n_  2=\n$.
The node must be counted as an extra puncture on each side, so the divisor $\eusm D_{\mathrm{sep}}$
corresponding to this degeneration (fig. \ref{divisors}(b)) is a copy of 
$\h\M_{\sg_1,\sn_1+1}\times \h\M_{\sg_2,\sn_2+1}$.  A quick calculation shows that again $\dim(\h\M_{\sg_1,\sn_1+1}\times \h\M_{\sg_2,\sn_2+1})=\dim\,\h\M_{\sg,\sn}-1$, so that the former
can appear as a divisor at infinity in the latter.

\begin{figure}
 \begin{center}
   \includegraphics[width=5.5in]{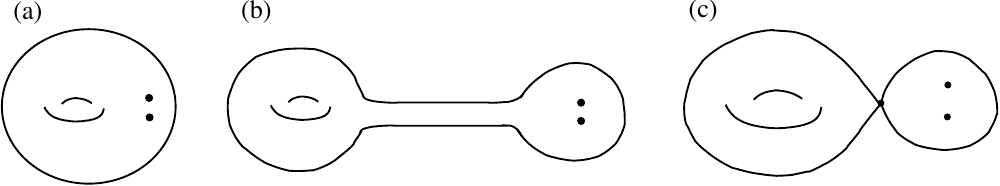}
 \end{center}
\caption{\small  
In the Deligne-Mumford compactification, a process in which two or more punctures on a Riemann surface $\Sigma_0$ approach each other as in (a), or equivalently are connected
to the rest of the surface through a long tube, as in  (b), leads to a separating divisor in which $\Sigma_0$ splits into two components, one of which has genus 0, as shown
in (c).  The number of punctures on the genus 0 component, counting the extra puncture at the node, is always at least 3.  In fact, we started in (a) with at least
2 punctures coming together, and there is 1 more where the two components of $\Sigma_0$ are glued together.  So the Deligne-Mumford compactification
can be constructed while never
allowing $\SIgma_0$ to have a genus 0 component with fewer than three punctures.   }
 \label{exceptional}
\end{figure}

The case that $\g_2$ (or $\g_1$) is 0 deserves special attention.  In compactifying the ordinary moduli space $\M_{\sg,\sn}$, why  is it necessary to include
a divisor in which $\Sigma_0$ splits into two components, one of which has genus 0?  The answer to this question (fig. \ref{exceptional}(a)) is that in compactifying $\M_{\sg,\sn}$
one needs to give a limit to a sequence in which the surface $\Sigma_0$ is kept fixed but 2 or 
more punctures in $\Sigma_0$ approach each other.  Let $s\geq 2$ be the number of punctures that are approaching each other. In the Deligne-Mumford compactification, the
limit of this sequence is represented, as in fig. \ref{exceptional}(c), by letting a genus 0 component split off from $\Sigma_0$.  This component has $s+1$ punctures, including an extra one from the node, so it has at least 3 punctures.  Thus, in constructing the Deligne-Mumford compactification, there is no need to allow
genus 0 components with fewer than 3 punctures.  

Another consequence is that $\h\M_{\sg,\sn}$ parametrizes Riemann surfaces that may have
singularities (nodes) but in which the $\n$ punctures are always distinct.  The limit of a sequence in which 2 or more points approach each other is
described by letting $\Sigma_0$ branch off a new component, not by actually taking the  points in question to be equal. 

\begin{figure}
 \begin{center}
   \includegraphics[width=3in]{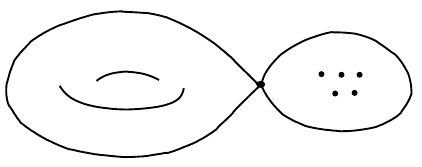}
 \end{center}
\caption{\small If $z_1,\dots,z_s$ (with $s\geq 2$) is any number of punctures in $\Sigma_0$, there is a divisor at infinity in $\h\M_{\sg,\sn}$ along which
$z_1,\dots,z_s$ are contained in a separate genus 0 component of $\Sigma_0$.  This locus is of complex codimension 1, and not of codimension $s-1$
as one might guess from the fact that naively it corresponds to the condition $z_1=\dots = z_s$.} 
 \label{exco}
\end{figure}

Now let us focus on the case $s>2$.  Consider a Riemann surface with nearby punctures $z_1,\dots,z_s$.  Naively, to
set $z_1=\dots=z_s$ is $s-1$ complex conditions and would appear to define a submanifold of moduli space of complex codimension $s-1$.
However, in the Deligne-Mumford compactification, in the limit that naively corresponds to $z_1=\dots=z_s$, we let $\Sigma_0$ split off a genus 0 component that contains the points $z_1,\dots,z_s$, which remain distinct (fig. \ref{exco}).  This
gives one of the separating divisors at infinity, here with $\g_1=\g$, $\g_2=0$, so the complex codimension is actually 1, not $s-1$.  What has happened is that
the naive codimension $s-1$ locus with $z_1=\dots= z_s$ has been ``blown up'' and replaced by a divisor.

\begin{figure}
 \begin{center}
   \includegraphics[width=4.5in]{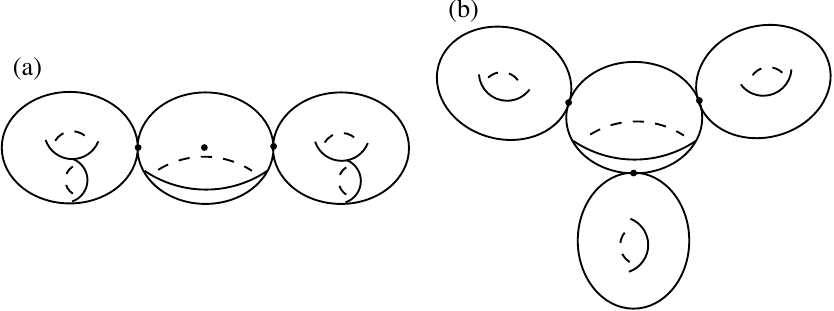}
 \end{center}
\caption{\small  
The divisors at infinity in the Deligne-Mumford compactification can intersect each other, leading to a Riemann surface $\Sigma_0$ that has more than 2 components
glued together at nodes.  Two examples are sketched here.   These examples were chosen to illustrate another point:  in the Deligne-Mumford compactification,
a genus 0 component of $\Sigma_0$ will always have at least 3 punctures, but there is no restriction on how many of those punctures may originate
from nodes.  In both (a) and (b), there is a genus 0 component with 3 punctures; in (a), 2 of these arise from nodes and in (b) all 3 do.}
 \label{illuminating}
\end{figure}

The description we have given of the divisors at infinity in $\h\M_{\sg,\sn}$ has the following implication.  
Since those divisors are themselves
compactified moduli spaces $\h\M_{\sg',\sn'}$ or products thereof, further degenerations are typically possible.  
Some examples are shown in figure \ref{illuminating}.  As long
as $\Sigma_0$ has a component of genus $\g'$ with $\n'$ punctures such that $\dim\,\h\M_{\sg',\sn'}>0$, further degeneration is always possible.  
Given the restrictions that we have placed on small values of $\g'$ and $\n'$,  the only case with $\dim\,\h\M_{\sg',\sn'}=0$ is $\g'=0$, $\n'=3$.  Accordingly,
3-punctured spheres play a special role.  A Riemann surface $\Sigma_0$ can degenerate until it is built by gluing together 3-punctured spheres.
Completely degenerate Riemann surfaces of this kind are dual to trivalent graphs.  They will be discussed in section \ref{applicdim}.

\begin{figure}
 \begin{center}
   \includegraphics[width=4.5in]{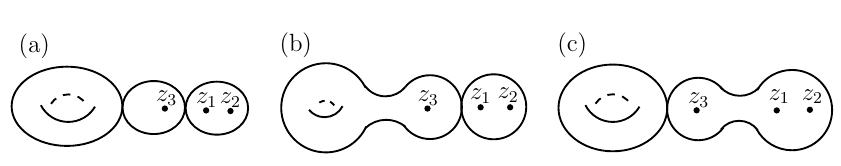}
 \end{center}
\caption{\small  The divisors $\eusm D_{12}$ and $\eusm D_{123}$ described in the text intersect on a locus in $\h\M_{\sg,\sn}$ that parametrizes singular
Riemann surfaces depicted in (a).  Surfaces of this type do correspond to intersection points of $\eusm D_{12}$ and $\eusm D_{123}$, since they can be
obtained by further degeneration from configurations of (b), which represent $\eusm D_{12}$ (since a genus zero component containing
$z_1$ and $z_2$ has split off), and from configurations of (c), which represent $\eusm D_{123}$ (since a genus zero component containing
all of $z_1,z_2,$ and $z_3$ has split off).
By contrast, for example, $\eusm D_{12}$ does not intersect $\eusm D_{13}$, since there is no picture that is a further degeneration both of (b),
which represents $\eusm D_{12}$, and of an equivalent picture with $z_2$ and $z_3$ exchanged, representing $\eusm D_{13}$.}
 \label{doublyx}
\end{figure}

We conclude this discussion with an example that may help one develop intuition about the Deligne-Mumford compactification.  In this
example, we consider a Riemann surface $\Sigma_0$ with three nearby punctures $z_1,z_2,z_3$.  We will keep $\Sigma_0$ fixed and vary only
the points $z_i$.  For each $i<j$, the Deligne-Mumford compactification has a divisor $\eusm D_{ij}$ that naively corresponds to $z_i=z_j$ (but which actually
is represented by the splitting off from $\Sigma_0$ of a genus 0 component that contains $z_i$ and $z_j$).  As we know from fig. \ref{exco}, 
$\h\M_{\sg,\sn}$ also has another divisor at infinity that we will call $\eusm D_{123}$, corresponding to the case that $\Sigma_0$ splits off a genus 0 component
that contains all  of $z_1,z_2,z_3$.  We want to understand the intersections of these divisors.  Naively, one may expect that the divisors $\eusm D_{12}$
and $\eusm D_{23}$, which naively correspond to  $z_1=z_2$ and to $z_2=z_3$, would intersect on a codimension 2 locus corresponding $z_1=z_2=z_3$.
But this cannot be right because the locus in $\h\M_{\sg,\sn}$ corresponding naively to $z_1=z_2=z_3$ is not of complex codimension 2; it is the divisor
$\eusm D_{123}$.  In fact, in $\h\M_{\sg,\sn}$, $\eusm D_{12}$ and $\eusm D_{23}$ do not intersect each other, but they both intersect
$\eusm D_{123}$.  This is indicated in fig. \ref{doublyx}.

\subsubsection{Normal Bundle To The Compactification Divisor}\label{normal}

A natural line bundle over a Riemann surface $\Sigma_0$ is its canonical bundle $K_{\SIgma_0}$.  We can use this line bundle to
define line bundles over the moduli space $\h\M_{\sg,\sn}$, which parametrizes
a Riemann surface $\Sigma_0$ with $\n$ punctures.  Taking the fiber of $K_{\SIgma_0}$ at the $\sigma^{th}$ puncture, for $\sigma=1,\dots,\n$,
we define a complex line bundle $\L_\sigma\to \M_{\sg,\sn}$.  These line bundles
are quite  nontrivial.  Their first Chern classes are the basic observables of two-dimensional topological gravity \cite{Wittentop}.
For our purposes, these line bundles are essential ingredients in describing the normal bundles to the compactification divisors
$\eusm D\subset \h\M_{\sg,\sn}$.  A basic understanding of these normal bundles is important for superstring perturbation theory.

For brevity in the notation, we will consider a divisor $\eusm D$ that describes 
the splitting of  $\Sigma_0$  into two components $\Sigma'$ and $\Sigma''$ joined at a pair of punctures.  The divisor $\eusm D$ is itself
a product $\eusm D=\h\M_{\sg_1,\sn_1+1}\times \h\M_{\sg_2,\sn_2+1}$, as explained before.  What we will say has an obvious analog for the
nonseparating degeneration.

We pick local coordinates $x$ and $y$ on  $\Sigma'$ and $\Sigma''$, respectively. The gluing of $\Sigma'$ and $\Sigma''$ at general points $x=a$ and $y=b$,
and the smoothing to make a smooth surface $\Sigma_0$, is described by a slight generalization of eqn. (\ref{onslo}):
\begin{equation}\label{brot} (x-a)(y-b)=q. \end{equation}
We interpret the parameters $a,b,$ and $q$ as all representing moduli of $\Sigma_0$.  We will work to first order in $x-a$, $y-b$, and $q$.

Describing the gluing by the explicit equation (\ref{brot}) has required a choice of local holomorphic coordinates $x-a$ and $y-b$ near the points $a\in \Sigma_1$
and $b\in\Sigma_2$.  The factors $x-a$ and $y-b$ in eqn. (\ref{brot}) are certainly not invariant under local reparametrization of $\Sigma_1$ and $\Sigma_2$.
However, since we work to first order in $x-a$ and in $y-b$, the factors $x-a$ and $y-b$ transform linearly under changes of coordinates
and in fact they transform like\footnote{For example, under $x\to\lambda x$, $a\to\lambda a$, we have $(x-a)\to \lambda (x-a)$, which
agrees with the transformation law $\d a\to \lambda \d a$.  In algebraic geometry, the cotangent bundle to a variety at a point is
defined as the space of functions vanishing at that point modulo those that vanish to second order. This motivates our assertion that $x-a$ tranforms
like $\d a$.}
the 1-forms $\d a$ and $\d b$.  So we see that $q$ must transform as the product $\d a\otimes \d b$.

A useful formulation of this is as follows. As in our discussion above, we define a line bundle $\L'\to \h\M_{\sg_1,\sn_1+1}$ whose fiber is the fiber
of the cotangent bundle $K_{\Sigma'}$ at $x=a$, and a line bundle $\L''\to \h\M_{\sg_1,\sn_1+1}$ whose fiber is the fiber of the cotangent bundle
$K_{\Sigma''}$ at $y=b$.  So $\d a\otimes \d b$ is naturally understood as a section of $\L'\otimes \L''$.  Hence the statement that $q$ transforms as $\d a\otimes \d b$
means that  we should interpret $q$ as a section of $\L'\otimes \L''$. 

Locally, we can view $q$ as a function on $\h\M_{\sg,\sn}$ that vanishes on the compactification divisor $\eusm D$.  However, this
formulation is not adequate globally.
The best way to formulate the statement that $q$ transforms as $\d a\otimes \d b$ is that $q$ should be viewed as a section of
the line bundle $\L'\otimes \L''\to \eusm D$.  This is a satisfactory global description of what $q$ means to linear order in $q$.  Beyond
linear order, the geometrical meaning of $q$ is more complicated (but usually not needed in practice).  

Equivalently, let $\eusmn_{\eusm D}$ be the normal bundle to $\eusm D$ in $\h\M_{\sg,\sn}$; it is a complex line bundle over $\eusm D$.
$q$ can be viewed as a linear function on $\eusmn_{\eusm D}$, so the fact that $q$ is a section of $\L'\otimes \L''$ means that there is a natural
isomorphism
\begin{equation}\label{malvento}\eusmn_{\eusm D}\cong\left( \L'\otimes \L''\right)^{-1}.\end{equation}
Later we will describe the super analog of this relation.

\subsection{Compactification Of Supermoduli Space}\label{compsuper}

\subsubsection{Neveu-Schwarz and Ramond Degenerations}\label{twogen}

The moduli space $\MM$ of super Riemann surfaces has a Deligne-Mumford compactification that is quite analogous to the corresponding 
compactification of the moduli space $\M$ of ordinary Riemann surfaces.\footnote{From the standpoint of algebraic geometry, the most detailed reference on this
topic is the  letter \cite{Deligne}, which unfortunately is unpublished. A number of details have been clarified in subsequent letters \cite{Deligneletters}. See \cite{BaS} for
another perspective via uniformization.}  Again compactification is achieved by adding divisors $\eusm D_\lambda$
at infinity that parametrize super Riemann surfaces with very simple singularities.  The singularities in question describe the collapse
of a narrow neck (or the growth of a long tube) and are quite analogous to the familiar singularity (\ref{onslo}) of an ordinary Riemann surface.  Perhaps the main difference 
from the bosonic case is that there are two types of tube that may collapse.  They were described
in section \ref{conftube} and
govern propagation of a closed string in the NS or Ramond sector.  Accordingly, in constructing the Deligne-Mumford compactification 
$\h\MM$ of $\MM$, we need to allow two different types of degeneration, which we will call Neveu-Schwarz (NS) and Ramond (R)
degenerations.

For an NS degeneration, we glue together two copies of $\C^{1|1}$, with local superconformal coordinates $x|\theta$ and $y|\psi$,
respectively, by the gluing formulas
\begin{align}\label{ondo}  xy & = -\varepsilon^2 \cr
                                           y\theta & = \varepsilon\psi \cr
                                            x\psi & =-\varepsilon\theta\cr
                                             \theta\psi&=0. \end{align}
                                             It is convenient to define \begin{equation}\label{delb}q_\NS=-\varepsilon^2.\end{equation}
The formulas (\ref{ondo}) obviously reduce to the bosonic gluing relation $xy=q$ if we set the odd coordinates $\theta,\psi$ to zero
and identify $q=q_\NS$.                                           
Eqn. (\ref{ondo}) defines a super Riemann surface, since
the transformation from $x|\theta$ to $y|\psi$ is superconformal.  One may verify this by showing that $D_\theta=(y/\varepsilon)D_\psi$,
where as usual $D_\theta=\partial_\theta+\theta\partial_x$, $D_\psi=\partial_\psi+\psi\partial_y$.  Alternatively, one may verify the superconformal nature of the mapping from $x|\theta$ to $y|\psi$  in a dual fashion by showing that $\d x-\theta\d\theta$ (whose kernel is generated by $D_\theta$)
is a multiple of $\d y-\psi\d\psi$
(whose kernel is generated by $D_\psi$):
\begin{equation}\label{golf}\d x-\theta\d\theta = \frac{\varepsilon^2}{y^2}\left(\d y-\psi\d\psi\right).\end{equation}

The super Riemann surface $\Sigma$ described by eqn. (\ref{ondo}) is smooth as long as $\varepsilon\not=0$, but it is singular when $\varepsilon$
vanishes.  The Deligne-Mumford compactification of $\MM$ is achieved precisely by allowing this kind of singularity, as well as an analogous
Ramond degeneration that we describe shortly.  As long as $\varepsilon\not=0$, $\Sigma$ can be identified with the NS supertube (\ref{zomy})
via $x=e^\varrho$, $\theta=e^{\varrho/2}\zeta$, $y=-
\varepsilon^2 e^{-\varrho}$, $\psi=-\varepsilon e^{-\varrho/2}\zeta$.  Just as in the bosonic case, we regard $\Sigma$ as a local
description of part of a compact super Riemann surface, valid for example for $|x|$, $|y|<1$.  The length of the supertube
is then $\tau=-\ln |q_\NS|^{-1}$, as in the bosonic case.  In the coordinates used in (\ref{ondo}), what happens at $\varepsilon=0$ 
is that $\SIgma$ splits up into two components, one parametrized by $x|\theta$ and one by $y|\psi$, and glued at $x=y=0$.  In the coordinate
system $\varrho|\zeta$, one says instead that as $\varepsilon\to 0$, the length of the supertube diverges.  

To describe  a Ramond degeneration, we again introduce two copies of $\C^{1|1}$ with local coordinates $x|\theta$ and
$y|\psi$.  But now we provide each copy of $\C^{1|1}$ with a  superconformal structure that degenerates at the divisor $x=0$ or
$y=0$, representing the Ramond punctures.  These singular superconformal structures are defined,
as in (\ref{dorg}), by the odd vector fields
\begin{equation}\label{omy}D^*_\theta =\frac{\partial}{\partial\theta}+\theta x\frac{\partial}{\partial x},~~D^*_\psi =\frac{\partial}{\partial\psi}+\psi y\frac{\partial}{\partial y}.\end{equation}
The gluing formulas are now simply
\begin{align}\label{tormo} xy& = q_\Ra \cr
                        \theta& =\pm \sqrt{-1}\psi.\end{align}
 Obviously, these formulas reduce to the bosonic gluing formula $xy=q$ if we set the odd variables to zero and set $q=q_\Ra$.
                        
This gluing defines a super Riemann surface $\Sigma$, since $D^*_\theta=\mp \sqrt{-1}D^*_\psi$.  Alternatively, in a dual language,
\begin{equation}\label{romox}\d x-x\theta \d\theta = -\frac{q_\Ra}{y^2}\left(\d y - y\psi\d\psi\right). \end{equation}
For $q_\Ra\not=0$, we can map $\Sigma$ to the Ramond supertube                         
(\ref{zoomy}) by setting $x=e^\varrho$, $y=q_\Ra e^{-\varrho}$, leaving $\theta$ unchanged.  So the behavior of $\Sigma$ for $q_\Ra\to 0$
can be described either as the collapse of a narrow neck or as the divergence of the length of a Ramond supertube.

\subsubsection{The Divisors At Infinity}\label{thed}

The purely bosonic analysis of sections \ref{dmcomp} extends to the super Riemann surface case, with just
one or two surprises.

In general we compactify the moduli space $\MM_{\sg,\sn_\NS,\sn_\Ra}$ that parametrizes
genus $\g$ super Riemann surfaces $\Sigma$  with $\n_  \NS$ Neveu-Schwarz punctures and 
$\n_  \Ra$ Ramond punctures, by allowing singularities of the types described in section \ref{twogen}.   
It is a rather fundamental fact that compactification
can be achieved by allowing singularities of only these types. 
We call the compactification $\h\MM_{\sg,\sn_\NS,\sn_\Ra}$. 

Compactification does not
affect the dimension of the moduli space, which is
\begin{equation}\label{fozzo}\dim\,\h\MM_{\sg,\sn_\NS,\sn_\Ra}=3\g-3+\n_\NS+\n_\Ra|2\g-2+\n_\NS+\frac{1}{2}\n_\Ra.\end{equation}
The divisors at infinity in $\h\MM_{\sg,\sn_\NS,\sn_\Ra}$  describe the possible
degenerations of $\Sigma$.  For similar reasons to the bosonic case, to develop a simple theory, we exclude the values $\g=0$, $\n_  \NS+\n_  \Ra
<3$ and $\g=1$, $\n_  \NS=\n_  \Ra=0$, at which $\Sigma$ would have continuous automorphisms.\footnote{The exceptional cases can of course be
treated by hand, but they are essentially not needed in 
superstring perturbation theory.  In any genus, to compute scattering amplitudes, one needs $n$-point functions with $n>0$.  The starting point of perturbation
theory is a superconformal field theory, which means that at the outset one knows what happens for $\g=0$, $\n_  \NS+\n_  \Ra\leq 2$.}  And similarly to the bosonic case,
in constructing $\h\MM$, all punctures remain distinct; sequences of super Riemann surfaces in which two or more punctures (of either type) approach each other are assigned limits in which $\Sigma$ splits off a genus zero component.  

As in the bosonic case,
$\Sigma$ can have  separating or nonseparating degenerations, and of course its degenerations are of NS and R type.  By obvious analogy with what we said in section \ref{dmcomp},
the divisor   in $ \h\MM_{\sg,\sn_\NS,\sn_\Ra}$ that parametrizes nonseparating degenerations of NS type is a copy of 
$\h\MM_{\sg-1,\sn_\NS+2,\sn_\Ra}$.   This is compatible with the dimension formula (\ref{fozzo}), which shows that
\begin{equation}\label{oformz}\dim\,\h\MM_{\sg,\sn_\NS,\sn_\Ra}=\dim\,\h\MM_{\sg-1,\sn_\NS+2,\sn_\Ra}+1|0,\end{equation}
so that $\h\MM_{\sg-1,\sn_\NS+2,\sn_\Ra}$ can be a divisor in $\h\MM_{\sg,\sn_\NS,\sn_\Ra}$.

Similarly, one might think at first that the divisor  $\eusm D\subset \h\MM_{\sg,\sn_\NS,\sn_\Ra}$ corresponding to a nonseparating degeneration of R type
would be a copy of $\h\MM_{\sg-1,\sn_\NS,\sn_\Ra+2}$. However, the dimension formula shows that this cannot be correct.  It gives
\begin{equation}\label{zform}\dim\,\h\MM_{\sg,\sn_\NS,\sn_\Ra}=\dim\,\h\MM_{\sg-1,\sn_\NS,\sn_\Ra+2}+1|1,\end{equation}
so that the two dimensions differ by $1|1$ and not by $1|0$ as would be the case if $\h\MM_{\sg-1,\sn_\NS,\sn_\Ra+2}$ were a divisor
in $\h\MM_{\sg,\sn_\NS,\sn_\Ra}$.   It turns out that the divisor $\eusm D$, rather than being isomorphic to $\h\MM_{\sg-1,\sn_\NS,\sn_\Ra+2}$, 
is a fiber bundle over that space with fibers of dimension $0|1$:
\begin{equation}\label{ofter}\begin{matrix}\C^{0|1}& \to &\eusm D \cr   & & \downarrow \cr & & \h\MM_{\sg-1,\sn_\NS,\sn_\Ra+2}.\end{matrix}\end{equation}
The structure group of this fibration acts on the fiber by  $\theta\to\pm \theta+\alpha$, with $\alpha$ an odd parameter.  (It is apparently not
known if the fibration has a holomorphic section.)

The origin of this fibration is as follows.  The divisor $\eusm D$ parametrizes singular super Riemann surfaces in which two Ramond divisors $\F$ and $\F'$
are glued together.  Locally we can describe a neighborhood of $\F$ by coordinates $x|\theta$ and superconformal structure
$D^*_\theta=\partial_\theta+\theta x\partial_x$, where $\F$ is defined by $x=0$; similarly, in suitable coordinates $y|\psi$ with
superconformal structure $D^*_\psi=\partial_\psi+\psi y\partial_y$, $\F'$ is defined by $y=0$.
Naively speaking, the gluing of $\F$ to $\F'$ is made by setting $\theta =\pm \sqrt{-1}\psi$.  (This is what eqn. (\ref{tormo}) amounts to at $q_\Ra=0$,
that is, along $\eusm D$.) However, a fact that was already important
in section \ref{pictures} was that when restricted to $\F$, a superconformal vector field may not vanish but may restrict to a multiple of
$\partial_\theta$, generating the symmetry $\theta\to \theta+\alpha$ of $\F$.  So once we have selected the two Ramond divisors $\F,\F'$ in a genus
$\g-1$ super Riemann surface $\Sigma'$ 
that we want to glue together to make a singular genus $\g$ super Riemann surface $\Sigma$, the gluing that makes $\Sigma$ is not
uniquely determined.  The general possible gluing is
\begin{equation}\label{olfom}\theta=-\alpha\pm \sqrt{-1}\psi,\end{equation}
where the fermionic parameter $\alpha$ parametrizes the fiber of the fibration (\ref{ofter}). (The meaning of the sign in $\pm \sqrt{-1}\psi$ will be explained in section
\ref{relgso}.)  

An important point is that the distinguished fermionic parameter $\alpha$ that enters the fibration (\ref{ofter}) only exists
when we restrict from the full moduli space to the divisor $\eusm D$.  Away from $q=0$, this parameter cannot be separated from the rest of the
odd and even moduli.  

This fermionic gluing parameter  has the following significance
for string theory.  We will describe this a little informally, as if it is only necessary to consider holomorphic degrees of freedom.
(In fact, the following comments apply to open superstrings, and with minor and standard modifications to closed superstrings.) 
In bosonic string theory, near a compactification divisor, there is a gluing parameter $q$.  The integral over $q$ gives the bosonic
string propagator
\begin{equation}\label{holdo}\frac{1}{L_0}.\end{equation}
With $L_0=p^2/2+N$, where $p$ is the momentum and $N$ has integer eigenvalues, the poles of $1/L_0$ lead to the singularities
of superstring scattering amplitudes.  Almost the same thing happens for the NS sector of superstrings.  Here the gluing parameter
is $\varepsilon=\pm \sqrt{-q_\NS}$.  The significance of the sign will be discused in section \ref{relgso}.  Leaving this aside for the moment,
the integration over $q_\NS$ again produces the propagator $1/L_0$, leading again to singularities of the scattering amplitudes due to bosonic
intermediate states.  

Now consider a Ramond degeneration.  The integral over $q_\Ra$ still gives a factor of $1/L_0$.  But now in integrating over $\eusm D$, we can integrate
first over the fibers of the fibration (\ref{ofter}), which means that we integrate first over the fermionic gluing parameter $\alpha$.  The integral over $\alpha$
gives a factor of $G_0$, the global worldsheet supercharge that obeys $G_0^2=L_0$.  The propagator for Ramond states is therefore
\begin{equation}\label{infeld}\frac{G_0}{L_0}=\frac{1}{G_0}.\end{equation}
$G_0$, which is sometimes called the Dirac-Ramond operator, was originally introduced in  \cite{ramond} as a stringy analog of the Dirac operator, and so $1/G_0$ is the analog for string
theory of the Dirac propagator of field theory.

We still must consider  separating degenerations, but there are no further surprises.  A divisor in $\h\M_{\sg,\sn_\NS,\sn_\Ra}$ representing a separating NS degeneration
is isomorphic to a product $\h\MM_{\sg_1,\sn_1,\sm_1}\times \h\MM_{\sg_2,\sn_2,\sm_2}$, with 
\begin{equation}\label{delfo}\g_1+\g_2=\g,~~\n_1+\n_2=\n_\NS+2,~~\m_1+\m_2=\n_\Ra,~~\m_1,\m_2\in 2\Z. \end{equation} This is as one would
expect by analogy with the bosonic case: a surface of genus $\g$ splits into two components of genera $\g_1$ and $\g_2$ that add to $\g$; the punctures originally
present are distributed between the two sides; and 1  extra NS puncture appears on each component, where they are glued together to make $\Sigma$.
By contrast, a divisor in $\h\MM_{\sg,n_\NS,n_\Ra}$ that represents a separating Ramond degeneration is a fiber bundle, with fibers of dimension $0|1$, over a product $\h\MM_{\sg_1,\sn_1,\sm_1}\times \h\MM_{\sg_2,\sn_2,\sm_2}$, now with 
\begin{equation}\label{delfog}\g_1+\g_2=\g,~~\n_1+\n_2=\n_\NS,~~\m_1+\m_2=\n_\Ra+2,~~\m_1,\m_2\in 2\Z. \end{equation}
The fiber of dimension $0|1$ enters the choice of gluing.

One moral of this story is that we must be careful in counting gluing parameters.  At an NS degeneration, there is from any point 
of view just one even bosonic gluing parameter $\varepsilon$ and
no odd ones.  So the space of gluing parameters is of dimension $1|0$.
At a Ramond degeneration, there are two slightly different but natural questions.  The locus in $\h\MM_{\sg,\n_\NS,\n_\Ra}$ that 
parametrizes super Riemann surfaces with a Ramond degeneration is a divisor, of codimension $1|0$,
with a single even gluing parameter $q_\Ra$.  But if one is interested in constructing this divisor from similar moduli spaces that parametrize
surfaces with smaller genus or more components, one must take
account also of the fermionic gluing parameter $\alpha$.  So for some purposes, it is better to think of the gluing parameters at a 
Ramond degeneration as being of dimension $1|1$.

\subsubsection{Gluing And The GSO Projection}\label{relgso}  

We have encountered an important minus sign for both NS and R degenerations.  Gluing of ordinary Riemann surfaces is described 
by a gluing parameter $q$ that
enters a formula $xy=q$.  In the super
Riemann surface version of this gluing, we need  in the NS case to take a square root, since the gluing parameter in eqn. (\ref{ondo}) is
 $\varepsilon =\pm\sqrt{-q_\NS}$.  Alternatively, in the Ramond case, the fermionic gluing  law (\ref{tormo}) directly involves a sign,  
$\theta=\pm \sqrt{-1}\psi$.   In each case, this minus sign appears in the gluing law between the fermionic coordinates
$\theta$ and $\psi$ on the two branches, via either $\theta=\pm \sqrt{-q_\NS}\psi/y$ or $\theta=\pm \sqrt{-1}\psi$.  

\begin{figure}
 \begin{center}
   \includegraphics[width=1.5in]{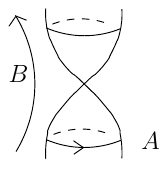}
 \end{center}
\caption{\small  Here we depict the local behavior of a super Riemann surface $\Sigma$ of genus $\g$
at a nonseparating degeneration.  We can pick an $A$-cycle of
$\Sigma$ that surrounds the node and a $B$-cycle that passes through the node, as shown here.  (The other $\g-1$ $A$-cycles and
$B$-cycles of $\Sigma$ can be chosen not to pass near the node. They play no role in 
the analysis.)}
 \label{doubly}
\end{figure}

This two-valuedness involves spin structures and the GSO projection.  Consider
a nonseparating degeneration, at which  $\Sigma$ can be constructed by gluing together two points or divisors in 
 a super Riemann surface $\Sigma'$ of genus $\g-1$. 
 The topological picture is the same as in the purely bosonic case (fig. \ref{divisors}(a)).  
 Prior to degenerating, $\Sigma$ has $2^{2\sg}$ spin structures, while $\Sigma'$ has only $2^{2(\sg-1)}$ spin structures.\footnote{In the presence
 of Ramond punctures on either $\Sigma$ or $\Sigma'$, the relevant objects are not spin structures but generalized spin structures as
 described in section \ref{compar}.  This  does not materially affect the following counting, 
 so we will use the more familiar
 language of spin structures.}  
We want to understand explicitly why $\Sigma$ has 4 times as many spin structures as $\Sigma'$.
  This can conveniently be described in terms of a 1-cycle
 $A$ on $\Sigma$ that encloses the node or narrow neck and another 1-cycle $B$ that passes through it, as depicted in fig. 
 \ref{doubly}.  In comparing spin structures on $\Sigma$ to those on $\Sigma'$, there are 2 factors of 2 that arise as follows:
  
(1) Spin structures on $\Sigma$ can be partially classified by whether they are of bounding or unbounding type when restricted to the cycle $A$
-- in other words whether the string propagating through the node is in the NS or Ramond sector.  The two cases correspond to NS and
R degenerations of $\Sigma$, respectively.  On $\Sigma'$, this distinction makes sense but is not part of the choice of spin structure.

(2) In addition, $\Sigma$ has pairs of spin structure that differ  in their ``type''  (NS or R) when restricted to $B$, but coincide in  their restriction to $A$
or to any 1-cycle that does not pass through the neck.  These pairs of spin structures are exchanged by including a minus
sign in the gluing between fermionic variables on the two sides of the neck.  In other words, they are exchanged if
one reverses the sign in the formulas  $\theta=\pm \sqrt{-q_\NS}\psi/y$ or $\theta=\pm \sqrt{-1}\psi$.  Locally (and in general
even globally\footnote{In the case of a Ramond degeneration, in the absence of Ramond punctures, one can make a distinguished choice of sign by asking
that the spin structure on $\Sigma$ should be even (or odd).  Our discussion in the text uses only what can be seen locally
without this sort of global information.}) there is no way to say which of these gluings is which, since one is free to reverse the sign of either $\theta$ or
$\psi$.
But there is a canonical operation on spin structures of exchanging this sign.    On $\Sigma'$, one does not make
this gluing so this factor of 2 in the number of spin structures is absent.

In general, as explained in section \ref{oddspin},
it does not make sense to say that two super Riemann surfaces are the same except for the choice of spin structure. 
A very limited but crucial exception\footnote{Another exception, though not relevant here, is that if $\Sigma$ is split, then it makes sense to change
its spin structure without changing anything else.}  comes from the two-valuedness we have just described: when
$q_\NS$ or $q_\Ra$ vanishes, there are pairs of spin 
structure that differ only by their ``type'' along a cycle $B$  that passes through the node.  
Singularities of superstring scattering amplitudes come from the contribution of an on-shell state passing
through the node.  So in evaluating such singularities,
one can sum over pairs of spin structures that differ only by their ``type'' along $B$ 
or in other words by the sign in the gluing between $\theta$ and $\psi$. (Moreover, this partial sum over spin structures
can be performed before the rest of the integral that determines a given singular contribution to a scattering amplitude.) 
To understand the implications of this for string theory,
consider a string state propagating through the narrow neck.  If we act on this string state with the operator $(-1)^F$ that counts world sheet
fermions mod 2, this will have the same effect as changing the sign of the fermionic coordinate in the gluing.   So averaging over the 
two possible signs of the gluing  amounts to  the action of
the GSO projection operator $(1+(-1)^F)/2$.
What we have just explained is the reason that only states that are invariant under the GSO projection contribute
singularities in superstring scattering amplitudes.

\subsubsection{Application To Type II Superstrings}\label{zna}

All of this has been stated purely holomorphically.  A Type II worldsheet is defined by, 
in a sense, combining holomorphic and antiholomorphic
super Riemann surfaces.  Its degenerations are described by applying what we have learned 
separately to the holomorphic and antiholomorphic
variables.  In particular, the degeneration of either the holomorphic or 
antiholomorphic variables can be of either NS or R
type, so there are four types of degeneration, which we can call NS-NS, NS-R, R-NS, and R-R.   
This matches the four types of punctures in Type II superstring theory.  
If $\Sigma$ develops a degeneration of, say, NS-R type, then after separating the two
branches to make a smooth surface (of lower genus or with more components) a puncture of NS-R 
type will appear on each branch.

 An important fact carries over
from the analysis of gluing parameters in section \ref{thed}.  An NS-NS 
degeneration has only its holomorphic and antiholomorphic
gluing parameters $q_\NS$ and $\tilde q_\NS$ (which are complex conjugates 
modulo the odd variables).  But an NS-R or R-NS puncture has,
in the same sense described in section \ref{thed}, a single fermionic gluing parameter. 
Integration over this gluing parameter causes
the propagator of an R-NS or NS-R string state to be Dirac-like.   And an R-R degeneration has a pair of
fermionic gluing parameters, one holomorphic and one antiholomorphic.  Integrating 
over these parameters has an effect that for massless
R-R fields can be described as follows (see, for example, \cite{Witten}, section 6.2.2). 
These fields are $p$-form fields in spacetime 
and the integration over the fermionic gluing parameters
ensures that in superstring perturbation theory, they couple only via their field strength.
                                           
The sign choice analyzed in section \ref{relgso} occurs separately for holomorphic 
and antiholomorphic variables, and leads to separate GSO projections
for holomorphic and antiholomorphic modes of the string. 

\subsection{The Canonical Parameters}\label{canonpar}

\subsubsection{Overview}\label{zongor}

The most delicate issues in superstring perturbation theory concern the behavior in the infrared region, near infinity in moduli space.  A key fact
in resolving them is that moduli space $\MM$ can be compactified by adding suitable divisors with a natural physical interpretation.

By definition a divisor is defined locally by vanishing of a single even parameter.  For an NS degeneration, this is the parameter $\varepsilon$
in  eqn. (\ref{ondo}), and for a Ramond degeneration, it is the parameter $q_\Ra$ in eqn. (\ref{tormo}).

We call $\varepsilon$ and $q_\Ra$ the canonical parameters, but they actually are properly understood not as complex-valued parameters
but as sections of certain line bundles, the duals of the normal bundles to the relevant compactification divisors  in the compactified moduli space $\h\MM$.
This was analyzed in the bosonic case in section \ref{normal}, and we describe the superanalog in section \ref{elgso}.  

Picking local superconformal coordinates trivializes the relevant line bundles and then, as in eqn. (\ref{ondo}) or (\ref{tormo}), $\varepsilon$ or $q_\Ra$
is a natural complex-valued parameter.   Under a change of trivialization, one would have $\varepsilon\to e^\phi\varepsilon$ for some function $\phi$
(and similarly for $q_\Ra$), but
not, for example, $\varepsilon\to \varepsilon+\eta_1\eta_2$ with odd moduli $\eta_1$ and $\eta_2$.    

The existence in this sense of a canonical parameter at infinity is key to resolving some questions that arose in the literature of the 1980's on
superstring perturbation theory.  It makes well-defined some integrals that are invariant under a change of variables $\varepsilon\to e^\phi\varepsilon$,
but would shift by a surface term if one were allowed to make
a change of variables $\varepsilon\to \varepsilon+\eta_1\eta_2$.   See \cite{Witten}.

\subsubsection{Example In Genus Zero}\label{ongo}

To gain experience, we will compute explicitly the canonical parameter in a simple special case.  Let $\Sigma$ be a super Riemann surface and
consider the degeneration in which two NS punctures approach each other.  In local superconformal coordinates $z|\theta$ on $\Sigma$,
we take the punctures to be at $z|\theta=z_1|\theta_1$ and $z|\theta=z_2|\theta_2$.  

We want to determine the canonical parameter for $z_1|\theta_1\to z_2|\theta_2$.  Even without any computation, one may suspect
that  $\varepsilon$ will be a function of the combination $z_1-z_2-\theta_1\theta_2$, which is invariant under the superconformal
transformation $z|\theta\to z-\alpha\theta|\theta+\alpha$.

\begin{figure}
 \begin{center}
   \includegraphics[width=4.5in]{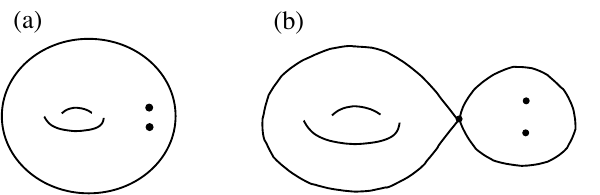}
 \end{center}
\caption{\small (a) A super Riemann surface $\Sigma$ with 2 nearby NS punctures labeled $p_1$ and $p_2$. (b) At infinity, $\Sigma$ 
splits as the union of a super Riemann
surface $\Sigma'$ that is a copy of $\Sigma$ with one less puncture, and a genus 0 component $\Sigma''$ that has three NS punctures,
namely $p_1$, $p_2$, and a third one at which $\Sigma''$ is glued to $\Sigma'$.}
 \label{depicted}
\end{figure}

$\Sigma$ has 2 even moduli and 2 odd moduli that will be relevant in this discussion, namely $z_1,z_2$ and $\theta_1,\theta_2$.  
At infinity, $\Sigma$ splits into a union of 2 components $\Sigma'$ and $\Sigma''$,
where $\Sigma'$ is simply $\Sigma$ with just one NS puncture that is relevant to our discussion (namely the one at which it is glued to $\Sigma''$; there may be some irrelevant ones far away), and $\Sigma''$ is a copy of $\CP^{1|1}$ with three NS punctures (counting
the node at which it is glued to $\Sigma'$).  This is depicted in fig. \ref{depicted}.  In terms of the gluing of $\Sigma'$ and $\Sigma''$, we count parameters as follows:
the one NS puncture of $\Sigma'$ depends on 1 odd and 1 even parameter; there will be 1 even gluing parameter $\varepsilon$; and $\CP^{1|1}$ with
3 NS punctures has 1 odd modulus.  Altogether the number of parameters is again $2|2$.

We describe $\Sigma''=\CP^{1|1}$ by superconformal coordinates $y|\psi$ (this description omits a divisor at $y=\infty$) and locate
the two punctures as follows:
\begin{align}\label{tobz} z|\theta=z_1|\theta_1 & ~\leftrightarrow ~y|\psi=1|0 \cr z|\theta= z_2|\theta_2 &~\leftrightarrow~ y|\psi = -1|0.\end{align}
 We glue $\Sigma'$ to $\Sigma''$ by identifying
the points $z|\theta=z_0|\theta_0$ with $y|\psi=0|\beta$, where $z_0|\psi_0$ are the moduli of the node in $\Sigma'$ and $\beta$ is the odd modulus
of $\Sigma''$.  The gluing formulas are obtained from (\ref{ondo}) upon replacing $x|\theta$ with superconformal coordinates
$z-z_0+\theta_0\theta|\theta-\theta_0$ and replacing $y|\psi$ with superconformal coordinates $y+\beta\psi|\psi-\beta$:
\begin{align}\label{zobz} (z-z_0+\theta_0\theta)(y+\beta\psi)& =-\varepsilon^2 \cr
                                         (y+\beta\psi)(\theta-\theta_0)& = \varepsilon (\psi-\beta)\cr
                                          (z-z_0+\theta_0\theta)(\psi-\beta)&=-\varepsilon(\theta-\theta_0)\cr
                                           (\theta-\theta_0)(\psi-\beta)&=0.\end{align}
                                           
In view of (\ref{tobz}), we determine $z_1|\theta_1$ by setting $y|\psi=1|0$ in (\ref{zobz}):
\begin{align}\label{jobz}z_1& = z_0-\varepsilon^2+\theta_0\varepsilon \beta \cr
                                      \theta_1&=\theta_0-\varepsilon\beta . \end{align}                                                                                      
                          Similarly, setting $y|\psi=-1|0$, we get                 
 \begin{align}\label{jobzo}z_2& = z_0+\varepsilon^2-\theta_0\varepsilon \beta \cr
                                      \theta_2&=\theta_0+\varepsilon\beta  .\end{align}   
                                      
Eliminating $z_0|\theta_0$ and solving for $\varepsilon$ in terms of $z_1|\theta_1$ and $z_2|\theta_2$, we get the expected result, up to a constant factor:
\begin{equation}\label{yffy}\varepsilon^2=-\frac{1}{2}\left(z_1-z_2-\theta_1\theta_2\right). \end{equation}          
This shows that $z_1-z_2-\theta_1\theta_2$ (or rather its square root, to account for the two ways of gluing the spin structures) is
the good parameter at infinity.  For example, this accounts for why in section \ref{zddn} the divisors $\eusm D_{ij}$ at infinity
were defined by $z_i-z_j-\theta_i\theta_j=0$.       

The factor $-1/2$ in (\ref{yffy}) has no particular significance, since it can certainly be removed
by $\varepsilon\to e^\phi\varepsilon$; what is significant is the combination $z_1-z_2-\theta_1\theta_2$.                

\subsubsection{Comparing The Different Types Of Degeneration}\label{ormal}

Before proceeding to a technical analysis of normal bundles, we will explain what
sort of information is actually most useful for superstring perturbation theory.

Let us compare the different types of oriented closed string degeneration.  For ordinary Riemann surfaces, we have the familiar
\begin{equation}\label{ozno} xy=q.\end{equation}
For Ramond degenerations of super Riemann surfaces, we have a minimal superextension of this
\begin{align}\label{oznox}xy& =q_\Ra \cr \theta& = \pm \sqrt{-1}\psi .\end{align}
  And the NS degeneration is more elaborate: 
\begin{align}\label{ondorp}  xy & = -\varepsilon^2\equiv q_\NS \cr
                                           y\theta & = \varepsilon\psi \cr
                                            x\psi & =-\varepsilon\theta\cr
                                             \theta\psi&=0. \end{align}
                                             
These formulas have an obvious similarity, and in a certain sense, modulo the odd variables,
 the supersymmetric degenerations (\ref{oznox}) and (\ref{ondorp}) reduce
to the bosonic degeneration (\ref{ozno}) with $q=q_\Ra$ or $q=q_\NS$.
The simplest explanation of what this means is just the following.  If $\SIgma$ is a super Riemann surface with a degeneration that in local coordinates
is described by (\ref{oznox}) or (\ref{ondorp}), and we use the reduced set of local coordinates on  the reduced space $\Sigma_\red$, then $\Sigma_\red$
 has a degeneration described by (\ref{ozno}) with $q=q_\Ra$ or $q=q_\NS$. 
 
It is better to give a formulation that does not depend on local coordinates.  For this, we should focus not on the parameters $q$, $q_\Ra$, and $\varepsilon$
but on the line bundles in which they take values.   These line bundles are dual to the normal bundles to the appropriate compactification divisors; the relevant line bundle was analyzed in the bosonic
case in section \ref{normal}.    Let us write $\eusm D$ for a compactification divisor in the bosonic case and $\eusm D_\Ra$ and $\eusm D_\NS$ for its two types of supersymmetric
cousin.  To be able to sensibly compare $q$, $q_\Ra$, and $\varepsilon$, we need a relation between the normal bundles $\eusmn_{\eusm D}$, $\eusmn_{\eusm D_\Ra}$, and $\eusmn_{\eusm D_\NS}$.
Since these are line bundles over different spaces, how can we compare them?    The answer is that the reduced spaces $\eusm D_{\Ra,\red}$ and $\eusm D_{\NS,\red}$
are locally isomorphic to $\eusm D$.  $\eusm D$ parametrizes a pair of Riemann surfaces joined at a point (or a Riemann surface with two points joined together) with some
additional punctures, while
$\eusm D_{\Ra,\red}$ and $\eusm D_{\NS,\red}$ describe the same data together with some discrete information (the labeling of the     punctures and a generalized spin structure)
that does not affect the analysis of the normal bundle.   If we simplify the notation by ignoring the extra discrete information, then the relation between the normal bundles is
that $\eusmn_{\eusm D_{\Ra}}$ when restricted to $\eusm D_{\Ra,\red}$ is naturally isomorphic to $\eusmn_{\eusm D}$
\begin{equation}\label{motto} \eusmn_{\eusm D_{\Ra}}|_{\eusm D_{\Ra,\red}}\cong \eusmn_{\eusm D},\end{equation}      
and similarly in the NS case                                 
\begin{equation}\label{mottop} \eusmn_{\eusm D_{\NS}}^2|_{\eusm D_{\NS,\red}}\cong \eusmn_{\eusm D}.\end{equation}     
These isomorphisms identify $q_\Ra$ or $q_\NS$ with $q$.

\subsubsection{Normal Bundle At Infinity: NS Case}\label{elgso}

Now we turn to the analysis of  the normal bundles to the compactification divisors in $\h\MM_{\sg,\sn_\NS,\sn_\Ra}$, or
equivalently, the global meaning of the gluing parameters.  
For brevity in the
notation, we consider divisors  that parametrize separating degenerations. All statements have immediate analogs for the
nonseparating case.
A super Riemann surface $\Sigma$ can degenerate to a pair of components $\Sigma'$ and $\Sigma''$ joined at a pair of either NS or R punctures. 
To distinguish the two cases, we will denote the divisors in question as $\eusm D_\NS$ and $\eusm D_\Ra$.
In the NS case, which we consider first, $\eusm D_\NS$ is the product of the moduli spaces that parametrize $\Sigma'$ and $\SIgma''$ with their 
punctures.  To minimize notation,
we write just $\eusm D_\NS=\h\MM(\Sigma')\times\h\MM(\SIgma'')$.

We will need the super analog of a definition explained at the beginning of section \ref{normal}.
A fundamental line bundle on a super Riemann surface $\Sigma$ is its Berezinian $\BBer(\Sigma)$.   
It has rank $0|1$; that is, its fibers are fermionic.
Consider the moduli space $\h\MM_{\sg,\sn_\NS,\sn_\Ra}$ that parametrizes a super Riemann surface $\Sigma$ 
with the given number of NS and R punctures.  
The NS punctures are simply points in $\Sigma$.  Taking the fiber of $\BBer(\Sigma)$ at the $\sigma^{th}$ NS 
puncture, for $\sigma=1,\dots,\n_\NS$, we get
a line bundle $\frakl_\sigma\to \h\MM_{\sg,\sn_\NS,\sn_\Ra}$, again with fermionic fibers.   (The $\frakl_\sigma$ do not quite have a close analog for Ramond punctures.)

The relation between $\frakl_\sigma$ and the corresponding line bundle $\L_\sigma$ that we defined in the bosonic case in section \ref{normal} is that
if we restrict to the reduced space of $\MM_{\sg,\sn_\NS,\sn_\Ra}$ and pull $\L_\sigma$ back from $\M_{\sg,\sn}$ to $\M_{\sg,\sn,\spin}$, then
\begin{equation}\label{trox}\frakl_\sigma^2|_{\MM_{\sg,\sn_\NS,\sn_\Ra,\red}}\cong \L_\sigma.\end{equation}
This just reflects the fact that if $\Sigma$ is split, then $\BBer(\Sigma)$, when restricted to $\Sigma_\red$, is $K_{\SIgma_\red}^{1/2}$, while $\L_\sigma$ is defined
using $K_\Sigma$.  Eqn. (\ref{trox}) will imply the desired
result (\ref{mottop}), once we express the normal bundle $\eusmn_{\eusm D_\NS}$ in
terms of the $\frakl$'s (eqn. (\ref{dolbo})).

Now let $x|\theta$ and $y|\psi$ be local superconformal coordinates on $\Sigma'$ and 
$\Sigma''$.  The gluing of $\Sigma'$ and $\Sigma''$ at general points 
$x|\theta=a|\alpha$ and $y|\psi=b|\beta$, and smoothing to make $\Sigma$, is described by a slight generalization of eqn. (\ref{ondo}): 
\begin{align}\label{blondo} (x-a+\alpha\theta)(y-b+\beta\psi)& =-\varepsilon^2 \cr
                                         (y-b+\beta\psi)(\theta-\alpha) & = \varepsilon(\psi-\beta) \cr
                                          (x-a+\alpha\theta)(\psi-\beta)&= -\varepsilon(\theta-\alpha)\cr
                                                              (\theta-\alpha)(\psi-\beta)& = 0.\end{align}
We have simply written (\ref{ondo}) in superconformal coordinates $ x-a+\alpha\theta|\theta-\alpha$   
and $y-b+\beta\psi|\psi-\beta$.
We view the even parameters $a,b,\varepsilon$ and the odd parameters $\alpha,\beta$ as  moduli of $\Sigma$. As in 
the the introductory remarks above, we write $\frakl'
\to \h\MM(\SIgma')$ for the line bundle whose fiber  at the point in $\hat\MM(\Sigma')$  corresponding to $\Sigma'$ 
is the fiber of $\BBer(\Sigma')$ at $x|\theta=a|\alpha$.  
Similarly, we write $\frakl''\to \h\MM(\SIgma'')$ for the line bundle defined by the fiber of $\BBer(\Sigma'')$ at $y|\psi=b|\beta$.

Describing the gluing by the explicit equation (\ref{blondo}) has required picking explicit superconformal coordinates 
$x-a+\alpha\theta|\theta-\alpha$ and $y-b+\beta\psi|\psi-\beta$
near, respectively,  $x|\theta=a|\alpha$ and $y|\psi=b|\beta$.         The normalization of $\varepsilon$ depends on 
this choice, and therefore, just as in section \ref{normal},
$\varepsilon$ is best understood as a section of a line bundle over $\eusm D_\NS$.   We claim that this line bundle is 
$\frakl'\otimes \frakl''$.  Equivalently, since $\varepsilon$ is
a linear function on the normal bundle $\eusmn_{\eusm D_\NS}$ to $\eusm D_\NS\subset\h\MM_{\sg,\sn_\NS,\sn_\Ra}$, we claim
\begin{equation}\label{dolbo}\eusmn_{\eusm D_\NS}\cong (\frakl'\otimes \frakl'')^{-1}.\end{equation}

To justify these claims, we observe that
under a change of superconformal coordinates near $x|\theta=a|\alpha$ , $x-a+\alpha\theta$ transforms
like $\d a -\alpha\d\alpha$.  (The basic case to consider is that $a|\alpha=0|0$ and the superconformal
change of coordinates is $x|\theta\to\lambda x|\lambda^{1/2}\theta$, with $\lambda\in \C$. Under this transformation, $x-a+\alpha\theta$
and $\d a-\alpha\d\alpha$ are both multiplied by $\lambda$.)  In view of the
relation of the line bundle $(\frakl')^2$ on $\h\MM(\SIgma')$ to 
the line bundle $\D^{-2}\cong \BBer(\Sigma')^2$ on 
$\SIgma'$,  and the interpretation (\ref{amurk}) of $\d a-\alpha\d\alpha$ as a section of $\D^{-2}$,
this means that $x-a+\alpha\theta$ can be interpreted near $\eusm D_\NS$ as  a section of $(\frakl')^2$.  Similarly 
$y-b-\beta\theta$ should
be interpreted 
as a section of $(\frakl'')^2$.  So from the
first equation in (\ref{blondo}), $\varepsilon^2$ should be interpreted as a section of $(\frakl'\otimes \frakl'')^2$.  The fact 
that $\varepsilon$ is  a section of $\frakl'\otimes \frakl''$
(rather than a more general square root of $(\frakl'\otimes \frakl'')^2$) follows from a similar consideration of the second 
and third equations in (\ref{blondo}).                                                

\subsubsection{The Ramond Case}\label{ramca}

Now let us consider the case of a Ramond degeneration, where Ramond punctures in $\Sigma'$ and $\Sigma''$ are 
glued and smoothed to make $\Sigma$.  The Ramond punctures really correspond to divisors $\F'\subset \Sigma'$ and
$\F''\subset \Sigma''$.  We pick local coordinates $x|\theta$ on $\Sigma'$ and $y|\psi$ on $\Sigma''$ such that  the superconformal
structures are defined by the vector fields $D^*_\theta$ and $D^*_\psi$ of eqn.  (\ref{omy}).  
The divisors $\F'$ and $\F''$ are defined respectively by $x=0$ and by $y=0$, and
the gluing is as in (\ref{tormo}):
\begin{align}\label{xormo} xy & = q_\Ra\cr    \theta&=\pm \sqrt{-1}\psi.\end{align}
The gluing of $\F'$ and $\F''$ by the second formula here gives a supermanifold of dimension $0|1$ that we will call $\F^\diamond$; it is isomorphic
to either $\F'$ or $\F''$.
The first equation makes it clear that in some sense $q_\Ra$ transforms as $\d x\otimes \d y$.  

In general, $\d x$ and $\d y$ take values in $T^*\Sigma$, which has rank $1|1$. Also, the gluing in (\ref{xormo}) happens at a divisor, not a point.
So we need more information if we want to identify
a vector space of dimension $1|0$ in which $q_\Ra$ takes values.  A splitting of $\Sigma$ gives one simple answer; if $\Sigma$ is split,
we can choose  local parameters $x$ and $y$ that are even under the $\Z_2$ symmetry defined by the splitting, and interpret $\d x$ and $\d y$ as cotangent vectors
to $\Sigma_\red$.  This is the same interpretation of $q_\Ra$ that we had in the bosonic case, so we  arrive at (\ref{motto}).

If $\Sigma$ is not split, we have to work harder to describe a $1|0$-dimensional vector space in which $q_\Ra$ takes values.
 $T^*\SIgma'$, when restricted to $\F'$, has a rank 1 subbundle $\U'$ consisting of
1-forms whose contraction with $D^*_\theta$ vanishes; similarly, $T^*\SIgma''$, when restricted to $\F''$, has a rank 1 subbundle
$\U''$ consisting of 1-forms whose contraction with $D^*_\psi$ vanishes.  Since $\d x$ and $\d y$ have vanishing contraction
with $D^*_\theta$ and $D^*_\psi$ at $x=y=0$,  they are naturally understood as sections of $\U'$ and $\U''$, respectively.
 So $\d x\otimes \d y$ is a section of $\U'\otimes \U''$.
 
 In general, a section of a line bundle over a supermanifold of dimension $0|1$, parametrized by an odd variable $\eta$,
 can be written $a+b\eta$, so such sections form a vector space of dimension $1|1$.  Taking the supermanifold to be $\F^\diamond$ and the line
 bundle to be $\U'\otimes \U''$, we would say based on this that $q_\Ra$ takes values in a vector space of dimension $1|1$. To reduce to a vector space of dimension $1|0$,
 we define the vector field $W=D^*_\theta\mp \sqrt{-1}D^*_\psi$. 
The object $\d x\otimes \d y$
 lies in a $1|0$-dimensional space $\U^*$ of sections of $\U'\otimes \U''\to \F^\diamond$ whose Lie derivative with respect to $W$ vanishes at $x=y=0$.
(Recall that the Lie derivative of a form $\omega$ with respect to a vector field $W$ is $\mathfrak L_W\omega =(\i_W\d+\d\i_W)\omega$, where $\i_W$ is contraction
with respect to $W$.  The Lie derivative of the symmetric tensor product $\d x\otimes \d y$ is $(\mathfrak L_W \d x)\otimes \d y+\d x\otimes (\mathfrak \L_W \d y)$.
 After setting $\psi=\mp \sqrt {-1}\theta$, we get $W=2\partial_\theta+\theta(x\partial_x-y\partial_y)$, and because of the minus
sign,  $\mathfrak L_W(\d x\otimes \d y)=0$, while $\mathfrak L_W(\theta \d x\otimes \d y)\not=0$.) 

$\U^*$ is the fiber of a line bundle  over the divisor $\eusm D_\Ra$
that parametrizes a Ramond degeneration. We just call this line bundle $\U^*$.
 We can interpret $q_\Ra$ as taking values in $\U^*$; the normal bundle to
$\eusm D_\Ra$ in $\h\MM$ is $\eusmn_\Ra=(\U^*)^{-1}$.

\subsection{Application To The Dimension Of Supermoduli Space}\label{applicdim}

\begin{figure}
 \begin{center}
   \includegraphics[width=5.5in]{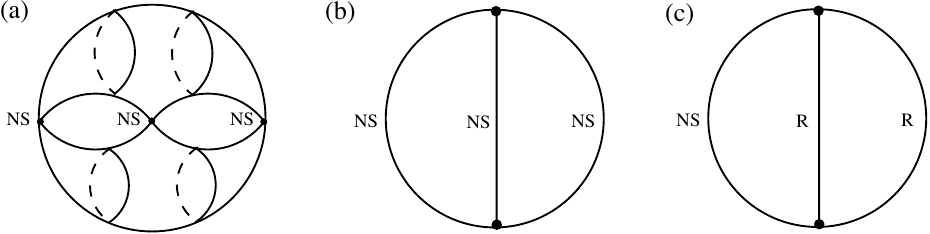}
 \end{center}
\caption{\small  In (a), we sketch the $\g=2$ case of a construction described in the text:
the gluing of two 3-punctured spheres (with NS punctures only) along 3 double points where they
meet.  Smoothing the double points into thin tubes that join at 3-holed (rather than 3-punctured) spheres
makes  a super Riemann surface of genus 2.  Sketched in (b) is a dual description of the same gluing.  We make a trivalent graph by collapsing each 3-holed sphere in (a) to a vertex
and expanding each double point where gluing takes place in (a) to a line between 2 vertices.  To get from the trivalent graph in (b) to
a Riemann surface, one thickens each line to a thin tube and each vertex to a 3-holed sphere where the tubes join smoothly.  (c) shows in the
same dual language another way to make a super Riemann surface of genus 2, this time with 2 $\NSthree$ vertices replaced by $\NSRR$ vertices. In (a) or (b),
we get 2 odd moduli from the vertices, while in (c), 2 odd moduli come from the lines labeled R. }
 \label{firstglue}
\end{figure}
Now that we understand how gluing works, we can give a new computation of the dimension of supermoduli space.  We will build
a super  Riemann surface $\Sigma$ of genus $\g$ by gluing some minimal building blocks -- spheres with 3 punctures.  
Since the number of Ramond punctures in a compact super Riemann surface
is always even, there are two possibilities: a sphere with 3 NS punctures and a moduli space
of dimension $0|1$, as
described in section \ref{addns}, and a sphere with 1 NS puncture, 2 Ramond punctures, and a moduli space of dimension $0|0$,
as described in section \ref{tryon}.  We will  refer to the first case as a 3-punctured sphere of type $\NSthree$, and the second case as a 3-punctured
sphere of  type $\NSRR$.  These objects are special because they have no bosonic moduli; a super Riemann surface built by gluing
together such objects has no bosonic moduli and hence cannot degenerate further.

First we  build a genus $\g$ super Riemann surface $\Sigma$ without punctures.  
One way to do so is  to start
with $2\g-2$  thrice-punctured spheres $\SIgma_1,\dots,\Sigma_{2\g-2}$, each of type $\NSthree$.  To build $\Sigma$, one 
glues together the
$\Sigma_i$ along their punctures to make a degenerate genus $\g$ curve with a total of $3\g-3$ double points, as shown 
in fig. \ref{firstglue}.  
(Each of the $\Sigma_i$ has 3 punctures, so altogether there are
 $3(2\g-2)$ punctures that one glues together pairwise at $3\g-3$ double points to build $\Sigma$. For what
 we are about to say, it does not matter how one pairs up the $3(2\g-2)$ punctures.)
Each of the $\Sigma_i$ has 1 odd modulus and no even ones, prior to gluing, making $2\g-2$ odd moduli in all.  Gluing and then smoothing the double points
adds 1 even gluing parameter and no odd ones at each
double point, making $3\g-3$ even parameters. So altogether  the dimension of the moduli space is $3\g-3|2\g-2$.

More generally, one can build a super Riemann surface of the same genus by taking $2\g-2-k$ of the $\Sigma_i$ to be of  type $\NSthree$  and $k$ to be of type $\NSRR$.  In the starting point, one has now only $2\g-2-k$ odd moduli, one for each 3-punctured sphere of 
type $\NSthree$.  But now one will have to glue $k$ pairs of Ramond punctures, and this will give back $k$ odd moduli
from the gluing parameters.
So one ends up with the same dimension for the moduli space. For an example, compare figs. \ref{firstglue}(b) and (c).

\begin{figure}
 \begin{center}
   \includegraphics[width=4.5in]{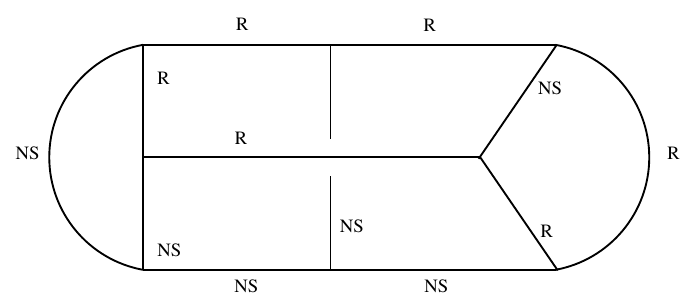}
 \end{center}
\caption{\small 
A trivalent graph with 5 loops.  Each line is labeled NS or R.
There are two kinds of vertex: an $\NSthree$ vertex that is the junction of  three NS lines, or an $\NSRR$ vertex
 that is the junction of an NS line
and 2 R lines.  From such a graph, one makes a super Riemann surface $\Sigma$ of genus 5 by thickening the lines
to tubes and the vertices to 3-holed spheres.  In this example, there are 2 $\NSthree$ vertices, each contributing $0|1$ to the dimension
of the supermoduli space, 6 NS lines, each contributing $1|0$, and 6 R lines, each contributing $1|1$.  So the dimension of the moduli
space of super Riemann surfaces of genus 5 is $12|8=3\g-3|2\g-2$.}
 \label{secondglue}
\end{figure}
As explained in figs. \ref{firstglue} and \ref{secondglue}, these gluing operations can be represented by trivalent graphs in which a vertex
represents a  3-punctured sphere, while  an edge or line between 2 vertices represents the gluing
of 2 punctures. (A line connecting a vertex to itself is also allowed.)
 Each line is labeled NS or R and represents the gluing of  NS or R punctures, respectively.
There are 2 types of vertex: $\NSthree$ vertices, at which 3 NS lines meet,
and $\NSRR$ vertices, at which 1 NS line and 2 R lines meet.   They represent the two kinds of 3-punctured sphere.
Given such a graph, one makes a topological type of super Riemann surface $\Sigma$
 by thickening the lines into tubes and the vertices into 3-holed spheres at which the tubes join.  
 (The spin structure of $\Sigma$ depends on choices made in the gluing.)
 The counting of moduli of $\Sigma$ is as follows:  the contribution of an $\NSthree$ vertex is $0|1$; the contribution from an $\NSRR$ vertex is $0|0$;
the contribution from an NS line is $1|0$; and the contribution from an R line is $1|1$.  A super Riemann surface of genus $\g$ can be represented
by any trivalent vertex constructed from these ingredients; irrrespective of the choice of the graph or its labeling, one always arrives
at the same result $3\g-3|2\g-2$ for the dimension of the moduli space.

\begin{figure}
 \begin{center}
   \includegraphics[width=5.5in]{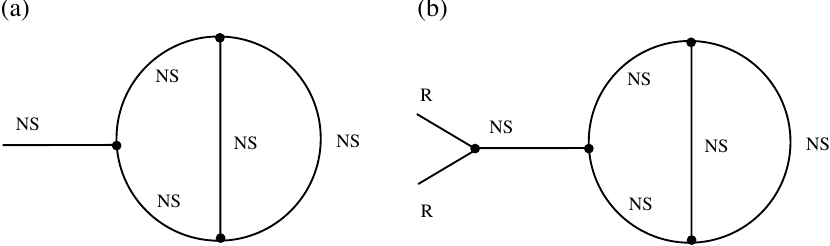}
 \end{center}
\caption{\small  
(a) A thickening of this graph gives a super Riemann surface of genus 2 with 1 NS puncture, corresponding to the external line in the graph.
The dimension of the moduli space is $4|3$, where we count $1|0$ for each internal NS line and $0|1$ for each $\NSthree$ vertex. (b) A thickening
of this graph gives a super Riemann surface of genus 2 with 2 R punctures, corresponding to the external lines.   The dimension of
the moduli space is $5|3$, with no contribution from the $\NSRR$ vertex.}
 \label{thirdglue}
\end{figure}

To include punctures in this description, we simply consider trivalent graphs with external lines. (An external line is a line that
has only one end attached to a vertex; the other end represents the puncture.  See fig. \ref{thirdglue} for simple
examples of graphs with external lines.)  A puncture of NS or R type is represented
by an external line labeled NS or R.  
In the presence of punctures, the dimension of the moduli space is computed by summing the contributions of vertices and internal lines
only.  The sum gives back the usual dimension formula (\ref{gendim}).

Graphs such as those we have drawn have an obvious analogy with Feynman diagrams 
in a field theory with cubic interactions (in fact, $(\NS)^3$
and $\NS\cdot \Ra^2$ interactions).  This analogy
is no coincidence; it is part of the mechanism by which string theory turns out to reproduce field 
theory at long distances.  In field theory, the ``leading
singularity'' of a perturbative scattering amplitude (see \cite{bcf} for a modern explanation and application)
is found by putting a maximal possible number of internal propagators on-shell.  In superstring theory,
this corresponds to the contribution of a maximally degenerate super Riemann surface, 
built by gluing 3-punctured spheres.  After picking a configuration of
3-punctured spheres, to compute the leading singularity, one still has to adjust loop momenta 
so that all momenta passing through nodes are on-shell.  In string theory, this 
will lead to an infinite sum over all the massive string states that can
flow through the various nodes.

\section{Open And/Or Unoriented Superstring Theories}\label{oundor}

So far, we have  considered only oriented superstring worldsheets without boundary.  These are appropriate for constructing perturbation
theory for oriented closed superstrings -- that is, for the heterotic and Type II superstring theories (without D-branes
or orientifold planes  in the case of Type II).

Type I superstring theory and more general orientifolds of Type II superstring theory are theories of unoriented strings.  In a theory of unoriented
strings, the string worldsheet is an unoriented and not necessarily orientable super Riemann surface.

Open as well as closed strings appear in Type I superstring theory and more generally in Type II superstring theory in the presence of D-branes. 
When open strings are present, the string worldsheet is a super Riemann surface with boundary.  Thus, in Type I superstring theory and in more general Type II
backgrounds with both orientifolds and D-branes, we must consider super Riemann surfaces that may be simultaneously unorientable and with boundary.

In this section, we explain what is required for these generalizations.
For brevity, we refer to a Riemann surface with boundary as an open Riemann surface.  A superstring worldsheet
is open or unoriented or unorientable if its reduced space is open or unoriented or unorientable.  
A  Riemann surface or superstring worldsheet without boundary is  said to be closed.

\subsection{The Classical Picture}\label{oundbor}

\subsubsection{The Closed Oriented Double Cover}\label{orco}

By an unoriented Riemann surface, we mean simply a two-manifold with a conformal structure but no choice of orientation.

Suppose that $\Sigma_0$ is an unoriented closed Riemann surface.  Then $\Sigma_0$ has a double cover $\Sigma_0'$ that
is  a closed  oriented Riemann surface.  A point in $\Sigma'_0$ is a point in $\Sigma_0$ together with a choice of orientation 
at that point.

If   
$\Sigma_0$ is orientable (but unoriented), this definition makes sense but is not terribly interesting.  In that case, $\Sigma'_0$ is just the disjoint union
of two oppositely oriented copies of $\Sigma_0$.  The interesting case is the case that $\Sigma_0$ is unorientable.
Then $\Sigma'_0$ is an ordinary connected and oriented Riemann surface.  A basic example is $\Sigma_0=\Bbb{RP}^2$.
The oriented double cover is $\Sigma'_0=S^2=\Bbb{CP}^1$.  The fact that, for a certain action of $\Z_2$ on $S^2$, the quotient
$S^2/\Z_2$ is $\Bbb{RP}^2$ is more or less the definition of $\Bbb{RP}^2$.

We can do something similar for the case that $\Sigma_0$ is an oriented Riemann surface with boundary.  Taking two
copies of $\Sigma_0$ with opposite orientation and gluing them together along their boundary, we produce a double cover
of $\Sigma_0$ (branched over its boundary) which is an oriented Riemann surface $\Sigma'_0$ without boundary.  A basic
example is that $\Sigma_0$ is the disc $|z|\leq 1$ in the complex plane.  Gluing two copies of $\Sigma_0$  along their boundary,
we build the double cover $\SIgma'_0=\CP^1$.

Finally, we can combine the two cases.  Suppose that $\Sigma_0$ is an unoriented Riemann surface with boundary. As before $\Sigma_0$
has an oriented double cover which we now call $\Sigma''_0$; a point in $\Sigma_0''$ is a point in $\Sigma_0$ with a choice of local
orientation.  The boundary of $\Sigma_0$ is a union of circles, each of which lifts in $\Sigma_0''$ to a pair of circles.  Gluing
together these pairs of circles by identifying corresponding points, 
we build a closed oriented Riemann surface that we call $\Sigma'_0$.  For an example, let $\Sigma_0$
be $\Bbb{RP}^2$ with an open disc removed, so that its boundary is a circle.  Then $\Sigma''_0$ is  a copy of $S^2$ with two open
balls removed; its boundary is a disjoint union of two circles.  Gluing these circles together, we build the closed oriented genus 1 Riemann surface
$\Sigma'_0$.

Let $\chi(X)$ denote the Euler characteristic of a space $X$.  In all instances of this construction,
\begin{equation}\label{chip}\chi(\SIgma'_0)= 2\chi(\Sigma_0).\end{equation} 
For example, if $\Sigma_0=\Bbb{RP}^2$, $\Sigma'_0=S^2$, we have $\chi(\Sigma_0)=1$, $\chi(\Sigma'_0)=2$.  
Alternatively, if $\Sigma_0$ is closed disc and $\Sigma'_0=S^2$, then again $\chi(\Sigma_0)=1$, $\chi(\Sigma'_0)=2$.  
As one last example, if $\Sigma_0$ is $\Bbb{RP}^2$ minus an open disc, and $\SIgma'_0$ is a closed Riemann surface of genus 1,
then $\chi(\Sigma_0)=\chi(\Sigma'_0)=0$.  

Eqn. (\ref{chip}) gives a convenient way to compute the genus $\g'$ of $\Sigma_0'$:
it obeys $1-\g'=\chi(\Sigma_0)$ or
\begin{equation}\label{obom}\g'=1-\chi(\Sigma_0).\end{equation}
The topological type of the oriented surface $\Sigma_0'$ is completely determined by $\g'$.  So all distinct types of $\Sigma_0$ with the same
Euler characteristic have closed oriented double covers that are topologically equivalent.

In all examples of this construction, $\Sigma_0'$  has 
 has an orientation-reversing (or equivalently, a complex structure reversing) symmetry $\rho$ of order 2
that exchanges pairs of points in $\Sigma_0'$ that correspond to the same point in $\Sigma_0$ with different local orientations.  
The fixed point set of $\rho$ consists of the circles in $\Sigma_0'$ 
that lie over the boundary of $\Sigma_0$.
$\rho$ generates a $\Z_2$ action on $\Sigma'_0$ and  $\Sigma_0=\SIgma'_0/\Z_2$.  A symmetry of order 2 is called an involution, so we call $\rho$ an
antiholomorphic involution of $\Sigma_0'$.  Since $\rho:\SIgma_0'\to
\Sigma_0'$ reverses the complex structure, we can think of it as an isomorphism between $\Sigma_0'$ and its complex conjugate $\bar\Sigma_0'$:
\begin{equation}\label{demof}\rho:\SIgma_0'\cong \bar\Sigma_0'.\end{equation}

\subsubsection{The Moduli Space}\label{themo}

The moduli space of an open and/or unorientable Riemann surface $\Sigma_0$ can be described as follows.

Suppose that the oriented double cover $\Sigma_0'$ has genus $\g'$.  If we ignore the relationship to $\Sigma_0$, then the deformations
of $\Sigma'_0$ are parametrized by the usual moduli space $\M_{\g'}$ of Riemann surfaces of genus $\g$.  

$\M_{\sg'}$ has an antiholomorphic involution $\varphi$ that maps the point in  $\M_{\sg'}$ corresponding to a surface $\Sigma_0'$ to the point
that corresponds to the complex conjugate $\bar\Sigma_0'$ of $\Sigma_0'$.  We usually describe this more briefly (but less precisely)  by saying that $\varphi$
maps $\Sigma_0'$ to $\bar\Sigma_0'$.  A fixed point of  $\varphi$ corresponds to a Riemann surface $\Sigma_0'$ such that $\bar\Sigma_0'$ is isomorphic
to $\Sigma_0'$.  So $\Sigma_0'$ corresponds to a fixed point if and only if there is an antiholomorphic involution:
\begin{equation}\label{zoboro}\rho:\SIgma_0'\cong \bar\Sigma_0'.\end{equation}

If $\rho$ exists, we define $\SIgma_0=\Sigma_0'/\Z_2$, where $\Z_2$ is generated by $\rho$.
Then $\Sigma_0$ is an open and/or unorientable Riemann surface, and $\Sigma_0'$ is its close oriented
double cover, as described in section \ref{orco}.
So the fixed point set $\M_{\sg'}^\varphi$ of $\varphi$ parametrizes surfaces $\Sigma_0'$ that are orientable double covers of some $\Sigma_0$.

The only constraint on $\Sigma_0$ is that its Euler characteristic must obey (\ref{obom}).   All topological types of open and/or unorientable Riemann surface
$\Sigma_0$ that obey this condition can arise. There are finitely many possibilities, parametrized by a set that we will call $ S$.
 So actually $\M_{\sg'}^\varphi$ is a union of finitely many components, one for each $\frak s\in S$:
 \begin{equation}\label{retrom}\M_{\sg'}^\varphi=\cup_{\frak s\in S}\Gamma_{\frak s}.\end{equation}
Here $\Gamma_{\frak s}$ is the moduli space of conformal structures on an open and/or unoriented Riemann surface $\Sigma_0$ of topological type $\frak s$.  

The union in (\ref{retrom}) is not a disjoint union; the $\Gamma_{\frak s}$ can intersect each other at points corresponding to surfaces $\Sigma_0'$ that 
admit more than one antiholomorphic involution.  These intersections are not important in superstring perturbation theory; one simply integrates over
one or more of the $\Gamma_{\frak s}$ (depending on which string theory one considers), ignoring the fact that they may intersect each other.

More important is the following.  $\M_{\sg'}$ is of course a noncompact manifold (or rather orbifold) of complex dimension $3\g'-3$.  As  a component of the fixed
point set of a real or antiholomorphic involution, $\Gamma_{\frak s}$ has real dimension $3\g'-3$.   It is noncompact, just llike $\M_{\sg'}$.  The closure of $\Gamma_{\frak s}$
in $\h\M_{\sg'}$ gives a compactification $\h\Gamma_{\frak s}$ of $\Gamma_{\frak s}$.  
In general, $\h\Gamma_{\frak s}$ is a manifold with boundary; typically, these manifolds with
boundary  join together pairwise on their common boundaries.  
This is relevant for string perturbation theory and will be described
in more detail in section \ref{undbe}.

\subsection{Open And/Or Unoriented  Superstring Worldsheets}\label{zordinary}

We now want to describe the analogous construction for open and/or unoriented superstring worldsheets.  It turns out that all the cases
can be treated together.  

The local structure of an open and/or unoriented superstring worldsheet will be the same as the local structure of a Type II superstring worldsheet,
as described in section \ref{tanalog}.  
However, it seems to be convenient to first describe the integration cycle $\varGamma_{\frak s}$ that in superstring perturbation theory
parametrizes open and/or unoriented superstring worldsheets of topological
type $\frak s$.     Here $\frak s$ is once again a topological type of open and/or unoriented
Riemann surface of Euler characteristic $1-\g'$.   Then we will describe the objects that $\varGamma_{\frak s}$  parametrizes.

\subsubsection{The Integration Cycle}\label{intcy}

We start with $\MM_{\sg'}$, the moduli space of closed oriented super Riemann surfaces of genus $\g'$.
Its reduced space is $\M_{\sg',\spin}$, which parametrizes ordinary Riemann surfaces of genus $\g'$ with a spin structure.  By forgetting the spin
structure, we get a projection $\pi:\M_{\sg',\spin}\to\M_{\sg'}$.  We define $\varGamma_{\frak s,\red}=\pi^{-1}(\Gamma_{\frak s})$, where as   in section 
\ref{themo}, $\Gamma_{\frak s}$ is the moduli space of conformal structures on an open and/or unoriented surface of topological type $\frak s$.  

$\varGamma_{\frak s,\red}$ parametrizes pairs\footnote{Such pairs have been considered recently in the context of superconformal field theory
on open and/or unoriented surfaces \cite{DFM}.}  consisting of an open and/or unoriented surface $\Sigma_0$ of topological type $\frak s$ and a spin structure
on the closed oriented double cover $\Sigma_0'$ of $\Sigma_0$. (The spin structure is not necessarily invariant under the
antiholomorphic involution of $\Sigma_0'$.)  It is middle-dimensional in $\MM_{\sg',\red}=\M_{\sg',\spin}$, and in particular it has
real dimension $3\g'-3$.  As usual in constructing integration cycles for superstring perturbation theory, 
we thicken $\varGamma_{\frak s,\red}$ in the fermionic directions to a smooth supermanifold
$\varGamma_{\frak s}\subset \MM_{\sg'}$ whose odd dimension is the same as that of $\M_{\sg'}$.  So $\varGamma_{\frak s}$ is of dimension
$3\g'-3|2\g'-2$.  $\varGamma_{\frak s}$ is the cycle over which we integrate to compute scattering amplitudes of open and/or unoriented supersymmetric strings
(or more precisely, the contribution to this scattering amplitude of surfaces of topological type $\frak s$). 
As in section \ref{colzo}, $\varGamma_{\frak s}$ is not defined uniquely but only up to homology -- that is, up to infinitesimal wiggling
of the fermionic directions.   

\subsubsection{The String Worldsheet}\label{ws}

Now we want to explain what sort of open and/or unoriented superstring worldsheets are parametrized by $\varGamma_{\frak s}$.
In a sense, we will construct an open and/or unoriented superstring worldsheet $\Sigma$ as an orientifold
(the quotient by an orientation-reversing orientation) of an ordinary Type II superstring worldsheet $\Sigma^*$.
As in section \ref{tanalog}, we begin with a product of super Riemann surfaces, but now we take these to be identical,\footnote{To follow the presentation
in sections \ref{colzo} and \ref{tanalog} more strictly, we should take the two factors in $\Y$ to be complex conjugate and then in the next paragraph
 embed  $\Sigma^*_\red$ in $\Y_\red$ by $x\to (x,\bar{\rho(x)})$ rather than $x\to (x,\rho(x))$.  The author finds the description given in the text to be less confusing.}
 so our starting point is
$\Y=\Sigma'\times \Sigma'$, where $\Sigma'$ is a holomorphic super Riemann surface of genus $\g'$.    Thus $\Y$ is a complex supermanifold of dimension $2|2$.
We define a holomorphic involution $\tau$ that exchanges the two factors in $\Y$.  

Suppose that $\Sigma'_{\red}$ is invariant under an antiholomorphic involution $\rho$. 
The automorphisms  $\rho$ and  $\tau$ of $\SIgma'_\red\times\SIgma'_\red$
generate two $\Z_2$ groups that we will call  $\Z_2^\rho$ and $ \Z_2^\tau$, respectively.
Let $\Sigma^*_\red$ be a copy of $\Sigma'_\red$ embedded
in $\Y_\red=\Sigma'_\red\times\Sigma'_\red$ by
\begin{equation}\label{zolx}x\to (x,\rho(x)).\end{equation}
$\tau$ acts on $\Sigma^*_\red$, mapping  $(x,\rho(x))$ to $(\rho(x),x)=(x',\rho(x'))$ where $x'=\rho(x)$.  So the action of $\tau$ on $\Y$ induces
the action of $\rho$ on $\Sigma^*_\red\cong \Sigma'_\red$:
\begin{equation}\label{olx} \Sigma^*_\red/\Z_2^\tau\cong \Sigma'_\red/\Z_2^\rho.\end{equation}

As usual, $\SIgma^*_\red$ can be thickened in the fermionic directions to give a smooth supermanifold $\Sigma^*\subset \Y$ of dimension $2|2$, 
in  a way
that is unique up to small wiggling in the fermionic directions.
Moreover, this can be done in a $\tau$-invariant fashion.  The supermanifold or rather superorbifold that we want
as the worldsheet of an open and/or unoriented string is $\Sigma=\Sigma^*/\Z^\tau_2$.
The reduced space of $\Sigma$ is $\Sigma'_\red/\Z_2^\rho$, in view of (\ref{olx}).  
  
The input to this construction was a super Riemann surface $\Sigma'$ whose reduced space is $\rho$-invariant.  A family of such objects
is parametrized by $\varGamma_\sf$.   
Given such an object, we have constructed  a smooth supermanifold (or, if $\rho$ has fixed points, a superorbifold)  $\Sigma$ of dimension $2|2$ parametrized by
$\varGamma_\sf$.   These are the superstring worldsheets parametrized by $\varGamma_\sf$.

The local structure of $\Sigma$, away from fixed points 
of the action of $\rho$ on $\Sigma'_\red$,
is the same as the local structure of an oriented closed Type II string worldsheet, as described in section \ref{tanalog}.
Locally, we can say that functions on the first factor in $\Y=\Sigma'\times\Sigma'$ restrict to antiholomorphic functions on 
$\Sigma$, and functions on the second
factor restrict to holomorphic functions.  Globally, the action of $\tau$ (or $\rho$) exchanges what we mean by 
``holomorphic'' and ``antiholomorphic,'' as one
expects for an open and/or unoriented worldsheet.    Similarly, locally $\Sigma$ has
antiholomorphic and holomorphic tangent bundles $T_L\Sigma$ and $T_R\Sigma$, with odd subbundles 
$\tilde\D$ and $\D$ that define its superconformal
structure; globally these are exchanged by $\tau$ or $\rho$.

As in sections \ref{colzo} and \ref{tanalog}, we can deform away from the choices that were made in this construction.
For example, we can move $\Sigma'$ slightly way from the locus where $\Sigma'_\red$ admits an antiholomorphic involution.  $\rho$ can then be deformed
(not uniquely) so that it still acts as an orientation-reversing involution, but it is no longer antiholomorphic.  The above construction still goes through, but 
holomorphic functions on $\Sigma_\red$ are no longer the complex conjugates of antiholomorphic ones.
This is analogous to what happens in sections \ref{colzo} and \ref{tanalog} if $\Sigma_{R,\red}$ is not the complex conjugate of $\Sigma_{L,\red}$.

What happens at the fixed points of $\rho$?
There is a general notion of a smooth supermanifold with boundary (for example, see section 3.5 of \cite{Wittennotes}).  But even if $\Sigma_\red$ has a boundary,
$\Sigma$ is not a smooth supermanifold with boundary by that definition; it is  a more general orbifold, a quotient by $\Z_2$ of an ordinary smooth
supermanifold $\Sigma^*$ without  boundary.  One can pick local coordinates $t^1,t^2|\theta^1,\theta^2$ on $\Sigma^*$ so that the $\Z_2$
action is $t^1,t^2|\theta^1,\theta^2\to t^1,-t^2|\theta^1,-\theta^2$.    For $\Sigma^*/\Z_2$ to satisfy the usual definition of a supermanifold with boundary,
the local model should be $t^1,t^2|\theta^1,\theta^2\to t^1,-t^2|\theta^1,\theta^2$, with a sign change only for one even coordinate and
no odd ones.   We will presently meet
supermanifolds with boundary in the conventional sense, namely the compactifications $\hat\varGamma_\sf$ of the integration cycles $\varGamma_\sf$.

\subsection{Punctures}\label{undbun}

Now let us discuss how this analysis generalizes in the presence of punctures.

\subsubsection{The Bosonic Case}\label{bosc}

We first incorporate punctures in the analysis of the classical moduli space in section \ref{themo}.
The moduli space $\M_{\sg',\sn}$ that parametrizes a Riemann surface $\Sigma_0'$ of genus $\g'$ with $\n$ punctures still
admits the antiholomorphic involution $\varphi$ that maps $\Sigma_0'$ to the complex conjugate space with the same punctures.
A fixed point of $\varphi$ now corresponds to a surface $\Sigma_0'$ with an antiholomorphic involution $\rho$ that maps the set of punctures
to itself.  This means that the punctures consist of pairs of points exchanged by $\rho$, or else fixed points of $\rho$.
If there are $\n_  0$ pairs and $\n_  1$ fixed points, then obviously
\begin{equation}\label{olfo} \n=2\n_  0+\n_  1. \end{equation}
In the quotient $\Sigma_0=\Sigma_0'/\Z_2$, pairs of punctures descend to interior points of $\Sigma_0$; we call these bulk punctures.
$\rho$-invariant punctures descend to boundary points of $\Sigma_0$; we call these boundary punctures.  In string theory, external closed strings
couple to bulk punctures and external open strings couple to boundary punctures.  

The components of the fixed point set $\M_{\sg',\sn}^\varphi$ are now labeled  by the topological type of $\Sigma_0$, the choice of nonnegative
integers $\n_  0$, $\n_  1$ obeying (\ref{olfo}), and the arrangement of the open-string punctures on the boundary components of $\Sigma_0$ (one must
specify which open-string puncture is on which boundary component and how they are cyclically ordered).  Let us write $S$
for the set of possible choices of all this data.  Then $\M_{\sg,\sn}^\varphi$ has a component $\Gamma_{\frak s}$ for every $\frak s\in S$ and
\begin{equation}\label{betro}\M^\varphi_{\sg',\sn}=\cup_{\frak s\in S}\Gamma_{\frak s}.\end{equation}
$\Gamma_{\frak s}$ parametrizes conformal structures on an open and/or unoriented surface $\Sigma_0$ with the indicated topology and set of punctures.

\subsubsection{Extension To Superstrings}\label{osc}

For the superstring analog, it is again easiest to first describe the integration cycle that parametrizes open and/or 
unoriented worldsheets with punctures,
and then to describe the worldsheets themselves.  We start with $\MM_{\sg',\sn_\NS,\sn_\Ra}$, the moduli space of super 
Riemann surfaces of genus $\g$
with the indicated numbers of NS and R punctures.  We set $\n=\n_\NS+\n_\Ra$.  The reduced space  $\MM_{\sg',\sn_\NS,\sn_\Ra,\red}$ parametrizes
a Riemann surface $\Sigma'_\red$ endowed with a spin structure and $\n$ punctures, each labeled as an NS or R puncture.

We want to focus on the case that $\Sigma'_\red$, forgetting its spin structures and the types of the punctures, admits an antiholomorphic involution $\rho$.
Topologically, there are many choices.  We must specify all the discrete data that were relevant in the bosonic case.  In addition, a $\rho$-invariant puncture
may be of NS or R type, so that there will be two types of open-string puncture,
 and a pair of punctures exchanged by $\rho$ may be of types NS-NS, NS-R, or R-R, so that there will be three types of closed-string puncture.  
 (For open and/or unoriented superstrings, there is no globally-defined distinction between NS-R and R-NS punctures.) Let $S^*$ be the whole collection
of discrete choices of the topology of $\Sigma'_\red/\Z_2^\rho$ and the types and arrangements of punctures.  

For every $\frak s\in S^*$, let $\varGamma_{\frak s,\red }$ be the subspace of $\MM_{\sg',\sn_\NS,\sn_\Ra,\red}$ defined by the condition that $\Sigma'_\red$ (ignoring its
spin structure) has an antiholomorphic involution of topological type $\frak s$.   In the usual fashion, we thicken $\varGamma_{\frak s,\red}$ to a smooth
supermanifold   $\varGamma_{\frak s}\subset \MM_{\sg',\sn_\NS,\sn_\Ra}$ with the same odd dimension as $\MM_{\sg',\sn_\NS,\sn_\Ra}$.  $\varGamma_{\frak s}$ is the parameter
space of open and/or unoriented super Riemann surfaces of type $\frak s$.

Moreover, the construction ensures, as usual, that the even dimension of  $\varGamma_{\frak s}$ is the complex dimension of  $\MM_{\sg',\sn_\NS,\sn_\Ra}$.
So one can write a dimension formula for $\varGamma_{\frak s}$.  It seems that the most useful way to record the information is to state
the contributions that different kinds of puncture make to the dimension of $\varGamma_{\frak s}$.
The contribution of an open-string puncture of NS or R type is
\begin{align}\label{vobost}\mathrm{NS}:&~~1|1.\cr
                                          \mathrm{R}:&~~1|\frac{1}{2}.\end{align} 
And the contribution of a closed-string puncture of one of the three types is
\begin{align}\label{zobost} \mathrm{NS-NS}:& ~~ 2|2 \cr
                                       \mathrm{NS-R}:&~~2|3/2\cr
                                        \mathrm{R-R}:&~~2|1. \end{align}
                                        
\subsubsection{The Worldsheet}\label{tws}

To construct the string worldsheets that are parametrized by $\varGamma_{\frak s}$, we basically repeat what has been said in section \ref{ws}
in the presence of punctures.  We start with the complex supermanifold $\Y=\Sigma'\times\SIgma'$, where $\SIgma'$ is now a super Riemann surface
of genus $\g'$ with $\n_  \NS$ Neveu-Schwarz and $\n_  \Ra$ Ramond punctures; we write $\tau$ for the involution of $\Y$ that exchanges the two factors.
We ask that $\Sigma'_\red$ should admit the action of an antiholomorphic involution $\rho$ of topological type $\frak s$, mapping the set of punctures to
itself (but perhaps permuting them).    We let $\Sigma^*_\red$ be
a copy of $\Sigma'_\red$, embedded in $\Y_\red$ by $x\to (x,\rho(x))$.  We thicken $\Sigma^*_\red$ to a smooth and $\tau$-invariant
 supermanifold $\Sigma^*\subset\Y$  of dimension $2|2$, and we define $\Sigma=\Sigma^*_\red/\Z_2^\tau$.   $\Sigma$ is an open and/or unorientable
 string worldsheet with punctures.  Its reduced space is $\Sigma'_\red/\Z_2^\rho$. 
                                        
\subsection{Compactification Of Moduli Spaces Of Open And/Or Unoriented Surfaces}\label{undbe}

\subsubsection{The Classical Case}\label{anex}

\begin{figure}
 \begin{center}
   \includegraphics[width=4.5in]{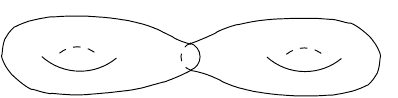}
 \end{center}
\caption{\small  
A closed oriented surface $\Sigma_0'$ that consists of two identical but oppositely oriented components joined at a narrow neck.  $\Sigma_0'$ has an
antiholomorphic involution that exchanges the two components.}
 \label{zandy}
\end{figure}
In section \ref{themo}, we defined the moduli space $\Gamma_{\frak s}$ that parametrizes conformal structures on an open and/or unoriented surface
$\Sigma_0$ of topological type $\frak s$.  $\Gamma_{\frak s}$ is a component of the fixed point set of the natural antiholomorphic involution 
$\varphi$ of $\M_{\sg'}$, the moduli space of Riemann surfaces of genus $\g'$.    We also remarked that $\Gamma_{\frak s}$ can be compactified
by taking its closure in $\hat\M_{\sg'}$.  We write $\hat\Gamma_{\frak s}$ for this compactification.   

Some of the important differences between the theory of open and/or unoriented strings and the theory of closed oriented strings come from
the fact that the compactification $\hat\Gamma_{\frak s}$ is a manifold (or orbifold) with boundary.  In superstring theory,
this leads to anomaly cancellation conditions
that have no close analog for closed oriented strings; the compactified moduli space of closed oriented Riemann surfaces has no boundary.
 
Here and in section \ref{opendeg}, we will describe the degenerations that are responsible for the boundary of $\hat\Gamma_{\frak s}$.  
In fig. \ref{zandy}, we sketch a closed oriented surface $\Sigma_0'$ that
is obtained by gluing at a narrow neck two identical but oppositely oriented components.  $\Sigma_0'$ has an antiholomorphic involution $\rho$
that exchanges the two components.

The behavior in the narrow neck is described by the usual gluing relation
\begin{equation}\label{zomber}x y = q. \end{equation}
$x$ is a local parameter on one side of the narrow neck in $\Sigma_0'$ and $y$ is a local parameter on the other side.
Now let us discuss how $\rho$ acts.  $\rho$ exchanges the two branches, so it exchanges $x$ and $y$.  It is antiholomorphic, so it must
map $x$ to a multiple of $\bar y$, and vice-versa.  We can define $x$ and $y$ so that
\begin{equation}\label{omber} \rho(x)=\bar y, ~~\rho(y)=\bar x.\end{equation}
With these choices of $x$ and $y$, the parameter $q$ in (\ref{zomber}) must be real so that the gluing relation is $\rho$-invariant. 

Now let us examine the quotient $\Sigma_0=\Sigma_0'/\Z^\rho_2$.  The boundary of $\SIgma_0$ consists of the fixed points of $\rho$.
A fixed point of $\rho$ obeys $y=\bar x$, so the gluing relation gives $x\bar x=q$.
So the fixed point set of $\rho$ consists of a circle if $q>0$, and is empty if $q<0$.

\begin{figure}
 \begin{center}
   \includegraphics[width=4.5in]{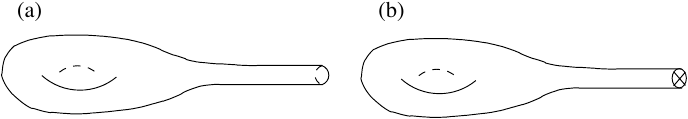}
 \end{center}
\caption{\small  
(a) For $q>0$, $\Sigma_0'/\Z_2$ has a long tube that ends on a boundary.  (b) For $q<0$, $\Sigma_0'/\Z_2$ has a long tube that ends on a crosscap.}
 \label{wandy}
\end{figure}
The quotient
$\Sigma_0'/\Z_2$ is depicted for the two cases in fig. \ref{wandy}.  It is convenient to use the conformal frame in which the gluing is described
by a long tube of length $\log |q|^{-1}/2$  (we divide the formula (\ref{proplength}) for the length of the tube by 2 because the tube is cut in half to make $\Sigma_0'/\Z_2$).
For $q>0$, $\Sigma_0'/\Z_2$ is a Riemann surface with boundary; the boundary, which is the end of the tube, is the circle $x\bar x=q$.
For $q<0$, there is no boundary, since $\rho$ has no fixed points.   The tube still ends at $x\bar x=|q|$, since a point with $x\bar x<|q|$ can be
mapped to $x\bar x>|q|$ by $ \rho:x\to \bar y=q/\bar x$.  For $q<0$ and $x\bar x=|q|$, the transformation $x\to q/\bar x$ is equivalent to $x\to -x$.
So opposite points of the circle $x\bar x=|q|$ are identified and $\Sigma_0'/\Z_2$ has no boundary.  One says in this case that the long tube ends on a crosscap,
a copy of $\RP^2$ with an open disc omitted.  

The topology of the surface $\Sigma_0'/\Z_2$ clearly depends on the sign of $q$.  We will call this quotient $\Sigma_0$ if $q>0$ and $\Sigma_1$ if $q<0$.
And we write $\Gamma_0$ and $\Gamma_1$ for the moduli spaces of conformal structures on $\Sigma_0$ or $\SIgma_1$.
The parameter $q$ is a real modulus of $\Sigma_0$ and $\Sigma_1$ and compactification of either $\Gamma_0$ or $\Gamma_1$ requires
us to allow the limiting value $q=0$.   The locus $q=0$ is a boundary component of the compactifications $\h\Gamma_0$ and $\h\Gamma_1$.

Since these moduli spaces have a common boundary, one can glue them together to make a manifold without boundary, different
parts of which parametrize open and/or unoriented surfaces with different topology.  Sometimes this is even useful in superstring theory \cite{GS}.
However, in general in superstring theory, it is not helpful to try to fit $\h\Gamma_0$ and $\h\Gamma_1$ together.  
For example, there are Type II orientifolds
without D-branes in which the string worldsheet has no boundary but need not be oriented; and there are Type II theories with D-branes but no
orientifold planes in which the string worldsheet is oriented but may not have a boundary.   In these examples, 
$\h\Gamma_0$ is relevant but not  $\h\Gamma_1$, or vice-versa,  so in
general one has to consider them separately.

The limits of $\Sigma_0$ and $\Sigma_1$ as $q\to 0$ cannot be deduced in a simple way from the way the double cover $\Sigma_0'$ degenerates
for $q\to 0$. $\Sigma_0'$
has only one limit for $q\to 0$ regardless of the direction from which one approaches $q=0$, but the objects which one wants to identify
as limits of $\Sigma_0$ and $\Sigma_1$ as $q$ approaches 0 from above or below are different.  

\begin{figure}
 \begin{center}
   \includegraphics[width=4.5in]{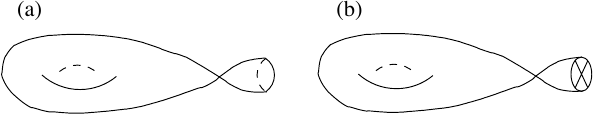}
 \end{center}
\caption{\small  
(a) For $q$ approaching zero from above, the long tube of fig. \ref{wandy}(a) is conformally equivalent to a collapsing neck and $\Sigma_0$ splits off a component consisting
of a disc with one puncture (the node).  (b) Similarly, for $q$ approaching zero from below, $\Sigma_1$
splits off a component that is conformally equivalent to $\Bbb{RP}^2$ with one puncture (again the node).}
 \label{xandy}
\end{figure}

It is not hard to understand what those objects should be.  In fig. \ref{wandy}(a) or (b), we see a long tube whose length diverges as $q$ approaches
0 from above or below.  We simply follow the general rule that the limit of a long tube is conformally equvalent to a collapsing neck. So
the limiting configurations corresponding to fig. \ref{wandy} for $q\to 0$ are as shown in fig. \ref{xandy}.  For $q\to 0$, $\Sigma_0$ or $\Sigma_1$
splits off a component that consists of a disc with one puncture (where it meets the node) or $\Bbb{RP}^2$ with one puncture (again
where it meets the node).  These are the natural limiting configurations at $q=0$ because the singularities they contain are
the usual closed-string degenerations, which are needed anyway in the Deligne-Mumford compactification so it
is unavoidable to allow them.  Moreover, they have a natural physical
interpretation.  

There are at least two lessons from all this for superstring theory:

(1) The compactified moduli space of open and/or unoriented Riemann surfaces is a manifold with boundary.  Similarly, the compactified moduli space
of open and/or unoriented super Riemann surfaces will be a supermanifold with boundary.  Integration on a supermanifold with boundary makes sense
provided the right structure is in place, but this is one of the issues that one has to consider.

(2) In compactifying the moduli space, we had to allow an open and/or unoriented Riemann surface to split off a component with continuous
symmetries.  In fact, a disc or $\Bbb{RP}^2$ with one closed-string puncture has a $U(1)$ group of symmetries -- the group
of rotations around the puncture. The components with symmetries turn out to be important in the study of
anomalies \cite{PolCai}, \cite{Witten}.  
By contrast, in compactifying the moduli space of oriented Riemann surfaces,
only components  without continuous symmetries are required.  

\subsubsection{Open-String Degenerations}\label{opendeg}

There is one more important example somewhat like the one studied in section \ref{anex}.  We return to the basic setting of fig. \ref{zandy} with a closed
oriented surface $\Sigma_0'$ that consists of two components joined at a narrow neck.  But now we assume that $\Sigma_0'$ has an antiholomorphic
involution $\rho$ that maps each component to itself, instead of exchanging them.

In terms of the gluing law
\begin{equation}\label{irtom}xy=q,\end{equation}
this means that the local parameters $x$ and $y$ are mapped to their own complex conjugates (up to constant multiples that we can absorb in the
definitions of $x$ and $y$):
\begin{equation}\label{obzo} \rho(x)=\bar x, ~~\rho(y)=\bar y.\end{equation}
Again, $\rho$-invariance of the gluing relation implies that $q$ must be real.

\begin{figure}
 \begin{center}
   \includegraphics[width=5.5in]{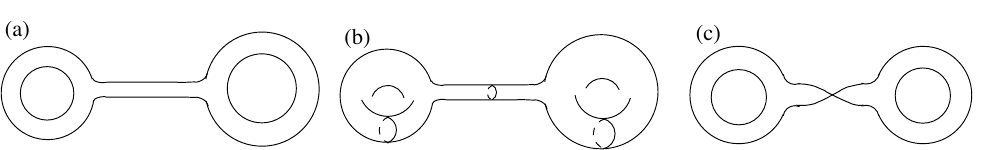}
 \end{center}
\caption{\small  
(a) A long strip connecting two components of a  Riemann surface $\Sigma_0$ with boundary. 
(b) The long strip of (a)  is simply the quotient by $\Z_2$ of a long tube in the associated closed oriented
double cover $\Sigma_0'$. The quotient is taken by flattening the closed oriented surface seen here
onto the page. (c) The  limit in which the strip in  (a)  becomes infinitely long is conformally equivalent
 to a degeneration in which $\Sigma_0$ splits
into a pair of components, joined at a pair of open-string punctures.  We call this an open-string degeneration. }
 \label{handy}
\end{figure}

Let us determine the fixed point set of $\rho$, which is the same as the boundary of the quotient $\Sigma_0'/\Z^\rho_2$.  Clearly the fixed point set is given
by the hyperbola $xy=q$ in the real $x\text{-}y$ plane.  The hyperbola consists topologically of a pair of lines. To describe $\Sigma_0$ explicitly, we go
back to complex variables and
 write $x=e^\varrho$, $y=qe^{-\varrho}$, with $\varrho\cong\varrho+2\pi\sqrt{-1}$.  $\rho$ acts on $\varrho $ by   $\varrho\to \bar\varrho$.   Imposing
 both relations $\varrho\cong\varrho+2\pi \sqrt{-1}$ and $\varrho\cong\bar\varrho$, we can parametrize
$\Sigma'_0/\Z_2$ by
\begin{equation}\label{noax}\varrho= s+i\vartheta,~~ 0\leq \vartheta \leq \pi,\end{equation}
and the endpoints $\vartheta=0,\pi$ are the boundaries of $\Sigma_0$.  With these restrictions on $\vartheta$, $\varrho$ parametrizes
a long strip (fig. \ref{handy}(a)).  This long strip is simply the quotient by $\Z_2$  of the long tube that is described by the equation $xy=q$ if we do not divide by $\Z_2$
(fig. \ref{handy}(b)).
Physically, the  long strip describes the propagation of an open string for a long proper time.

To compactify the moduli space $\Gamma$ that parametrizes $\Sigma'_0/\Z_2$ for positive (or negative) $q$, we need to introduce
a limit of the long strip at $q=0$. It is not difficult to see what this must be.   We use the fact that a long strip is conformally equivalent to a strip
with a narrow neck.  In the limit $q\to 0$, $\Sigma'_0/\Z_2$ decomposes into a union of two components glued together at a pair of open-string punctures
(fig. \ref{handy}(c)).  This is what we will call an open-string degeneration.  In string theory, infrared singularities due to on-shell open strings arise
from such open-string degenerations.   In this language, the degenerations studied in section \ref{infinity} might be called closed-string degenerations.

\begin{figure}
 \begin{center}
   \includegraphics[width=4in]{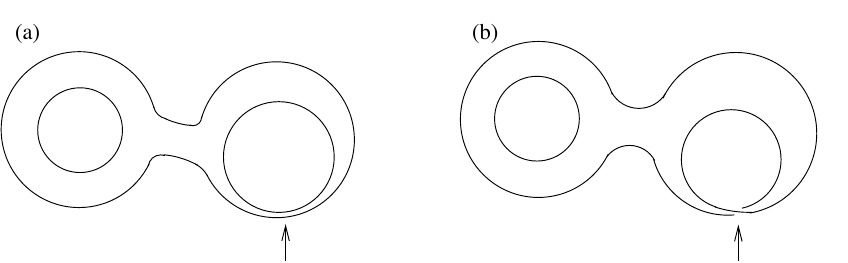}
 \end{center}
\caption{\small  
A nonseparating open-string degeneration (a) and its twisted version (b), both marked with arrows.  (a) and (b) can be exchanged by cutting
the indicated narrow neck, twisting it, and regluing. This corresponds to changing the sign of $q$.}
 \label{opennon}
\end{figure}

  The two cases of positive and negative $q$ differ by a ``twist''  of the strip, as sketched in
fig. \ref{opennon}.   
This happens as follows.  The two boundary components of the strip  can be distinguished by the sign of $x$, and also by the sign of $y$.
For $q\to 0$, $x$ parametrizes the boundary of one component of $\Sigma_0$, and $y$ parametrizes the boundary of the other.
The relation between positivity of $x$ and positivity of $y$ is reversed if we change the sign of $q$, and this introduces the twist.

Just like closed-string degenerations, open-string degenerations can be either separating (fig. \ref{handy}) or nonseparating (fig. \ref{opennon}).  In general $\Sigma'_0/\Z_2$ has different topology for positive or negative $q$.  For example, in fig. 
\ref{opennon},   $\Sigma'_0/\Z_2$ is unorientable for one sign of $q$ though it is orientable for the other sign.
In this case, the compactified moduli space $\h\Gamma_{\frak s}$ clearly must be understood as a manifold with boundary, with $q=0$ defining one of the boundary
components.   

Even when the topology of $\Sigma'_0/\Z_2$ is not affected by the twist, its conformal structure generically jumps discontinuously in crossing
$q=0$. So one should think of $q=0$ as a boundary of the moduli space $\h\Gamma_{\frak s}$, even if the limits as $q$ approaches
0 from above and below represent two different
parts of the boundary of the same $\h\Gamma_{\frak s}$.

\subsubsection{Superanalogs}\label{zondo}

Compactifying the parameter space $\varGamma_\sf$ 
of a super Riemann surface $\Sigma$ does not involve many new ingredients. 
The reduced space $\varGamma_{\sf,\red} \subset \MM_{\sg',\red}$  
  has a natural compactification 
$\h\varGamma_{\sf,\red}$ obtained by taking its closure in $\h\MM_{\sg',\red}$.  It can then be thickened in the fermionic
directions to give the compactified integration cycle $\h\varGamma_\sf$.

This compactification parametrizes superanalogs of the usual closed string degenerations, but this involves
no particular novelties relative to what we analyzed in section \ref{infinity}.  More interestingly, the
compactification of $\varGamma_\sf$ also describes 
superanalogs of the special degenerations that we considered in sections \ref{anex} and \ref{opendeg}, namely the
ones responsible for the fact that the compactified classical parameter space
$\h\Gamma_\sf$ of an open and/or unoriented Riemann surface $\SIgma_0$ is a manifold with boundary.  
It turns out that, because of the existence of these special degenerations, $\h\varGamma_\sf$ is a (smooth)
supermanifold with boundary.  This requires some discussion, since the proper definition of a supermanifold
with boundary, such that integration makes sense, is rather subtle (see for example \cite{Wittennotes}, sections 3.4 and 3.5).
We want to make sure that the
appropriate structure is present in the case of $\varGamma_\sf$.

To describe the special degenerations explicitly in the superstring context,
one replaces the classical gluing law $xy=q$ by its familiar
Neveu-Schwarz and Ramond analogs, namely
\begin{align}\label{bondo}  xy & = -\varepsilon^2 \cr
                                           y\theta & = \varepsilon\psi \cr
                                            x\psi & =-\varepsilon\theta\cr
                                             \theta\psi&=0. \end{align}
and 
\begin{align}\label{tormov} xy& = q_\Ra \cr
                        \theta& =\pm \sqrt{-1}\psi.\end{align}      
The degenerations that we want to focus on here  are the ones for which, classically, the gluing parameter $q$ is real.  
We will call them real degenerations.                                        
In the superstring context, at a real degeneration,  $q_\Ra$ and $q_\NS=-\varepsilon^2$ are real 
(and $\varepsilon$ is real or a real multiple of $\sqrt{-1}$) modulo the odd variables.   We say that a real degeneration is of NS type
or R type depending on whether it is described by the NS gluing formulas (\ref{bondo}) or their
Ramond counterparts (\ref{tormov}).

Real degenerations describe the propagation through a long tube or strip of 
either (1) a  closed string whose worldsheet ends on a boundary
or crosscap, as in section \ref{anex}; or (2)  an open string, as in section \ref{opendeg}.  
A real degeneration is of NS (R) type if the closed-string state in (1) is of NS-NS (R-R) type, or the open-string state
in (2) is of NS (R) type.  In a real degeneration, the sum over the sign of the fermionic
gluing (the sign of $\varepsilon$ or the sign in $\theta=\pm \sqrt{-1}\psi$) enforces the GSO projection.
In case (1), the sum over the sign of the fermionic gluing ensures that the closed-string state 
created by a boundary or crosscap is GSO even;
in case (2), it ensures that  open-string states that create infrared singularities by 
propagating for a long proper time through a long strip are GSO even.

Now let us discuss the boundary of  $\h\varGamma_{\frak s}$.  
The case of a real Ramond degeneration is more obvious so we consider it first.
An open-string degeneration of R type leads to a boundary component of
$\h\varGamma_{\frak s}$; as in the bosonic case, the two cases $q_\Ra\geq 0$ and 
$q_\Ra\leq 0$  represent different moduli 
spaces, typically parametrizing superstring worldsheets with different topology.  
Either of these two moduli spaces has a boundary at $q_\Ra=0$.

In general, to define  a supermanifold $M$ with boundary in such a way that integration is possible, one needs some additional structure.  
Locally, one requires
an even function $f$ that defines the boundary, in the sense that it is positive definite or negative definite 
(when reduced modulo odd variables)
in the interior of $M$ and has a first-order zero on the boundary of $M$.  To be a little more precise, 
we need only an equivalence
class of such functions $f$ modulo
\begin{equation}\label{irito} f\to e^\phi f,\end{equation}
where $\phi$ is a function on $M$ that is real modulo the odd variables.  
 Note that an allowed transformation  of $f$
does not include, for example, $f\to f+\eta_1\eta_2$, where $\eta_1$ and $\eta_2$ are odd functions.
The role of the function $f$ in integration theory is explained, for example,
in \cite{Wittennotes}, sections 3.4 and 3.5. 

The structure just described is precisely what we have near the boundary of $\h\varGamma_\sf$ 
associated to a Ramond degeneration.  
For the function $f$ we can
take $q_\Ra$.   Actually, $q_\Ra$ is really only well-defined to first order near the boundary of 
$\h\varGamma_\sf$;  and even to first order, it is not a
function but a section of a certain real\footnote{A line bundle over a smooth supermanifold is real if its transition functions are real modulo the odd variables.} line 
bundle, the dual of the normal 
bundle to 
the locus of real Ramond degenerations in $\h\varGamma_\sf$.    Rather as in section \ref{normal},
a  line bundle appears  because of the dependence
of $q_\Ra$ on the choice of the local parameters $x$ and $y$ in eqn. (\ref{tormov}).  
If we change the local parameters, $q_\Ra$ charges by $q_\Ra\to e^\phi q_\Ra$, for some function $\phi$ (which
in general may depend on all the other moduli of $\Sigma$).
Hence the indeterminacy of the function $q_\Ra$
 is precisely of the form of eqn. (\ref{irito}), with $q_\Ra$ understood as the distinguished function that vanishes
 on the boundary.  Therefore the structure of $\h\varGamma_\sf$ near its boundary
is precisely such that integration of a smooth measure (a smooth section of the Berezinian) 
over $\h\varGamma_\sf$ makes sense.  There is more to say because the measures that one has to integrate in superstring theory 
have singularities at the boundary; those issues are discussed in \cite{Witten}.  But the fact that one would be able 
to integrate a smooth measure over $\h\varGamma_\sf$
is certainly part of the input to the fact that perturbation theory makes sense for open and/or unoriented superstrings.  

A real NS degeneration does not actually contribute a 
boundary component to $\h\varGamma_\sf$, because the two choices of sign of $\varepsilon$
match smoothly at $\varepsilon=0$.    Nonetheless, 
to make sense of the integrals that appear in superstring
perturbation theory, one needs the existence of a distinguished parameter 
$\varepsilon$ (subject to $\varepsilon\to e^\phi\varepsilon$)
associated to a real NS degeneration.  That is because of the tachyon singularity at $\varepsilon=0$.
The sum over the two signs of $\varepsilon$ is needed to implement the 
GSO projection and eliminate the tachyon.  Roughly speaking, after eliminating the tachyon, the locus $\varepsilon=0$
is effectively a boundary, and again the integral over a supermanifold with boundary 
 only makes sense because of the existence of a distinguished parameter that vanishes at the boundary,
namely $\varepsilon$.  

We conclude by explaining the analog of section \ref{thed} for open-string degenerations, or
in other words by describing in the case of an open-string degeneration
the locus in $\h\varGamma_\sf$ with $q_\Ra=0$ or $\varepsilon=0$. (Real closed-string degenerations 
are treated in section \ref{comcont}.)
When a worldsheet $\Sigma$ undergoes an open-string degeneration, it can be compared 
to the smooth surface $\Sigma^*$ (its ``normalization'') that results from detaching the two
branches at the node.  ($\Sigma^*$ may or may not be connected, depending on whether 
the degeneration is separating.)
The locus in $\h\varGamma_{\frak s}$ that parametrizes the open-string degeneration can be compared, 
just as we did in section \ref{thed} for closed strings,
to the parameter space $\h\varGamma_{\frak s}^*$ of $\Sigma^*$.  For an open-string degeneration of NS type, 
the degeneration locus in $\h\varGamma_{\frak s}$
is a copy of $\h\varGamma^*_{\frak s}$, but for one of R type, it is a fiber bundle  over $\h\varGamma^*_{\frak s}$
with fibers of dimension $0|1$.  As in section \ref{thed}, the fibers parametrize
the different ways to glue together Ramond divisors on the two  sides. 
One can summarize this by saying that while at an open-string NS node there is only the even gluing parameter 
$\varepsilon$, an open-string Ramond node
has a fermionic gluing parameter as well as the even parameter $q_\Ra$.  (After integrating 
over this parameter, the open-string Ramond propagator
is $G_0/L_0$, in contrast to the open-string NS propagator, which is $1/L_0$.)  
All this is closely analogous to what happens for closed strings.

\subsubsection{Components With Continuous Symmetries}\label{comcont}

The compactified parameter space $\h\varGamma_{\frak s}$ describes many possible 
ways that $\Sigma$ can degenerate by reducing its genus (understood as
the genus of the closed oriented double cover $\Sigma'$) or splitting into multiple components.  
For the same reasons as in the bosonic discussion
of section \ref{anex}, an example of what can happen is that $\Sigma$ can split off a component 
that has a symmetry group $\G$ of positive even dimension.
This will be a component  whose reduced space is a disc with a single closed-string puncture, or 
$\Bbb{RP}^2$ with a single
puncture.  Even though these sound like  special cases, they are important in understanding superstring 
anomalies, so we will look more closely.

For either the once-punctured disc or once-punctured $\RP^2$, 
the single puncture is of NS-NS type or of R-R type (it cannot be of type NS-R 
for then on the closed oriented double
cover, there would be a single R puncture, which is impossible).
 It is instructive to work out the dimension formula for the dimension of
 moduli space for a punctured Riemann surface of one of these types. The disc and 
 $\RP^2$ can be considered together.   First we assume that the puncture is of NS-NS type.
 The dimension formula with only  NS-NS punctures reads
 \begin{equation}\label{yrze} \dim\,\varGamma-\dim\,\G =-3\chi(\Sigma)+2\n_  {\mathrm{NS-NS}}|
 -2\chi(\Sigma)+2\n_  {\mathrm{NS-NS}}.\end{equation}
 In evaluating this formula, it helps to bear in mind that, for either a disc or $\RP^2$ with one NS-NS puncture,
  the associated closed oriented super Riemann surface $\Sigma'$   is $\CP^{1|1}$ with two NS punctures.
This surface has no moduli, so $\varGamma$ is a point
and  $\dim\,\varGamma=0|0$.  Both the disc and $\Bbb{RP}^2$ have $\chi=1$.  
So the dimension formula with $\n_  {\mathrm{NS-NS}}=1$  implies that 
$\dim\,\G=1|0$.  Indeed, the bosonic part of the automorphism group is $U(1)$, consisting of rotations of the disc or $\Bbb{RP}^2$ around the puncture.  And the odd
dimension of the automorphism group is 0, because $\Sigma'$ has no odd automorphisms.

Now suppose that the puncture is of R-R type.  The dimension formula with only R-R punctures reads
\begin{equation}\label{byrze}\dim\,\varGamma-\dim\,\G =-3\chi(\Sigma)+2\n_  {\mathrm{R-R}}|-2\chi(\Sigma)+\n_  {\mathrm{R-R}}.\end{equation}
$\varGamma$ is still a point, with $\dim\,\varGamma=0|0$, just as before.  The dimension formula with $\n_  {\mathrm{R-R}}=1$ now
gives $\dim\, \G=1|1$.  The bosonic part of the automorphism group is $U(1)$,  as before.  To understand why $\G$ has odd dimension 1,
observe that the associated closed oriented super Riemann surface $\Sigma'$ is now of genus 0 with 2 Ramond punctures.  This surface has a one-dimensional space
of odd superconformal vector fields, as we explained in discussing eqn. (\ref{trymo}). 

We conclude by discussing the consequences of the automorphism group $\G$ for gluing.  We start with
bosonic strings.    Let us return  to fig. \ref{xandy} of section \ref{anex}.  The degeneration depicted
in this figure is a closed-string degeneration, which normally is controlled by a complex gluing
parameter $q$. Yet in the particular
case that one of the components meeting at the double point is a disc or $\RP^2$ (with no punctures except the double point), one may restrict to real nonnegative\footnote{In fig. \ref{zandy}, 
a real degeneration corresponds
to real $q$, which may be positive or negative.  The two signs correspond to cases (a) and (b) in fig. \ref{xandy}, 
so in those pictures, there is no longer a choice of sign.}
values of $q$.   We have already given one explanation of this in section \ref{anex}.   
Another  follows from the existence of the automorphism group.  Ordinarily,
when two Riemann surfaces $\Sigma_1$ and $\Sigma_2$ are joined by the narrow neck $xy=q$, 
the parameter $\mathrm{Arg}\,q$ has the following interpretation: a shift
in $\mathrm{Arg}\,q$ by an angle $\phi$ has the same effect as cutting the narrow neck, 
rotating one side relative to the other by the angle $\phi$, and gluing the two sides
back together again.  When $\Sigma_1$ or $\Sigma_2$ has a symmetry of rotating around the neck, 
the parameter $\mathrm{Arg}\,q$ can be absorbed in rotating $\Sigma_1$ or
$\Sigma_2$ and becomes irrelevant.  
This is the case precisely when  $\Sigma_1$ or $\Sigma_2$ is a disc or $\RP^2$ with just one puncture.

What we have said so far relies on the bosonic part of the automorphism group, and is equally valid for bosonic strings or for NS-NS and R-R degenerations of superstrings.  Now we consider the fermionic part of the automorphism
group, which is present only in the R-R superstring case.  Usually, at an R-R closed-string degeneration, there are two fermionic gluing parameters, one for holomorphic degrees of freedom and one
for antiholomorphic degrees of freedom.  However, if one of the two branches, say $\Sigma_2$, is a disc or $\RP^2$ with one puncture, then it has a fermionic symmetry
that can be used to transform away one of the fermionic gluing parameters.  This happens because the fermionic symmetry of the closed oriented double cover of $\Sigma_2$ acts
nontrivially on the Ramond divisors (this is shown in eqn. (\ref{mogro})) and hence shifts the fermionic gluing data.
This is quite analogous to what happened to $\mathrm{Arg}\,q$.  The disappearance of 
one of the two fermionic gluing
parameters is part of the mechanism by which an R-R  degeneration of the particular types shown in fig. \ref{xandy}  can lead to anomalies.  This is described in 
\cite{Witten}.  

\section{Contour Integrals And Super Period Matrix}\label{contourintegrals}

Our goal in this section is to describe the analog on a super Riemann surface of the periods of a holomorphic differential on an ordinary Riemann surface.
We will also describe the super period matrix constructed from the periods, and demonstrate its basic properties.
Finally, we describe an application \cite{DPhthree} to superstring perturbation theory.

\subsection{Basics}\label{peribasics}

We will need to consider a super Riemann surface $\Sigma$
as a smooth supermanifold, as
described in section \ref{sufsmooth}.  However,  antiholomorphic odd variables would add nothing.  So as in most
of section \ref{sufsmooth}, we take $\Sigma$ to be a heterotic string worldsheet.  This means that $\Sigma$
is a smooth supermanifold of dimension $2|1$ embedded in $\Sigma_L\times \Sigma_R$, where $\Sigma_L$
is an ordinary Riemann surface and $\Sigma_R$ is a super Riemann surface (as usual, we can assume that $\Sigma_L$
is the complex conjugate of $\Sigma_{R,\red}$ and $\Sigma_\red$ is diagonally embedded in $\Sigma_L\times
\SIgma_{R,\red}$).  We will write $\BBer(\Sigma)$ for the Berezinian of the holomorphic cotangent
bundle to $\Sigma$ (that is, for the restriction to $\Sigma$ of $\BBer(\Sigma_R)$).

Suppose that $\Sigma $ is split from a holomorphic point of view (in the sense that $\Sigma_R$ is a split
super Riemann surface).  
Then its reduced space $\Sigma_\red$ is naturally embedded in $\Sigma$, with a projection $\pi:\SIgma\to\Sigma_\red$.
Even if $\Sigma$ is not holomorphically split, 
the structure theorem of smooth supermanifolds (reviewed for instance in 
section 2.2 of \cite{Wittennotes}) says that $\Sigma_\red$
can be embedded in $\Sigma$, again with a projection $\pi:\Sigma\to\Sigma_\red$.  
The embedding and projection are not completely natural, but they are natural up to
homology (in fact, up to infinitesimal homologies involving the fermionic directions), 
and this will be good enough for us.\footnote{Actually,  on a super Riemann surface, as opposed
to a more general $1|1$ complex supermanifold, once an embedding of $\Sigma_\red$ in $\Sigma$ 
is given, there is a natural projection $\pi:\Sigma\to\Sigma_\red$.
The subbundle $\D\subset T\Sigma$, when restricted to $\Sigma_\red$, is a fiber 
bundle over $\Sigma_\red$ with fibers of rank $0|1$. This bundle is
naturally embedded in $\Sigma$ and its total space is $\Sigma$. So $\Sigma$ is a fiber bundle over $\Sigma_\red$
and this gives the projection $\pi$.  We do not often need this fact,
but it will be relevant at the end of appendix \ref{zelfo}. } 

We begin by describing the cycles on which we will compute periods.
Let $\gamma_\red$ be an ordinary oriented closed 1-cycle in $\Sigma_\red$, with 
$\Sigma_\red$ embedded in $\Sigma$ as above.  (We take $\gamma_\red$
to be closed as we do not want to consider contours with boundary.)  Then after embedding 
$\Sigma_\red$ in $\Sigma$, we can view $\gamma_\red$ as a submanifold of $\Sigma$ of dimension $1|0$.  If $\mu$ is
a 1-form on $\Sigma$, then we can integrate it over $\gamma_\red$ in the usual way. 
For this, we parametrize $\gamma_\red$ by an angular variable $s$; when restricted to $\gamma_\red$,
$\mu$ is an ordinary 1-form $f(s)\,\d s$, and we define
\begin{equation}\label{zonny}\oint_{\gamma_\red} \mu=\oint_{\gamma_\red} f(s)\,\d s.\end{equation}
If $\mu$ is closed, then $\oint_{\gamma_\red}\mu$ depends only on the homology class of 
$\gamma_{\red}$ in $\Sigma$, not on the embedding of $\gamma_\red$ in $\Sigma_\red$
or of $\Sigma_\red$ in $\Sigma$.
A difference from ordinary compact Riemann surfaces (and  from compact Kahler manifolds of any dimension) 
is that even if $\mu$ is holomorphic, it need
not be closed.  Indeed if $\Sigma$ is split and $a(z)\,\d z$ is a holomorphic 1-form on $\Sigma_\red$,
then $a(z)\varpi$ is a globally-defined and holomorphic but not closed 1-form on $\Sigma$; here
\begin{equation}\label{ygro}\varpi=\d z-\theta\d\theta\end{equation}
was introduced in the discussion of eqn. (\ref{zomurk}).

There is another type of object that is appropriate for contour-like integrals.   Let $\sigma$ be
a holomorphic section of $\BBer(\Sigma)$.  Then $\sigma$ can be regarded 
as an integral form on $\Sigma$ of codimension 1,
so it can be naturally integrated over a  submanifold  $\gamma\subset \Sigma$ of 
real codimension $1|0$ and hence of dimension $1|1$.  
This is explained, for example, in \cite{Wittennotes}, section 3.3.4.
The integral, which we denote as $\oint_{\gamma}\sigma$,  only depends on the homology 
class of $\gamma$.  As an example of a codimension $1|0$ cycle in $\Sigma$,
we can take  $\gamma=\pi^{-1}(\gamma_{\red})$
with $\gamma_\red$ as before.  

Up to homology, all 
of the relevant $1|0$ and $1|1$ cycles $\gamma_\red$ or $\gamma$ in $\Sigma$ arise in this 
way from ordinary 1-cycles in $\Sigma_\red$.  So all periods of a 1-form on $\Sigma$
are associated with $A$-cycles or $B$-cycles on the ordinary Riemann surface $\Sigma_\red$.  
There are thus $2\g$ periods in all.

A difference from ordinary Riemann surface theory is that in general there are more independent 
objects that one might want to integrate than there are independent cycles to
integrate them over.  Suppose for example that $\Sigma$ is split; it is then constructed from its reduced
space $\Sigma_\red$ and a square root $K^{1/2}$ of the canonical bundle $K$ of $\Sigma_\red$.  In this case, a 
holomorphic section $\sigma$ of $\BBer(\Sigma)$ takes the form
$\sigma=(\alpha(z)+\theta b(z))[\d z|\d\theta]$, with $\alpha\in H^0(\Sigma_\red,K^{1/2})$ and 
$b\in H^0(\Sigma_\red,K)$.  
If $\sigma$ is even (as we will generally assume), 
then, since the measure $[\d z|\d\theta]$ is odd, $\alpha$ is odd and $b$ is even.  
We can reduce an integral over $\gamma$ to an integral over $\gamma_\red$
by first integrating over $\theta$.  Using the properties of the Berezin integral, this gives
\begin{equation}\label{cocyc}\oint_{\gamma} \alpha(z)[\d z|\d\theta]=0,\end{equation}
and
\begin{equation}\label{ocry}\oint_{\gamma}\theta b(z)[\d z|\d\theta]=\oint_{\gamma_\red} b(z) \d z.\end{equation}

Thus, in the split case, if there is a holomorphic section $\alpha(z)$ of $H^0(\Sigma_\red,K^{1/2})$, 
there is a holomorphic section $\sigma=\alpha(z)[\d z|\d\theta]$ of $\BBer(\Sigma)$ whose periods 
all vanish (and similarly, there is a holomorphic 1-form $\mu=\d(\theta\alpha)$ whose periods all vanish).
The theory of periods and contour integration on $\Sigma$ is more simple if this does not occur, since for example then $\sigma$ is uniquely 
determined by its $A$-periods, just as on an ordinary Riemann surface.   In practice, if $K^{1/2}$ defines an odd spin structure, the dimension of $H^0(\Sigma_\red,K^{1/2})$ is always
odd and never zero.  If $K^{1/2}$ defines an even spin structure, then the dimension of $H^0(\SIgma_\red,K^{1/2})$ is always even and this cohomology group vanishes
for a generic choice of the complex structure of $\Sigma_\red$.  That is the situation in which we will discuss the super period matrix.  (At the end of section \ref{formula},
 we will see that the super period matrix develops a pole when $H^0(\SIgma_\red,
K^{1/2})$ becomes nonzero.)

An important technical point is that the hypothesis that $H^0(\SIgma_\red,K^{1/2})=0$ ensures that no jumping
occurs in the cohomology groups $H^k(\Sigma,\BBer(\Sigma))$, $k=0,1$, as a function of the odd moduli.\footnote{See the end
of section \ref{ryon} for a simple example showing that such jumping can occur for $H^0(\Sigma_\red,K^{1/2})\not=0$.}  
 In general, just as on an ordinary Riemann surface, for
any line bundle (or vector bundle) $\V$, the difference in graded dimension between $H^0(\Sigma,\V)$ and
$H^1(\Sigma,\V)$ is a topological invariant (determined by the Riemann-Roch theorem), not subject to jumping.
So absence of jumping of $H^1(\Sigma,\BBer(\SIgma))$ will imply absence of jumping of $H^0(\SIgma,\BBer(\Sigma))$.  
 By Serre duality,
$H^1(\SIgma,\BBer(\SIgma))$ is dual to $\Pi H^0(\Sigma,\O)$, where $\O$ is a trivial line bundle.  So it suffices
to show that there is no jumping in $H^0(\Sigma,
\O)$, which is the space of global holomorphic functions on $\Sigma$.  For $\Sigma$ split, a global holomorphic function $w$
has an expansion $w=w_0+\theta w_1$, where  $w_0$ is a global holomorphic function on $\Sigma_\red$ and
therefore a constant,
while $w_1$ is an element of $H^0(\SIgma_\red,K^{1/2})$.  The assumption that $H^0(\Sigma_\red,K^{1/2})=0$ 
thus implies
that $w_1=0$.  So we have learned that $H^0(\Sigma,\O)$ is generated in the split case by the constant function 1.
Certainly, this constant function makes sense as an element of $H^0(\SIgma,\O)$ even when we turn on odd
moduli, so there is no jumping of $H^0(\Sigma,\O)$ as a function of odd moduli, and hence also none in $H^1(\Sigma,
\BBer(\Sigma))$ or $H^0(\Sigma,\BBer(\Sigma))$.  
 Jumping of $H^0(\SIgma,\BBer(\SIgma))$ occurs only
when we vary the even moduli of
$\SIgma_\red$ so that $H^0(\SIgma_\red,K^{1/2})$ becomes nonzero.  

We have already seen that in the split
case, with $H^0(\Sigma_\red,K^{1/2})=0$, $H^0(\Sigma,\BBer(\SIgma))$ is naturally isomorphic to 
$H^0(\Sigma_\red,K)$, and in particular has dimension $\g|0$, where $\g$ is the genus of $\SIgma$.  
Absence of jumping means that the dimension remains $\g|0$ when we turn on the odd moduli.

  Periods of holomorphic sections of $\BBer(\Sigma)$ are 
  best described by a super period matrix \cite{RSV,DPhagain}.  
On the reduced space $\Sigma_\red$, pick a basis of $A$-cycles and $B$-cycles $A_\red^i$, $B_{j,\red}$, $i,j=1,\dots,g$
with
\begin{equation}\label{zono}A_\red^i\cap A_\red^j=0=B_{i,\red}\cap B_{j,\red},~~ A^i_\red\cap B_{j,\red}=\delta^i_j.\end{equation}
We thicken these in the usual way to cycles $A^i,\,B_j\subset\Sigma$ of codimension $1|0$ or dimension $1|1$.
Assuming that $H^0(\Sigma_\red,K^{1/2})=0$, define a basis  
$\sigma_j$, $j=1\dots g$ of holomorphic sections of $\BBer(\Sigma)$ by requiring that the
integrals over $A$-cycles  are
\begin{equation}\label{zinco}\oint_{A^{i}}\sigma_j = \delta^i_j.\end{equation}
(The no-jumping result  and the existence and uniqueness of the
$\sigma_j$ in the split case ensures that the $\sigma_j$ exist and are unique even after we turn on the
odd moduli.)
Then define
\begin{equation}\label{rinco}\hat \Omega_{ij}=\oint_{B_j}\sigma_i.\end{equation}
This definition does not make it immediately clear that $\hat\Omega_{ij}$ is symmetric.  Proving that it is  will be a goal of section \ref{symmetry}.

When $\Sigma$, is split, $\hat\Omega_{ij}$ coincides with the classical period matrix $\Omega_{ij}$.
Indeed,
from (\ref{ocry}), we see that in the split case, the periods of a holomorphic section $\sigma=b(z)\theta[\d z|\d\theta]$
of $\BBer(\Sigma)$ are the same as the periods of the ordinary differential
$b(z)\,\d z$ on the reduced space $\Sigma_\red$.  When $\Sigma$ is not split, $\hat\Omega_{ij}$ no longer
coincides with $\Omega_{ij}$.  Computing the difference between them when $\Sigma$ is not split \cite{DPhagain} will be the goal of section \ref{formula}. 

\subsection{Symmetry of the Super Period Matrix}\label{symmetry}

To show that the super period matrix is symmetric, we will follow \cite{RSV} and first show
an equivalence between the two types of contour integration on a super Riemann surface.  By an elementary though at first sight somewhat
mysterious formula, one can map a holomorphic section $\sigma=\phi(z|\theta)[\d z|\d\theta]$ of $\BBer(\Sigma)$ to a closed holomorphic 1-form $\mu$ in such a way that
\begin{equation}\label{kilfred}\oint_{\gamma_\red}\mu=\oint_{\gamma}\sigma. \end{equation}
The formula is as follows.  If $\sigma=\phi(z|\theta)[\d z|\d\theta],$ then
\begin{equation}\label{firred}\mu=\d\theta\phi(z|\theta) +\varpi D_\theta\phi(z|\theta).\end{equation}
It is not difficult to verify that $\d\mu=0$ (the formula (\ref{horsefly}) for the exterior derivative is helpful).      Actually, from the formula (\ref{firred}), one can verify directly\footnote{\label{zasp} A general holomorphic 1-form $\mu = s(z|\theta) \d z+ t(z|\theta)\d \theta$ (where
we will consider $s$ even and $t$ odd),
with $s(z|\theta)=s_0(z)+\theta s_1(z)$, $t(z|\theta)=t_0(z)+\theta t_1(z)$, is  closed if and only if
$t_1=0$, $s_1=\partial_z t_0$, so $\mu=(s_0+\theta\partial_z t_0) \d z+t_0 \d\theta$.   This has the
form of eqn. (\ref{firred}) if and only if $\phi(z|\theta)=t_0+s_0\theta$.  } that the map from
$\sigma$ to $\mu$ is a 1-1 map from holomorphic volume forms (that is holomorphic sections of $\BBer(\Sigma)$) to closed holomorphic 1-forms.
The somewhat mysterious formula (\ref{firred}) has a more conceptual explanation \cite{Beltwo}, as will be described in appendix \ref{zelfo}.
This explanation shows that the passage from $\sigma\in H^0(\Sigma,\BBer(\Sigma))$ to $\mu\in H^0(\Sigma,T^*\Sigma)$ does not
depend on the choice of local superconformal coordinates $z|\theta$.  

Explicitly, if $\phi(z|\theta)=\alpha(z)+\theta b(z)$, we compute $\mu$ from eqn. (\ref{firred}):
\begin{equation}\label{pilfred}\mu= b(z)\d z+\d(\theta\alpha).\end{equation}  
In computing $\oint_{\gamma_\red}\mu$, the  exact form $\d(\theta\alpha)$ can be dropped because  actually, one can always pick a  superconformal coordinate
system valid everywhere on the circle\footnote{The sheaf cohomology class that obstructs  splitting of $\Sigma$ 
is trivial when restricted to a small neighborhood of a circle. So such a neighborhood is isomorphic to ${\Pi} T^{1/2}\Sigma_0$ (with one of the two possible choices of $T^{1/2}\SIgma_0$, corresponding to NS and R
spin structures) where
$\Sigma_0$ is an ordinary annulus in the complex $z$-plane.}   $\gamma_\red$ so that the computation leading to eqn. (\ref{pilfred}) is valid throughout 
$\gamma_\red$.
We also have the Berezin integral over $\theta$, giving $\oint_{\gamma}\sigma=\oint_{\gamma_\red} b(z)\d z$.
So
\begin{equation}\label{zilfof}\oint_{\gamma}\sigma =\oint_{\gamma_\red} \mu.\end{equation}
For a more conceptual explanation of this result, see appendix \ref{zelfo}. 

Eqn. (\ref{zilfof}) has the following interesting corollary.  
The formula (\ref{firred}) implies that
 $\mu$ is exact -- it is of the form $\d s$ for a globally-defined 
 holomorphic function $s$ -- if and only if  $\sigma=D_\theta s$.  So 
\begin{equation}\label{domoxot}\oint_\gamma [\d z|\d\theta]D_\theta s=\oint_{\gamma_\red}\d s=0,\end{equation}
giving an alternative proof of an assertion of section \ref{whatfor}.

Since our first step in analyzing the periods of $\sigma\in H^0(\Sigma,\BBer(\SIgma))$ has been to equate them to periods of $\mu\in H^0(\Sigma,T^*\Sigma)$,
one may ask why we do not merely start with the latter.  One answer is that as $\BBer(\Sigma)$ is a line bundle, while $T^*\SIgma$ has rank $1|1$,
the description by sections of $\BBer(\Sigma)$ is often more straightforward.  For an example of this, see section \ref{formula}, which would be more
complicated if expressed in terms of $\mu$.

If $p$ and $q$ are any closed 1-forms on the ordinary Riemann surface $\Sigma_\red$, one has the fact of classical topology (essentially
Riemann's bilinear relation)
\begin{equation}\label{itzo}\int_{\Sigma_\red}p\wedge q=\sum_i\left(\oint_{A^i_\red}p\oint_{B_{i,\red}}q-\oint_{B_{i,\red}}p\oint_{A^i_\red}q\right).\end{equation}
If $p$ and $q$ are holomorphic 1-forms -- and thus of type $(1,0)$ -- then $p\wedge q=0$, as a result of which eqn. (\ref{itzo})  simplifies to
\begin{equation}\label{tzo}\oint_{A^i_\red}p\oint_{B_{i,\red}}q-\oint_{B_{i,\red}}p\oint_{A^i_\red}q =0.\end{equation}
The classical period matrix is defined by introducing holomorphic 1-forms $\lambda_i$ such that $\oint_{A^i_\red}\lambda_j=\delta^i_j$ and
then setting $\Omega_{ij}=\oint_{B_{i\,\red}}\lambda_j$.  Its symmetry $\Omega_{ij}=\Omega_{ji}$ follows immediately from (\ref{tzo}) with $p=\lambda_i$,
$q=\lambda_j$.

We want to imitate this proof on a super Riemann surface.  We have already taken the first step -- defining a basis of sections $\sigma_j\in H^0(\Sigma,\BBer(\Sigma))$
with canonical $A$-periods (\ref{zinco}).  The above procedure maps them to a basis of closed holomorphic 1-forms $\mu_j$ also with canonical
$A$-periods:
\begin{equation}\label{tryop}\oint_{A^i_\red}\mu_j=\delta^i_j.\end{equation}
However, in imitating the classical proof, we run into a snag.  On a super Riemann surface, two holomorphic 1-forms $\mu_i$ and $\mu_j$ may have a non-zero
wedge product $\mu_i\wedge \mu_j$; for example, $\d z\wedge\d\theta$ is nonzero, and $\d\theta$, being even, can be raised to any power.  So it is not immediately obvious that $\int_{\Sigma_\red}\mu_i\wedge\mu_j=0$.
It is true that $\mu_i\wedge\mu_j$ is of type $(2,0)$ on $\Sigma$, and it is also true that on the ordinary
Riemann surface $\Sigma_\red$, a $(2,0)$-form would vanish. But
 if $\Sigma $ is not  holomorpically split, the embedding of $\Sigma_\red$ in $\Sigma$ cannot be chosen to be
 holomorphic, so the restriction  to $\Sigma_\red$ of a $(2,0)$-form on $\Sigma$ is not necessarily a form of type
 $(2,0)$ on $\Sigma_\red$.
 
A cure for this difficulty was explained in \cite{RSV}.  Though a holomorphic 2-form on  $\Sigma$ need not vanish, such a form, if closed,
is always exact. This statement is not true on a general complex supermanifold of dimension $1|1$; the proof uses the superconformal structure of $\Sigma$.   A  general holomorphic 2-form on $\Sigma$ is $\Psi=(\d\theta)^2 p(z|\theta)+\d\theta\varpi \rho(z|\theta)$, for  holomorphic functions $p$ and $\rho$.
The condition that $\d\Psi=0$ gives $\rho=D_\theta p$ (again it helps here to use (\ref{horsefly})), from which it follows that $\Psi=\d f$ with $f=-p\,\varpi$. $f$
does not depend on the local superconformal coordinates that were used for this computation,\footnote{To prove this, recall from eqn. (\ref{zomurk})
that there is a natural projection  $\kappa:T^*\Sigma\to \D^{-1}$.  So there is a natural projection $\kappa^{\otimes 2}:T^*\Sigma\otimes T^*\Sigma\to \D^{-2}$.
The object $p\varpi$ is a section of $\D^{-2}\subset T^*\SIgma$ that can be globally defined as $\kappa^{\otimes 2}(\Psi)$.}
 so it is globally-defined.  

So now if $\mu_i$ are closed holomorphic 1-forms normalized as in (\ref{tryop}), then $\mu_i\wedge\mu_j$ is exact and thus
$\int_{\SIgma_\red}\mu_i\wedge \mu_j=0$.  Hence if in (\ref{itzo}) we take $p=\mu_i$, $q=\mu_j$, we get finally
\begin{equation}\label{tronx}\hat\Omega_{ij}=\hat\Omega_{ji},\end{equation}
with $\hat\Omega_{ij}=\oint_{B^j}\sigma_i$.

\subsection{Formula For The Super Period Matrix}\label{formula}

Essentially following \cite{DPhagain}, we will now compute the dependence on odd moduli of the super period matrix of a super Riemann surface.
We begin with a split super Riemann surface $\Sigma$, with an even spin structure and $H^0(\Sigma_\red,K^{1/2})=0$. 
We will perturb the complex structure of $\Sigma$ in the framework of section \ref{defsmooth} and compute how the period matrix varies.
We make the perturbation in the framework of eqn. (\ref{zorb}), except that we are only interested in varying with respect to odd moduli,
since we consider the classical period matrix as a function of even moduli to be a familiar quantity.  Of course, the calculation we will perform has a close
analog  to study the dependence of the classical period matrix on bosonic moduli.  

We use the description of a variation of complex structure of $\Sigma$ given in eqn. (\ref{zorb}), except that we drop the metric perturbation $h_{\t z}^z$ and
keep only the gravitino perturbation $\chi_{\t z}^\theta$.  We define the perturbation by saying that after the perturbation, a holomorphic function is
annihilated by
\begin{equation}\label{irod}\partial'_{\t z}=\partial_{\t z}+\chi_{\t z}^\theta\left(\partial_\theta-\theta\partial_z\right).\end{equation}
We take the gravitino field $\chi_{\t z}^\theta$ to be
\begin{equation}\label{rodo} \chi^\theta_{\t z}=\sum_{a=1}^n\eta_a f_{a\,\t z}^\theta,\end{equation}
where the $\eta_a$ are odd moduli and the $f_a$ are even $(0,1)$-forms valued in $T\Sigma_\red^{1/2}$.  It is natural to take
 $n$ to be  the odd dimension of $H^1(\Sigma_\red,T\Sigma_\red^{1/2})$ -- which coincides with the odd dimension of the supermoduli space $\MM$ -- and the $f_a$,
 $a=1,\dots ,n$, to 
 represent a basis of this space.  
 
 What we have described in eqn. (\ref{irod}) is a $0|n$-dimensional family of super Riemann surfaces that  is transverse to the split locus in $\MM$.  
 We have not given any instructions on how to pick a natural family; in general, we do not know any nice way to do so.   The family of super Riemann
 surfaces  determined by the $f_a$ depends 
 not only on
 their cohomology classes but on
 the actual $(0,1)$-forms $f_a$; if we make a gauge transformation on one of the $f_a$, this will in general transform some of the others into the metric
 perturbations by $h_{\t z}^z$ that were present in the more general deformation of eqn. (\ref{zorb}).   After performing the calculation, we will
 explain how to include the metric deformations and get a more natural formalism.

We start with a holomorphic section $\sigma$ of $\BBer(\Sigma)$.  In local superconformal coordinates $z|\theta$, 
$\sigma=\phi(z|\theta)[\d z|\d \theta]$, with $\phi$ a holomorphic function.
When we perturb the complex structure of $\Sigma$, $\sigma$ will have to be modified so as to be still a holomorphic 
section of $\BBer(\Sigma)$.  It will not suffice to modify the
function $\phi(z|\theta)$; we also have to change the symbol $[\d z|\d\theta]$.  The reason is that $\BBer(\Sigma)$ is defined so that a 
typical element at a given point in $\Sigma$
is a symbol $[u|\zeta]$ where $u|\zeta$ is a basis of forms of type $(1,0)$ at the given point.   When we perturb the 
complex structure of $\Sigma$ as in (\ref{irod}), the 1-forms
$\d z$ and $\d \theta$ cease to be of type $(1,0)$ and need corrections.

The condition for a 1-form to be of type $(1,0)$ is that its contraction with the vector fields $\partial'_{\t z}$ defined in 
eqn. (\ref{irod}) should vanish.  This condition is satisfied by
$\d z+\chi_{\t z}^\theta\theta \d \t z$ and by $\d\theta+\chi_{\t z}^\theta\d\t z$.  So a general section of $\BBer(\Sigma)$ after the perturbation takes the form
\begin{equation}\label{kofo} \h\sigma =\h\phi(\t z;\neg z|\theta) \left[\left.\d z+\chi_{\t z}^\theta\theta \d \t z\right|\d\theta+\chi_{\t z}^\theta\d\t z\right].\end{equation}
We have to determine the condition on $\h\phi$ that will make $\h\sigma$ holomorphic.  We assume that $\h\sigma$
is even and $\h\phi$ is odd.

For this, it is convenient to view a section of $\BBer(\Sigma)$ as an integral form on $\Sigma$ of codimension 1.  In turn, such a form is a function on ${\Pi} T\Sigma$,
meaning locally that it is a function of the coordinates $\t z, \,z$, and $\theta$ of $\Sigma$ and of a new set of fiber coordinates 
$\d\t z$, $\d z$, $\d\theta$ of reversed parity.
(See for example \cite{Wittennotes}, section 3.2, for an explanation of this formalism.)  As a function on ${\Pi} TM$, $\h\sigma$ is
\begin{align}\label{zofo}\h\sigma=&\h\phi(\t z;\neg z|\theta)(\d z+\chi_{\t z}^\theta\theta \d \t z)\delta(\d\theta+\chi_{\t z}^\theta\d\t z)\cr =&
\h\phi(\t z;\neg z|\theta)\left(\d z+\chi_{\t z}^\theta\theta \d \t z\right)\left(\delta(\d\theta)+\delta'(\d\theta)\chi_{\t z}^\theta\d\t z\right).\end{align}

What we have written in eqn. (\ref{zofo}), as a function on ${\Pi} T\Sigma$,  
is the most general section of $\BBer(\Sigma)$ for the family of super Riemann
surfaces defined by the operator $\partial'_{\t z}$ of eqn. (\ref{irod}).  
Now we have to find the condition for such a section to be holomorphic.   A small short cut is available
here.  On an ordinary complex manifold $X$ of complex dimension $m$, the condition 
for an $(m,0)$-form to be holomorphic is simply that it should be closed.
Similarly on a complex supermanifold, a section of $\BBer(\Sigma)$ understood 
as an integral form is holomorphic if and only if it is closed.  In the present context,
this means that $\h\sigma$ should be annihilated by
\begin{equation}\label{zurfo}\d =\d \t z\partial_{\t z}+\d z\partial_z+\d\theta\partial_\theta.\end{equation}
A small computation reveals that
\begin{equation}\label{urfo}\d\h\sigma=-\d\t z\,\d z\,\delta(\d\theta)
\left(\partial_{\t z}\h\phi-\partial_z(\h\phi \chi_{\t z}^\theta\theta)
+\partial_\theta(\h\phi\chi_{\t z}^\theta)\right)
.\end{equation}

Let $\h\phi(\t z;\neg z|\theta)=\h \alpha(\t z;\neg z)+\theta \h b(\t z;\neg z)$.  From (\ref{urfo}), the condition that $\d\h\sigma=0$ becomes
a pair of equations
\begin{align}\label{menz} \partial_{\t z}\h\alpha +\h b \chi_{\t z}^\theta & = 0 \cr
                                         \partial_{\t z}\h b-\partial_z\left(\h \alpha \chi_{\t z}^\theta\right)&=0.\end{align}
These equations have been obtained in \cite{DPhagain}.                                         
                                         
To understand these formulas better, let $\gamma_\red$ be a 1-cycle in $\Sigma_\red$ and $\gamma$ its pullback to $\Sigma$.  Let
us compute $\oint_{\gamma}\h\sigma$.  We integrate first over $\theta$ and $\d\theta$.  Among functions of the form
$\theta^a\delta^{(b)}(\d\theta)$ (where $\delta^{(b)}$ is the $b^{th}$ derivative of a delta function), the only
nonzero integral is $\int \D(\theta,\d\theta)\,\theta\delta(\d\theta)=1$.  
      Integrating over $\theta$ and $\d\theta$ with the help of this fact, we reduce to
an ordinary integral over $\gamma_\red$:
\begin{equation}\label{hedry}   \int_{\gamma}\h\sigma=\int_{\gamma_\red} \h\rho ,\end{equation}                              
where $\h\rho$ is a 1-form on $\Sigma_\red$:
\begin{equation}\label{yre}\hat\rho=\hat b\,\d z+ \hat\alpha\chi_{\t z}^\theta\d\t z  .\end{equation}
The second equation in (\ref{menz}) says that $\d\h\rho=0$, and therefore $\int_{\gamma_\red}\h\rho$ and 
$\int_{\gamma}\h\sigma$ only depend on the homology
class of $\gamma_\red$.

Because of our assumption that $H^0(\Sigma_\red,K^{1/2})=0$, the first equation in (\ref{menz}) has a unique solution for $\h\alpha$ in terms of $\h b$.
We let $S(z,z')$ be the propagator of the Dirac operator.  This is a holomorphic section of the line bundle\footnote{$K^{1/2}\boxtimes K^{1/2}$ is defined
as the tensor product $K_{\Sigma_\red}^{1/2}\otimes K_{\SIgma'_\red}^{1/2}$ of the square roots of the canonical bundles of $\Sigma_\red$
and $\Sigma'_\red$.  In writing formulas, we use a local complex coordinate $z$ and trivialize $K$ and $K^{1/2}$ by
$\d z$ and $(\d z)^{1/2}$.} $K^{1/2}\boxtimes K^{1/2}$ over what we will call 
$\Sigma_\red\times\Sigma'_\red$ (the product of two copies of $\Sigma_\red$)
with a delta function source on the diagonal:
\begin{equation}\label{lable}\partial_{\t z} S(z,z')=2\pi\delta^2(z,z') .\end{equation}
(We define $\delta^2(z,z')$ by $\int\d^2 z \,\delta^2(z,z')=1$ where $\d^2z =-i\,\d\t z\, \d z$.  Thus if $z=(u+iv)/\sqrt 2$, $\t z=(u-iv)/\sqrt 2$, then
$\d^2z=\d u \,\d v$.  The normalization ensures that $S(z,z')\sim 1/(z-z')$ and has been chosen to agree with
\cite{DPh,DPhagain}.)
$S(z,z')$ exists because the kernel of the Dirac operator is trivial, in other words because $H^0(\SIgma_\red,K^{1/2})=0$.  With the aid of $S(z,z')$, we have\footnote{The integral over $\SIgma_\red'$ makes sense because the quantity being integrated is naturally a $(1,1)$-form.  In their dependence on $z'$,
$\chi_{\t z'}^\theta(\t z';\neg z')\d\t z'$ is a $(0,1)$-form
valued in $T\Sigma_\red^{1/2}=K^{-1/2}$, $S(z,z')$ is a section of $K^{1/2}$, and $\h b(\t z';z') \,\d z'$ is $(1,0)$-form.}
\begin{equation}\label{beklin} \h\alpha(\t z;\neg z)=-\frac{1}{2\pi}\int_{\Sigma'_\red} S(z,z')  \chi_{\t z'}^\theta(\t z';\neg z')\,\h b(\t z';\neg z')\d^2 z'.\end{equation}
And hence we can write an equation for $\h b$ only:
\begin{equation}\label{eklin}\partial_{\t z}\h b(\t z;\neg  z)=-\frac{1}{2\pi}\partial_z \int_{\Sigma_\red'} \chi_{\t z}^\theta(\t z;\neg z)  S(z,z') 
\chi_{\t z'}^\theta(\t z';\neg z')\h b(\t z';\neg z') \d^2 z' \end{equation}

Now let us compute the super period matrix.  For this, we start at $\eta_a=0$ with the holomorphic sections $\sigma_i=\theta b_i(z)[\d z|\d\theta]$ of $\BBer(\Sigma)$ that are normalized
so that $\oint_{A^{i }}\sigma_j=\delta^i_j$.  Equivalently, the ordinary holomorphic 1-form  $\mu_j=b_j\d z$ on $\Sigma_\red$ obeys
\begin{equation}\label{nonnx}\oint_{A^i_\red}\mu_j=\delta^i_j.\end{equation}
The deformation of $\sigma_i$ as a function of the $\eta_a$ is not uniquely determined, since when we expand $\hat\sigma_i$ as a polynomial in the $\eta_a$, at each stage of the expansion we could add a monomial in the $\eta$'s
times one of the  $\sigma_j$'s.  We fix this ambiguity by requiring that
\begin{equation}\label{bonx}\oint_{A^{i}}\hat\sigma_j=\delta^i_j ,\end{equation}
independent of the $\eta$'s.  Equivalently, setting $\h\rho_i=\h b_i\d z+\h\alpha_i \chi_{\t z}^\theta \d\t z$,
we require
\begin{equation}\label{zonx}\oint_{A^i_\red}\hat\rho_j=\delta_{ij}.\end{equation}
Alternatively, we define $\hat\rho_i'=\hat\rho_i-\mu_i$, so that $\hat\rho'_i=0$ at $\eta=0$. 
In terms of $\hat\rho'_i$, the condition is
\begin{equation}\label{wonix}\oint_{A^i_\red}\hat\rho'_j=0.\end{equation}

We want to compute
\begin{equation}\label{wonx}\hat\Omega_{ij}=\oint_{B_i}\hat\sigma_j=\oint_{B_{i,\red}}\hat\rho_j.\end{equation}
Equivalently, we want to compute
\begin{equation}\label{onx}\hat\Omega_{ij}-\Omega_{ij}=\oint_{B_{j,\red}}\left(\hat\rho_i-\mu_i\right)=\oint_{B_{j,\red}}\h\rho_i'.\end{equation}
Applying the topological relation (\ref{itzo}) with $p=\mu_j$, $q=\hat\rho'_i$, and remembering that $\mu_j$ is of type $(1,0)$ on the ordinary Riemann surface $\Sigma_\red$, so that
the $(1,0)$ part of  $\hat\rho'_i$ does not contribute to $\mu_j\wedge\hat\rho'_i$,
we learn that 
\begin{equation}\label{monx}\hat\Omega_{ij}-\Omega_{ij}=\int_{\Sigma_\red} \mu_j\wedge\hat\rho'_i
=\int_{\SIgma_\red}\mu_j\wedge \hat\alpha_i \chi_{\t z}^\theta\,\d\t z.\end{equation}
In turn, we can use (\ref{beklin}) to eliminate $\hat\alpha_i$:
\begin{equation}\label{delfom}\hat\Omega_{ij}-\Omega_{ij}=-\frac{1}{2\pi}\int_{\Sigma_\red\times \SIgma'_\red}\mu_j(z)\chi_{\t z}^\theta(\t z;z)\d\t z\,  S(z,z')
 \chi_{\t z}^\theta(\t z';\neg z')\,\h b_i(\t z';\neg z')\d^2 z'.\end{equation}
When $\hat\Omega_{ij}-\Omega_{ij}$ is expanded in powers of the $\eta_a$'s, the lowest order term, which we call $\hat\Omega^{(2)}_{ij}$,
 is quadratic.  It
can found by just replacing $\hat b_i(\t z';\neg z')\d z'$ in the last formula by $b_i$.  To exhibit the symmetry of 
$\hat\Omega^{(2)}_{ij}$, we use $\d^2z=-i\d \t z\d z$ so that $b_i\d^2z =-i\d \t z\mu_i$ with $\mu_i=b_i \d z$:
\begin{equation}\label{golly} \hat\Omega^{(2)}_{ij} =\frac{i}{2\pi}\int_{\Sigma_\red\times\Sigma'{}_\red}\mu_j(z)\chi_{\t z}^\theta(\t z;z)\d\t z\,S(z,z')\chi_{\t z'}^\theta(z')\d\t z'\,\mu_i(z').\end{equation}
This formula was obtained in  \cite{DPhagain} (where our $\chi$ is called $\chi/2$).
 Higher order terms can be evaluated by using (\ref{eklin}) to express $\hat b_i$ as
a polynomial in the $\eta$'s.

To make the expression of $\hat\Omega^{(2)}_{ij}$ even more explicit, we write $\chi_{\t z}^\theta=\sum_a\eta_a f_{a\,\t z}^\theta$ as in (\ref{rodo}), and we find
\begin{equation}\label{molly}
\hat\Omega^{(2)}_{ij}=\frac{i}{2\pi}\sum_{a,b=1}^n\eta_a\eta_b\int_{\Sigma_\red\times\Sigma'{}_\red}\mu_i(z)f_{a\,\t z}^\theta(\t z;z)\d\t z \,S(z,z')f_{b\,\t z'}^\theta(\t z';z')\d \t z'\,\mu_j(z').\end{equation} 
It is interesting now to make a gauge transformation on one of the $(0,1)$-forms $f_{a\,\t z}^\theta$,
\begin{equation}\label{olly} f^\theta_{a\,\t z}\to f^\theta_{a\,\t z}+\partial_{\t z}c_a^\theta,\end{equation}
where $c_a^\theta$ is a section of $T\Sigma_\red^{1/2}$.  Upon integrating by parts, one finds  that $\hat\Omega^{(2)}_{ij}$ is invariant under this
gauge transformation of $f_{a\,\t z}^\theta$ if and generically only if the support of $c_a^\theta$ is disjoint from the support of all the  $(0,1)$-forms
$f_{b\,\t z}^\theta$, $b\not=a$.  The meaning of this is that a gauge transformation of $f_{a\,\t z}^\theta$ will, if the support of the gauge parameter is not restricted,
rotate the perturbation by $f_{b\,\t z}^\theta$ into a metric perturbation, namely the perturbation by $h_{\t z}^z$ in (\ref{zorb}). 
Differently put, as was explained before we began the computation, the particular family of super Riemann surfaces for which we have evaluated the super period matrix depends on the one-forms
$f_{a\,\t z}^\theta$ and not only on their cohomology classes.

To get a somewhat more intrinsic formula, we should include metric perturbations as well as fluctuations in the gravitino field.  We will just
briefly sketch how to do this.  In doing so, we will specialize to the case that there are only two odd parameters $\eta_a$, $a=1,2$.  The natural
thing to do is to consider a general deformation of the super Riemann surface $\Sigma$ over the ring $\C[\eta_1,\eta_2]$ that is generated over the
base field $\C$ by the two odd parameters $\eta_1,\eta_2$.   Such a deformation is characterized by the gravitino deformation 
$\chi_{\t z}^\theta=\sum_a\eta_a f_{a\,\t z}^\theta$, but also by a metric perturbation, that is a deformation of the complex structure of $\Sigma_\red$
by a $(0,1)$-form $\eta_1\eta_2 h_{\t z}^z$ that is valued in $T\Sigma_\red$.  The relevant formalism here has actually been described in detail in 
section 3.5.1 of \cite{DWtwo}, which the reader should consult for a fuller explanation.  The three fields $(f_1,f_2,h)$  (which are called $\chi^1,\,\chi^2,\,h^{12}$ in \cite{DWtwo})
are subject to the gauge invariance
\begin{align}\label{zolb} f_{a\,\t z}^\theta & \to f_{a\,\t z}^\theta+\partial_{\t z} c_a^\theta \cr h_{\t z}^z & \to h_{\t z }^z+\partial_{\t z}w^z+c_1^\theta f_{2\,\t z}^\theta-c_2^\theta f_{1\,\t z}^\theta, \end{align}  where $w^z$ is a section of $T\Sigma_\red.$
This collection of fields and gauge invariance describes the most general deformation of a split super Riemann surface over the ring $\C[\eta_1,\eta_2]$.  
The super period matrix as a function of the parameters $\eta_1,\eta_2$ is given by the formula
\begin{align}\label{moolly}
\hat\Omega^{(2)}_{ij}=&\frac{i}{2\pi}\eta_1\eta_2\int_{\Sigma_\red\times\Sigma'{}_\red}\left(\mu_i(z)f_{1\,\t z}^\theta(\t z;z)\d\t z \,S(z,z')f_{2\,\t z'}^\theta(\t z';z')\d \t z'\,\mu_j(z')+i\leftrightarrow j\right) \cr &+2\eta_1\eta_2\int_{\Sigma_\red}\d\t z\, h_{\t z}^z\,b_i b_j \d z.\end{align} 
The term proportional to $h_{\t z}^z$
 is the classical expression for the response of the ordinary period matrix to a metric perturbation by
$\eta_1\eta_2 h_{\t z}^z$. (To describe this term more intrinsically, the $T\Sigma_\red$-valued $(0,1)$-form $\d\t z h_{\t z}^z\partial_z$
is multiplied by the quadratic differential $\mu_i\mu_j$  to make a $(1,1)$-form that is then integrated.)
 The reader can verify that gauge invariance is now satisfied.

Going back to eqn. (\ref{molly}), we can now inquire about what happens to the super period matrix as $\Sigma_\red$ approaches the locus in its moduli space
at which $H^0(\Sigma_\red,K^{1/2})$ becomes nonzero.  At that point, the Dirac propagator $S$ becomes divergent because of a zero-mode contribution; it
varies as $1/\epsilon$ where $\epsilon$ is the small eigenvalue of the Dirac equation.
The formula (\ref{molly}) shows that the super period matrix develops a pole along this locus. The residue of the pole
is bilinear in odd moduli.

\subsection{Super Period Matrix In Superstring Perturbation Theory}\label{apinf}

The super period matrix has played a striking role in the computation of the superstring vacuum 
amplitude at two-loop order \cite{DPhthree}.  This receives a potential contribution only from a genus 2 surface with even spin structure.  Here
we will describe this application.

If $\Sigma_0$ is a Riemann surface of genus 2 with an even spin structure, then $H^0(\Sigma_0,K^{1/2})$ vanishes always, not just generically.
This is not true for any higher genus.  For example, if $\Sigma_0$ has genus 3, again with an even spin structure, then $H^0(\Sigma_0,K^{1/2})$ can be
nonzero if (and only if) $\Sigma_0$ is hyperelliptic.\footnote{These assertions are proved as follows.  
As $K^{1/2}$ is an even spin structure, $H^0(\Sigma_0,K^{1/2})$ has even dimension. If $\Sigma_0$ has genus 2, then $K^{1/2}$ has degree
$\g-1=1$.  In general, a line bundle of degree 1 on a compact Riemann surface $\Sigma_0$ of positive genus can have at most a one-dimensional space
of  holomorphic sections.  A nonnegative even integer that is at most 1 must be 0, so
so $H^0(\Sigma_0,K^{1/2})=0$.
  If $\Sigma_0$ has genus 3, then $K^{1/2}$ has degree $\g-1=2$.  In general, a compact Riemann surface $\Sigma_0$ of positive genus has a line bundle of degree 2
with a two-dimensional space of holomorphic sections if and only if $\Sigma_0$ is hyperelliptic. (One uses the ratio of the two sections to define
a map of $\Sigma_0$ to $\Bbb{CP}^1$, and the fact that the degree of the line bundle is 2 means that the map is a double cover.)
 If $\Sigma_0$ is a hyperelliptic curve of genus 3, it can be written
as a double cover $y^2=\prod_{i=1}^8(x-e_i)$ of $\Bbb{CP}^1$.  Consider the holomorphic differential  $\omega=\d x /y$, whose zeroes
are a pair of double zeroes at $x=\infty$, $y=\pm x^4$.   $\Sigma_0$ has an even spin structure defined by   a square root $K^{1/2}$ of the canonical bundle $K$ that has a two-dimensional space of holomorphic
sections, spanned by $\omega^{1/2}$ and $x\omega^{1/2}$.}  

Now let $\Sigma$ be a super Riemann surface of genus 2 with an even spin structure.  Because of the fact mentioned in the last paragraph, the super period
matrix $\hat\Omega_{ij}$ of $\Sigma$ is always defined.

Another important fact is that every $2\times 2$ symmetric complex matrix $\Omega_{ij}$ with positive definite imaginary part is the
period matrix of a genus 2 Riemann surface that is unique up to isomorphism.  (The moduli space of genus 2 surfaces has complex dimension $3\g-3=3$,
which coincides with the dimension of the space of symmetric $2\times 2$ matrices.) This actually remains true for genus 3, but for $\g\geq 4$, a generic $\Omega_{ij}$
is not a period matrix.

From what has been said in the last paragraph, if $\Sigma$ is a super Riemann surface of genus 2 with an even spin structure, then there is an ordinary
Riemann surface $\Sigma_0$ of genus 2 whose period matrix $\Omega_{ij}$ coincides with the super period matrix $\hat\Omega_{ij}$ of $\Sigma$.  $\Sigma$ is a slightly
thickened and deformed version of $\Sigma_0$ (thickened by its fermionic dimension and deformed by its 
odd moduli), so up to homology there is a natural map from $\Sigma$ to $\Sigma_0$.
This means that it makes sense to endow $\Sigma_0$ with the same spin structure as $\Sigma$.

The map from $\Sigma$ to $\Sigma_0$ is a holomorphic map $\pi$ from $\MM_{2,+}$, the moduli space of super Riemann surfaces of genus 2 with even spin structure,
to $\M_{2,\mathrm{Spin+}}$, the moduli space of ordinary Riemann surfaces of genus 2 with an even spin structure.  But $\M_{2,\mathrm{Spin+}}$ is the reduced
space of $\MM_{2,+}$, and so is naturally embedded in $\MM_{2,+}$ as its subspace that parametrizes
split super Riemann surfaces.
The map $\pi:\MM_{2,+}\to \M_{2,\mathrm{Spin+}}$ is the identity when restricted to split super Riemann surfaces
(since  $\h\Omega_{ij}=\Omega_{ij}$ if the odd moduli are zero), so it
is a holomorphic splitting of $\MM_{2,+}$.
Given such a holomorphic splitting, we can attempt to calculate superstring scattering amplitudes by integrating first over the odd moduli -- that is over the fibers
of $\pi$ -- to reduce to an integral over the purely bosonic moduli space $\M_{2,\mathrm{Spin+}}$.  This approach has turned out to be very powerful, as reviewed in  \cite{DPhthree}.

As one can learn from that reference and from additional papers cited there, it has also been possible to compute certain scattering amplitudes -- as well as the
vacuum amplitude -- in genus 2.  At the present level of understanding, this was a rather difficult calculation that has not yet been fully elucidated in terms of supergeometry.

\section{$\N=2$ Super Riemann Surfaces and Duality}\label{duality}

So far our super Riemann surfaces have all been $\N=1$ super Riemann surfaces, which provide a natural framework for perturbative Type I, 
Type II, and heterotic superstrings.
Here we will describe some rather pretty facts \cite{Delone,DRS,BR,RoRa} about
$\N=2$ super Riemann surfaces, complex supermanifolds of dimension $1|1$, and duality.  The reader  will hopefully  help explain
what these facts mean for string theory.\footnote{The papers \cite{BV,BVtwo} may provide a clue.}

\subsection{Definition of an $\N=2$ Super Riemann Surface}\label{deltwo}

First we explain the definition of an $\N=2$ super Riemann surface \cite{Cohn}. An $\N=2$ super Riemann surface is a complex supermanifold
$W$ of dimension $1|2$ with some additional structure.  Its tangent bundle $TW$ is endowed with 2 subbundles $\D_+$ and $\D_-$,
each of rank $0|1$, with the following properties.  $\D_+$ is integrable in the sense that if $D_+$ is a nonzero section of $\D_+$,
then $\{D_+,D_+\}=2D_+^2$ is a multiple of $D_+$,
\begin{equation}\label{tonic}D_+^2= uD_+,\end{equation}
for some function $u$. Thus, ``integrability'' means that the sections of $\D_+$ form a graded Lie algebra, without needing to introduce $D_+^2$ as an independent
generator.  Similarly $\D_-$ is integrable, in the sense that if $D_-$ is a nonzero section of $\D_-$, then
\begin{equation}\label{onic}D_-^2= vD_-,\end{equation}
again for some function $v$.  However, the direct sum $\D=\D_+\oplus \D_-$ is nonintegrable, the obstruction being that, for any
nonzero sections $D_+$ and $D_-$ of $\D_+$ and $\D_-$, the anticommutator $\{D_+,D_-\}$ is linearly independent of $D_+$ and $D_-$.
Thus, $D_+$, $D_-$, and $\{D_+,D_-\}$ form a basis of $TW$.  From this it follows that the quotient $TW/(\D_+\oplus \D_-)$ is naturally
isomorphic to $\D_+\otimes \D_-$, so that there is an exact sequence
\begin{equation}\label{zonic}0\to \D_+\oplus \D_-\to TW\to \D_+\otimes \D_-\to 0.\end{equation}
Reasoning as in appendix \ref{obz}, one can deduce that $\BBer(W)$ is canonically trivial.  This is relevant 
among other things for writing Lagrangians.

By arguments similar to those in section \ref{holp}, one can show that locally, there exist superconformal coordinates $z|\theta^+,\theta^-$ 
in which $\D_+$ and $\D_-$ are generated by
\begin{align}\label{crest}D_+ & = \frac{\partial}{\partial \theta^+}+\theta^{-}\frac{\partial}{\partial z} \cr D_-& = 
\frac{\partial}{\partial\theta^-}+\theta^+\frac
{\partial}{\partial z}.\end{align}
These obey
\begin{equation}\label{nox}D_+^2=D_-^2=0,~~~\{D_+,D_-\}=2\partial_z.\end{equation}
Most of what we have described in section \ref{bintro} has an analog for $\N=2$.  For instance, one can
define a sheaf $\S$ of vector fields that preserve the $\N=2$ structure (its sections are explicitly described
in eqns. (\ref{moxie}) and (\ref{zoxie})), and a general $\N=2$ super Riemann
surface can be constructed by gluing together open sets in which the $\N=2$ structure takes the standard form described
in  eqn. (\ref{crest}).

Since $D_-^2=0=D_+^2$, one can look for a holomorphic function that is annihilated by $D_-$ or by $D_+$.  Since we can write
\begin{align}\label{yugo} D_-&=\exp(-\theta^-\theta^+\partial_z)\frac{\partial}{\partial\theta^-}\exp(\theta^-\theta^+\partial_z)\cr 
D_+&=\exp(\theta^-\theta^+\partial_z)\frac{\partial}{\partial\theta^+}\exp(-\theta^-\theta^+\partial_z),\end{align}
a function annihilated by $D_-$ is concretely of the form
\begin{equation}\label{reft}\exp(-\theta^-\theta^+\partial_z)f(z|\theta^+)=f(z-\theta^-\theta^+|\theta^+)\end{equation}
and a function annihilated by $D_+$ is concretely of the form
\begin{equation}\label{zeft}\exp(\theta^-\theta^+\partial_z)g(z|\theta^-)=g(z+\theta^-\theta^+|\theta^-).\end{equation}
Functions annihilated by $D_-$ or by $D_+$ are called chiral or antichiral, respectively.

Geometrically, one can think of a chiral function as a function on a space $X$ of dimension $1|1$ parametrized by
$z-\theta^-\theta^+|\theta^+$,  and an antichiral function as a function on another space $X'$ of dimension $1|1$ parametrized
by $z+\theta^-\theta^+|\theta^-$.    Here is a slightly different way to define $X$ and $X'$ that does not depend on a choice of
coordinates.  Integrability of $\D_-$ means that $W$ is fibered by the orbits generated by any nonzero section $D_-$ of $\D_-$; those
orbits are independent of the choice of $D_-$.  $X$ is the space of these orbits,\footnote{Roughly, integrability is important
here because if  $D_-^2$ is not a multiple of $D_-$, then to make a Lie supergroup with $D_-$ in its
Lie algebra, we would have to include also $D_-^2$ as a generator of the Lie algebra, and then the orbits would not have dimension $0|1$.  One could still define
curves in $W$ generated by $D_-$, but these curves would not be fibers of a fibration.  See section \ref{donw}.} which we call simply the orbits of $\D_-$.
$X'$ is similarly the space of orbits of $\D_+$.

Thus an $\N=2$ super Riemann surface $W$ has maps to two different $1|1$-dimensional supermanifolds $X,$ $X'$ with the properties
that functions on $X$ or $X'$ correspond to chiral or antichiral functions on $W$:
\begin{equation}\label{kaboo}  
\begin{array}{lcl}
&W&\\
\; \; \swarrow&&\searrow \\
 X&&\; \; X'.
\end{array} 
 \end{equation}
It turns out that individually $X$ and $X'$ have no special structure at all, beyond being complex supermanifolds of dimension $1|1$.
But there is a very special relationship between them, which is the subject of section \ref{scduality}.

A quick way (see  \cite{Kac} and \cite{DRS}) to show that $X$ (or  $X'$) has no structure beyond being a complex supermanifold of dimension $1|1$ is
to consider the action on $X$ of the Lie algebra of $\N=2$ superconformal vector fields.  A superconformal vector field on the $\N=2$
super Riemann surface $W$ is a holomorphic vector field that preserves the two subbundles $\D_+$ and $\D_-$ of $TW$.   Eqn. (\ref{reft})
implies that  in the coordinate system $w|\theta^+,\theta^-$
where $w=z-\theta^-\theta^+$,  the subbundle $\D_-$ is generated by
\begin{equation}\label{doft}D_-=\frac{\partial}{\partial\theta^-} .\end{equation}
 Similarly, in these coordinates, $\D_+$ is generated by
\begin{equation}\label{woft}D_+=\frac{\partial}{\partial\theta^+}+2\theta^-\partial_w. \end{equation}
The even superconformal vector fields on $W$ take the form
\begin{align}\label{moxie} V_g& = g(w)\partial_w+\frac{\partial_wg}{2}\left(\theta^+\partial_{\theta^+}+\theta^-\partial_{\theta^-}\right)\cr
v_k&=k(w)\left(\theta^-\partial_{\theta^-}-\theta^+\partial_{\theta^+}\right)  .\end{align}
The odd ones are
\begin{align}\label{zoxie}  \nu_{\alpha^+}& = \alpha^+(w) \partial_{\theta^+}\cr
   \nu_{\alpha^-} & = \left(\alpha^-(w) +2\partial_w\alpha^-(w)\theta^-\theta^+\right)\partial_{\theta^-}-2\alpha^-(w) \theta^+\partial_w. \end{align}
 Here $g, k$, and $\alpha^\pm$ are functions of $w$ only.  The vector fields in (\ref{moxie}) and (\ref{zoxie}), which are sections of the
 sheaf $\S$ of superconformal vector fields, generate what
 is usually called the $\N=2$ superconformal algebra.  To find the action of this algebra on $X$, we simply drop all terms 
 proportional to $\partial_{\theta^-}$, since $X=W/\D_-$ is the space of orbits of $\partial_{\theta^-}$, which generates $\D_-$.
 So as vector fields on $X$, the $\N=2$ generators are
 \begin{align}\label{oxie} V_g& = g(w)\partial_w+\frac{\partial_wg}{2}\theta^+\partial_{\theta^+}\cr
v_k&=-k(w)\theta^+\partial_{\theta^+}\cr    
 \nu_{\alpha^+}& = \alpha^+(w) \partial_{\theta^+}\cr
   \nu_{\alpha^-} & =-2\alpha^-(w) \theta^+\partial_w.   \end{align}
 An arbitrary holomorphic vector field $f(w|\theta^+)\partial_w+\rho(w|\theta^+)\partial_{\theta^+}$ can be written as a linear
 combination of these $\N=2$ generators in a unique fashion, so the $\N=2$ superconformal symmetries are simply the automorphisms
 of $X$ as a $1|1$ supermanifold.  
 
 {\it A priori}, the sheaf $\S$ of superconformal vector fields on $W$ is a sheaf of Lie algebras over $W$, but not a sheaf of $\O_W$ modules
 (a superconformal vector field cannot be multiplied by a holomorphic function on $W$ to give another superconformal vector field).
 However, we can proceed rather as we did for $\N=1$ super Riemann surfaces in section \ref{reconsider}.
 Viewing  $\S$ as a subsheaf of the sheaf $TW$ of all holomorphic vector fields on $W$, we can
 use the exact sequence (\ref{zonic}) and project $\S$ to $\D_+\otimes \D_-$.  A short computation using the above formulas reveals that this
 projection is an isomorphism, so $\S$ is isomorphic as a sheaf to the sheaf of sections of the line bundle $\D_+\otimes \D_-\to W$.

\subsection{$1|1$ Supermanifolds and Duality}\label{scduality}

Let $X$ be any complex supermanifold of dimension $1|1$.  Locally, we can parametrize $X$ by holomorphic coordinates $z|\psi$.

Now we want to consider minimal divisors in $X$.  By a minimal divisor we mean a subvariety of $X$ of dimension $0|1$ whose intersection with the
reduced space $X_\red$ is a single point.   Such a divisor is defined by an equation of the form
\begin{equation}\label{muthro} z =y+\eta\psi,\end{equation}
where  $y$ and $\eta$ parametrize the divisor.   So the space of  such minimal divisors is a $1|1$-dimensional supermanifold $X'$, locally
parametrized by $y|\eta$.

Actually, the relation between $X$ and $X'$ is symmetrical.  With $y$ and $\eta$ viewed as parameters, eqn. (\ref{muthro}) defines a minimal divisor in $X$,
but if instead we view $z$ and $\psi$ as parameters, then eqn. (\ref{muthro}) defines a minimal divisor in $X'$.  So $X$ parametrizes minimal
divisors in $X'$, and vice-versa.

Another way to explain the symmetric nature of the relationship between $X$ and $X'$ is as follows.  Suppose that $X'$ is the space of minimal divisors
in $X$. To a point $z|\psi$ in $X$, we associate the space of all minimal divisors that pass through this point.  This is a $0|1$-dimensional family of divisors,
so it represents a divisor in $X'$.  (To show this explicitly, we would go back to the equation (\ref{muthro}) which in local coordinates describes the family
of minimal divisors that pass through $z|\psi$.)  So again, just as a point in $X'$ determines a minimal divisor in $X$, so a point in $X$ determines
a minimal divisor in $X'$.

Now we will define a supermanifold $W$ of dimension $1|2$.  One way to define $W$ is 
that a point in $W$ is a pair consisting of a point in $X$ and a minimal
divisor through that point.  Because points in $X$ correspond to minimal divisors in $X'$ and vice-versa, 
we can equally well say that a point in $W$ is a point
in $X'$ together with a divisor through that point.  If we use local coordinates $z|\psi$ on $X$ and 
local coordinates $y|\eta$ on $X'$ as described above
so that the divisor in $X$ corresponding to $y|\eta$ is described by the equation (\ref{muthro}), 
then a point in $W$ is simply described by the whole
collection of coordinates $y,z|\eta,\psi$, subject to the condition (\ref{muthro}).   In other words, 
locally $W$ is defined as the submanifold of  $C^{2|2}$ defined
by $G=0$ where $G=z-y-\eta\psi$.

So now we have a complex supermanifold $W$ of dimension $1|2$ with maps to two 
different supermanifolds $X$ and $X'$, each of dimension $1|1$.
In fact, $W$ has the natural structure of an $\N=2$ super Riemann surface.  Abstractly, the subbundles $\D_+$ and $\D_-$ of $TW$ are the tangents
to the fibers of the  fibrations $W\to X$ and $W\to X'$.  More concretely, we define $\D_-$ to 
be spanned by vector fields on $W$ that commute with $z|\psi$ (the local coordinates of $X$).
A section of $\D_-$ can be represented by a vector field on $\C^{2|2}$ that 
commutes with $z,\psi,$ and $G$.  Such a vector field is a multiple of
\begin{equation}\label{incxo}D_-=\frac{\partial}{\partial\eta}-\psi\frac{\partial}{\partial y}.\end{equation}
Similarly, a section of $\D_+$ can be represented by a vector 
field on $\C^{2|2}$ that commutes with $y,\eta,$ and $G$. Such a vector field is a multiple of
\begin{equation}\label{zincxo} D_+=\frac{\partial}{\partial\psi}-\eta\frac{\partial}{\partial z}.\end{equation}
This exhibits the structure of an $\N=2$ super Riemann surface: 
we have $D_-^2=D_+^2=0$, while $\{D_-,D_+\}=-\partial_y-\partial_z$ is everywhere linearly
independent of $D_-$ and $D_+$.

So every $1|1$ supermanifold $X$ canonically determines an $\N=2$ super Riemann 
surface $W$. And conversely, we can reconstruct
$X$ from $W$ as roughly $W/\D_-$, as described in section \ref{deltwo}.

This construction has an amusing variant that we will describe briefly.  Starting with $X$, 
we introduce ${\Pi} TX$, the tangent bundle with statistics
reversed on the fiber.  This object was studied, for example, in \cite{Wittennotes}, section 3.2.  
If we describe $X$ by local coordinates $z|\psi$, then
to describe ${\Pi} TX$, we introduce additional coordinates $\d\psi$ and $\d z$, where 
$\d\psi$ is even and $\d z$ is odd.  On ${\Pi} TX$ there is a natural
odd vector field
\begin{equation}\label{izvo}
\d=\d z\frac{\partial}{\partial z}+\d\theta\frac{\partial}{\partial \theta}.\end{equation}
It is usually called the exterior derivative on $X$.  It obeys $\d^2=0$.

There is a scaling symmetry of ${\Pi} TX$ that rescales the fiber coordinates, 
$\d\theta|\d z\to \lambda\d\theta|\lambda\d z$, where $\lambda$ is even and nonzero.
We define $W=\Bbb{P}({\Pi} TX)$ to be the projectivization  of ${\Pi} TX$.  This projectivization 
is obtained by constraining the even fiber coordinates (in our present problem, only $\d\theta$) to be 
not all zero and then dividing by $\C^*$.  Locally, we can use the $\C^*$ action to map $\d\theta$ to 1, 
after which $W$ is a fiber
bundle over $X$ with fibers of rank $0|1$ parametrized by $\d z$.  
We define $\D_-$ to be the subbundle of $TW$ generated by
\begin{equation}\label{jhdf}D_-=\frac{\partial}{\partial \d z}.\end{equation}
Clearly $D_-^2=0$, so $\D_-$ is integrable.  The quotient $W/\D_-$ is just $X$.

The vector field $\d$ on ${\Pi} TX$ is not invariant under scaling, so it does not descend to a vector 
field on $W$ in a natural way.  However, scaling
of $\d$ does not affect the subbundle of ${\Pi} TX$ generated by $\d$, so that subbundle 
 does descend in a natural way to a subbundle $\D_+$ of $TW$ of rank $0|1$.  Locally,
$\D_+$ is generated by what we get from $\d$ upon using the scaling to set $\d\theta=1$.  Thus, 
$\D_+$ is generated by
\begin{equation}D_+=\d z\frac{\partial}{\partial z}+\frac{\partial}{\partial\theta}.\end{equation}
Again we see the structure of an $\N=2$ super Riemann surface, with $D_-^2=D_+^2=0$, $\{D_-,D_+\}=\partial_z$.

The relation of this description of $W$ to the description in terms of minimal divisors in $X$ is simply the following.
At a point  in $p\in W$, a nonzero vector  $\mu\in \Pi TX|_p$ (where $\Pi TX|_p$ is the fiber of $\Pi TX$ at
$p$) determines a minimal divisor through $p$, namely the one obtained by displacing $p$ in the $\mu$ direction.
This divisor only depends on $\mu$ up to scaling, so the space of minimal divisors through $p$ is the fiber of $\Bbb{P}(\Pi TX)$.

\subsection{Self-Duality Of $\N=1$ Super Riemann Surfaces}\label{donw}

Since we have found a natural duality for every $1|1$-dimensional complex supermanifold $X$, 
the question now arises of what happens
if $X$ is actually a super Riemann surface.

Given any odd vector field $\nu$ on $X$, we regard $\nu$ as generating a superdiffeomorphism of $X$.  Starting at any point and acting
with $\exp(\alpha\nu)=1+\alpha\nu$, where $\alpha$ is an odd parameter, we generate a $0|1$-dimensional submanifold of $X$ that passes
through the given point.  This much holds for any odd vector field, but in going farther there are two fundamentally
different cases: $\nu$ may be integrable or non-integrable.
  Integrability means  that (after possibly multiplying
$\nu$ by a nonzero function, an operation that does not affect the curves it generates) $\nu^2=0$.  If so, we can find local coordinates $z|\theta$ on $X$ with
$\nu=\partial_\theta$.   Then $\nu$ has a family of orbits of dimension $1|0$, parametrized by $z$, and $X$ is fibered over this orbit space.

A super Riemann surface $X$ is at the opposite extreme. Locally, the superconformal structure determines,
 up to multiplication by a nonzero function (which does not affect the following)
a natural odd vector field on $X$, namely
\begin{equation}\label{zonky} D_\theta=\frac{\partial}{\partial\theta}+\theta\frac{\partial}{\partial z}.\end{equation}
But $D_\theta$ is everywhere non-integrable, that is,  $D_\theta^2$ is everywhere nonzero and  linearly independent from $D_\theta$.
Given any point $z|\theta=z_0|\theta_0$ in $X$,
the curve generated by $D_\theta$ passing through this point is given in parametric form as
\begin{align}\label{prot} \theta & = \theta_0+\alpha \cr
                                                z & = z_0+\alpha\theta\end{align}
                                                with odd parameter $\alpha$.
Equivalently, it is given by an equation
\begin{equation}\label{irot}z=z_0-\theta_0\theta.\end{equation}
These formulas  were already introduced in section \ref{definitions}.

What has happened as a result of the non-integrability of $\D$ is that these $0|1$-dimensional varieties  do not fit together into a fibration of $X$.  
There are too many of them; the family of divisors in (\ref{irot}) has dimension $1|1$, with parameters $z_0$ and $\theta_0$,
and not $1|0$, as in the integrable case.

 We see in (\ref{irot})  that a point $z_0|\theta_0\in X$
naturally determines a minimal divisor in $X$, and every such divisor occurs for a unique $z_0|\theta_0$.  This means that $X$ is self-dual under
the duality of $1|1$ supermanifolds: $X$ parametrizes its own minimal divisors.  
More specifically, the holomorphic isomorphism $\varphi:X\cong X'$ is such that for $z_0|\theta_0\in X$, the divisor in $X$ corresponding to $\varphi(z_0|\theta_0)$, namely the divisor defined in (\ref{irot}),  passes
through $z_0|\theta_0$.

Conversely, suppose that $X$ is a $1|1$-dimensional complex manifold and that we are given an isomorphism $\varphi:X\cong X'$, where $X'$ parametrizes
minimal divisors in $X$ and for $z_0|\theta_0\in X$, the divisor $\varphi(z_0|\theta_0)$ passes through $z_0|\theta_0$.  Then we can define a subbundle $\D\subset TX$ of rank $0|1$
by saying that the fiber of $\D$ at $z_0|\theta_0$ is the tangent space to $\varphi(z_0|\theta_0)$.  $\D$ is everywhere nonintegrable or $\varphi$ would not be an isomorphism.  (We have
just seen that if $\D$ is integrable, it does not have enough orbits for $\varphi$ to be an isomorphism.)
So $X$ has a natural super Riemann surface structure.

\vskip 1cm
\noindent {\bf Acknowledgments} Research  supported in part by NSF grant  PHY-0969448.  I thank P. Deligne for 
much valuable advice
over the years on  supergeometry and super Riemann surfaces, and especially for explaining 
to me many of the matters that are summarized in section 6, as well
as some details in section 5.
Helpful comments on the manuscript have also been provided by K. Becker, J. Distler, G. Moore,
and especially D. Robbins and B. Safdi.

\appendix
  
\section{More On Berezinians}\label{obz}
In section \ref{holvol}, we established an isomorphism between $\BBer(\Sigma)$ and $\D^{-1}$.  
Here we explain an alternative proof of this isomorphism that requires 
less computation and more knowledge about Berezinians.

First, recall that for every vector space $V$ of dimension $a|b$, one defines a 
one-dimensional vector space $\Ber(V)$, the Berezinian of $V$, of statistics $(-1)^b$.
As explained for instance in section 3.1 of \cite{Wittennotes}, to every basis 
$e_1\dots|\dots\rho_b$ of $V$, there is a corresponding
basis element $[e_1\dots|\dots \rho_b]$ of $\Ber(V)$.  If a new basis  
$e_1'\dots|\dots \rho'_b$ is obtained from $e_1\dots|\dots\rho_b$ by a linear
transformation $W$, then $[e_1'\dots|\dots\rho'_q]=\Ber(W)[e_1\dots|\dots\rho_b]$.  (Here $\Ber(W)$ is the
Berezinian of the linear transformation $W$, the superanalog of the determinant.)
If $V$ varies as the fibers of a vector bundle $R\to X$ for some
supermanifold $X$, then by applying this construction fiberwise, we make a line bundle over 
$X$ that we call $\Ber(R)$, the Berezinian of $R$.
If $R$ is a holomorphic vector bundle over a complex supermanifold, then $\Ber(R)$ 
is a holomorphic line bundle, and in that case we denote it as $\BBer(R)$.

We will phrase the argument that follows using a slightly different definition of the Berezinian 
of a vector bundle.  (This alternative definition is presented
here for variety, but will actually be slightly less precise than what was said in the last 
paragraph, since -- in the form described here -- we will
define the Berezinian line bundle only up to isomorphism.)  We leave it to the reader to relate the two definitions. 
Picking an open cover $U_\alpha$ of $X$, suppose that the vector bundle $R$ can be defined by 
transition functions $r_{\alpha\beta}$ on intersections
$U_\alpha\cap U_\beta$, obeying the usual conditions such as the requirement that
\begin{equation}\label{plod} r_{\alpha\beta}r_{\beta\gamma}r_{\gamma_\alpha}=1 \end{equation}
on triple intersections $U_\alpha\cap U_\beta \cap U_\gamma$.  Then we define the 
line bundle $\Ber(R)$ via the transition
functions $r^*_{\alpha\beta}=\Ber(r_{\alpha\beta})$.  The multiplicative property of the 
Berezinian of a matrix together with (\ref{plod})
ensures that $r^*_{\alpha\beta}r^*_{\beta\gamma}
 r^*_{\gamma\alpha}=1$ in $U_\alpha\cap U_\beta \cap U_\gamma$, so the objects 
 $r^*_{\alpha\beta}$ really are transition functions
of a line bundle, which we will call $\Ber(R)$.   If $R$ has purely bosonic fibers, $\Ber(R)$ 
reduces to what is usually called the determinant
line bundle of $R$.  If $R$ is a holomorphic vector bundle over a complex supermanifold, 
then $\Ber(R)$ is obviously a holomorphic line bundle.

Now suppose that we are given an exact sequence of vector bundles over $X$,
\begin{equation}\label{proxx} 0 \to Q\to R\to S\to 0.\end{equation}
This means that transition functions of $R$ can be put in the triangular form
\begin{equation}\label{roxx} r_{\alpha\beta}=\begin{pmatrix} q_{\alpha\beta} & c_{\alpha\beta}\cr 0 & s_{\alpha\beta}\end{pmatrix},\end{equation}
where $q_{\alpha\beta}$ and $s_{\alpha\beta}$ are, respectively, transition functions for $Q$ and for $S$.
($q$, $r$, and $s$ all act in general on both odd and even variables; we have not tried to display this in (\ref{roxx}).)
From this triangular form, it follows that 
\begin{equation}\label{zolow}\Ber(r_{\alpha\beta})=\Ber(q_{\alpha\beta})\Ber(s_{\alpha\beta}). \end{equation}
$\Ber(q_{\alpha\beta})$ and $\Ber(s_{\alpha\beta})$ are transition functions for the  line bundles $\Ber(Q)$ and $\Ber(S)$,
so their products $\Ber(q_{\alpha\beta})\Ber(s_{\alpha\beta})$ are transition functions for the tensor product $\Ber(Q)\otimes \Ber(S)$.
Thus (\ref{zolow}) gives
a natural isomorphism $\Ber(R)\cong \Ber(Q)\otimes \Ber(S)$ for every exact sequence (\ref{proxx}).  This generalizes the analogous
isomorphism for determinant line bundles in the purely bosonic case.  The isomorphism $\Ber(R)\cong \Ber(Q)\otimes \Ber(S)$ is clearly
holomorphic in the case of an exact sequence of holomorphic vector bundles.

The cotangent bundle of a super Riemann surface appears in the exact sequence  (\ref{zomurk}):
\begin{equation}\label{tomurk} 0 \to \D^{-2}\to T^*\Sigma\to \D^{-1}\to 0.\end{equation}
This implies a natural isomorphism
\begin{equation}\label{trex}\BBer(T^*\Sigma)\cong \BBer(\D^{-2})\otimes \BBer(\D^{-1}). \end{equation}
Now if $\lambda$ is a nonzero complex number which we regard as a $1\times 1$ matrix acting on a vector space $E$ of dimension $1|0$,
then $\Ber\,\lambda=\lambda$, but if $E$ has dimension $0|1$, then $\Ber\,\lambda=\lambda^{-1}$.  This follows directly from the definition
of the Berezinian of a matrix.  For a line bundle $\L\to X$, considering that the passage from $\L$ to $\Ber(\L)$ involves taking the
Berezinians of $1\times 1$ matrices (namely the transition functions of $\L$), we see that $\Ber(\L)\cong \L$ if $\L$ has bosonic fibers,
but $\Ber(\L)\cong \L^{-1}$ if $\L$ has fermionic fibers.  In (\ref{trex}), $\D^{-2}$ has bosonic fibers, while $\D^{-1}$ is fermionic,
so finally we get  $\BBer(T^*\Sigma)\cong \D^{-2}\otimes \D\cong \D^{-1}$.  But $\BBer(T^*\Sigma)$ is what we usually call  $\BBer(\Sigma)$.

\section{More On Pictures}\label{morep}

Here we will make a few remarks on physical states or vertex operators in superstring theory for different values of the picture number.  The goal is to make
contact with observations of section \ref{pictures} about how the moduli space of super Riemann surfaces can be modified to accommodate different picture numbers.

The commuting ghosts of superstring theory are a pair of free fields $\beta$ and $\gamma$ of conformal dimension $3/2$ and $-1/2$, respectively.  (They appear in eqn.
(\ref{ghosts}) along with their anticommuting partners.)  Their mode expansions read
\begin{equation}\label{helom} \beta(z)=\sum_n z^{-n-3/2}\beta_n,~~~\gamma(z)=\sum_n z^{-n+1/2}\gamma_n. \end{equation}
$n$ takes values in $\Z+\epsilon$, where $\epsilon=1/2$  in the NS sector, and 0 in the Ramond sector.    The ghost vacuum $|q\rangle$ with picture number $q$ is defined by the conditions 
\begin{align}\label{elom}  \beta_n|q\rangle & = 0,~~~n>-q-3/2  \cr
                                              \gamma_n|q\rangle & = 0,~~~n\geq q+3/2.  \end{align}
Superstring theory also has anticommuting ghosts $b$ and $c$ of conformal dimensions 2 and $-1$, respectively, and expansions
\begin{equation}\label{zelom} b(z)=\sum_n z^{-n-2}b_n,~~~c(z)=\sum_n z^{-n+1}c_n. \end{equation}                                              
                                                
We want to consider a physical state $|\Psi\rangle$ of the combined system for which the commuting ghosts are in the state $|q\rangle$.  
We also assume that 
\begin{align}\label{belom} c_n|\Psi\rangle=&0,~~n\geq 1 \cr
                                                 b_n|\Psi\rangle=& 0,~~~n \geq 0. \end{align}
(For free fermions as opposed to free bosons, there is no close analog of picture number, so we do not introduce a free parameter here.)                                                 
Thus the dependence of $|\Psi\rangle$ on the superghosts is completely fixed, but so far we have said nothing about the matter part of the state.
                                                 
A physical state must be annihilated by the BRST operator 
\begin{equation}\label{urtu}Q=\sum_{n\in\Z+\epsilon}  \gamma_n G_{-n}+\sum_{m\in\Z}c_mL_{-m}+\dots\end{equation}
where $G_n$ and $L_m$ are respectively the odd and even generators of the super Virasoro algebra.  These operators can be realized by the 
superconformal vector fields
\begin{equation}\label{burtu} G_n=z^{n+1/2}(\partial_\theta-\theta\partial_z),~~ L_m=-\left(z^{m+1}\partial_z+\frac{m+1}{2}z^m\theta\partial_\theta\right),\end{equation}
from which one can read off their commutation relations.    The conditions for a state of the form described above to be annihilated by $Q$  are
\begin{align}\label{nurtu}  G_n|\Psi\rangle = & 0 , ~~~n\geq -q-1/2\cr
                                               L_n|\Psi\rangle= & 0 , ~~~ n\geq 0.\end{align}
So, setting $r=n+1/2$, the state $|\Psi\rangle$ is annihilated by the operators that represent the vector fields
\begin{equation}\label{wurtu} z^r(\partial_\theta-\theta\partial_z), ~~ r\geq -q,\end{equation}                                                
as well as
\begin{equation}\label{purtu} z^{m+1}\partial_z +\frac{m+1}{2}z^m\theta\partial_\theta,~~m\geq 0.\end{equation}

We see that this construction will not work unless $q<0$.  If $q=0$, the constraints (\ref{nurtu}) include $G_{-1/2}|\Psi\rangle=0$,
which implies that $|\Psi\rangle$ is annihilated by $L_{-1}=G_{-1/2}^2$; when added to the constraints (\ref{nurtu}), this becomes too strong
and forces the matter part of the state $|\Psi\rangle$ to be trivial.  Things are even worse if $q>0$. Then we would get the constraint generated
by $G_{-3/2}$ and by the time we close the constraints on a Lie algebra, we would find  that $|\Psi\rangle$ is annihilated by the entire super Virasoro algebra.

However, the above construction makes perfectly good sense if $q<0$.  In fact, in section \ref{pictures}, we made a proposal for what should be  the sheaf $\S'$ of superconformal vector
fields in the presence of a puncture associated to a vertex operator of picture number $q<0$.  The proposal was precisely that sections of $\S'$ should have the behavior given
in eqns. (\ref{wurtu}) and (\ref{purtu}) near $z=\theta=0$.

Since the constraints obeyed by the vertex operators that we have just introduced are those in eqn. (\ref{nurtu}),
the vector fields that we should regard as symmetry generators when we construct a moduli space of super Riemann surfaces  in which those vertex operators are inserted are those in 
eqns. (\ref{wurtu}) and (\ref{purtu}).  The corresponding supermoduli spaces make perfect sense and were described in section \ref{pictures}. 
These supermoduli spaces coincide with the standard ones if and only if $q$ has the canonical value $-1$ for the NS sector,
or $-1/2$ for the R sector.   Precisely in these
cases, the appropriate vertex operators of the matter sector are the usual superconformal primaries.

For $q$ less than the canonical values  $-1$ and $-1/2$, the conditions obeyed by the vertex operators are different.  
We lose some constraints on $|\Psi\rangle$, but we gain some gauge-invariances
$|\Psi\rangle\to |\Psi\rangle+G_n|\chi\rangle$, $0\leq n\leq -q-3/2$.  The net effect is that the space of physical states is the same. For example, see \cite{HMM,BeZw} for proofs
of this assertion.  For $q<-1$ or $q<-1/2$, the extra  states $G_n|\chi\rangle$ that
should decouple from superstring scattering amplitudes are projected out when one integrates over the extra  odd moduli that exist for the given
value of $q$.

What about vertex operators with picture number $q\geq 0$?  They certainly exist as well; inevitably, in view of what we have explained, they take a more complicated form than the
operators described above.
In fact, in \cite{FMS}, a natural picture-changing operation was described that increases $q$ by 1.   This operation is very useful in computing perturbative superstring amplitudes
of low orders.  In \cite{EHV}, it was interpreted as expressing the result of integrating over one of the odd moduli.  (The odd modulus in question is represented by a gravitino
with delta function support at the position of the vertex operator.)  In genus zero, one can certainly calculate with vertex operators of any picture number.  It seems unlikely -- at least
to the author -- that in general there is a natural recipe to calculate genus $\g$ scattering amplitudes using vertex operators of $q\geq 0$, because there is no good way in general to remove
an odd modulus from the supermoduli space $\MM$ defined at a canonical picture number.

\section{Picture-Changing On A Super Riemann Surface}\label{zelfo}

In section \ref{symmetry}, we made use of a map described in \cite{RSV} from a 
holomorphic section $\sigma$ of $\BBer(\Sigma)$
to a closed holomorphic 1-form $\mu$.  In local superconformal coordinates $z|\theta$, 
this map takes  $\sigma=\phi(z|\theta) [\d z|\d\theta]$
to the closed holomorphic 1-form 
\begin{equation}\label{pogo}\mu=\phi(z|\theta)\d\theta+(\d z-\theta\d\theta)D_\theta\phi.\end{equation}
(We take $\sigma$ to be even and $\phi(z|\theta)$ to be odd.) 

It is not immediately obvious why this map does not depend on the choice of local superconformal coordinates, or why $\mu$ is closed.
However, a natural explanation has been given in \cite{Beltwo}.  First of all, since $\sigma$ is an integral form while $\mu$ is a differential form,
the map from $\sigma$ to $\mu$ involves ``picture-changing.''  Thus, we can represent $\sigma$ as a function on ${\Pi} T\Sigma$
(this formalism is reviewed in section 3.2 of \cite{Wittennotes}):
\begin{equation}\label{incor}\sigma=\phi(z|\theta)\d z\delta(\d\theta).\end{equation}
Similarly, $\mu$ as already written in eqn. (\ref{pogo}) is naturally understood as a function on ${\Pi} T\Sigma$.  But these are different
kinds of function; $\mu$ has a polynomial dependence on $\d\theta$ while $\sigma$ in its dependence on $\d\theta$ is a delta function
supported at the origin.  A transformation from one type of function to the other is analogous to picture-changing in superstring perturbation theory.

A general picture-changing operation on a supermanifold $M$ has been defined in \cite{Beltwo} and was reviewed in \cite{Wittennotes}, section 4.2.
The operation can be defined whenever one is given a rank $0|1$ subbundle $L$ of $TM$.  So a natural example is the case that $M$ is a super Riemann
surface $\SIgma$, with $L$ being the rank $0|1$ subbundle $\D\subset T\Sigma$ that defines the superconformal structure.
Roughly, picture-changing is defined by integration over the orbits generated by sections of $L$.
Any nonzero local section of $L$ can be used to generate the orbits and evaluate the picture-changing
transformation.  In the super Riemann surface case, we can work in superconformal coordinates $z|\theta$ and pick
the usual section $D_\theta=\partial_\theta+\theta\partial_z$ of $\D$.  
The coordinate transformation generated by $D_\theta$ is
\begin{align}\label{zembo} z & \to z+\eta\theta \cr
                                             \theta & \to \theta+\eta \end{align}
                                         with a fermionic parameter $\eta$.             This change of coordinates maps $\sigma$ to
 \begin{align}\label{hope}\sigma^*=&
 \phi(z+\eta\theta|\theta+\eta)\d(z+\eta\theta) \delta(\d\theta+\d\eta)\cr=
 &(\phi+\eta D_\theta\phi)(\d z-\eta \d\theta+\d\eta\cdot
 \theta)
 \delta(\d\theta+\d\eta).\end{align}
To complete the picture-changing operation, we now define $\mu$ by integrating  
over the new odd and even variables $\eta$ and $\d\eta$:
\begin{equation}\label{oico}\mu=\int\D(\eta,\d\eta) \,\sigma^*.\end{equation}                                             
Since $\mu$ is obtained by integrating over a single odd variable and its differential, 
in general its picture number differs by 1 from that of $\sigma$.
In the present example, the integral over the even variable $\d\eta$ is performed using
using the delta function $\delta(\d\theta+\d\eta)$ in eqn. (\ref{hope}), and the integral over $\eta$ 
is a Berezin integral that
picks out the coefficient of the term linear in $\eta$.   Upon carrying out these steps, one arrives at
the formula (\ref{pogo}) for $\mu$.  General arguments show that this picture-changing operation maps 
a closed form (such as $\sigma$)
to a closed form of picture number greater by 1 (such as $\mu$).

Thus, as explained in \cite{Beltwo}, the possibly mysterious-looking transformation from $\sigma$ 
to $\mu$ is a special case
of a general notion of picture-changing on supermanifolds. 

The important result (\ref{zilfof}) of section \ref{symmetry}
\begin{equation}\label{zilfob}\int_{\gamma}\sigma =\int_{\gamma_\red} \mu,\end{equation}
is an immediate corollary of this explanation of the relation between $\sigma$ and $\mu$.  Given
the 1-cycle $\gamma_\red\subset \Sigma$, we can thicken it to a $1|1$-cycle $\gamma\subset\Sigma$
by simply moving $\gamma_\red$ in the direction of a nonzero section of $\D$.  Thus, we can assume that $\gamma$
fibers over $\gamma_\red$ with fibers generated by $\D$.  The result (\ref{zilfob}) comes by evaluating the
integral on the left hand side by first integrating over the fibers of $\gamma\to\gamma_\red$ to reduce
to an integral over $\gamma_\red$.  Almost by definition, this operation transforms $\sigma$ into $\mu$ and
$\int_\gamma\sigma$ into $\int_{\gamma_\red}\mu$.

\subsection{Picture-Changing In The Presence Of A Ramond Divisor}\label{picram}

It is useful to extend this analysis slightly to the case that the superconformal structure of $\Sigma$ has singularities
along Ramond divisors.  As usual, the local model near a Ramond divisor $\F$ at $z=0$ is that the superconformal
structure is generated in local coordinates $z|\theta$ by
\begin{equation}\label{urit} D_\theta^* =\partial_\theta +z\theta\partial_z. \end{equation}
Equivalently, it is defined by the 1-form
\begin{equation}\label{nurit} \varpi^*=\d z-z\theta\d \theta. \end{equation}

The picture-changing procedure can still be used to map a holomorphic section $\sigma=\phi(z|\theta)[\d z|\d\theta]$ of
$\BBer(\Sigma)$ to a closed holomorphic 1-form $\mu$.  One way to do this is to work away from $z=0$ with superconformal
coordinates $z|\psi$, where $\psi=\sqrt z \theta$.  In these coordinates, we can simply use eqn. (\ref{pogo}), which can then
be transformed back to the coordinates $z|\theta$, which are regular at $z=0$.
However, it is just as easy and more instructive to repeat the above derivation.  Eqn. (\ref{hope}) becomes 
 \begin{align}\label{ope}\sigma^*=&
 \phi(z+z\eta\theta|\theta+\eta)\d(z+z\eta\theta) \delta(\d\theta+\d\eta)\cr=
 &(\phi+\eta D^*_\theta\phi)(\d z(1+\eta\theta)-z\eta \d\theta+z\d\eta\cdot
 \theta)
 \delta(\d\theta+\d\eta).\end{align}
 The Berezin integral over $\eta$ and $\d\eta$ now gives
 \begin{equation}\label{izzo}\mu=D_\theta^*\phi(\d z-z\theta\d\theta )+\phi(\d z \,\theta+z\, \d\theta).      \end{equation}
 
This formula has the important property that the coefficient of $\d\theta$ vanishes at $z=0$.  In other words, if $\sigma$
is an arbitrary holomorphic section of $\BBer(\Sigma)$, then $\mu$ vanishes when restricted to the Ramond divisor $\F$
at $z=0$, though it is otherwise an arbitrary closed holomorphic 1-form.
 (The restriction of a 1-form to the divisor $z=0$ is defined by setting $z=\d z=0$.)  Thus in the presence
of a Ramond divisor, the picture-changing map from a holomorphic section of $\BBer(\Sigma)$ to a closed holomorphic 1-form
may fail to be a $1-1$ map. 

To understand why this happens, let us consider the special case that $\Sigma$ is split, with reduced space $\Sigma_\red$.  In the absence of Ramond divisors,
$\Sigma$ is the total space of the fibration $\Pi T^{1/2}\to \Sigma_\red$, where $T^{1/2}$ is a square root of the tangent bundle of $\Sigma_\red$.
In the presence of Ramond punctures at points $p_1,\dots, p_{\sn_\Ra}\in\Sigma_\red$,   corresponding to divisors $\F_1,\dots,\F_{\sn_\Ra}\in
\Sigma$, $T^{1/2}$ is replaced by a line bundle $\RR$ with an isomorphism
\begin{equation}\label{zonkers}\RR^2\cong T\otimes \O(-p_1-\dots -p_{\sn_\Ra}). \end{equation}
An odd section of $\BBer(\Sigma)$ has the form $\alpha(z)[\d z|\d\theta]$, where now $\alpha$ is a section of $K\otimes \RR\to\Sigma_\red$ (as usual,
$K=T^*\Sigma_\red$).  Adding Ramond punctures makes $\RR$ more negative, making it more difficult to find an odd holomorphic
section of $\BBer(\Sigma)$.  On the other hand, a closed holomorphic 1-form which is odd takes the form $\d(\theta f(z))$, where
now $f(z)$ is a section of $\RR^{-1}$.  Adding Ramond punctures makes $\RR^{-1}$ more positive, increasing the supply of closed
holomorphic 1-forms which are odd.   So as soon as $\RR $ is more negative than $T^{1/2}$, there is no
correspondence between closed holomorphic 1-forms and holomorphic sections of $\BBer(\Sigma)$.

In the presence of Ramond punctures, to establish an isomorphism between holomorphic volume-forms $\sigma$ and
closed holomorphic 1-forms $\mu$, we need to modify the definitions.  There are two ways to do this: {\it (i)} we can require
$\mu$ to vanish when restricted to a Ramond divisor $\F$; {\it (ii)} we can allow $\sigma$ to have a pole along $\F$.  The first possibility is clear; let us examine the second one, in the context of the local model (\ref{urit}). A short calculation shows that if $\sigma=(1/z)[\d z|\d\theta]$, then $\mu=\d\theta$.   On the other hand, if $\sigma=(\theta/z)[\d z|\d\theta]$, then $\mu=\d z/z$
has a pole at $z=0$.  So we will generate, locally, an arbitrary closed holomorphic 1-form that is regular at $z=0$ if we allow
$\sigma$ to have a pole whose residue is independent of $\theta$ or in other words is constant along the Ramond divisor.

The condition that the residue of the pole is independent of $\theta$ has a natural interpretation.  On a complex manifold or 
supermanifold $\Sigma$ of any dimension, a section of $\BBer(\Sigma)$
that has a simple pole along a divisor $\F$ has a residue that is a section of $\BBer(\F)$; the residue does not depend on a choice of local
coordinates.  In dimension $1|1$, with $\F$ defined by the condition $z=0$, the residue of  
\begin{equation}\label{undo} \sigma=\frac{a(z|\theta)}{z}[\d z|\d\theta] ,  \end{equation}
where $a(z|\theta)$ is a local holomorphic function, is the section 
\begin{equation}\label{wumdo}\lambda=  a(0|\theta) [\d\theta]\end{equation}
of $\BBer(\F)$.  The condition that $a(0|\theta)$ is independent of $\theta$ is equivalent to saying that $\lambda$ is exact.
In one direction, this follows from the fact that if $a(0|\theta)$ is a constant $c$, then $\lambda=c[\d\theta]=\d(-c\theta \delta'(\d\theta))$.
In the opposite direction, if $a(0|\theta)$ has a non-trivial dependence on $\theta$, then $\int_\F\lambda\not=0$ (for 
example, $\int_\F\theta[\d\theta]=1$), so $\lambda$ is
not exact.  

In short, in the presence of a Ramond divisor $\F$, we can restore the $1-1$ correspondence between a section 
$\sigma$ of $\BBer(\Sigma)$ and a closed holomorphic
1-form $\mu$ by either requiring $\mu$ to vanish along $\F$ or allowing $\sigma$ to have a simple pole along $\F$ with exact
residue.   More generally, we can make a mixture of these two choices.  Suppose that 
 $\F=\cup_{i=1}^{\sn_\Ra} \F_i$, where the $\F_i$ are minimal
Ramond divisors modeled by eqn. (\ref{urit}).  If $\Sigma$ is split, the $\F_i$ correspond to points $p_i\in \Sigma_\red$.
Along each of the $\F_i$, we can separately decide to allow $\sigma$ to have a pole or to require $\mu$ to vanish.
So in general, we  make any decomposition $\F=\F'\cup \F''$, where $\F'$ is the union of some of the $\F_i$ and $\F''$ is the union of the others.
Then we define $\BBer\,'(\Sigma)$ to be the sheaf of sections of $\BBer(\Sigma)$ that may have simple poles along $\F'$ with
exact residues.  Picture-changing  gives a $1-1$ map from sections of $\BBer\,'(\Sigma)$ to closed holomorphic
1-forms with vanishing restriction to $\F''$.
\bibliographystyle{unsrt}

\end{document}